\documentclass[twocolumn]{aastex63}

\usepackage{aas_macros}
\usepackage{natbib}
\usepackage{hyperref}
\usepackage{amsmath}
\usepackage{xspace}
\usepackage{graphicx}
\usepackage{enumitem}
\usepackage{amssymb}
\usepackage{xifthen}
\usepackage{longtable}
\usepackage{soul} 
\usepackage{amsmath}
\usepackage{amssymb}
\usepackage{xspace}
\usepackage{xifthen}

\newcommand{\unit}[1]{\ensuremath{\mathrm{\,#1}}\xspace}

\newcommand{\mas}{\unit{mas}}

\newcommand{\km}{\unit{km}}
\newcommand{\kms}{\km \second^{-1}}
\newcommand{\pc}{\unit{pc}}
\newcommand{\kpc}{\unit{kpc}}
\newcommand{\second}{\unit{s}}

\newcommand{\Msun}{\unit{M_\odot}}

\newcommand{\e}{\unit{e^{-}}}

\newcommand{\yr}{\unit{yr}}
\newcommand{\masyr}{\unit{\mas \yr^{-1}}}

\newcommand{\COMMENT}[3]{{}}

\newcommand{\degree}{\ensuremath{{}^{\circ}}\xspace}

\received{\today}
\revised{\today}
\accepted{\today}
\submitjournal{ApJ}

\shorttitle{Proper Motions and Orbits of  MW dSphs}
\shortauthors{Pace, Erkal, \& Li}

\graphicspath{{./}{figures/}}

\begin{document}

\title{Proper Motions, Orbits, and Tidal Influences of Milky Way Dwarf Spheroidal Galaxies}

\correspondingauthor{Andrew B. Pace}
\email{apace@andrew.cmu.edu}

\author[0000-0002-6021-8760]{Andrew B. Pace}
\affiliation{McWilliams Center for Cosmology, Carnegie Mellon University, 5000 Forbes Ave, Pittsburgh, PA 15213, USA }

\author[0000-0002-8448-5505]{Denis~Erkal}
\affiliation{Department of Physics, University of Surrey, Guildford GU2 7XH, UK}

\author[0000-0002-9110-6163]{Ting~S.~Li}
\affiliation{Department of Astronomy and Astrophysics, University of Toronto, 50 St. George Street, Toronto ON, M5S 3H4, Canada}

\begin{abstract}
{We combine {\it Gaia} EDR3 astrometry with accurate  photometry  and utilize a probabilistic mixture model to measure the systemic proper motion of 52  dwarf spheroidal (dSph) satellite galaxies of the Milky Way (MW). }
For the 46 dSphs  with literature line-of-sight velocities we compute orbits in both a MW and a combined MW + Large Magellanic Cloud (LMC) potential and identify {Car~II, Car~III,  Hor~I, Hyi~I, Phx~II, and Ret~II as} likely LMC satellites. 
{40\% of our dSph sample has a $>25\%$ change in  pericenter and/or apocenter with the MW + LMC potential.}
For these orbits, we Monte Carlo sample over the observational uncertainties for each dSph {and} the uncertainties in the MW and LMC potentials. 
{We predict that {Ant~II,  Boo~III, Cra~II, Gru~II, and Tuc~III} should be be tidally disrupting by comparing each dSph’s average density relative to the MW density at its pericenter.}
dSphs with large ellipticity {(CVn~I, Her, Tuc~V, UMa~I, UMa~II, UMi, Wil~1)} show a preference for their orbital direction to align with their major axis even {for dSphs with large pericenters}. 
We compare the {dSph} radial orbital phase to subhalos in MW-like N-body simulations and infer that there is not an excess of satellites near their pericenter. 
With {projections of} future {\it Gaia} data releases, we find {dSph} orbital precision  will be limited by uncertainties in {the} distance and/or  MW potential rather than proper motion precision. 
Finally, we provide our membership {catalogs} to enable community follow-up.
\end{abstract}


\section{Introduction} \label{sec:intro}

The Milky Way (MW) dwarf spheroidal (dSph) satellite galaxies are a diverse set of galaxies spanning a wide range of stellar masses, sizes,  dynamical masses, star formation histories, and orbital histories \citep[e.g.][]{McConnachie2012AJ....144....4M, Simon2019ARA&A..57..375S}.
dSphs are near enough that their 6D phase space can be measured although the tangential motion is the most difficult, generally requiring space based astrometry. {For many years, this was the exclusive domain of the Hubble Space Telescope} \citep[e.g.,][]{Piatek2007AJ....133..818P, Sohn2017ApJ...849...93S}.

The tangential and orbital motion of MW dSphs  has been revolutionized by astrometry from the {\it Gaia} mission. 
With the release of the first proper motion {\it Gaia}  catalogs (i.e., {\it Gaia} DR2), the measurement of systemic proper motion of nearly all the MW dSphs has been possible \citep[e.g., ][]{GaiaHelmi2018A&A...616A..12G, Simon2018ApJ...863...89S, Fritz2018A&A...619A.103F, Pace2019ApJ...875...77P, McConnachie2020AJ....160..124M} and has led to the determination of their orbital motion within the MW  \citep[e.g., ][]{GaiaHelmi2018A&A...616A..12G, Simon2018ApJ...863...89S, Fritz2018A&A...619A.103F}.
This has also led to measurements of the orbital anisotropy of the dSph system \citep{Riley2019MNRAS.486.2679R}, satellite infall times \citep{Fillingham2019arXiv190604180F}, updates on potential planes of satellites in the MW \citep[also known as, the vast polar structure, ][]{Fritz2018A&A...619A.103F}, a potential excess of dSph near their orbital pericenter \citep{Simon2018ApJ...863...89S, Fritz2018A&A...619A.103F}, and  measurements of the mass of the MW  \citep{Callingham2019MNRAS.484.5453C, Li2020ApJ...894...10L, Fritz2020MNRAS.494.5178F}

The recent discovery of MW satellites in the Dark Energy Survey (DES) and other southern surveys has revealed a new population of dSphs and several are likely LMC/SMC satellites \citep[e.g.,][]{Koposov2015ApJ...805..130K, Bechtol2015ApJ...807...50B, Torrealba2018MNRAS.475.5085T}. 
With radial velocities and {\it Gaia} proper motions a handful of satellites have been associated with the LMC \citep{Kallivayalil2018ApJ...867...19K,Erkal2020MNRAS.495.2554E, Patel2020ApJ...893..121P}.
In addition to bringing in its own satellite population, the LMC also perturbs the orbits of dSphs in the MW \citep[e.g.][]{Gomez2015ApJ...802..128G,Erkal2020MNRAS.495.2554E, Patel2020ApJ...893..121P}. While satellites which pass close to the LMC are directly accelerated, satellites with more distant passages can be indirectly affected by the reflex motion of the MW \citep[e.g. Leo I and Antlia 2,][]{Erkal2020MNRAS.498.5574E, Ji2021ApJ...921...32J,CorreaMagnus2022MNRAS.511.2610C}. This reflex motion of $\sim40\kms$ has also been measured in the MW's stellar halo \citep[][]{Erkal2021MNRAS.506.2677E,Petersen2021NatAs...5..251P}.

With {\it Gaia} EDR3 the 
systemic proper motions of the dSphs have significantly improved \citep[e.g.][]{McConnachie2020RNAAS...4..229M, Li2021ApJ...916....8L, MartinezGarcia2021MNRAS.505.5884M, Vitral2021MNRAS.504.1355V, Battaglia2022A&A...657A..54B}.
This is due to both the reduced statistic errors with an additional year of data and the reduction of the systematic errors by roughly a factor of two relative to DR2. 
As a result, the orbital properties have improved \citep{Li2021ApJ...916....8L, Battaglia2022A&A...657A..54B} and the internal rotation has been observed in a few dSphs with {\it Gaia} EDR3 data \citep{MartinezGarcia2021MNRAS.505.5884M}.

Here we measure the systemic proper motion and identity candidate members of 52 dwarfs (46 with line-of-sight velocities),  compute the orbits both with and without the influence of the LMC, and discuss the tidal influences of the MW. 
In Section~\ref{section:data}, we give an overview of the astrometric {\it Gaia} EDR3 data, the photometric data sets we complement the {\it Gaia} data with, and describe our initial quality selection and color-magnitude selection. 
In Section~\ref{section:methods}, we present our methodology for measuring the systemic proper motions and for computing orbits.
In Section~\ref{section:results}, we present the systemic proper motions and the orbital properties.  
In Section~\ref{section:discussion}, we compare the orbital properties to indicators of tidal influences, discuss the orientation and orbital direction of the dSphs, discuss  LMC association, discuss the potential excess of satellites near pericenter, and make projections for errors of the orbital parameters of future measurements. 
We summarize and conclude in Section~\ref{section:conclusion}.

\section{Data}
\label{section:data}

We list the 54 MW dwarf galaxies and ultra-faint dwarf galaxies analyzed in this work along with  relevant properties  in Table~\ref{tab:overview}. 
We restrict our sample to MW satellites, the most distant being Eri~II. Throughout this analysis we will refer to  the objects as dwarf spheroidal galaxies (dSphs) even though several objects do not have spectroscopic confirmation or have an ambiguous classification (e.g., Dra~II,  Sgr~II, Tuc~III). 
We do not include the recently discovered dSphs Eridanus~IV and Pegasus~IV as similar methods were used derive the systemic proper motion and only Pegasus~IV has a line-of-sight velocity measurement \citep{Cerny2021ApJ...920L..44C, Cerny2022arXiv220311788C}.
We will refer to the dSphs by their shorted acronyms throughout the paper which are listed  in Table~\ref{tab:overview} along with their   full  names.

We group the satellites with $M_V<-7.7$ as `bright' satellites and fainter satellites as ultra-faints (UFDs) following \citet{Simon2019ARA&A..57..375S}.  This groups the more recently discovered satellites, Ant~II, Cra~II, and CVn~I with the traditionally labelled classical satellites (Cra, Dra, For, Leo~I, Leo~II, Scl, Sxt, UMi). 
Until {the advent of large CCD photometric surveys}, the former escaped detection due to their low surface brightness \citep[e.g.,][]{Zucker2006ApJ...643L.103Z, Koposov2008ApJ...686..279K, Torrealba2016MNRAS.459.2370T}.

\subsection{Astrometric Data}

We use the astrometric {\it Gaia} EDR3 catalog \citep{Gaia_Brown_2021A&A...649A...1G} for our systemic proper motion measurements.  
We consider two samples with different quality cuts which we refer to as `clean' and `complete.' The former is more restrictive and selects higher quality astrometry which we will use for our sytemic proper motion measurements whereas the latter is more inclusive and we will  to maximize the number of candidate stars.

The  quality selection for the clean sample is as follows \citep{Gaia_Lindegren_2021A&A...649A...2L, Gaia_Riello2021A&A...649A...3R}:
\begin{itemize}
    \item $\texttt{astrometric\_params\_solved} >3$
    \item $G<G_{\rm max},$
    \item $\texttt{astrometric\_excess\_noise\_sig} <2,$
    \item $ \texttt{ruwe} <1.3,$
    \item $|C^*| \le 3 \sigma_{C^*}(G),$
    \item $\texttt{ipd\_frac\_multi\_peak} <2,$
    \item $\texttt{ipd\_gof\_harmonic\_amplitude} <0.1,$ however this cut is only applied to some of the dSphs\footnote{For two-thirds of the satellites analyzed, this cut  removes $\sim2$\% of stars within 1$^{\circ}$ of the satellite.  However, for satellites with fewer visibility periods, this cut  removes a large portion of the  stars ($\sim5-32$\% for \texttt{visibility\_periods\_used}$<15$).  In particular, this cut would remove all known spectroscopic members in Aquarius~II and removes $\sim30$\% of the stars around the Crater~II and Sextans dSphs. The following dwarfs do not have the \texttt{ipd\_gof\_harmonic\_amplitude} cut applied: Aqu~II, Cet~III, CB~I, Cra~II, Leo~I, Leo~II, Leo~IV, Leo~V, Peg~III, Psc~II, Sgr~II, Seg~1, Seg~2, Sxt, Tri~II, and Vir~I. },
    \item $\varpi - 3 \times \sigma_\varpi < 0,$
    \item $v_{\rm tangential} - 3 \times \sigma_{v_{\rm tangential}} < v_{\rm escape},$
    \item $\texttt{visibility\_periods\_used}>10,$
    \item $\texttt{duplicated\_source}=\mathrm{False}.$
\end{itemize}
$G_{\rm max}$ is determined based on the approximate magnitude where 90\% of stars have an astrometric solution and varies between $G_{\rm max}=20.35-20.85$ for the dSph sample. 
{Here $|C^*|$ is  the corrected BP and RP flux excess factor \citep[see Equation 6 of][]{Gaia_Riello2021A&A...649A...3R}.}
We compute the tangential velocity ($v_{\rm tangential}$)   of each star by converting the proper motions into Galactic coordinates in the Galactic Standard of Rest (GSR) frame after accounting for the Sun's reflex motion, assuming $({\rm U_\odot,\, V_\odot, \, W_\odot}) = (11.1, 12.24, 7.25)\kms$, a circular velocity of $220\kms$ \citep{Schonrich2010MNRAS.403.1829S}, and assume each star is at the satellite's heliocentric distance.
$v_{\rm esc}$ is computed with the potential \texttt{MWPotential2014} (with a slightly increased halo mass, $M_{\rm vir}=1.6\times 10^{12} \Msun$) from \texttt{galpy} \citep{Bovy2015ApJS..216...29B}.
The escape velocity is a conservative cut to remove high proper motion stars that are nearby foreground MW stars. 
We remove AGN/QSOs galaxies from the sample with the {\it Gaia} catalog \texttt{gaiaedr3.agn\_cross\_id}.

For the complete sample, the following cuts are modified to be less restrictive:  
$G_{\rm max}=21$, \texttt{ruwe} $<1.4$,  $\varpi - 3.5 \times \sigma_\varpi < 0$, and $v_{\rm tangential} - 3.5 \times \sigma_{v_{\rm tangential}} < v_{\rm escape}$.
We remove the selection cuts on \texttt{ipd\_gof\_harmonic\_amplitude},   \texttt{ipd\_frac\_multi\_peak}, and \texttt{visibility\_periods\_used}.

We compute the systematic proper motion errors following \citet{Gaia_Lindegren_2021A&A...649A...2L}.
The proper motion covariance function is:

\begin{equation}
\label{eq:systematic}
    V_{\mu}(\theta) = 292 \exp{\left(-\theta /12\degree \right)}+258 \exp{\left(-\theta /0.25\degree \right)} \, {\rm \mu as^2 ~ yr^{-2}},
\end{equation}

\noindent where $\theta$ is the angular separation between data points. 
We treat $\sigma_{\mu, {\rm sys}}$=$\sqrt{V_{\mu}(\theta)}$ as the systematic error for each dSph. We use the half-light radius ($r_h$) of each dSph as the characteristic angular scale. We list $\sigma_{\mu, {\rm sys}}$ values in Table~\ref{tab:results}.
For our sample the proper motion systematic errors varied between $\sim 16-23~{\rm \mu as ~ yr^{-1}}$ and the median value is $22~{\rm \mu as ~ yr^{-1}}$. 
An alternative form is presented in \citet{Vasiliev2021MNRAS.505.5978V} that has values of $V_{\mu}(\theta)$ that are 15\% to 40\% larger than Equation~\ref{eq:systematic}.

\subsection{Photometry}

We utilize several different photometric catalogs to improve the separation between dSph member stars and MW interlopers. 
This is primarily composed of Dark Energy Camera \citep[DECam;][]{Flaugher2015AJ....150..150F} based data  and Pan-STARRS1 DR1 (PS1) \citep{Chambers2016arXiv161205560C} in the Northern sky.
For the `bright' dSphs  we  use {\it Gaia} EDR3  $G$, $G_{RP}$  photometry. 
The DECam based catalogs include: the Dark Energy Survey (DES) DR2 \citep{DES2021ApJS..255...20A},  the Dark Energy Camera Legacy Survey (DECaLs) DR9 \citep{Dey2019AJ....157..168D}, Survey of the MAgellanic Stellar History DR2 \citep{Nidever2021AJ....161...74N}, DECam Local Volume Exploration Survey DR1 \citep{DrlicaWagner2021ApJS..256....2D}, and the NOIRLab Source Catalog  (NSC) DR2 \citep{Nidever2021AJ....161..192N}. 
{At faint magnitudes ($G\gtrsim 20$), the {\it Gaia} $G-G_{RP}$ color has significant errors and utilizing accurate auxiliary photometry significantly improves the separation between MW foreground and the dSph \citep[see Figure~1 of][]{Pace2019ApJ...875...77P}.}

We opt to not apply any star/galaxy separation from the photometric surveys and instead use the {\it Gaia} astrometry as our stellar selection.
In particular, there are several bright stars in  DES DR2 that are considered galaxies (\texttt{extended\_class\_coadd}=3, e.g. the brightest member in Tuc~III, \citealt{Hansen2017ApJ...838...44H} and a bright candidate member in Cet~II) and the inclusion of these stars are key to determine the  systemic proper motion. 

\begin{figure*}
\includegraphics[width=\textwidth]{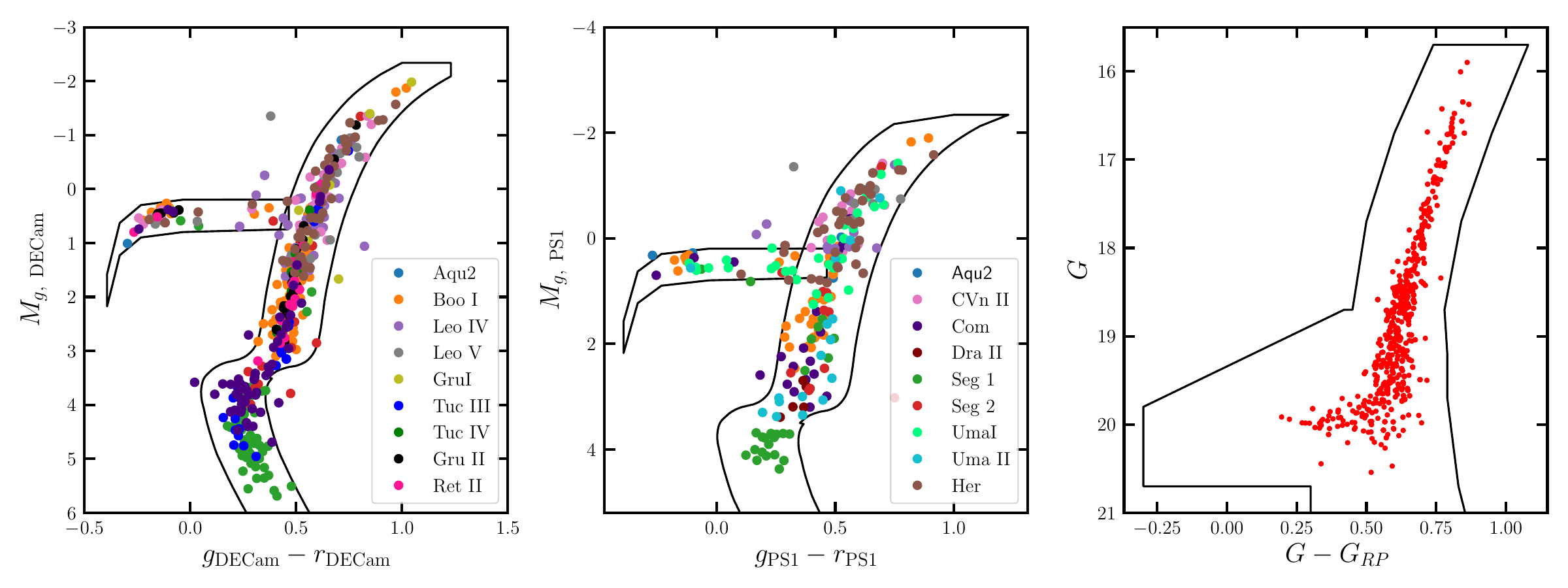}
\caption{Color-magnitude selection based on spectroscopic members (see Tables~\ref{tab:overview},\ref{tab:future_targets} for citations). 
(Left) DECam based selection. The same filter is used for most of the DECam dSphs. (middle)  PS1 based selection. The  filter is expanded at faint magnitudes based on photometric errors. (right)  {\it Gaia} based selection for Dra. A similar selection is made based on spectroscopic members for the other bright dSphs. 
}
\label{fig:cmd_selction}
\end{figure*}

We apply empirical isochrone based filters in the color-magnitude diagrams to improve  dSph  member selection.  
The isochrone filter is created based on spectroscopic members (see citations in Tables~\ref{tab:overview} \&~\ref{tab:future_targets}) and starts from an old, metal-poor isochrone. 
The  DECam based selection is similar to \citet{Pace2019ApJ...875...77P} but includes the red horizontal branch.  
This selection is based primarily on a g-r color selection of 0.12-0.15 around a [Fe/H]=$-$2 and age = 12 Gyr Dartmouth isochrone \citep{Dotter2008ApJS..178...89D} and around the ridgeline of the M92 globular cluster for  horizontal branch stars (since Dartmouth isochrone does not contain a horizontal branch).
For the DECam based photometry we do not increase the filter due to photometric errors as they are generally small at the limiting  {\it Gaia} magnitude.
The PS1 isochrone filter is created in a similar manner except we use PARSEC isochrone from \citet{Bressan2012MNRAS.427..127B}.  
In contrast to the DECam based selection, we increase the width of the filter at faint magnitudes with an additional error term  based on the median errors at a given magnitude, added in quadrature with the constant width of 0.15-0.16 in g-r color. 
For the bright satellites, we construct a wide {\it Gaia} $G-R_{RP}$-$G$ color-magnitude box based on spectroscopic members.  The box width increases with magnitude to account for the large errors in color for fainter  {\it Gaia} stars.
We show the spectroscopic selection in Figure~\ref{fig:cmd_selction} along with the spectroscopic members used to construct the filter. 

\startlongtable
\begin{longrotatetable}
\begin{deluxetable*}{l cc ccc  cccc c}
\tablewidth{0pt}
\tablecaption{dSph  Properties
\label{tab:overview}
}
\tablehead{\colhead{Name (Abbreviation)} &  \colhead{RA} & \colhead{DEC} &  \colhead{$r_h$} & \colhead{$\epsilon$} & \colhead{$\theta$} & \colhead{$d$} & \colhead{$M_V$} & \colhead{$v_{\rm los}$} & \colhead{$\sigma_{\rm los}$} & \colhead{Ref}\\
&  deg & deg & arcmin & & & kpc & & $\kms$&   $\kms$ &
}
\startdata
Antlia II(Ant II) & 143.8868 & -36.7673 & $ 76.20 \pm 7.20 $ &  $ 0.38 \pm 0.08 $ &  $ 156.0 \pm 6.0 $ &  $ 132.0 \pm 6.0 $ &  $ -9.03 \pm 0.15 $ &  $ 290.7\pm0.5 $ &  $5.71\pm1.08$ &     a \\
Aquarius II(Aqu II) & 338.4813 & -9.3274 & $ 5.10 \pm 0.80 $ &  $ 0.39 \pm 0.09 $ &  $ 121.0 \pm 9.0 $ &  $ 107.9 \pm 3.3 $ &  $ -4.36 \pm 0.14 $ &  $ -71.1\pm2.5 $ &  $5.4\pm2.15$ &     b \\
Bo\"{o}tes I(Boo I) & 210.0200 & 14.5135 & $ 9.97 \pm 0.27 $ &  $ 0.3 \pm 0.03 $ &  $ 6.0 \pm 3.0 $ &  $ 66.0 \pm 3.0 $ &  $ -6.02 \pm 0.25 $ &  $ 101.8\pm0.7 $ &  $4.6\pm0.7$ &     c,d,e \\
Bo\"{o}tes II(Boo II) & 209.5141 & 12.8553 & $ 3.17 \pm 0.42 $ &  $ 0.25 \pm 0.11 $ &  $ -68.0 \pm 27.0 $ &  $ 42.0 \pm 2.0 $ &  $ -2.94 \pm 0.74 $ &  $ -117.0\pm5.2 $ &  $10.5\pm7.4$ &     f,e,g \\
Bo\"{o}tes III(Boo III) & 209.3000 & 26.8000 & $ 33.03 \pm 2.50 $ &  $ 0.33 \pm 0.085 $ &  $ -81.0 \pm 8.0 $ &  $ 46.5 \pm 2.0 $ &  $ -5.75 \pm 0.5 $ &  $ 197.5\pm3.8 $ &  $14.0\pm3.2$ &     h,i,j \\
Bo\"{o}tes IV(Boo IV) & 233.6890 & 43.7260 & $ 7.60 \pm 0.80 $ &  $ 0.64 \pm 0.05 $ &  $ 3.0 \pm 4.0 $ &  $ 209.0 \pm 19.0 $ &  $ -4.53 \pm 0.22 $ &  $  $ &   &    k \\
Canes Venatici I(CVn I) & 202.0091 & 33.5521 & $ 7.12 \pm 0.21 $ &  $ 0.44 \pm 0.03 $ &  $ 80.0 \pm 2.0 $ &  $ 210.0 \pm 6.0 $ &  $ -8.8 \pm 0.06 $ &  $ 30.9\pm0.6 $ &  $7.6\pm0.4$ &     l,e,m \\
Canes Venatici II(CVn II) & 194.2927 & 34.3226 & $ 1.52 \pm 0.24 $ &  $ 0.4 \pm 0.13 $ &  $ 9.0 \pm 15.0 $ &  $ 160.0 \pm 4.5 $ &  $ -5.17 \pm 0.32 $ &  $ -128.9\pm1.2 $ &  $4.6\pm0.8$ &     n,e,m \\
Carina(Car) & 100.4065 & -50.9593 & $ 10.10 \pm 0.10 $ &  $ 0.36 \pm 0.01 $ &  $ 60.0 \pm 1.0 $ &  $ 105.6 \pm 5.4 $ &  $ -9.43 \pm 0.05 $ &  $ 222.9\pm0.1 $ &  $6.6\pm1.2$ &     o,e,p \\
Carina II(Car II) & 114.1066 & -57.9991 & $ 8.69 \pm 0.75 $ &  $ 0.34 \pm 0.07 $ &  $ 170.0 \pm 9.0 $ &  $ 37.4 \pm 0.4 $ &  $ -4.57 \pm 0.1 $ &  $ 477.2\pm1.2 $ &  $3.4\pm1.0$ &     q,r \\
Carina III(Car III) & 114.6298 & -57.8997 & $ 3.75 \pm 1.00 $ &  $ 0.55 \pm 0.18 $ &  $ 150.0 \pm 14.0 $ &  $ 27.8 \pm 0.6 $ &  $ -2.4 \pm 0.2 $ &  $ 284.6\pm3.25 $ &  $5.6\pm3.2$ &     q,r \\
Centaurus I(Cen I) & 189.5850 & -40.9020 & $ 2.90 \pm 0.45 $ &  $ 0.4 \pm 0.1 $ &  $ 20.0 \pm 11.0 $ &  $ 116.3 \pm 1.1 $ &  $ -5.55 \pm 0.11 $ &  $  $ &   &    s \\
Cetus II(Cet II) & 19.4700 & -17.4200 & $ 1.90 \pm 0.75 $ &  $ 0.0 \pm 0.0 $ &  $ 0.0 \pm 0.0 $ &  $ 30.0 \pm 3.0 $ &  $ 0.0 \pm 0.68 $ &  $  $ &   &    t \\
Cetus III(Cet III) & 31.3310 & -4.2700 & $ 1.23 \pm 0.30 $ &  $ 0.76 \pm 0.07 $ &  $ 101.0 \pm 5.5 $ &  $ 251.0 \pm 17.5 $ &  $ -2.45 \pm 0.565 $ &  $  $ &   &    u \\
Columba I(Col I) & 82.8570 & -28.0425 & $ 2.20 \pm 0.20 $ &  $ 0.3 \pm 0.1 $ &  $ 24.0 \pm 9.0 $ &  $ 183.0 \pm 10.0 $ &  $ -4.2 \pm 0.2 $ &  $ 153.7\pm4.9 $ &   &    v,w \\
Coma Berenices(CB) & 186.7454 & 23.9069 & $ 5.64 \pm 0.30 $ &  $ 0.37 \pm 0.05 $ &  $ -57.0 \pm 4.0 $ &  $ 42.0 \pm 1.5 $ &  $ -4.38 \pm 0.25 $ &  $ 98.1\pm0.9 $ &  $4.6\pm0.8$ &     e,x,m \\
Crater II(Cra II) & 177.3100 & -18.4130 & $ 31.20 \pm 2.50 $ &  $ 0.0 \pm 0.0 $ &  $ 0.0 \pm 0.0 $ &  $ 117.5 \pm 1.1 $ &  $ -8.2 \pm 0.1 $ &  $ 87.5\pm0.4 $ &  $2.7\pm0.3$ &     y,z \\
Draco(Dra) & 260.0684 & 57.9185 & $ 9.67 \pm 0.09 $ &  $ 0.29 \pm 0.01 $ &  $ 87.0 \pm 1.0 $ &  $ 75.8 \pm 5.4 $ &  $ -8.71 \pm 0.05 $ &  $ -290.7\pm0.75 $ &  $9.1\pm1.2$ &     aa,e,ab \\
Draco II(Dra II) & 238.1983 & 64.5653 & $ 3.00 \pm 0.60 $ &  $ 0.23 \pm 0.15 $ &  $ 76.0 \pm 27.0 $ &  $ 21.5 \pm 0.4 $ &  $ -0.8 \pm 0.7 $ &  $ -342.5\pm1.15 $ &   &    ac \\
Eridanus II(Eri II) & 56.0925 & -43.5329 & $ 2.31 \pm 0.12 $ &  $ 0.48 \pm 0.04 $ &  $ 72.6 \pm 3.3 $ &  $ 366.0 \pm 17.0 $ &  $ -7.1 \pm 0.3 $ &  $ 75.6\pm1.3 $ &  $6.9\pm1.05$ &     ad,ae \\
Fornax(For) & 39.9583 & -34.4997 & $ 19.90 \pm 0.06 $ &  $ 0.29 \pm 0.02 $ &  $ 42.7 \pm 0.3 $ &  $ 147.2 \pm 8.4 $ &  $ -13.46 \pm 0.14 $ &  $ 55.2\pm0.1 $ &  $12.1\pm0.2$ &     e,af,p,ag \\
Grus I(Gru I) & 344.1660 & -50.1680 & $ 4.16 \pm 0.64 $ &  $ 0.44 \pm 0.09 $ &  $ 153.0 \pm 7.5 $ &  $ 127.0 \pm 6.0 $ &  $ -4.3 \pm 0.3 $ &  $ -140.5\pm2.0 $ &   &    ah,ai,aj \\
Grus II(Gru II) & 331.0250 & -46.4420 & $ 5.90 \pm 0.50 $ &  $ 0.0 \pm 0.0 $ &  $ 0.0 \pm 0.0 $ &  $ 55.0 \pm 2.0 $ &  $ -3.5 \pm 0.3 $ &  $ -110.0\pm0.5 $ &   &    ai,ak \\
Hercules(Her) & 247.7722 & 12.7852 & $ 5.63 \pm 0.46 $ &  $ 0.69 \pm 0.03 $ &  $ -73.0 \pm 2.0 $ &  $ 130.6 \pm 6.1 $ &  $ -5.83 \pm 0.17 $ &  $ 45.0\pm1.1 $ &  $5.1\pm0.9$ &     e,al,m \\
Horologium I(Hor I) & 43.8813 & -54.1160 & $ 1.46 \pm 0.07 $ &  $ 0.16 \pm 0.06 $ &  $ 69.0 \pm 11.0 $ &  $ 79.0 \pm 4.0 $ &  $ -3.4 \pm 0.1 $ &  $ 112.8\pm2.55 $ &  $4.9\pm1.85$ &     am,an,j \\
Horologium II(Hor II) & 49.1077 & -50.0486 & $ 2.09 \pm 0.42 $ &  $ 0.52 \pm 15.0 $ &  $ 127.0 \pm 11.0 $ &  $ 78.0 \pm 8.0 $ &  $ -2.6 \pm 0.25 $ &  $  $ &   &    ao \\
Hydra II(Hyd II) & 185.4251 & -31.9860 & $ 1.70 \pm 0.25 $ &  $ 0.01 \pm 0.1 $ &  $ 28.0 \pm 37.5 $ &  $ 151.0 \pm 8.0 $ &  $ -5.1 \pm 0.3 $ &  $ 303.1\pm1.4 $ &   &    ap,aq,ar \\
Hydrus I(Hyi I) & 37.3890 & -79.3089 & $ 7.42 \pm 0.58 $ &  $ 0.21 \pm 0.11 $ &  $ 97.0 \pm 14.0 $ &  $ 27.6 \pm 0.5 $ &  $ -4.71 \pm 0.08 $ &  $ 80.4\pm0.6 $ &  $2.7\pm0.45$ &     as \\
Leo I(Leo I) & 152.1146 & 12.3059 & $ 3.65 \pm 0.03 $ &  $ 0.3 \pm 0.1 $ &  $ 78.0 \pm 1.0 $ &  $ 258.2 \pm 9.5 $ &  $ -11.78 \pm 0.28 $ &  $ 282.9\pm0.5 $ &  $9.2\pm0.4$ &     at,e,au \\
Leo II(Leo II) & 168.3627 & 22.1529 & $ 2.52 \pm 0.03 $ &  $ 0.07 \pm 0.01 $ &  $ 38.0 \pm 8.0 $ &  $ 233.0 \pm 15.0 $ &  $ -9.74 \pm 0.04 $ &  $ 78.5\pm0.6 $ &  $7.4\pm0.4$ &     av,e,aw \\
Leo IV(Leo IV) & 173.2405 & -0.5453 & $ 2.54 \pm 0.27 $ &  $ 0.17 \pm 0.09 $ &  $ -28.0 \pm 38.0 $ &  $ 151.4 \pm 4.4 $ &  $ -4.99 \pm 0.26 $ &  $ 132.3\pm1.4 $ &  $3.3\pm1.7$ &     ax,e,m \\
Leo V(Leo V) & 172.7857 & 2.2194 & $ 1.00 \pm 0.32 $ &  $ 0.43 \pm 0.22 $ &  $ -71.0 \pm 26.0 $ &  $ 169.0 \pm 4.4 $ &  $ -4.4 \pm 0.36 $ &  $ 173.0\pm0.9 $ &  $3.2\pm1.55$ &     ay,ax,e \\
Pegasus III(Peg III) & 336.1074 & 5.4150 & $ 1.67 \pm 0.23 $ &  $ 0.37 \pm 0.085 $ &  $ 83.0 \pm 7.5 $ &  $ 215.0 \pm 12.0 $ &  $ -4.17 \pm 0.205 $ &  $ -222.9\pm2.6 $ &  $5.4\pm2.75$ &     az,ba \\
Phoenix II(Phx II) & 354.9960 & -54.4115 & $ 1.50 \pm 0.30 $ &  $ 0.4 \pm 0.1 $ &  $ 156.0 \pm 13.0 $ &  $ 84.1 \pm 8.0 $ &  $ -2.7 \pm 0.4 $ &  $ 32.4\pm3.75 $ &   &    w,bb \\
Pictor I(Pic I) & 70.9475 & -50.2831 & $ 0.90 \pm 0.09 $ &  $ 0.46 \pm 0.08 $ &  $ 58.0 \pm 6.0 $ &  $ 125.9 \pm 5.0 $ &  $ -3.1 \pm 0.3 $ &  $  $ &   &    am,j \\
Pictor II(Pic II) & 101.1800 & -59.8970 & $ 3.80 \pm 1.25 $ &  $ 0.13 \pm 17.5 $ &  $ 14.0 \pm 63.0 $ &  $ 45.0 \pm 4.5 $ &  $ -3.2 \pm 0.45 $ &  $  $ &   &    bc \\
Pisces II(Psc II) & 344.6345 & 5.9526 & $ 1.12 \pm 0.16 $ &  $ 0.34 \pm 0.1 $ &  $ 78.0 \pm 20.0 $ &  $ 183.0 \pm 15.0 $ &  $ -4.22 \pm 0.38 $ &  $ -226.5\pm2.7 $ &  $5.4\pm3.1$ &     ap,e,bd \\
Reticulum II(Ret II) & 53.9203 & -54.0513 & $ 6.30 \pm 0.40 $ &  $ 0.6 \pm 0.1 $ &  $ 68.0 \pm 2.0 $ &  $ 31.4 \pm 1.4 $ &  $ -3.1 \pm 0.1 $ &  $ 64.3\pm1.2 $ &  $3.6\pm0.85$ &     bb,be \\
Reticulum III(Ret III) & 56.3600 & -60.4500 & $ 2.40 \pm 0.85 $ &  $ 0.0 \pm 0.0 $ &  $ 0.0 \pm 0.0 $ &  $ 92.0 \pm 13.0 $ &  $ -3.3 \pm 0.29 $ &  $ 274.2\pm7.45 $ &   &    t,w \\
Sagittarius II(Sgr II) & 298.1687 & -22.0681 & $ 1.60 \pm 0.10 $ &  $ 0.0 \pm 0.0 $ &  $ 0.0 \pm 0.0 $ &  $ 70.2 \pm 5.0 $ &  $ -5.2 \pm 0.1 $ &  $ -177.2\pm0.55 $ &  $1.7\pm0.5$ &     bf,bb \\
Sculptor(Scl) & 15.0183 & -33.7186 & $ 11.17 \pm 0.05 $ &  $ 0.33 \pm 0.01 $ &  $ 92.0 \pm 1.0 $ &  $ 83.9 \pm 1.5 $ &  $ -10.82 \pm 0.14 $ &  $ 111.4\pm0.1 $ &  $9.2\pm1.1$ &     bg,e,p \\
Segue 1(Seg 1) & 151.7504 & 16.0756 & $ 3.62 \pm 0.42 $ &  $ 0.33 \pm 0.1 $ &  $ 77.0 \pm 15.0 $ &  $ 23.0 \pm 2.0 $ &  $ -1.3 \pm 0.73 $ &  $ 208.5\pm0.9 $ &  $3.7\pm1.25$ &     bh,e,bi \\
Segue 2(Seg 2) & 34.8226 & 20.1624 & $ 3.76 \pm 0.28 $ &  $ 0.22 \pm 0.07 $ &  $ 164.0 \pm 14.0 $ &  $ 36.6 \pm 2.45 $ &  $ -1.86 \pm 0.88 $ &  $ -40.2\pm0.9 $ &   &    bj,bk,e \\
Sextans(Sxt) & 153.2628 & -1.6133 & $ 16.50 \pm 0.10 $ &  $ 0.3 \pm 0.01 $ &  $ 57.0 \pm 1.0 $ &  $ 92.5 \pm 2.5 $ &  $ -8.72 \pm 0.06 $ &  $ 224.3\pm0.1 $ &  $7.9\pm1.3$ &     e,bl,p \\
Triangulum II(Tri II) & 33.3252 & 36.1702 & $ 2.50 \pm 0.30 $ &  $ 0.3 \pm 0.1 $ &  $ 73.0 \pm 17.0 $ &  $ 28.4 \pm 1.6 $ &  $ -1.6 \pm 0.4 $ &  $ -381.7\pm1.1 $ &   &    v,bm \\
Tucana II(Tuc II) & 342.9796 & -58.5689 & $ 12.89 \pm 1.85 $ &  $ 0.39 \pm 0.15 $ &  $ 107.0 \pm 18.0 $ &  $ 58.0 \pm 3.0 $ &  $ -3.8 \pm 0.1 $ &  $ -129.1\pm3.5 $ &  $8.6\pm3.55$ &     am,aj \\
Tucana III(Tuc III) & 359.1075 & -59.5833 & $ 5.10 \pm 1.20 $ &  $ 0.2 \pm 0.1 $ &  $ 25.0 \pm 38.0 $ &  $ 22.9 \pm 0.9 $ &  $ -1.3 \pm 0.2 $ &  $ -102.3\pm0.4 $ &   &    bb,bn \\
Tucana IV(Tuc IV) & 0.7170 & -60.8300 & $ 9.30 \pm 1.15 $ &  $ 0.39 \pm 0.085 $ &  $ 27.0 \pm 8.5 $ &  $ 47.0 \pm 4.0 $ &  $ -3.0 \pm 0.35 $ &  $ 15.9\pm1.75 $ &  $4.3\pm1.35$ &     ak \\
Tucana V(Tuc V) & 354.3470 & -63.2660 & $ 2.10 \pm 0.50 $ &  $ 0.51 \pm 0.135 $ &  $ 29.0 \pm 11.0 $ &  $ 55.0 \pm 5.5 $ &  $ -1.1 \pm 0.55 $ &  $ -36.2\pm2.35 $ &   &    ak \\
Ursa Major I(UMa I) & 158.7706 & 51.9479 & $ 8.31 \pm 0.35 $ &  $ 0.59 \pm 0.03 $ &  $ 67.0 \pm 2.0 $ &  $ 97.3 \pm 5.85 $ &  $ -5.13 \pm 0.38 $ &  $ -55.3\pm1.4 $ &  $7.0\pm1.0$ &     bo,e,m \\
Ursa Major II(UMa II) & 132.8726 & 63.1335 & $ 13.80 \pm 0.50 $ &  $ 0.56 \pm 0.03 $ &  $ -76.0 \pm 2.0 $ &  $ 34.7 \pm 2.1 $ &  $ -4.25 \pm 0.26 $ &  $ -116.5\pm1.9 $ &  $6.7\pm1.4$ &     bp,e,m \\
Ursa Minor(UMi) & 227.2420 & 67.2221 & $ 18.30 \pm 0.11 $ &  $ 0.55 \pm 0.01 $ &  $ 50.0 \pm 1.0 $ &  $ 76.2 \pm 4.2 $ &  $ -9.03 \pm 0.05 $ &  $ -247.0\pm0.4 $ &  $8.6\pm0.3$ &     bq,e,br \\
Virgo I(Vir I) & 180.0380 & -0.6810 & $ 1.76 \pm 0.45 $ &  $ 0.59 \pm 0.13 $ &  $ 62.0 \pm 10.5 $ &  $ 91.0 \pm 6.5 $ &  $ -0.33 \pm 0.81 $ &  $  $ &   &    u \\
Willman 1(Wil 1) & 162.3436 & 51.0501 & $ 2.51 \pm 0.22 $ &  $ 0.47 \pm 0.06 $ &  $ 73.0 \pm 4.0 $ &  $ 38.0 \pm 7.0 $ &  $ -2.53 \pm 0.74 $ &  $ -12.8\pm1.0 $ &  $4.5\pm0.9$ &     e,bs,bt \\
\enddata
\tablecomments{ Citations:
(a) \citep{Torrealba2019MNRAS.488.2743T}
(b) \citep{Torrealba2016MNRAS.463..712T}
(c) \citep{DallOra2006ApJ...653L.109D}
(d) \citep{Koposov2011ApJ...736..146K}
(e) \citep{Munoz2018ApJ...860...66M}
(f) \citep{Koch2009ApJ...690..453K}
(g) \citep{Walsh2008ApJ...688..245W}
(h) \citep{Carlin2009ApJ...702L...9C}
(i) \citep{Carlin2018ApJ...865....7C}
(j) \citep{Moskowitz2020ApJ...892...27M}
(k) \citep{Homma2019PASJ...71...94H}
(l) \citep{Kuehn2008ApJ...674L..81K}
(m) \citep{Simon2007ApJ...670..313S}
(n) \citep{Greco2008ApJ...675L..73G}
(o) \citep{Karczmarek2015AJ....150...90K}
(p) \citep{Walker2009AJ....137.3100W}
(q) \citep{Li2018ApJ...857..145L}
(r) \citep{Torrealba2018MNRAS.475.5085T}
(s) \citep{Mau2020ApJ...890..136M}
(t) \citep{DrlicaWagner2015ApJ...813..109D}
(u) \citep{Homma2018PASJ...70S..18H}
(v) \citep{Carlin2017AJ....154..267C}
(w) \citep{Fritz2019A\string&A...623A.129F}
(x) \citep{Musella2009ApJ...695L..83M}
(y) \citep{Caldwell2017ApJ...839...20C}
(z) \citep{Torrealba2016MNRAS.459.2370T}
(aa) \citep{Bonanos2004AJ....127..861B}
(ab) \citep{Walker2015MNRAS.448.2717W}
(ac) \citep{Longeard2018MNRAS.480.2609L}
(ad) \citep{Crnojevic2016ApJ...824L..14C}
(ae) \citep{Li2017ApJ...838....8L}
(af) \citep{Pietrzynski2009AJ....138..459P}
(ag) \citep{Wang2019ApJ...881..118W}
(ah) \citep{Cantu2021ApJ...916...81C}
(ai) \citep{MartinezVazquez2019MNRAS.490.2183M}
(aj) \citep{Walker2016ApJ...819...53W}
(ak) \citep{Simon2020ApJ...892..137S}
(al) \citep{MutluPakdil2020ApJ...902..106M}
(am) \citep{Koposov2015ApJ...805..130K}
(an) \citep{Koposov2015ApJ...811...62K}
(ao) \citep{Kim2015ApJ...808L..39K}
(ap) \citep{Kirby2015ApJ...810...56K}
(aq) \citep{Martin2015ApJ...804L...5M}
(ar) \citep{Vivas2016AJ....151..118V}
(as) \citep{Koposov2018MNRAS.479.5343K}
(at) \citep{Mateo2008ApJ...675..201M}
(au) \citep{Stetson2014PASP..126..616S}
(av) \citep{Bellazzini2005MNRAS.360..185B}
(aw) \citep{Spencer2017ApJ...836..202S}
(ax) \citep{Medina2018ApJ...855...43M}
(ay) \citep{Jenkins2021ApJ...920...92J}
(az) \citep{Kim2016ApJ...833...16K}
(ba) \citep{Richstein2022arXiv220401917R}
(bb) \citep{MutluPakdil2018ApJ...863...25M}
(bc) \citep{DrlicaWagner2016ApJ...833L...5D}
(bd) \citep{Sand2012ApJ...756...79S}
(be) \citep{Walker2015ApJ...808..108W}
(bf) \citep{Longeard2021MNRAS.503.2754L}
(bg) \citep{MartinezVazquez2015MNRAS.454.1509M}
(bh) \citep{Belokurov2007ApJ...654..897B}
(bi) \citep{Simon2011ApJ...733...46S}
(bj) \citep{Boettcher2013AJ....146...94B}
(bk) \citep{Kirby2013ApJ...770...16K}
(bl) \citep{Okamoto2017MNRAS.467..208O}
(bm) \citep{Kirby2017ApJ...838...83K}
(bn) \citep{Simon2017ApJ...838...11S}
(bo) \citep{Garofalo2013ApJ...767...62G}
(bp) \citep{DallOra2012ApJ...752...42D}
(bq) \citep{Bellazzini2002AJ....124.3222B}
(br) \citep{Spencer2018AJ....156..257S}
(bs) \citep{Willman2006astro.ph..3486W}
(bt) \citep{Willman2011AJ....142..128W}
}
\end{deluxetable*}
\end{longrotatetable}

\section{Methods}
\label{section:methods}

To measure the systemic proper motions of the MW satellites we construct proper motion based mixture models that build on the methodology presented in \citet{Pace2019ApJ...875...77P} and \citet{McConnachie2020AJ....160..124M}.
Briefly, we model the proper motions and spatial positions of stars near a satellite as a mixture of a satellite and MW foreground/background components: 

\begin{equation}
\mathcal{L} = (1 - f_{\rm MW}) \mathcal{L}_{\rm satellite} + f_{\rm MW} \mathcal{L}_{\rm MW} \, , \\ 
\end{equation}

\noindent where each individual component is decomposed into spatial and proper motion (PM) terms:

\begin{equation}
\mathcal{L}_{\rm satellite/MW} = \mathcal{L}_{\rm spatial} \mathcal{L}_{\rm PM} \, . \\ 
\end{equation}

For the satellite spatial term, we assume a projected Plummer stellar distribution \citep[][]{Plummer1911MNRAS..71..460P}:

\begin{equation}
\Sigma (R_e) = \frac{1}{\pi r_p^2 (1-\epsilon)}(1 + R_e^2/r_p^2)^{-2} \,, 
\end{equation}

\noindent where $r_p$,  $\epsilon$, and $R_e$ are the Plummer half-light radius, the ellipticity, and the elliptical radius, respectively.
In contrast to \citet{Pace2019ApJ...875...77P}, we vary the spatial parameters,  $r_p$, $\epsilon$, and $\theta$.
We assume Gaussian priors on these parameters based on literature measurements (listed in Table~\ref{tab:overview}).
For the MW spatial component we assume that the MW component is spatially constant in the small regions around the satellite.

For satellite proper motion component we model the proper motions as a multivariate distribution \citep[e.g.,][]{Pace2019ApJ...875...77P}.  We include the covariance in the proper motion errors (the cross term in the proper motion error) and instrinsic proper motion dispersion terms. 
While the proper motion errors have improved in EDR3, they are not precise enough to measure internal dispersions and we fix intrinsic dispersion terms for the satellite component as they are smaller than the proper motion uncertainties.
For most bright satellites we fix  $\sigma_\mu=10 \kms$ and for all UFDs (including Ant~II and Cra~II) we fix $\sigma_\mu=5 \kms$.

For the MW proper motion distribution, we utilize two models.
The first `fixed' background model, is created from the proper motion distribution of stars  at radii much larger than the target \citep{McConnachie2020AJ....160..124M}.
At a large distance from the target dSph, this sample will  only include MW stars.
Especially, we select stars between $R_{\rm max, sat} < R < R_{\rm max, BG}$ and set $R_{\rm max, sat}$ based on $r_h$.  For the smaller UFDs,  the maximum radius varies between $25\arcmin <R_{\rm max, sat}<60\arcmin$ and for the bright satellites it varies between $60\arcmin <R_{\rm max, sat}<210\arcmin$.
We set the  limiting radius of the background model to  $R_{\rm max, BG}=2.5^\circ$ for the UFDs and for the bright satellites (including Boo~III) we use $R_{\rm max, BG}=R_{\rm max,~sat}+1^\circ$.
We apply the same color-magnitude selection and astrometric filtering to the distant MW sample to mimic  the selection for candidate satellite stars.  
We note for several satellites we removed other known stellar systems, e.g., Boo~II from the Boo~I background (and vice versa), Palomar~3 and Sextans~A from the Sxt background. 
The second `Gaussian' background model, uses a multi-variate Gaussian distribution with free proper motion dispersion terms \citep[e.g.,][]{Pace2019ApJ...875...77P}. 
Other models have been explored in the literature including multiple Gaussian distributions \citep[e.g.,][]{Pace2019ApJ...875...77P, Vasiliev2021MNRAS.505.5978V}
or a Pearson VII distribution \citep{Vitral2021MNRAS.504.1355V}.

To determine the membership of stars we compute the relative likelihood between the satellite and total likelihood: $p = (1- f_{\rm MW})\mathcal{L}_{\rm satellite}/[(1- f_{\rm MW}) \mathcal{L}_{\rm satellite} + f_{\rm MW}\mathcal{L}_{\rm MW}]$ \citep[e.g.,][]{Martinez2011ApJ...738...55M}.  We take the median value to be the star's membership probability (which we refer to as $p_i$, for the $i$-th star) and compute the membership error (based on 16\% and 84\% confidence intervals) which we use to assess the confidence of the population  assignment for individual stars and any potential signal from  dSphs with few candidate members.

\subsection{Orbit Methods} \label{sec:orbit_method}

In order to simulate the orbits of these dwarfs, we account for the potential of the MW and the LMC. This is done using the technique of \cite{Erkal2020MNRAS.495.2554E} where the MW and the LMC are treated as individual particles sourcing their respective potentials. For the Milky Way potential we use the results of \cite{McMillan2017MNRAS.465...76M} where the MW consists of an NFW halo, a bulge, and four disks (thin, thick, HI, and H$_2$). In order to account for the uncertainties in the MW potential, we sample from the posterior chains in \cite{McMillan2017MNRAS.465...76M} in our fiducial setup. For the LMC, we use a Hernquist profile as in \cite{Erkal2020MNRAS.495.2554E}, with a mass of $1.38\pm0.255 \times10^{11} M_\odot$ \citep[from][]{Erkal2019MNRAS.487.2685E} and a scale radius chosen to match the enclosed mass at 8.7 kpc \citep{vanderMarel2014ApJ...781..121V}. In the fiducial setup, we also account for the observed uncertainties in the radial velocity \citep{van_der_Marel2002}, proper motion \citep{Kallivayalil2013ApJ...764..161K}, and distance to the LMC \citep{Pietrzynski2019Natur.567..200P}. Each dwarf is then rewound in the presence of the Milky Way and LMC for 10 Gyr to estimate its orbital properties. We note that these models include the Milky Way's reflex motion in response to the LMC which was initially highlighted in  \cite{Gomez2015ApJ...802..128G} and measured in \citet{Erkal2021MNRAS.506.2677E, Petersen2021NatAs...5..251P}.

\section{Results}
\label{section:results}
\subsection{Proper Motions}

\begin{deluxetable*}{l cc ccc ccc c}
\tablewidth{0pt}
\tablecaption{Systemic Proper Motion Measurements
\label{tab:results}
}
\tablehead{Dwarf & N & $N_{mem, F}$ & $\overline{\mu_{\alpha \star}}_{F}$ &  $\overline{\mu_{\delta}}_{F}$ & $N_{mem, G}$ & $\overline{\mu_{\alpha \star}}_{G}$ &  $\overline{\mu_{\delta}}_{G}$ & $\sigma_{\mu, {\rm sys}}$
}
\startdata
Ant II & 4889 & $558.2_{-25.2}^{+25.6}$ & $-0.093_{-0.008}^{+0.008}$ & $0.100_{-0.009}^{+0.009}$ &  $414.7_{-24.6}^{+24.9}$ &  $-0.090_{-0.009}^{+0.009}$ & $0.100_{-0.010}^{+0.010}$ & 0.016 \\
Aqu II & 51 & $16.3_{-1.8}^{+1.8}$ & $-0.170_{-0.119}^{+0.113}$ & $-0.466_{-0.095}^{+0.096}$ &  $14.3_{-1.7}^{+2.1}$ &  $-0.183_{-0.122}^{+0.121}$ & $-0.446_{-0.096}^{+0.099}$ & 0.022 \\
Boo I & 373 & $167.9_{-3.6}^{+3.5}$ & $-0.385_{-0.017}^{+0.017}$ & $-1.068_{-0.013}^{+0.013}$ &  $170.0_{-4.5}^{+4.5}$ &  $-0.387_{-0.016}^{+0.017}$ & $-1.064_{-0.013}^{+0.013}$ & 0.021 \\
Boo II & 88 & $20.9_{-0.9}^{+1.0}$ & $-2.426_{-0.077}^{+0.080}$ & $-0.414_{-0.061}^{+0.061}$ &  $20.2_{-1.0}^{+1.1}$ &  $-2.419_{-0.080}^{+0.078}$ & $-0.413_{-0.061}^{+0.061}$ & 0.022 \\
Boo III & 1073 & $73.6_{-5.7}^{+6.0}$ & $-1.176_{-0.019}^{+0.019}$ & $-0.890_{-0.015}^{+0.015}$ &  $90.8_{-6.8}^{+7.1}$ &  $-1.168_{-0.018}^{+0.018}$ & $-0.890_{-0.014}^{+0.014}$ & 0.018 \\
Boo IV & 43 & $4.2_{-0.8}^{+0.5}$ & $0.469_{-0.244}^{+0.180}$ & $0.489_{-0.255}^{+0.256}$ &  $4.1_{-1.3}^{+0.5}$ &  $0.445_{-0.433}^{+0.195}$ & $0.500_{-0.295}^{+0.311}$ & 0.021 \\
CVn I & 322 & $122.5_{-1.4}^{+1.4}$ & $-0.096_{-0.031}^{+0.030}$ & $-0.116_{-0.020}^{+0.020}$ &  $120.5_{-1.8}^{+1.8}$ &  $-0.093_{-0.030}^{+0.030}$ & $-0.114_{-0.020}^{+0.020}$ & 0.021 \\
CVn II & 15 & $11.1_{-0.6}^{+0.6}$ & $-0.124_{-0.115}^{+0.117}$ & $-0.254_{-0.080}^{+0.082}$ &  $11.8_{-0.5}^{+1.1}$ &  $-0.116_{-0.109}^{+0.111}$ & $-0.264_{-0.079}^{+0.080}$ & 0.023 \\
Car & 10273 & $2043.4_{-11.4}^{+11.5}$ & $0.532_{-0.006}^{+0.007}$ & $0.127_{-0.006}^{+0.006}$ &  $1952.8_{-11.7}^{+11.8}$ &  $0.534_{-0.007}^{+0.007}$ & $0.124_{-0.006}^{+0.006}$ & 0.020 \\
Car II & 5033 & $60.4_{-3.7}^{+3.7}$ & $1.885_{-0.019}^{+0.018}$ & $0.133_{-0.019}^{+0.019}$ &   &   &  & 0.021 \\
Car III & 5033 & $9.5_{-1.0}^{+1.3}$ & $3.095_{-0.041}^{+0.040}$ & $1.395_{-0.045}^{+0.045}$ &   &   &  & 0.022 \\
Cen I & 282 & $19.0_{-1.8}^{+1.8}$ & $-0.074_{-0.065}^{+0.062}$ & $-0.199_{-0.055}^{+0.054}$ &  $17.9_{-1.7}^{+1.8}$ &  $-0.063_{-0.065}^{+0.063}$ & $-0.198_{-0.056}^{+0.056}$ & 0.022 \\
Cet II & 151 & $4.9_{-0.1}^{+0.1}$ & $2.844_{-0.059}^{+0.061}$ & $0.474_{-0.063}^{+0.064}$ &  $5.0_{-0.0}^{+0.0}$ &  $2.845_{-0.060}^{+0.060}$ & $0.475_{-0.064}^{+0.063}$ & 0.023 \\
Col I & 54 & $5.7_{-0.3}^{+0.3}$ & $0.169_{-0.073}^{+0.071}$ & $-0.400_{-0.079}^{+0.079}$ &  $5.6_{-0.3}^{+0.3}$ &  $0.168_{-0.073}^{+0.071}$ & $-0.400_{-0.081}^{+0.081}$ & 0.023 \\
CB & 265 & $35.5_{-0.9}^{+0.9}$ & $0.423_{-0.027}^{+0.026}$ & $-1.721_{-0.024}^{+0.024}$ &  $35.7_{-1.1}^{+1.0}$ &  $0.423_{-0.025}^{+0.026}$ & $-1.720_{-0.024}^{+0.024}$ & 0.022 \\
Cra II & 9310 & $390.3_{-13.4}^{+13.8}$ & $-0.072_{-0.020}^{+0.020}$ & $-0.112_{-0.013}^{+0.013}$ &  $371.2_{-12.9}^{+13.2}$ &  $-0.053_{-0.021}^{+0.020}$ & $-0.103_{-0.013}^{+0.013}$ & 0.018 \\
Dra & 5678 & $1517.6_{-4.0}^{+4.0}$ & $0.044_{-0.006}^{+0.005}$ & $-0.188_{-0.006}^{+0.006}$ &  $1506.3_{-4.7}^{+4.7}$ &  $0.046_{-0.006}^{+0.006}$ & $-0.188_{-0.006}^{+0.006}$ & 0.021 \\
Dra II & 247 & $20.0_{-0.8}^{+0.6}$ & $1.027_{-0.065}^{+0.067}$ & $0.887_{-0.072}^{+0.072}$ &  $19.5_{-1.1}^{+0.8}$ &  $1.030_{-0.068}^{+0.069}$ & $0.889_{-0.072}^{+0.077}$ & 0.022 \\
Eri II & 23 & $19.5_{-0.5}^{+0.5}$ & $0.125_{-0.100}^{+0.101}$ & $0.013_{-0.127}^{+0.123}$ &  $19.5_{-0.8}^{+2.1}$ &  $0.136_{-0.100}^{+0.098}$ & $0.003_{-0.121}^{+0.125}$ & 0.023 \\
For & 17007 & $16222.9_{-5.6}^{+5.6}$ & $0.381_{-0.001}^{+0.001}$ & $-0.359_{-0.002}^{+0.002}$ &  $16198.2_{-9.8}^{+9.8}$ &  $0.381_{-0.001}^{+0.001}$ & $-0.358_{-0.002}^{+0.002}$ & 0.019 \\
Gru I & 74 & $9.3_{-0.4}^{+0.3}$ & $0.069_{-0.050}^{+0.051}$ & $-0.248_{-0.072}^{+0.071}$ &  $9.4_{-0.4}^{+0.3}$ &  $0.070_{-0.050}^{+0.050}$ & $-0.246_{-0.072}^{+0.073}$ & 0.022 \\
Gru II & 204 & $32.8_{-3.1}^{+3.2}$ & $0.384_{-0.033}^{+0.033}$ & $-1.484_{-0.040}^{+0.039}$ &  $34.4_{-3.4}^{+3.5}$ &  $0.384_{-0.032}^{+0.033}$ & $-1.478_{-0.040}^{+0.038}$ & 0.022 \\
Her & 184 & $40.9_{-1.5}^{+1.4}$ & $-0.035_{-0.042}^{+0.042}$ & $-0.339_{-0.036}^{+0.035}$ &  $41.0_{-1.4}^{+1.3}$ &  $-0.031_{-0.041}^{+0.040}$ & $-0.334_{-0.034}^{+0.034}$ & 0.022 \\
Hor I & 50 & $19.1_{-0.4}^{+0.5}$ & $0.847_{-0.035}^{+0.034}$ & $-0.607_{-0.035}^{+0.035}$ &  $19.2_{-0.6}^{+0.6}$ &  $0.846_{-0.034}^{+0.034}$ & $-0.606_{-0.036}^{+0.036}$ & 0.023 \\
Hor II & 40 & $3.9_{-0.3}^{+0.3}$ & $0.967_{-0.171}^{+0.173}$ & $-0.771_{-0.230}^{+0.220}$ &  $3.8_{-0.4}^{+0.3}$ &  $0.976_{-0.179}^{+0.177}$ & $-0.762_{-0.237}^{+0.233}$ & 0.023 \\
Hyd II & 82 & $17.5_{-0.5}^{+0.4}$ & $-0.394_{-0.140}^{+0.140}$ & $0.000_{-0.104}^{+0.103}$ &  $17.3_{-0.6}^{+0.5}$ &  $-0.395_{-0.142}^{+0.139}$ & $0.001_{-0.106}^{+0.106}$ & 0.023 \\
Hyi I & 1801 & $102.4_{-4.0}^{+4.0}$ & $3.781_{-0.016}^{+0.016}$ & $-1.496_{-0.015}^{+0.015}$ &  $92.3_{-4.0}^{+3.9}$ &  $3.783_{-0.016}^{+0.016}$ & $-1.495_{-0.015}^{+0.015}$ & 0.021 \\
Leo I & 1031 & $920.7_{-1.1}^{+1.1}$ & $-0.050_{-0.014}^{+0.014}$ & $-0.120_{-0.010}^{+0.010}$ &  $920.6_{-1.8}^{+1.8}$ &  $-0.047_{-0.014}^{+0.014}$ & $-0.118_{-0.010}^{+0.010}$ & 0.022 \\
Leo II & 343 & $264.4_{-0.5}^{+0.5}$ & $-0.109_{-0.028}^{+0.028}$ & $-0.150_{-0.026}^{+0.026}$ &  $263.9_{-0.8}^{+1.0}$ &  $-0.108_{-0.028}^{+0.028}$ & $-0.149_{-0.027}^{+0.027}$ & 0.023 \\
Leo IV & 11 & $6.2_{-0.1}^{+0.2}$ & $-0.009_{-0.152}^{+0.152}$ & $-0.279_{-0.112}^{+0.115}$ &  $6.2_{-0.2}^{+0.8}$ &  $-0.021_{-0.150}^{+0.152}$ & $-0.279_{-0.111}^{+0.110}$ & 0.023 \\
Leo V & 6 & $6.0_{-0.0}^{+0.0}$ & $0.113_{-0.215}^{+0.219}$ & $-0.391_{-0.153}^{+0.155}$ &  $6.0_{-0.0}^{+0.0}$ &  $0.115_{-0.217}^{+0.213}$ & $-0.391_{-0.154}^{+0.151}$ & 0.023 \\
Peg III & 25 & $3.9_{-0.4}^{+0.3}$ & $-0.030_{-0.210}^{+0.210}$ & $-0.580_{-0.208}^{+0.213}$ &  $3.8_{-0.4}^{+0.3}$ &  $-0.019_{-0.218}^{+0.212}$ & $-0.567_{-0.217}^{+0.217}$ & 0.023 \\
Phx II & 45 & $9.5_{-0.3}^{+0.3}$ & $0.507_{-0.048}^{+0.047}$ & $-1.199_{-0.057}^{+0.058}$ &  $9.5_{-0.3}^{+0.3}$ &  $0.507_{-0.048}^{+0.047}$ & $-1.198_{-0.059}^{+0.058}$ & 0.023 \\
Pic I & 68 & $8.3_{-0.3}^{+0.3}$ & $0.153_{-0.088}^{+0.086}$ & $0.096_{-0.114}^{+0.118}$ &  $8.2_{-0.4}^{+0.3}$ &  $0.150_{-0.087}^{+0.087}$ & $0.097_{-0.117}^{+0.119}$ & 0.023 \\
Pic II & 455 & $6.1_{-1.5}^{+3.5}$ & $1.091_{-0.423}^{+0.113}$ & $1.179_{-0.087}^{+0.116}$ &   &   &  & 0.022 \\
Psc II & & 3 & $0.681_{-0.307}^{+0.309}$ & $-0.645_{-0.209}^{+0.215}$ &  &  & & 0.022 \\
Ret II & 465 & $50.2_{-1.4}^{+1.4}$ & $2.377_{-0.024}^{+0.023}$ & $-1.379_{-0.025}^{+0.026}$ &  $49.4_{-1.5}^{+1.4}$ &  $2.375_{-0.023}^{+0.023}$ & $-1.378_{-0.026}^{+0.027}$ & 0.021 \\
Ret III & 67 & $5.7_{-1.5}^{+0.8}$ & $0.260_{-0.144}^{+0.140}$ & $-0.502_{-0.226}^{+0.222}$ &  $4.9_{-2.2}^{+1.1}$ &  $0.260_{-0.173}^{+0.163}$ & $-0.524_{-0.318}^{+0.330}$ & 0.023 \\
Sgr II & 769 & $65.2_{-1.3}^{+1.3}$ & $-0.769_{-0.035}^{+0.035}$ & $-0.903_{-0.023}^{+0.022}$ &  $63.1_{-1.3}^{+1.3}$ &  $-0.771_{-0.035}^{+0.036}$ & $-0.902_{-0.023}^{+0.023}$ & 0.023 \\
Scl & 7362 & $6184.2_{-3.6}^{+3.5}$ & $0.100_{-0.002}^{+0.002}$ & $-0.158_{-0.002}^{+0.002}$ &  $6195.5_{-6.5}^{+6.6}$ &  $0.101_{-0.003}^{+0.003}$ & $-0.156_{-0.002}^{+0.002}$ & 0.020 \\
Seg 1 & 302 & $17.9_{-1.9}^{+1.9}$ & $-2.102_{-0.051}^{+0.051}$ & $-3.375_{-0.046}^{+0.044}$ &  $16.5_{-1.9}^{+1.9}$ &  $-2.099_{-0.054}^{+0.053}$ & $-3.375_{-0.047}^{+0.047}$ & 0.022 \\
Seg 2 & 201 & $16.4_{-0.7}^{+0.7}$ & $1.446_{-0.059}^{+0.059}$ & $-0.322_{-0.050}^{+0.049}$ &  $15.9_{-0.7}^{+0.7}$ &  $1.445_{-0.061}^{+0.059}$ & $-0.321_{-0.051}^{+0.050}$ & 0.022 \\
Sxt & 4359 & $1361.0_{-6.5}^{+6.5}$ & $-0.409_{-0.008}^{+0.009}$ & $0.037_{-0.009}^{+0.009}$ &  $1333.0_{-8.2}^{+8.2}$ &  $-0.409_{-0.009}^{+0.009}$ & $0.041_{-0.009}^{+0.009}$ & 0.019 \\
Tri II & 799 & $10.7_{-1.3}^{+1.5}$ & $0.575_{-0.060}^{+0.060}$ & $0.112_{-0.067}^{+0.069}$ &  $11.1_{-1.3}^{+1.5}$ &  $0.571_{-0.058}^{+0.058}$ & $0.109_{-0.067}^{+0.065}$ & 0.023 \\
Tuc II & 277 & $40.6_{-4.6}^{+4.7}$ & $0.911_{-0.026}^{+0.024}$ & $-1.280_{-0.029}^{+0.029}$ &  $42.5_{-5.0}^{+5.0}$ &  $0.905_{-0.026}^{+0.024}$ & $-1.277_{-0.030}^{+0.029}$ & 0.020 \\
Tuc III & 881 & $46.8_{-4.7}^{+5.1}$ & $-0.048_{-0.036}^{+0.035}$ & $-1.638_{-0.039}^{+0.039}$ &  $54.3_{-5.7}^{+5.9}$ &  $-0.040_{-0.034}^{+0.034}$ & $-1.629_{-0.038}^{+0.038}$ & 0.022 \\
Tuc IV & 344 & $11.0_{-1.9}^{+2.2}$ & $0.534_{-0.053}^{+0.050}$ & $-1.707_{-0.055}^{+0.054}$ &  $12.1_{-2.1}^{+2.3}$ &  $0.540_{-0.051}^{+0.049}$ & $-1.697_{-0.055}^{+0.053}$ & 0.021 \\
Tuc V & 62 & $4.6_{-4.6}^{+1.7}$ & $-0.161_{-0.176}^{+0.087}$ & $-1.157_{-0.195}^{+0.150}$ &  $5.6_{-3.1}^{+1.2}$ &  $-0.152_{-0.068}^{+0.057}$ & $-1.151_{-0.079}^{+0.074}$ & 0.023 \\
UMa I & 122 & $44.0_{-1.0}^{+0.9}$ & $-0.401_{-0.036}^{+0.036}$ & $-0.613_{-0.042}^{+0.040}$ &  $42.9_{-1.3}^{+1.2}$ &  $-0.398_{-0.038}^{+0.036}$ & $-0.614_{-0.043}^{+0.042}$ & 0.021 \\
UMa II & 812 & $47.0_{-2.3}^{+2.3}$ & $1.731_{-0.021}^{+0.021}$ & $-1.906_{-0.025}^{+0.024}$ &  $42.6_{-2.6}^{+2.6}$ &  $1.734_{-0.022}^{+0.022}$ & $-1.902_{-0.025}^{+0.025}$ & 0.020 \\
UMi & 5113 & $1909.1_{-6.9}^{+6.9}$ & $-0.120_{-0.005}^{+0.005}$ & $0.071_{-0.005}^{+0.005}$ &  $1890.2_{-8.1}^{+8.0}$ &  $-0.119_{-0.005}^{+0.005}$ & $0.072_{-0.005}^{+0.005}$ & 0.019 \\
Wil 1 & 76 & $7.7_{-0.6}^{+0.9}$ & $0.255_{-0.087}^{+0.077}$ & $-1.110_{-0.091}^{+0.095}$ &  $7.6_{-0.5}^{+1.1}$ &  $0.241_{-0.085}^{+0.082}$ & $-1.108_{-0.096}^{+0.096}$ & 0.023 \\
\enddata
\end{deluxetable*}

\begin{figure*}
\includegraphics[width=\textwidth]{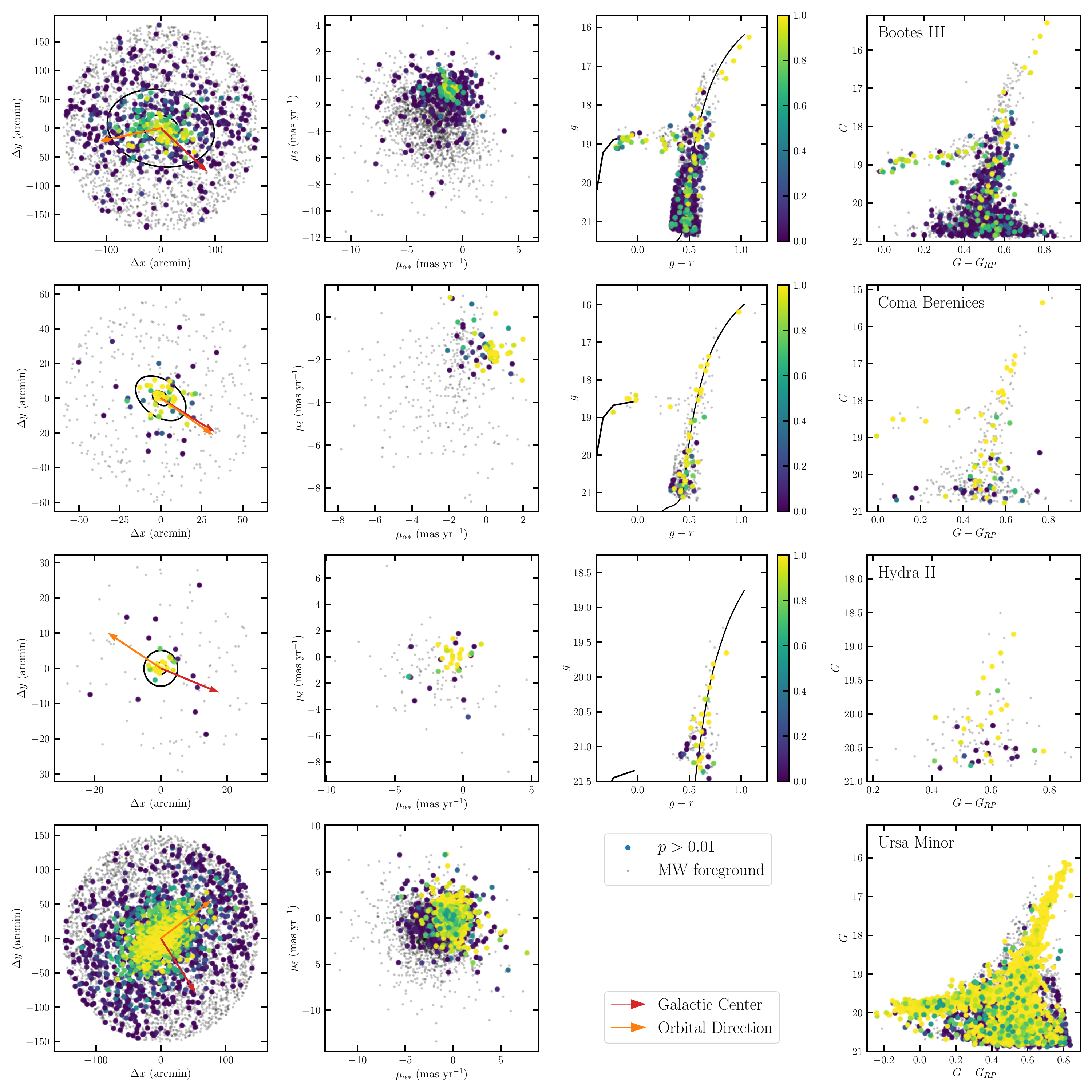}
\caption{Results of the mixture model for 4 dSphs, from top to bottom: Boo~III, CB, Hyd~II, and UMi. 
The four rows are the spatial distribution (tangent plane), the proper motion (vector point diagram), a DECam based color-magnitude diagram (not included for UMi), and a {\it Gaia} color-magnitude diagram. Points with membership probability $p > 0.01$ are colored according to their probability; the rest are considered  MW foreground stars and are shown as grey points. The red arrow points toward the Galactic center and the orange arrow is the direction of the reflex-corrected proper motion which is approximately equal to the orbital motion. Two ellipses are included in the spatial distribution at one and three times the half-light radius. 
}
\label{fig:example_diagnostic}
\end{figure*}

We are able to measure the systemic proper motion of 52 of our 54 dSph sample of which 46 have line-of-sight velocities.   
{We have the first confident detection of the systemic proper motion of Peg~III and do not detect a signal in Cet~III or Vir~I.}
We identify between $\sim4$ and $\sim16200$ members in the 52 dSphs.
To demonstrate the ability of our model to identify dwarf members, we show example results for four dSphs in Figure~\ref{fig:example_diagnostic}. The identified member stars cluster spatially, cluster in proper motion space, and cluster along metal-poor isochrones in color-magnitude space. 
In Table~\ref{tab:results}, we list our results for the systemic proper motion of the 52 dwarfs. We include our measurements and number of  members  with both the fixed and Gaussian background models  with the clean sample. 
The systemic proper motions are in excellent agreement between the two background models, with differences $\lesssim 0.01 \masyr$. 

For the majority of the UFDs there is excellent agreement in the    total membership with both background models.  Only in three UFDs, Hyi~I, Boo~III, and Tuc~III, are there differences with $\Delta N \geq 5$. 
For the brighter dSphs, most show differences in total membership between the two background models. 
Ant~II and Car in particular have large differences of $\sim 140$ and $\sim 110$ stars while other bright dwarfs have differences on the order $\sim 5-10$ stars. 
The individual stars with a large difference in membership  between the two background models tend to be fainter stars with large proper motion errors.
We note that for Car~II, Car~III, and Pic~II we do not have results from the Gaussian background model, this is discussed in more detail in  Appendix~\ref{appendix:comments}. 
We consider the results from the fixed background model  as our default model.

We  find that the two data samples, `clean' and `complete,' have excellent agreement between them.  
Results with the  same background model finds similar membership for stars that overlap between the `clean' and `complete' samples. 
We base our primary results on the `clean' sample in this analysis and provide  membership for both samples and both background models in Appendix~\ref{appendix:membership}. This will enable future spectroscopic follow-up. 
To enable the search for distant members, we recompute our membership analysis without the spatial component and only use the  proper motion posterior \citep[e.g.,][]{Chiti2020ApJ...891....8C, Qi2022MNRAS.512.5601Q} and include these membership probabilities in Appendix~\ref{appendix:membership}.

We are not able to  measure the systemic motion of Cet~III or Vir~I and are only able to measure a signal in Psc~II when spectroscopic information is included (see Appendix~\ref{appendix:comments}). Neither Cet~III nor Vir~I have any spectroscopic follow-up and there are no stars with high membership probability. 
In order to measure the systemic proper motion with {\it Gaia} astrometry, members will need to be identified  beforehand (i.e., with spectroscopy) and these systems may be faint enough that there are no stars above the {\it Gaia} magnitude limit.
Boo~IV, Hor~II, Peg~III, and Tuc~V have the most uncertain detections with $\sim4$ members in each.  Only Tuc~V has {\it Gaia} candidate stars that are spectroscopically confirmed.

Systematic proper motion errors computed using Equation~\ref{eq:systematic} are  presented in Table \ref{tab:results}. There are 15 dSphs where the systematic proper motion error is larger than the statistical error.
This includes all `bright' dSphs except for Leo~II, and five UFDs,  Boo~I, Boo~III, Car~II, Hyi~I, and Sgr~II.
Excluding Leo~II, dSphs with more than $\sim60$ members are  dominated by systematic errors whereas systems with fewer members are dominated by  statistical errors.  We note that because of Leo~II's  distance it has a larger statistical error than other systems with a similar number of stars.

\begin{figure}
\includegraphics[width=\columnwidth]{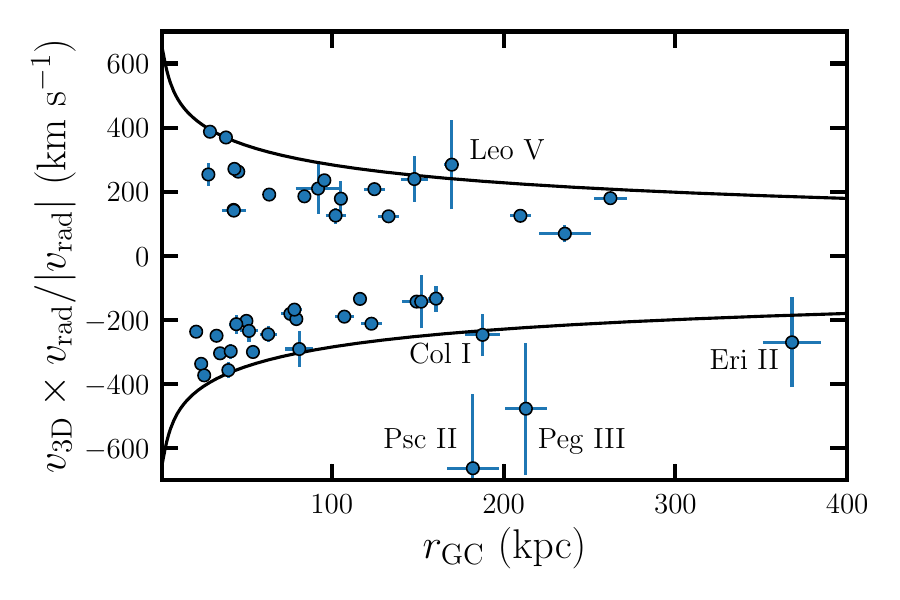}
\caption{Phase-space diagram for the MW  dSphs where the x-axis shows the distance to the Galactic center (GC) and the y-axis shows the total velocity in 3D, where positive (negative) indicates that the dSph is moving away from (towards) the GC.  The black lines represent the escape velocity of the MW.
}
\label{fig:phase_space}
\end{figure}

In  Figure~\ref{fig:phase_space}, we show the phase-space diagram for the 46 MW dSphs with systemic proper motion measurements and line-of-sight velocity data. 
For comparison we include the escape velocity of the MW.  The 5 dSphs outside of the MW escape velocity are labeled but we note that they have the largest tangential velocity errors and their total velocity may be overestimated.

\subsection{Orbits}

\begin{figure*}
\begin{center}
 \includegraphics[width=0.8\textwidth]{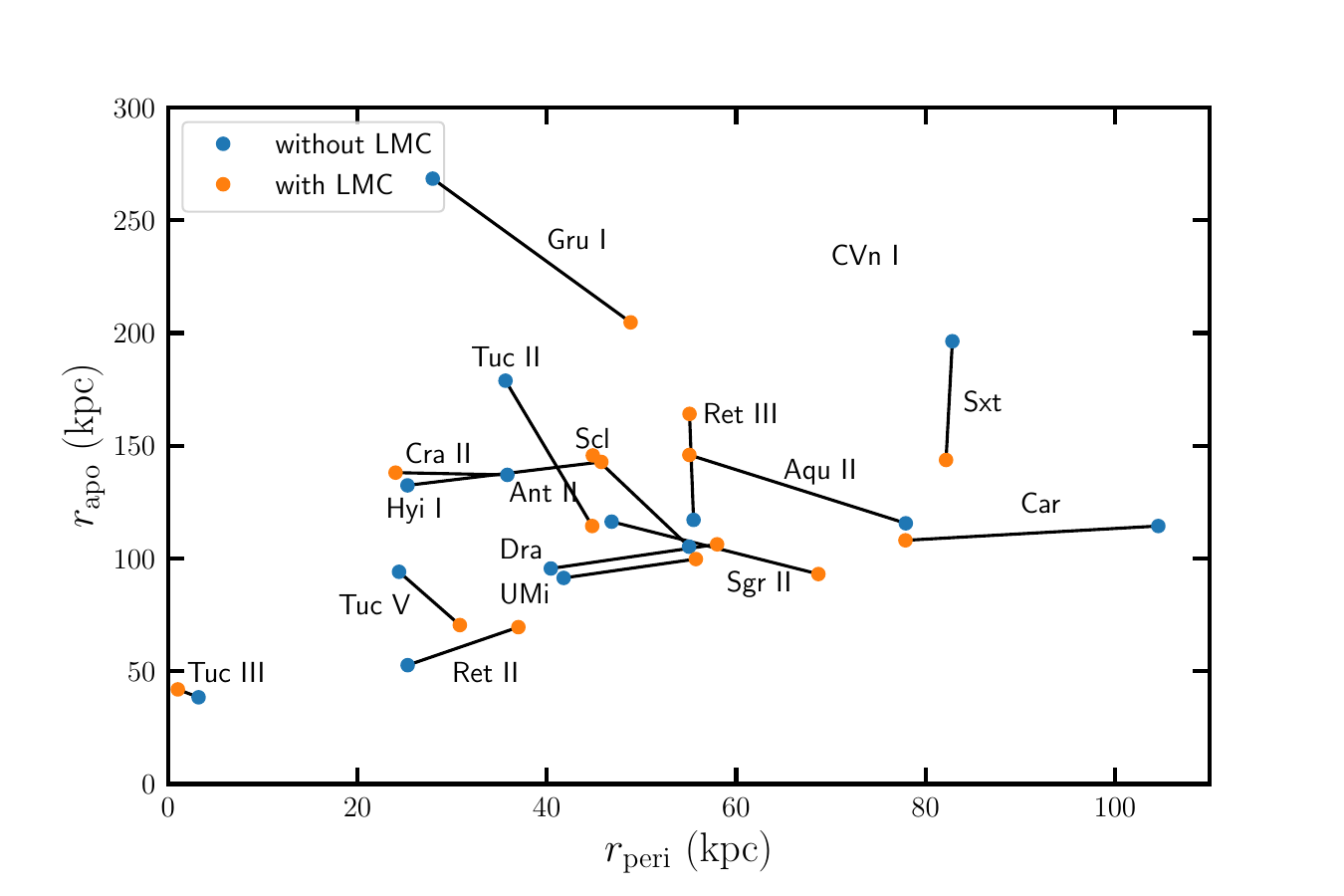}
\caption{Comparison of the orbital pericenter and apocenter of dSphs with a large relative change ($>25\%$ in either) in Milky Way potentials with  (orange points) and without (blue points) the presence of the LMC.  
}
\label{fig:orbit_differences}
\end{center}
\end{figure*}

Next, we explore the orbits of these dwarfs in the Milky Way as described in Section~\ref{sec:orbit_method}. In our fiducial setup, we include the LMC and Monte Carlo over the uncertainties in the observational properties of each dwarf (i.e. proper motions, radial velocity, and distance), as well as the potential parameters of the Milky Way and the LMC (see Sec.~\ref{sec:orbit_method} for more details). This Monte Carlo process is repeated 10,000 times for each dwarf to sample the uncertainties. In addition to this fiducial run, we also have an `nL' run where the LMC's effect is not included. This allows us to see how much the orbital properties are influenced by the LMC. We note that we have also repeated the same suite several times with different assumptions about the observational, Milky Way potential, and LMC potential errors in order to explore the dominant contribution to the error. We discuss this analysis in Section~\ref{sec:future_improvements}. 

\begin{deluxetable*}{l cc ccc ccc cc c }
\tablewidth{0pt}
\tablecaption{dSph Orbital Properties
}
\tablehead{\colhead{Dwarf} & \colhead{$r_{\rm peri}$} & \colhead{$r_{\rm apo}$} & \colhead{ecc} & \colhead{$f_{\rm peri}$} & \colhead{$r_{\rm peri}^{\rm nL}$} & \colhead{$r_{\rm apo}^{\rm nL}$} & \colhead{ecc$^{\rm nL}$} & \colhead{$f_{\rm peri}^{\rm nL}$} & \colhead{$r_{\rm peri}/r_{\rm peri}^{\rm nL}$}  & \colhead{$r_{\rm apo}/r_{\rm apo}^{\rm nL}$} & \colhead{$p_{\rm LMC}$}  \\
& kpc & kpc &  & & kpc & kpc & & & & &  
}
\startdata
Ant II & $38.2_{-7.8}^{+10.0}$ & $137.2_{-6.8}^{+7.6}$ & $0.56_{-0.07}^{+0.06}$ & $0.96_{-0.05}^{+0.04}$ & $50.7_{-9.7}^{+12.9}$ & $144.7_{-7.7}^{+9.2}$ & $0.48_{-0.07}^{+0.07}$ &  $0.88_{-0.07}^{+0.03}$ & 0.75  &  0.95 & 0.00 \\
Aqu II & $55.1_{-32.7}^{+40.8}$ & $145.9_{-17.4}^{+53.6}$ & $0.49_{-0.13}^{+0.23}$ & $0.57_{-0.46}^{+0.18}$ & $77.9_{-41.2}^{+23.2}$ & $115.6_{-7.8}^{+81.3}$ & $0.31_{-0.13}^{+0.24}$ &  $0.75_{-0.72}^{+0.20}$ & 0.71  &  1.26 & 0.00 \\
Boo I & $37.9_{-6.8}^{+7.5}$ & $71.7_{-5.5}^{+8.3}$ & $0.31_{-0.04}^{+0.05}$ & $0.77_{-0.17}^{+0.09}$ & $35.3_{-6.1}^{+7.1}$ & $80.9_{-7.1}^{+11.0}$ & $0.40_{-0.03}^{+0.04}$ &  $0.62_{-0.15}^{+0.09}$ & 1.07  &  0.89 & 0.00 \\
Boo II & $35.8_{-2.5}^{+2.9}$ & $176.3_{-56.0}^{+122.5}$ & $0.65_{-0.10}^{+0.10}$ & $0.03_{-0.02}^{+0.02}$ & $39.0_{-1.9}^{+1.9}$ & $203.0_{-75.4}^{+178.1}$ & $0.68_{-0.11}^{+0.10}$ &  $0.00_{-0.00}^{+0.00}$ & 0.92  &  0.87 & 0.00 \\
Boo III & $7.8_{-2.0}^{+2.3}$ & $97.9_{-10.9}^{+17.1}$ & $0.86_{-0.03}^{+0.03}$ & $0.42_{-0.07}^{+0.05}$ & $7.5_{-1.9}^{+2.2}$ & $108.1_{-12.8}^{+21.4}$ & $0.87_{-0.03}^{+0.03}$ &  $0.38_{-0.07}^{+0.05}$ & 1.04  &  0.91 & 0.00 \\
CVn I & $84.5_{-37.2}^{+53.6}$ & $229.8_{-13.2}^{+31.7}$ & $0.46_{-0.13}^{+0.18}$ & $0.87_{-0.28}^{+0.08}$ & $68.7_{-31.0}^{+42.7}$ & $256.5_{-16.0}^{+34.2}$ & $0.58_{-0.12}^{+0.15}$ &  $0.76_{-0.20}^{+0.08}$ & 1.23  &  0.90 & 0.10 \\
CVn II & $47.5_{-29.7}^{+46.8}$ & $234.0_{-27.1}^{+92.5}$ & $0.66_{-0.14}^{+0.19}$ & $0.62_{-0.29}^{+0.13}$ & $46.0_{-27.6}^{+45.5}$ & $201.1_{-15.0}^{+55.2}$ & $0.64_{-0.16}^{+0.20}$ &  $0.75_{-0.29}^{+0.09}$ & 1.03  &  1.16 & 0.01 \\
Car & $77.9_{-17.9}^{+24.1}$ & $108.1_{-5.7}^{+7.9}$ & $0.18_{-0.12}^{+0.10}$ & $1.00_{-0.09}^{+0.00}$ & $104.6_{-22.5}^{+7.5}$ & $114.4_{-11.8}^{+49.7}$ & $0.10_{-0.07}^{+0.11}$ &  $0.10_{-0.10}^{+0.89}$ & 0.74  &  0.94 & 0.28 \\
Car II & $29.2_{-0.6}^{+0.6}$ & $176.0_{-43.5}^{+170.7}$ & $0.72_{-0.07}^{+0.13}$ & $0.06_{-0.03}^{+0.03}$ & $28.9_{-0.6}^{+0.6}$ & $227.1_{-55.4}^{+143.4}$ & $0.78_{-0.06}^{+0.08}$ &  $0.05_{-0.02}^{+0.02}$ & 1.01  &  0.78 & 1.00 \\
Car III & $28.8_{-0.6}^{+0.6}$ & $230.1_{-44.3}^{+121.2}$ & $0.78_{-0.04}^{+0.07}$ & $0.00_{-0.00}^{+0.00}$ & $28.7_{-0.6}^{+0.6}$ & $185.7_{-50.3}^{+122.5}$ & $0.74_{-0.08}^{+0.09}$ &  $0.00_{-0.00}^{+0.00}$ & 1.00  &  1.24 & 1.00 \\
Col I & $165.5_{-57.2}^{+15.0}$ & $303.0_{-112.3}^{+248.4}$ & $0.24_{-0.11}^{+0.15}$ & $0.20_{-0.15}^{+0.77}$ & $175.0_{-65.1}^{+17.5}$ & $232.9_{-102.2}^{+253.9}$ & $0.25_{-0.09}^{+0.14}$ &  $0.26_{-0.22}^{+0.67}$ & 0.95  &  1.30 & 0.01 \\
CB & $42.5_{-1.6}^{+1.6}$ & $68.1_{-11.0}^{+17.1}$ & $0.23_{-0.07}^{+0.09}$ & $0.03_{-0.01}^{+0.03}$ & $42.4_{-1.6}^{+1.5}$ & $80.4_{-14.3}^{+23.4}$ & $0.31_{-0.08}^{+0.10}$ &  $0.02_{-0.01}^{+0.02}$ & 1.00  &  0.85 & 0.00 \\
Cra II & $24.0_{-5.2}^{+5.6}$ & $138.1_{-4.9}^{+7.9}$ & $0.71_{-0.05}^{+0.05}$ & $0.81_{-0.06}^{+0.04}$ & $35.8_{-6.8}^{+8.4}$ & $137.1_{-4.5}^{+7.0}$ & $0.59_{-0.06}^{+0.06}$ &  $0.80_{-0.07}^{+0.04}$ & 0.67  &  1.01 & 0.36 \\
Dra & $58.0_{-9.5}^{+11.4}$ & $106.3_{-13.1}^{+20.4}$ & $0.30_{-0.04}^{+0.04}$ & $0.37_{-0.21}^{+0.16}$ & $40.4_{-5.4}^{+6.5}$ & $95.6_{-9.1}^{+12.0}$ & $0.41_{-0.02}^{+0.03}$ &  $0.65_{-0.12}^{+0.08}$ & 1.43  &  1.11 & 0.55 \\
Dra II & $21.4_{-1.1}^{+1.7}$ & $90.8_{-14.3}^{+28.5}$ & $0.62_{-0.04}^{+0.06}$ & $0.04_{-0.02}^{+0.02}$ & $19.9_{-0.5}^{+0.5}$ & $80.5_{-10.8}^{+19.1}$ & $0.60_{-0.04}^{+0.06}$ &  $0.06_{-0.02}^{+0.02}$ & 1.07  &  1.13 & 0.96 \\
Eri II & $114.4_{-67.6}^{+80.9}$ & $440.9_{-47.5}^{+158.3}$ & $0.57_{-0.19}^{+0.22}$ & $0.85_{-0.15}^{+0.07}$ & $212.9_{-59.1}^{+69.1}$ & $454.5_{-68.0}^{+160.5}$ & $0.64_{-0.16}^{+0.19}$ &  $0.75_{-0.13}^{+0.08}$ & 0.54  &  0.97 & 0.03 \\
For & $76.7_{-27.9}^{+43.1}$ & $152.7_{-9.1}^{+9.7}$ & $0.33_{-0.18}^{+0.17}$ & $0.96_{-0.07}^{+0.02}$ & $85.2_{-29.3}^{+38.6}$ & $160.0_{-12.3}^{+24.7}$ & $0.31_{-0.10}^{+0.14}$ &  $0.87_{-0.32}^{+0.07}$ & 0.90  &  0.95 & 0.00 \\
Gru I & $48.9_{-22.9}^{+27.0}$ & $204.7_{-23.9}^{+58.1}$ & $0.62_{-0.09}^{+0.14}$ & $0.48_{-0.21}^{+0.12}$ & $28.0_{-13.4}^{+15.9}$ & $268.4_{-45.0}^{+99.1}$ & $0.82_{-0.07}^{+0.08}$ &  $0.40_{-0.12}^{+0.10}$ & 1.75  &  0.76 & 0.00 \\
Gru II & $27.2_{-6.4}^{+8.4}$ & $64.6_{-4.1}^{+5.3}$ & $0.41_{-0.08}^{+0.08}$ & $0.62_{-0.14}^{+0.08}$ & $24.8_{-4.8}^{+5.1}$ & $72.4_{-6.9}^{+10.6}$ & $0.50_{-0.03}^{+0.04}$ &  $0.54_{-0.12}^{+0.09}$ & 1.10  &  0.89 & 0.00 \\
Her & $67.4_{-16.1}^{+15.5}$ & $253.8_{-53.1}^{+115.6}$ & $0.60_{-0.04}^{+0.06}$ & $0.31_{-0.15}^{+0.17}$ & $56.8_{-15.0}^{+15.8}$ & $237.5_{-40.0}^{+84.3}$ & $0.63_{-0.05}^{+0.06}$ &  $0.37_{-0.16}^{+0.14}$ & 1.19  &  1.07 & 0.00 \\
Hor I & $67.6_{-14.6}^{+13.5}$ & $81.3_{-4.4}^{+5.5}$ & $0.09_{-0.07}^{+0.11}$ & $0.84_{-0.75}^{+0.10}$ & $68.0_{-16.8}^{+10.5}$ & $91.2_{-11.0}^{+35.8}$ & $0.19_{-0.04}^{+0.10}$ &  $0.49_{-0.40}^{+0.37}$ & 0.99  &  0.89 & 0.60 \\
Hyd II & $99.2_{-55.7}^{+30.6}$ & $237.3_{-57.6}^{+191.2}$ & $0.56_{-0.10}^{+0.14}$ & $0.41_{-0.31}^{+0.35}$ & $80.4_{-51.8}^{+27.5}$ & $214.6_{-73.9}^{+208.6}$ & $0.61_{-0.08}^{+0.11}$ &  $0.34_{-0.23}^{+0.32}$ & 1.23  &  1.11 & 0.06 \\
Hyi I & $45.8_{-6.0}^{+16.1}$ & $142.8_{-46.9}^{+191.5}$ & $0.46_{-0.15}^{+0.22}$ & $0.00_{-0.00}^{+0.00}$ & $25.3_{-0.5}^{+0.5}$ & $132.4_{-29.2}^{+63.6}$ & $0.68_{-0.07}^{+0.09}$ &  $0.00_{-0.00}^{+0.00}$ & 1.81  &  1.08 & 1.00 \\
Leo I & $47.5_{-24.0}^{+30.9}$ & $401.5_{-45.8}^{+83.2}$ & $0.79_{-0.09}^{+0.10}$ & $0.61_{-0.14}^{+0.10}$ & $42.9_{-23.2}^{+28.9}$ & $532.8_{-86.3}^{+109.6}$ & $0.86_{-0.06}^{+0.06}$ &  $0.39_{-0.08}^{+0.08}$ & 1.11  &  0.75 & 0.01 \\
Leo II & $61.4_{-34.7}^{+62.3}$ & $230.0_{-17.1}^{+17.6}$ & $0.58_{-0.28}^{+0.22}$ & $1.00_{-0.00}^{+0.00}$ & $54.3_{-31.6}^{+55.7}$ & $240.1_{-15.3}^{+16.7}$ & $0.63_{-0.25}^{+0.20}$ &  $0.98_{-0.03}^{+0.01}$ & 1.13  &  0.96 & 0.03 \\
Leo IV & $66.8_{-44.1}^{+60.7}$ & $153.7_{-8.8}^{+87.1}$ & $0.44_{-0.20}^{+0.30}$ & $1.00_{-0.77}^{+0.00}$ & $81.8_{-51.6}^{+73.2}$ & $153.7_{-5.6}^{+87.0}$ & $0.38_{-0.26}^{+0.34}$ &  $1.00_{-0.99}^{+0.00}$ & 0.82  &  1.00 & 0.05 \\
Leo V & $165.8_{-49.2}^{+5.8}$ & $189.1_{-30.5}^{+264.3}$ & $0.35_{-0.22}^{+0.25}$ & $0.38_{-0.37}^{+0.62}$ & $137.9_{-55.4}^{+6.5}$ & $0.0_{-35.0}^{+253.7}$ & $0.40_{-0.16}^{+0.21}$ &  $0.41_{-0.39}^{+0.51}$ & 1.20  &   & 0.01 \\
Peg III & $141.0_{-79.3}^{+87.8}$ & $251.5_{-34.6}^{+234.0}$ & $0.25_{-0.16}^{+0.31}$ & $0.79_{-0.79}^{+0.15}$ & $162.6_{-75.4}^{+49.4}$ & $252.6_{-64.6}^{+267.4}$ & $0.35_{-0.07}^{+0.26}$ &  $0.68_{-0.49}^{+0.23}$ & 0.87  &  1.00 & 0.01 \\
Phx II & $84.6_{-35.6}^{+91.3}$ & $174.2_{-95.1}^{+166.5}$ & $0.06_{-0.05}^{+0.55}$ & $0.00_{-0.00}^{+0.39}$ & $76.5_{-10.0}^{+6.6}$ & $181.3_{-106.9}^{+250.7}$ & $0.42_{-0.20}^{+0.20}$ &  $0.02_{-0.02}^{+0.18}$ & 1.11  &  0.96 & 0.93 \\
Pis II & $130.5_{-72.3}^{+70.1}$ & $265.7_{-72.7}^{+304.3}$ & $0.27_{-0.14}^{+0.31}$ & $0.53_{-0.53}^{+0.32}$ & $147.3_{-47.8}^{+26.2}$ & $248.6_{-116.3}^{+250.8}$ & $0.39_{-0.07}^{+0.10}$ &  $0.32_{-0.25}^{+0.48}$ & 0.89  &  1.07 & 0.01 \\
Ret II & $37.0_{-5.3}^{+2.9}$ & $69.6_{-20.9}^{+51.1}$ & $0.28_{-0.09}^{+0.18}$ & $0.00_{-0.00}^{+0.00}$ & $25.3_{-3.1}^{+2.8}$ & $52.7_{-7.8}^{+12.1}$ & $0.36_{-0.02}^{+0.04}$ &  $0.27_{-0.11}^{+0.14}$ & 1.46  &  1.32 & 0.96 \\
Ret III & $55.1_{-32.7}^{+32.8}$ & $164.1_{-40.3}^{+127.4}$ & $0.59_{-0.11}^{+0.16}$ & $0.37_{-0.30}^{+0.24}$ & $55.5_{-37.5}^{+30.0}$ & $117.1_{-27.2}^{+124.7}$ & $0.47_{-0.13}^{+0.22}$ &  $0.54_{-0.49}^{+0.29}$ & 0.99  &  1.40 & 0.00 \\
Sgr II & $68.7_{-9.8}^{+9.6}$ & $93.1_{-15.6}^{+24.3}$ & $0.16_{-0.07}^{+0.09}$ & $0.00_{-0.00}^{+0.11}$ & $46.9_{-9.6}^{+8.5}$ & $116.3_{-29.5}^{+63.7}$ & $0.44_{-0.04}^{+0.09}$ &  $0.24_{-0.14}^{+0.20}$ & 1.47  &  0.80 & 0.00 \\
Scl & $44.9_{-3.9}^{+4.3}$ & $145.7_{-14.2}^{+25.2}$ & $0.54_{-0.03}^{+0.04}$ & $0.39_{-0.10}^{+0.08}$ & $55.0_{-5.2}^{+5.5}$ & $105.3_{-5.9}^{+11.4}$ & $0.32_{-0.02}^{+0.02}$ &  $0.58_{-0.16}^{+0.11}$ & 0.82  &  1.38 & 0.00 \\
Seg 1 & $19.8_{-4.8}^{+4.2}$ & $47.9_{-10.0}^{+20.1}$ & $0.44_{-0.03}^{+0.06}$ & $0.29_{-0.16}^{+0.19}$ & $19.5_{-4.9}^{+4.3}$ & $48.5_{-10.8}^{+20.3}$ & $0.45_{-0.03}^{+0.06}$ &  $0.29_{-0.16}^{+0.20}$ & 1.02  &  0.99 & 0.00 \\
Seg 2 & $18.0_{-3.1}^{+3.8}$ & $48.2_{-3.2}^{+3.5}$ & $0.45_{-0.06}^{+0.05}$ & $0.82_{-0.05}^{+0.03}$ & $17.4_{-2.9}^{+3.9}$ & $46.0_{-3.0}^{+3.3}$ & $0.45_{-0.06}^{+0.05}$ &  $0.88_{-0.04}^{+0.03}$ & 1.04  &  1.05 & 0.00 \\
Sxt & $82.2_{-4.3}^{+3.8}$ & $143.7_{-25.7}^{+48.2}$ & $0.27_{-0.07}^{+0.11}$ & $0.22_{-0.11}^{+0.20}$ & $82.8_{-4.0}^{+3.7}$ & $196.4_{-38.5}^{+82.4}$ & $0.41_{-0.08}^{+0.12}$ &  $0.11_{-0.06}^{+0.08}$ & 0.99  &  0.73 & 0.00 \\
Tri II & $12.2_{-1.3}^{+1.5}$ & $85.6_{-7.7}^{+12.7}$ & $0.75_{-0.02}^{+0.03}$ & $0.31_{-0.05}^{+0.04}$ & $12.6_{-1.1}^{+1.1}$ & $100.1_{-11.8}^{+21.8}$ & $0.78_{-0.02}^{+0.03}$ &  $0.25_{-0.05}^{+0.04}$ & 0.97  &  0.86 & 0.00 \\
Tuc II & $44.8_{-10.1}^{+12.3}$ & $114.4_{-20.6}^{+35.8}$ & $0.45_{-0.05}^{+0.05}$ & $0.13_{-0.13}^{+0.15}$ & $35.6_{-4.8}^{+4.5}$ & $178.9_{-48.0}^{+104.1}$ & $0.67_{-0.05}^{+0.08}$ &  $0.13_{-0.06}^{+0.07}$ & 1.26  &  0.64 & 0.00 \\
Tuc III & $1.0_{-0.4}^{+0.4}$ & $42.0_{-2.7}^{+3.3}$ & $0.95_{-0.02}^{+0.02}$ & $0.49_{-0.03}^{+0.02}$ & $3.2_{-0.2}^{+0.2}$ & $38.5_{-2.1}^{+2.5}$ & $0.85_{-0.01}^{+0.01}$ &  $0.51_{-0.03}^{+0.02}$ & 0.32  &  1.09 & 0.00 \\
Tuc IV & $32.1_{-12.6}^{+18.5}$ & $52.7_{-4.2}^{+5.7}$ & $0.25_{-0.18}^{+0.18}$ & $0.59_{-0.59}^{+0.15}$ & $28.7_{-9.1}^{+9.4}$ & $60.4_{-10.6}^{+22.3}$ & $0.38_{-0.04}^{+0.08}$ &  $0.50_{-0.27}^{+0.21}$ & 1.12  &  0.87 & 0.25 \\
Tuc V & $30.8_{-9.4}^{+14.3}$ & $70.5_{-9.5}^{+16.6}$ & $0.40_{-0.07}^{+0.09}$ & $0.52_{-0.28}^{+0.13}$ & $24.4_{-8.5}^{+10.4}$ & $94.1_{-20.4}^{+47.6}$ & $0.61_{-0.04}^{+0.07}$ &  $0.39_{-0.20}^{+0.17}$ & 1.26  &  0.75 & 0.00 \\
UMa I & $49.9_{-15.6}^{+46.2}$ & $103.5_{-7.4}^{+7.6}$ & $0.35_{-0.20}^{+0.13}$ & $0.98_{-0.08}^{+0.02}$ & $46.5_{-16.0}^{+26.0}$ & $102.1_{-5.8}^{+6.2}$ & $0.37_{-0.18}^{+0.15}$ &  $1.00_{-0.00}^{+0.00}$ & 1.07  &  1.01 & 0.00 \\
UMa II & $41.4_{-3.6}^{+3.4}$ & $83.3_{-19.2}^{+31.7}$ & $0.34_{-0.08}^{+0.11}$ & $0.00_{-0.00}^{+0.04}$ & $39.3_{-2.6}^{+2.3}$ & $102.1_{-28.8}^{+55.4}$ & $0.44_{-0.11}^{+0.14}$ &  $0.02_{-0.01}^{+0.03}$ & 1.05  &  0.82 & 0.00 \\
UMi & $55.7_{-7.0}^{+8.4}$ & $99.8_{-8.4}^{+13.0}$ & $0.29_{-0.03}^{+0.03}$ & $0.51_{-0.20}^{+0.13}$ & $41.8_{-4.5}^{+5.3}$ & $91.4_{-5.8}^{+7.4}$ & $0.37_{-0.03}^{+0.03}$ &  $0.74_{-0.10}^{+0.06}$ & 1.33  &  1.09 & 0.65 \\
Wil 1 & $16.2_{-3.0}^{+5.2}$ & $41.9_{-6.6}^{+6.7}$ & $0.42_{-0.09}^{+0.08}$ & $1.00_{-0.00}^{+0.00}$ & $18.7_{-3.7}^{+6.4}$ & $43.1_{-7.0}^{+7.1}$ & $0.38_{-0.09}^{+0.08}$ &  $0.99_{-0.01}^{+0.00}$ & 0.86  &  0.97 & 0.00 \\
\enddata
\tablecomments{\label{tab:orbits} Columns with the superscript `nL' are orbital parameters without the influence of the LMC. }
\end{deluxetable*}

For the 46 dwarfs with line-of-sight velocities we compute their orbital properties with and without the presence of the LMC. For the 16 dwarfs whose orbits are significantly affected ($>25\%$ change in either pericenter or apocenter), we show the orbital properties with and without the LMC in Figure~\ref{fig:orbit_differences}. This shows that in order to get precise orbits, the LMC must be accounted for.

In Table~\ref{tab:orbits},  we list the pericenter, apocenter, eccentricity, and probability of being an LMC satellite. We include  the ratio of pericenter and apocenter  with and without the influence of the LMC  to highlight which dwarfs are significantly affected by the inclusion of the LMC. We note that we define the pericenter and apocenter respectively as the first local minimum and maximum in the distance from the Milky Way during the backwards rewinding of each satellite. This is motivated by the results of \cite{DSouza2022MNRAS.512..739D} who showed that while the most recent pericenter and apocenter can be reliably determined during backwards integration, subsequent pericenters and apocenters are more poorly constrained. As a result, if a satellite is unbound from the Milky Way, it may not have a pericenter or an apocenter. We note that we only compute orbital uncertainties in pericenter and apocenter for the subset of realizations which respectively reach their pericenter and apocenter during the integration. We find that  $\sim$2\%, 6\%, 15\%,  34\%, 46\%, 77\%,  87\%, 88\%, 89\%, and 89\%  of the samples in Psc~II, Peg~III, Eri~II,  Col~I, Leo~V, Hyd~II,  Leo~IV, CVn~II, Phx~II, and Ret~II respectively, reach their pericenters and/or apocenters during our orbit integration and these samples might not be bound to the MW. 
The first 5 of these dSphs have the largest proper motion errors of the sample. 
The reminder of satellites have pericenters and apocenters for $>90\%$ of the sample.

The following dwarfs have a significant change ($>25\%$) to their orbital pericenter due to the presence of the  LMC: Ant~II, Aqu~II, Cvn~I, Car, Cra~II, Dra, Eri~II, Gru~I, Hyi~I, Ret~II, Sgr~II, Tuc~III, and UMi. These updated orbits may have a significant effect on the tidal disruption of these satellites. Indeed, for Ant~II and Cra~II, the effect of these updated orbits on the dwarf's tidal disruption has already been studied \citep[i.e.][]{Ji2021ApJ...921...32J}.

Several dwarfs  have a significant change to their orbital apocenter due to the presence of the LMC: Aqu~II, Col~I, Leo~V, Ret~II, Ret~III, Scl, Sxt, and Tuc~II. While some of these are believed to be LMC satellites which would naturally affect their apocenters (i.e. Ret~II), the change in the orbits of the remaining dwarfs may significantly affect models of when they were accreted.
Col~I and Leo~V meet this criteria, but Col~I contains a much larger pericenter and apocenter than the other dSphs and Leo~V is unbound without the presence of the LMC. 

For the probability of being an LMC satellite, we use the approach of \cite{Patel2020ApJ...893..121P}: for each dwarf we determine whether it was within the escape velocity of the LMC at its most recent closest approach to the LMC. We note that we also tried the method of \cite{Erkal2020MNRAS.495.2554E} who instead evaluated whether the satellite was energetically bound to the LMC 5 Gyr ago. Since we are sampling over a wide range of Milky Way potentials and LMC masses, as opposed to \cite{Erkal2020MNRAS.495.2554E} who used a single Milky Way potential and a discrete set of LMC masses, this method does not seem to be as robust as the approach of \cite{Patel2020ApJ...893..121P}. This also agrees with the results of \cite{DSouza2022MNRAS.512..739D} who show that the accuracy of orbits decrease with increased lookback time. We discuss the LMC connection of the satellites in more detail in  Section~\ref{sec:lmc_connection}. 

\section{Discussion}
\label{section:discussion}

\subsection{Tidal influence of the Milky Way}

\begin{figure*}
\includegraphics[width=\textwidth]{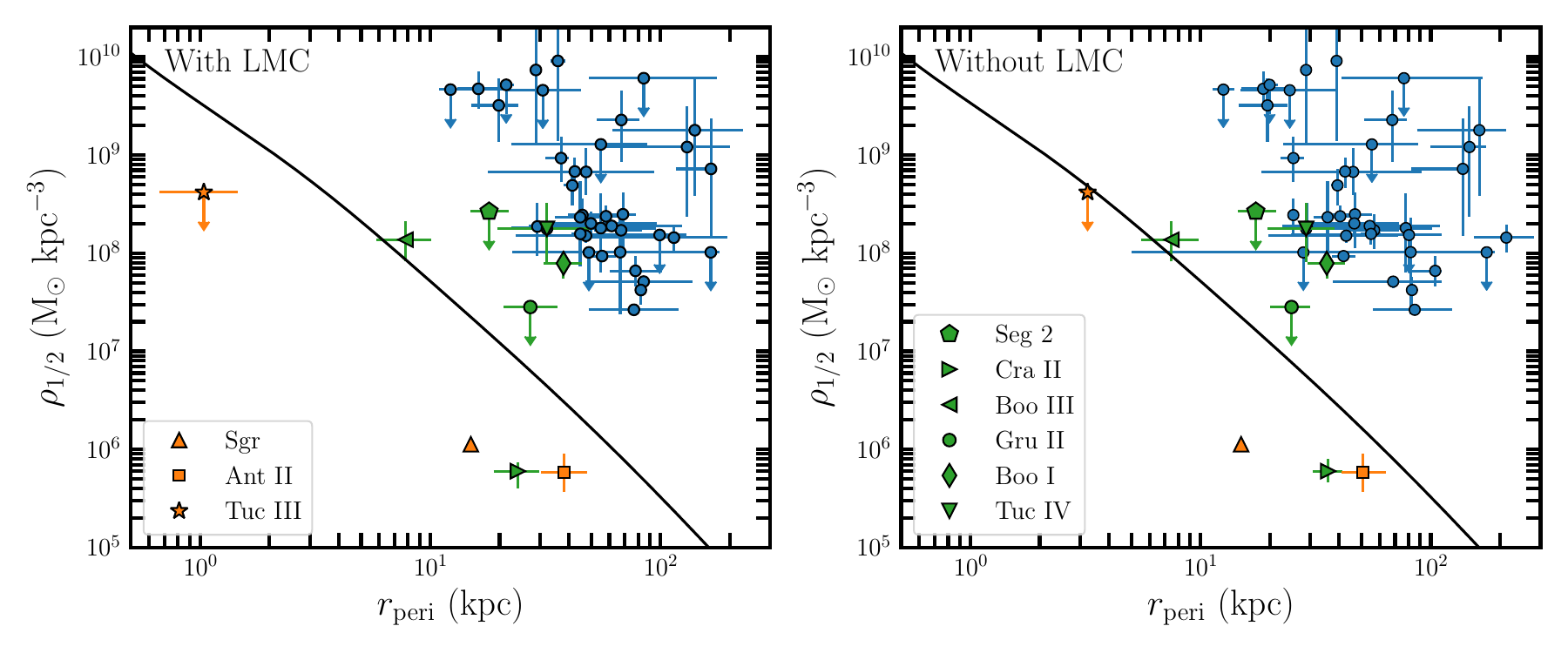}
\caption{Comparison of the pericenter ($r_{\rm peri}$) with and without the LMC influence (left panel and right panel, respectively) versus the average density within the half-light radius calculated from the stellar kinematics ($\rho_{1/2}$).  The black line shows  twice the enclosed MW density as a function of radius. If the satellite sits below this line, its Jacobi radius will be larger than the half-light radius and it will likely be tidal disrupting. Satellites near or below the curve are labeled.  
Orange symbols denote dSphs which are clearly tidally disrupting: Ant II,  Sgr, and Tuc III.  Whereas green symbols denote dSphs that are potentially undergoing tidal disruption and near the MW average density: Boo I, Boo III, Cra~II, Gru~II, Seg~2, and Tuc~IV (see text for details). 
}
\label{fig:density_pericenter}
\end{figure*}

\begin{figure*}
\includegraphics[width=\textwidth]{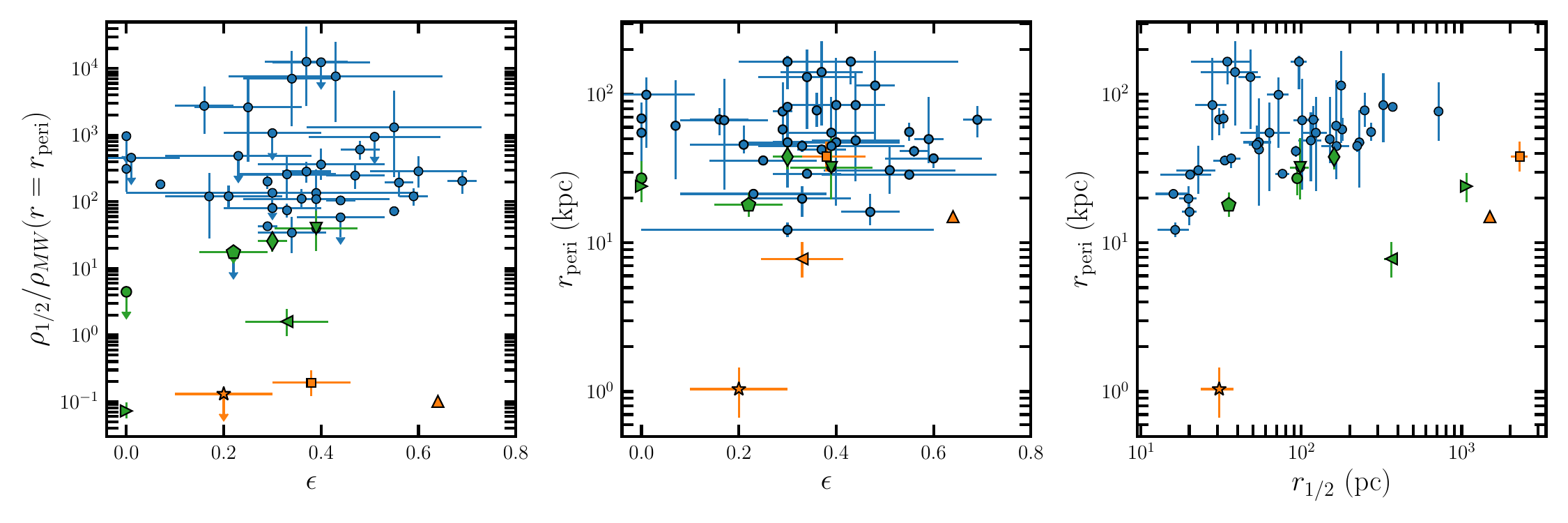}
\caption{Correlation plots of dSphs in terms of their densities, ellipticies, pericenters, and half-light radii. Symbols and colors are the same as Figure~\ref{fig:density_pericenter}. \textbf{Left panel:} Ellipticity ($\epsilon$) versus the average dSph density divided by the average MW density at the dSph's pericenter ($\rho_{1/2}/\rho_{MW}(r=r_{\rm peri})$). \textbf{Center panel:} Ellipticity ($\epsilon$) versus pericenter ($r_{\rm peri}$). \textbf{Right panel:} Half-light radius ($r_{ 1/2}$) vs pericenter ($r_{\rm peri}$).  
There is no clear trend for pericenter or the average density ratio with ellipticity (left and middle panel). The dSphs with evidence of tidal disruption and the smallest $\rho_{1/2}/\rho_{MW}(r=r_{\rm peri})$ ratios have larger half-light radii at a fixed pericenter than the general dSph population (right panel). 
}
\label{fig:other_tidal}
\end{figure*}

There has been extensive discussion on the tidal influence of the MW on its dSph population and here we compare our orbital parameters to some commonly used diagnostics. 
To directly address whether a satellite can be tidally influenced, we compare the  average dSph density within the half-light radii to twice the average MW density at the orbital pericenter in Figure~\ref{fig:density_pericenter}. This follows from the tidal radius assuming a flat rotation curve for the Milky Way and that the dwarf is on a circular orbit \citep{King1962AJ.....67..471K}:
\begin{equation}
r_t = r \Big( \frac{m}{2M(<r)} \Big)^{\frac{1}{3}} \,, \end{equation}
where $r_t$ is the tidal radius, $r$ is the distance from the Milky Way, $m$ is the mass of the dwarf, and $M(<r)$ is the enclosed Milky Way's mass within $r$. Re-arranging this and re-calling that we expect strong disruption when the half-light radius is similar to the tidal radius, the condition is:
\begin{equation}
\frac{m_{1/2}}{r_{1/2}^3} =  \frac{2M(<r)}{r^3} 
\end{equation}
which implies our condition of $\rho_{1/2} = 2 \bar{\rho}_{\rm MW}$. 
To calculate $m_{1/2}$ for the dSphs, we use the dynamical mass estimator from \citet{Wolf2010MNRAS.406.1220W}. 
We  additionally include the Sagittarius (Sgr) dSph \footnote{Sgr is not in our primary sample and the systemic proper motion and orbital motion was not derived in this work. For reference, we use $r_{\rm peri}\sim 16~\kpc$ \citep{Vasiliev2021MNRAS.501.2279V}, $\sigma \sim 15~\kms$ \citep{Vasiliev2020MNRAS.497.4162V}, and $r_{h}\sim 2500~\pc$ and $\epsilon \sim 0.64$ \citep{McConnachie2012AJ....144....4M}. }  in  Figure~\ref{fig:density_pericenter} and our tidal disruption analysis. Sgr is undergoing tidal disruption \citep[e.g.,][]{Vasiliev2021MNRAS.501.2279V} and    was excluded from our mixture model analysis due to its large angular size and low Galactic latitude.

In Figure~\ref{fig:density_pericenter}, there are a total of 11 dSphs with $\rho_{1/2}/\rho_{MW}(r=r_{\rm peri}) \lesssim 10$ that could have tidal influences. Two of these, Car~II and Hyi~I,  are likely LMC satellites (see Section~\ref{sec:lmc_connection})  and we exclude them from this discussion as their past dynamical evolution has primarily been influenced  by the LMC and they are on near their  first pericenter in the MW.
Three of the dSphs below the average MW density are clearly tidally disrupting based on independent literature analysis  (Ant~II, Sgr, and Tuc~III) and we  denote them with orange symbols in  Figure~\ref{fig:density_pericenter}.
Tuc~III has clear tidal tails extending $\sim2\degree$ from the satellite  \citep{DrlicaWagner2015ApJ...813..109D, Shipp2018ApJ...862..114S} and there is  velocity gradient along the tidal tails \citep{Li2018ApJ...866...22L}. 
Ant~II has a velocity gradient that aligns with the orbital direction  and there is qualitative agreement between tidal stripping models and Ant~II's kinematic and spatial properties  \citep{Ji2021ApJ...921...32J, Vivas2022ApJ...926...78V}.

We denote the other six dSphs with $\rho_{1/2}/\rho_{MW}(r=r_{\rm peri}) \lesssim 10$ as potentially disrupting (Boo~I, Boo~III, Cra~II, Gru~II, Seg~2, and Tuc~IV) and   more observational evidence and/or detailed dynamical modeling is required to confidently assess the tidally disrupting scenario. 
Boo~III has been argued to be tidally disrupting based on its large velocity dispersion \citep{Carlin2009ApJ...702L...9C}, its small pericenter ($r_{\rm peri} \sim 12 ~\kpc$)  and possible connection to the Styx stream \citep{Carlin2018ApJ...865....7C}.
\citet{Simon2020ApJ...892..137S} noted that the tidal radius of Gru~II is just larger than its physical size and may be vulnerable to tidal stripping. 
There is a tentative velocity gradient in Cra~II and  the tidal radius is less than the half-light radius but the predicted tidal features are beyond the range of current spectroscopic samples   \citep{Ji2021ApJ...921...32J}.
For Boo~I there is a potential velocity gradient \citep{Longeard2021arXiv210710849L} and there are several blue horizontal branch star candidates at large distances, outside the King limiting radius \citep{Filion2021ApJ...923..218F} which are both  consistent with tidal stripping models \citep{Longeard2021arXiv210710849L, Filion2021ApJ...923..218F}.
Tuc~IV  has had a recent, direct ($\Delta d \sim 4 ~\kpc$) collision with the LMC \citep{Simon2020ApJ...892..137S}. Tidal stripping has been used as an explanation for why  is offset from the stellar mass-metallicity relation  \citep{Kirby2013ApJ...770...16K}. 
No detailed tidal stripping models have been carried out for Boo~III,  Gru~II,  Seg~2, or Tuc~IV.
These dSphs are prime targets for searches for direct evidence of tidal disruption and/or detailed dynamical modeling.

There are other satellites that have small pericenters ($r_{\rm peri} <30~\kpc$; Car~III, Dra~II, Seg~1, Tri~II, and Wil~1), but they all have larger average densities and are therefore  resilient to the tidal influence of the MW. 
We note that if the velocity dispersion  was over-estimated these satellites could be undergoing tidal disruption by the MW  \citep[e.g., from unresolved binaries][or from small sample sizes]{Minor2019MNRAS.487.2961M}. 
Some of the dSphs have upper limits on their velocity dispersion and if we assume a value of  $\sigma_{\rm los}\sim 1 \kms$,  we would infer   density ratios of $\rho_{1/2}/\rho_{MW}(r=r_{\rm peri})\sim2$, 8, 9 for Gru~II, Seg~2, and Tri~II, respectively and they would be considered prime candidates for tidal influence.

A large ellipticity has previously been used as evidence for tidal disruption \citep[e.g.,][]{Munoz2010AJ....140..138M, Kupper2017ApJ...834..112K}.
In Figure~\ref{fig:other_tidal}, we  compare the ellipticity, the stellar half-light radius,  the ratio of average dSph density to MW density, and the orbital pericenters.  
We see no clear trend with the average density ratio  or pericenter  with the ellipticity. This agrees with conclusions from $N$-body simulations that high ellipticity does not imply tidal disruption \citep{Munoz2008ApJ...679..346M}. 
It is interesting that roughly half the dSphs with low density ratios are nearly spherical (Cra~II, Gru~II, Seg~2, Tuc~III) while the other half are  elongated (Ant~II, Boo~I, Boo~III, Sgr, Tuc~IV).
Clearly, there is some additional dependence on orbital phase for whether a large ellipticity would be observed in a disrupting satellite.
The right-hand panel of Figure~\ref{fig:other_tidal} compares the spherically averaged half-light radius and the pericenter.  At a fixed pericenter, dSphs that have some indications (Boo~III and Cra~II) or are  likely tidally disrupting  (Ant~II, Sgr, and Tuc~III) have larger sizes than the general dSph population.  The exceptions to this trend  (Boo~I, Gru~II, Seg~2, and Tuc~IV)  may be in an earlier stage of disruption than the other likely disrupting dwarfs. 
We similarly  examined the mass-to-light ratio of the dSphs and did not see any trends when comparing to orbital properties and direct tidal indicators.

\begin{figure*}
\includegraphics[width=\textwidth]{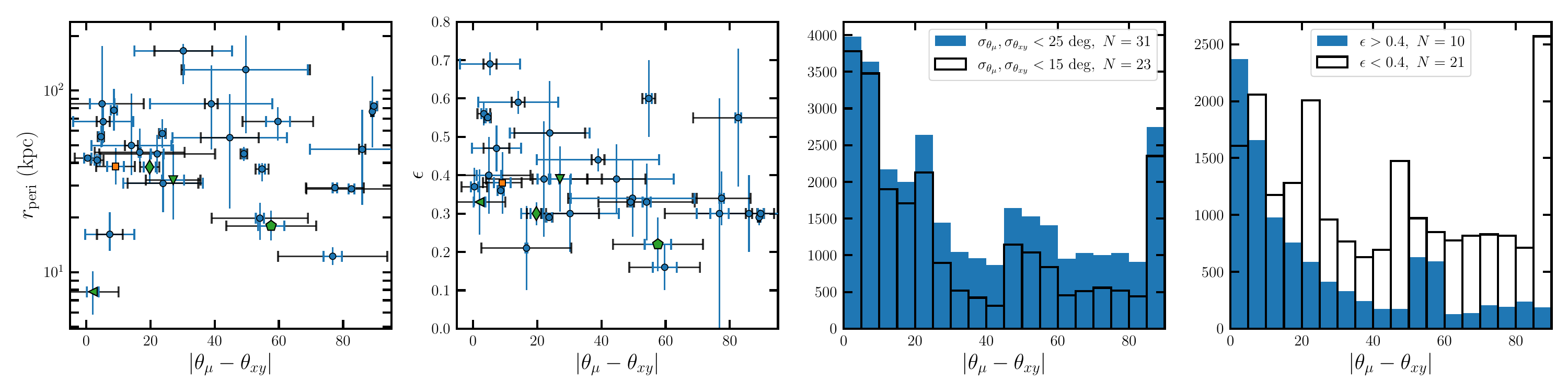}
\caption{Comparison between the orbital direction and spatial orientation of the dSphs. Symbols and colors follow Figure~\ref{fig:density_pericenter}.
(left) Difference between the reflex-corrected proper motion (i.e. the orbital direction) and the position angle (i.e. spatial orientation) $| \theta_{\mu}-\theta_{\rm xy} |$ versus the pericenter. When $| \theta_{\mu}-\theta_{\rm xy} | \sim 0$ the orbit and major axis of the dSph are aligned. The blue error bar is from the reflex corrected proper motion and the black error bar is due to the distribution in the position angle. (center left) $| \theta_{\mu}-\theta_{\rm xy} |$ versus the ellipticity of the dSph.  
(center right) Histogram of $| \theta_{\mu}-\theta_{\rm xy} |$ for the same sample.  To simulate errors we draw from each dSph's  $| \theta_{\mu}-\theta_{\rm xy} |$  distribution 1000 times.  The black bins are a subset of dSphs with smaller uncertainties. There is an excess of dSphs aligned with their orbital motion. 
(right) Histogram of $| \theta_{\mu}-\theta_{\rm xy} |$ but separated by ellipticity.  The blue bins are the elongated sample  ($\epsilon >0.4$).  There is a clear preference for systems with large ellipticity to align with their orbital direction. 
}
\label{fig:angular_difference}
\end{figure*}

\subsection{Is Orientation a Signature of Tidal Disruption?}
Next, we explore the relationship between the direction of the orbital motion and the orientation of each satellite.
Specially, we compute the difference  between the position angle ($\theta_{\rm xy}$) and the direction of the reflex-corrected proper motion (i.e., the orbital direction, $\theta_{\mu}$). 
In the left and center left panels of Figure~\ref{fig:angular_difference}, we compare $| \theta_{\mu}-\theta_{\rm xy} |$ to the pericenter and ellipticity, respectively. 
We have excluded dSphs where the orbital direction and/or position angle  are poorly measured ($\sigma_{\theta_\mu}, \sigma_{\theta_{\rm xy}} > 25\degree$).

The right-hand panels of Figure~\ref{fig:angular_difference} show the $| \theta_{\mu}-\theta_{\rm xy} |$ distribution of the dSph sample.  To construct the global satellite $| \theta_{\mu}-\theta_{\rm xy} |$, we have sampled each dSph's  $| \theta_{\mu}-\theta_{\rm xy} |$ distribution  1000 times with an error determined by adding the error from the position angle and reflex-corrected proper motion in  quadrature.
The center right panel shows the  dSph sample  
($\sigma_{\theta_\mu}, \sigma_{\theta_{\rm xy}} < 25\degree$) and  a subset (black bins) with more precise measurements ($\sigma_{\theta_\mu}, \sigma_{\theta_{\rm xy}} < 15\degree$).
There is an excess of satellites whose shape is aligned with their orbital motion and this correlation becomes more significant when  poorly measured dSphs are removed.
The right hand panel splits the same sample by ellipticity (at $\epsilon=0.4$).
dSphs with large ellipticties  are in general aligned with their orbital motion whereas less elliptical dSphs have  uniform orientations. 
In particular, there are six dSphs with $| \theta_{\mu}-\theta_{\rm xy} |<15\degree$ (Her, Phx~II, UMa~I, UMa~II, UMi, and Wil~1). 
The  two dSphs with high ellipticity that are not aligned, Car~III and Ret~II, are both highly likely to be LMC satellites and the LMC association likely affects their orientation relative to the MW.

Several of the disrupting dSphs are excluded from the orientation sample due to their sphericity (Cra~II, Gru~II, and Tuc~III) and most of the elliptical disrupting dSphs are aligned with the orbital motion.
The exception in Figure~\ref{fig:angular_difference}, is Seg~2 which has a low ellipticity ($\epsilon \sim 0.2$).
The core of Tuc~III is spherical ($\epsilon \sim 0.2$), and the Tuc~III tidal tails are aligned with the reflex-corrected proper motion \citep{Shipp2019ApJ...885....3S}.
Ant~II, Boo~I, Boo~III, and Tuc~IV all have the orbital motion aligned with the major axis.
Similarly, the orientation of the Sgr dSph is aligned with its orbital motion though it is not in our nominal sample \citep[e.g.,][]{delPino2021ApJ...908..244D}.

We have further compared the orientation of each satellite to the direction of the Galactic center and the orbital direction with the direction to the Galactic center. The only trend is an excess of satellites {with an orbital direction that is perpendicular to}   the direction to the  Galactic center. If the likely LMC satellites are removed this excess is removed. 

Thus, the only alignment that we find is that satellites with large ellipticity tend to be oriented along their orbit.  While tidal disruption would be a natural explanation for this, there is no corresponding trend with small pericenters or low average density compared to the average MW density. 
A possible explanation for this alignment is  tidal torques. 
Based on numerical simulations, there is an expected radial alignment between satellite orientation and the Galactic center due to tidal torques \citep{Pereira2008ApJ...672..825P}. However, the orientation changes throughout orbit and we do not have a complete sample of MW satellites.
Previous work has found that the MW dSphs share a common  orientation and that it may be related to the Vast Polar Orbital structure  \citep{Sanders2017MNRAS.472.2670S}. 
The Vast Polar Orbital structure  and orbital poles alignment may be caused by the  LMC \citep{GaravitoCamargo2021arXiv210807321G} but  \citet{Pawlowski2021arXiv211105358P} show the magnitude of the LMC perturbation is to small to fully explain the Vast Polar Orbital structure.

\subsection{Comparison to Previous Results and Spectroscopic Catalogs}

\begin{deluxetable*}{l cc ccc ccc}
\tablewidth{0pt}
\tablecaption{Summary of known spectroscopic members and potential targets of UFD galaxies
\label{tab:future_targets}
}
\tablehead{Dwarf & $N_{mem, F}$ & $N_{\rm targets}$ & $N_{\rm expected}$ & $N_{\rm mem, Gaia}$ & $N_{\rm mem, total}$ &  Spectroscopy citations
}
\startdata
Aqu II & $17.5_{-2.0}^{+2.1}$ & 23 & 13.8 & 3 & 9 & a \\
Boo I & $187.6_{-3.9}^{+3.8}$ & 202 & 140.0 & 46 & 100 & b,c,d,e \\
Boo II & $23.2_{-1.2}^{+1.2}$ & 23 & 16.7 & 6 & 6 & f,g \\
Boo III & $114.6_{-9.4}^{+10.0}$ & 256 & 129.7 & 13 & 20 & h \\
Boo IV & $6.7_{-1.5}^{+0.6}$ & 7 & 6.8 & 0 & 0 &  \\
CVn II & $16.3_{-0.5}^{+0.4}$ & 4 & 2.3 & 14 & 25 & i \\
Car II & $65.1_{-4.0}^{+4.0}$ & 72 & 39.4 & 17 & 18 & j,k \\
Car III & $10.0_{-1.4}^{+1.9}$ & 9 & 4.6 & 5 & 5 & j,k \\
Cen I & $28.1_{-2.3}^{+2.3}$ & 37 & 26.6 & 0 & 0 &  \\
Cet II & $7.9_{-0.6}^{+0.5}$ & 9 & 7.6 & 0 & 0 &  \\
Col I & $7.6_{-0.3}^{+0.4}$ & 3 & 2.9 & 5 & 9 & l \\
CB & $42.2_{-1.2}^{+1.2}$ & 38 & 28.4 & 14 & 59 & i \\
Dra II & $24.2_{-1.2}^{+1.0}$ & 19 & 14.3 & 10 & 14 & m,n \\
Eri II & $21.6_{-0.5}^{+0.5}$ & 10 & 7.6 & 14 & 92 & o,p \\
Gru I & $12.6_{-0.6}^{+0.6}$ & 6 & 5.5 & 7 & 7 & q \\
Gru II & $40.1_{-3.6}^{+3.7}$ & 40 & 16.3 & 19 & 21 & r \\
Her & $46.0_{-2.1}^{+1.9}$ & 27 & 16.3 & 28 & 59 & i \\
Hor I & $18.9_{-0.5}^{+0.5}$ & 17 & 12.8 & 6 & 6 & s,t \\
Hor II & $4.0_{-0.3}^{+0.3}$ & 4 & 3.9 & 0 & 3 & l \\
Hyd II & $21.5_{-0.8}^{+0.7}$ & 17 & 15.0 & 6 & 13 & u \\
Hyi I & $118.5_{-4.8}^{+4.8}$ & 133 & 77.8 & 31 & 31 & v \\
Leo IV & $8.5_{-0.2}^{+0.3}$ & 3 & 1.5 & 7 & 25 & i,e \\
Leo V & $8.4_{-0.2}^{+0.2}$ & 0 & 0.0 & 9 & 15 & w,x,e \\
Peg III & $3.9_{-0.4}^{+0.3}$ & 4 & 3.9 & 0 & 7 & y \\
Phx II & $12.7_{-0.5}^{+0.5}$ & 8 & 6.6 & 6 & 7 & l \\
Pic I & $8.2_{-0.4}^{+0.4}$ & 9 & 8.2 & 0 & 0 &  \\
Pic II & $6.4_{-2.5}^{+3.5}$ & 8 & 6.0 & 0 & 0 &  \\
Ret II & $56.1_{-1.6}^{+1.7}$ & 32 & 22.2 & 29 & 29 & z,aa,s \\
Ret III & $7.1_{-1.6}^{+1.0}$ & 6 & 4.0 & 2 & 3 & l \\
Sgr II & $69.3_{-1.5}^{+1.5}$ & 55 & 42.0 & 24 & 39 & ab \\
Seg 1 & $26.7_{-3.0}^{+3.0}$ & 9 & 3.1 & 12 & 72 & ac,ad \\
Seg 2 & $19.7_{-0.9}^{+0.9}$ & 2 & 0.3 & 12 & 26 & ae,af \\
Tri II & $11.8_{-1.5}^{+1.7}$ & 8 & 4.1 & 7 & 14 & ag,ah \\
Tuc II & $43.3_{-5.2}^{+5.2}$ & 53 & 19.0 & 22 & 22 & q,ai,aj \\
Tuc III & $56.3_{-5.3}^{+5.8}$ & 72 & 33.9 & 22 & 52 & ak,al \\
Tuc IV & $12.2_{-2.2}^{+2.5}$ & 12 & 4.0 & 7 & 11 & r \\
Tuc V & $6.5_{-2.3}^{+1.2}$ & 6 & 2.8 & 3 & 3 & r \\
UMa I & $50.5_{-1.3}^{+1.3}$ & 32 & 24.2 & 24 & 40 & am,c,i \\
UMa II & $51.8_{-2.5}^{+2.5}$ & 53 & 34.0 & 14 & 29 & c,i \\
Wil 1 & $9.0_{-0.7}^{+1.1}$ & 2 & 0.5 & 9 & 44 & c,an \\
\enddata
\tablecomments{
$N_{mem, F}$--total membership of each dSph with the complete sample, $N_{\rm targets}$--number of unobserved stars with $p>0.1$, $N_{\rm expected}$--expected number of members if all targets are observed, $N_{\rm mem, Gaia}$--number of known members with astrometric solutions,  $N_{\rm mem, total}$--total number of spectroscopic numbers.  
Citations: 
(a) \citep{Torrealba2016MNRAS.463..712T}
(b) \citep{Munoz2006ApJ...650L..51M}
(c) \citep{Martin2007MNRAS.380..281M}
(d) \citep{Koposov2011ApJ...736..146K}
(e) \citep{Jenkins2021ApJ...920...92J}
(f) \citep{Koch2009ApJ...690..453K}
(g) \citep{Ji2016ApJ...817...41J}
(h) \citep{Carlin2009ApJ...702L...9C}
(i) \citep{Simon2007ApJ...670..313S}
(j) \citep{Li2018ApJ...857..145L}
(k) \citep{Ji2020ApJ...889...27J}
(l) \citep{Fritz2019A\string&A...623A.129F}
(m) \citep{Martin2016MNRAS.458L..59M}
(n) \citep{Longeard2018MNRAS.480.2609L}
(o) \citep{Li2017ApJ...838....8L}
(p) \citep{Zoutendijk2021A\string&A...651A..80Z}
(q) \citep{Walker2016ApJ...819...53W}
(r) \citep{Simon2020ApJ...892..137S}
(s) \citep{Koposov2015ApJ...811...62K}
(t) \citep{Nagasawa2018ApJ...852...99N}
(u) \citep{Kirby2015ApJ...810...56K}
(v) \citep{Koposov2018MNRAS.479.5343K}
(w) \citep{Collins2017MNRAS.467..573C}
(x) \citep{MutluPakdil2019ApJ...885...53M}
(y) \citep{Kim2016ApJ...833...16K}
(z) \citep{Walker2015ApJ...808..108W}
(aa) \citep{Simon2015ApJ...808...95S}
(ab) \citep{Longeard2020MNRAS.491..356L}
(ac) \citep{Norris2010ApJ...723.1632N}
(ad) \citep{Simon2011ApJ...733...46S}
(ae) \citep{Kirby2013ApJ...770...16K}
(af) \citep{Belokurov2009MNRAS.397.1748B}
(ag) \citep{Martin2016ApJ...818...40M}
(ah) \citep{Kirby2017ApJ...838...83K}
(ai) \citep{Chiti2018ApJ...857...74C}
(aj) \citep{Chiti2021NatAs...5..392C}
(ak) \citep{Simon2017ApJ...838...11S}
(al) \citep{Li2018ApJ...866...22L}
(am) \citep{Kleyna2005ApJ...630L.141K}
(an) \citep{Willman2011AJ....142..128W}
}
\end{deluxetable*}

Overall, we find excellent agreement between our results and other {\it Gaia} EDR3 proper motion results\footnote{We have also compared our proper motion results to previous {\it Gaia} DR2 results \citep{Torrealba2019MNRAS.488.2743T, Chakrabarti2019ApJ...886...67C, McConnachie2020AJ....160..124M, Kallivayalil2018ApJ...867...19K, Fritz2018A&A...619A.103F, GaiaHelmi2018A&A...616A..12G, Simon2018ApJ...863...89S, Carlin2018ApJ...865....7C, Massari2018A&A...620A.155M, Mau2020ApJ...890..136M, Pace2019ApJ...875...77P, Fritz2019A&A...623A.129F, Fu2019ApJ...883...11F, Walker2019MNRAS.490.4121W, Longeard2018MNRAS.480.2609L, Simon2021ApJ...908...18S, Simon2020ApJ...892..137S, Gregory2020MNRAS.496.1092G, MutluPakdil2019ApJ...885...53M, Longeard2020MNRAS.491..356L, Massari2018NatAs...2..156M, Chiti2021NatAs...5..392C, Pace2020MNRAS.495.3022P} and to non-{\it Gaia} proper motion measurements \citep{Piatek2003AJ....126.2346P, Walker2008ApJ...688L..75W, Pryor2015AJ....149...42P, CasettiDinescu2016MNRAS.461..271C, Sohn2017ApJ...849...93S, Piatek2002AJ....124.3198P, Dinescu2004AJ....128..687D, Piatek2007AJ....133..818P, Mendez2011AJ....142...93M, Sohn2013ApJ...768..139S, Lepine2011ApJ...741..100L, Piatek2016AJ....152..166P, Piatek2006AJ....131.1445P, Fritz2018ApJ...860..164F, CasettiDinescu2018MNRAS.473.4064C, Piatek2005AJ....130...95P}. We include more details of this comparison in Appendix~\ref{appendix:membership}.} \citep{McConnachie2020RNAAS...4..229M, Li2021ApJ...916....8L, MartinezGarcia2021MNRAS.505.5884M, Vitral2021MNRAS.504.1355V, Battaglia2022A&A...657A..54B, Qi2022MNRAS.512.5601Q}. 
In particular,  \citet{McConnachie2020RNAAS...4..229M, Battaglia2022A&A...657A..54B} apply similar mixture models based on spatial position and proper motion with an additional  color-magnitude component based on {\it Gaia} photometry. Similar to \citet{Pace2019ApJ...875...77P} with {\it Gaia} DR2 data, we advocate for the use of auxiliary photometry especially for faint stars to assist with the identification of dSph members but acknowledge that the addition of a color-magnitude likelihood term is valuable to identify dSph stars. The distribution of MW stars is not uniform in color-magnitude space and including that information in the mixture model is valuable. 
\citet{McConnachie2020RNAAS...4..229M} include a prior on the systemic proper motions that requires the corresponding tangential velocity  to be bound to the MW. 
For {\it Gaia} DR2 measurements this affected a large number of dwarfs relative to other measurements \citep{McConnachie2020AJ....160..124M}.   
With EDR3, this prior generally only affects more distant satellites ($>100~\kpc$) with a low number of members (e.g, Boo~IV, Leo~IV, Leo~V, Psc~II, Peg~III) and  some distant dSphs have smaller proper motion errors than our results due to this prior.  
Previous {\it Gaia} DR2 results are commonly offset from the EDR3 results due to zero-point proper motion systematics in DR2 that have roughly  decreased by a factor of two in EDR3 \citep{GaiaCollaboration_Luri2021A&A...649A...7G}. The EDR3 proper motions are more precise than previous {\it HST} measurements for Leo~I and Leo~II in contrast to the DR2 results.

Ultra-faint dwarfs have spectroscopic samples varying between 3 and $\sim70$ members.  
We have compared current spectroscopic samples to our membership catalogs to assist in validating the method (i.e. to check we are correctly identifying known dSph members and MW foreground stars) and to identify the most promising targets for future followup. 
Here we consider a candidate as any stars with $p>0.1$. 
In Table~\ref{tab:future_targets}, we show the results of this exercise. 
In particular we list the number of expected members ($N_{\rm expected}$) if all stars with $p>0.1$ are targeted ($N_{\rm targets}$).
For  almost all dSphs, we find excellent agreement between known spectroscopic dSph members and a high mixture model membership probability and a  corresponding agreement between known  spectroscopic MW foreground stars and a low or  zero membership probability from our mixture model.  
One object with disagreement is Wil~1 as several previously identified spectroscopic members are identified as MW stars based on {\it Gaia} astrometry.  This disagreement is partly due to the difficultly in identifying spectroscopic members as the Wil~1 line-of-sight velocity overlaps with the MW distribution.
Based on the total membership, there are several dwarfs where future spectroscopic observations with {\it Gaia} selected observations can double or triple the sample sizes (e.g., Aqu~II, Boo~II, Boo~III, Dra~II, Hyi~I, Phx~II, Ret~II, Sgr~II, UMa~II).
In addition, there are a number of bright candidates ($g<18.5$) that are excellent targets for high resolution spectroscopic follow-up for detailed chemical abundance studies.

\subsection{Association with the Large Magellanic Cloud}
\label{sec:lmc_connection}

\begin{figure*}
\includegraphics[width=\textwidth]{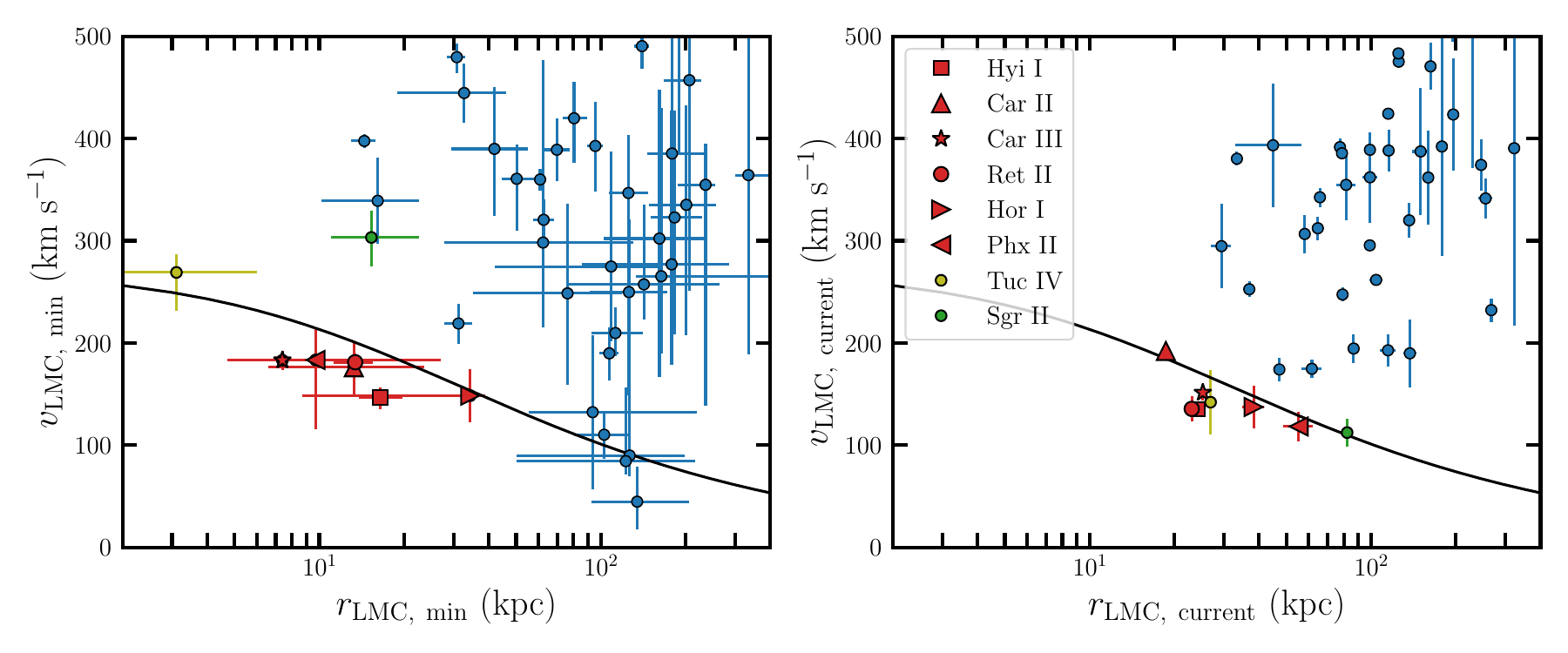}
\caption{
The distance to the LMC versus the relative velocity at each  satellite's previous closest LMC approach (left) and the current distance versus the current relative velocity (right). Red symbols denote the candidate LMC satellite population (Car~II, Car~III, Hyi~I, Hor~I, Phx~II, Ret~II). The black line is the LMC escape velocity curve with $M_{\rm LMC} = 13.8\times 10^{10}~M_{\odot}$. The satellite with the closest approach is Tuc~IV (gold circle) which is not considered an LMC satellite due to its large relative velocity. 
Currently, Sgr~II (green symbol) is closer to the MW than the LMC and is more likely to be associated with the MW. 
The five satellites at large distances with low relative velocities (they are Dra, Dra~II, UMi, Cra~II, Car~I) are not bound to the LMC. Most of them are closer to the MW at the time of closest approach. 
}
\label{fig:lmc_rad_dist}
\end{figure*}

With full phase space information, it is possible to determine which dSphs were previously associated with the LMC prior to their infall into the MW 
\citep{Deason2015MNRAS.453.3568D, Jethwa2016MNRAS.461.2212J, Sales2017MNRAS.465.1879S, Kallivayalil2018ApJ...867...19K, Fritz2019A&A...623A.129F, Patel2020ApJ...893..121P, Battaglia2022A&A...657A..54B, Erkal2020MNRAS.495.2554E, SantosSantos2021MNRAS.504.4551S, CorreaMagnus2022MNRAS.511.2610C}. 
To determine LMC association, we compute the fraction of orbits, $p_{\rm LMC}$, where a dSph's relative velocity ($v_{\rm LMC,~min}$) at its most recent approach to the LMC ($r_{\rm LMC,~min}$) is less than the LMC's escape velocity. 
We include the $p_{\rm LMC}$ values for each dSph in Table~\ref{tab:orbits}. We show each dSph's  $r_{\rm LMC,~min}$ and $v_{\rm LMC,~min}$ in Figure~\ref{fig:lmc_rad_dist}, along with each dSphs current position and velocity relative to the LMC.  

Based on our orbit modeling, we identify Car~II, Car~III,  Hor~I, Hyi~I, Phx~II, and Ret~II as   likely  LMC satellites ($p_{\rm LMC}>0.5$).
While five of the six dSphs we identify as LMC satellites are currently within the LMC's escape velocity (Car~II is currently $<10~\kms$ outside of the escape velocity), we note that the LMC's Jacobi radius is likely lower today than in the past due to the proximity of the MW \citep{Battaglia2022A&A...657A..54B}.  
The same six dSphs have been identified as likely  LMC satellites in other studies with different methodology for determining association \citep{Kallivayalil2018ApJ...867...19K,Erkal2020MNRAS.495.2554E, Patel2020ApJ...893..121P, Battaglia2022A&A...657A..54B}. 
For example, \citet{Erkal2020MNRAS.495.2554E} determine LMC association by computing the binding energy relative to the LMC after rewinding for 5 Gyr, or when the LMC reaches apocenter if that is earlier. 
Similarly, \citet{CorreaMagnus2022MNRAS.511.2610C} define a LMC satellite as one which was energetically bound to the LMC at some point between 1 to 3 Gyr ago. 
\citet{Battaglia2022A&A...657A..54B} determine LMC association if the satellite was inside the LMC Jacobi radius at the time of closest approach. 
However, we note \citet{CorreaMagnus2022MNRAS.511.2610C} only consider  five of the six to be likely associated and find  Ret~II to have a low probability of being associated. 
The different methodology used to identify LMC satellites generally does not make a difference for these six objects (Car~II, Car~III, Hor~I, Hyi~I, Phx~II, and Ret~II) and they were likely previously associated with the LMC.

We find that Tuc~IV has had a close encounter ($r_{\rm LMC,~min}\sim3\kpc$) with the LMC\footnote{Tuc~IV has possibly undergone a three-body interaction with LMC and SMC \citep{Simon2020ApJ...892..137S}.} with a large relative velocity   and is currently within the LMC escape velocity. 
Several studies have considered Tuc~IV to have a low probability of being associated with the LMC \citep{Simon2020ApJ...892..137S, Battaglia2022A&A...657A..54B}. 
Similar to Tuc~IV, we find that Sgr~II, Tuc~III, and Tuc~V have had close encounters ($r_{\rm LMC,~min}\sim 10-20~\kpc$) but with much larger relative velocities ($v_{\rm LMC,~min}\sim 300-400~\kms$). Of these, Sgr~II is currently near the LMC's escape velocity, however it is unlikely to be associated as it is closer to the MW.

While Dra, Dra~II, and UMi also pass our association criteria based on their relative velocity being less than the LMC escape velocity at their closest approach, they were closer to the MW at this time and their Milky Way apocenters are less than $r_{\rm min,~LMC}$.  Thus, the low relative velocity is just fortuitous. 
Dra~II has been noted to have a potential LMC association if the LMC was on its  second pericenter, a scenario which is increasingly unlikely \citep{Kallivayalil2018ApJ...867...19K}. 
Similarly, in some orbits the relative velocity of Car and Cra~II are less than the LMC escape velocity at $r_{\rm LMC,~min}$ but both are more distant than $100\kpc$ from the LMC at this time and the MW has had a  larger influence on them. 
Finally, Gru~II is the next closest satellite in phase space relative to the LMC, both at the present day and during its closest approach. It is considered a recently captured satellite by \citet{Battaglia2022A&A...657A..54B}.

\subsection{On the Excess of Satellites near Pericenter}

\begin{figure}
\includegraphics[width=\columnwidth]{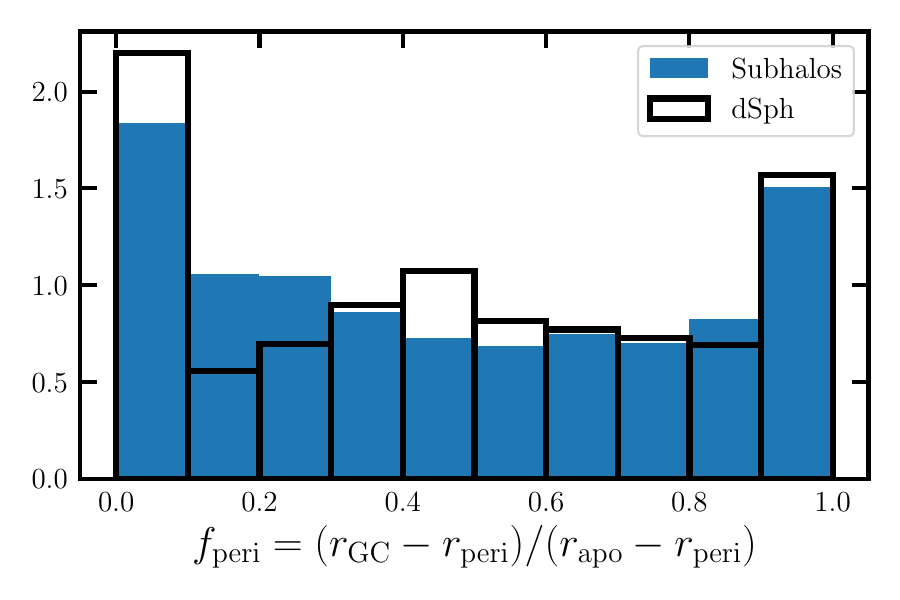}
\caption{Pericenter fraction ($f_{\rm peri} = (r_{GC} - r_{\rm peri} )/ (r_{\rm apo} - r_{\rm peri}$)  of observed dSphs (black) and subhalos (blue) from six MW-like $N$-body cosmological simulations. 
We exclude LMC dSphs and dSphs with large uncertainties from the observed sample (see text).
For the observed sample, the bins are made from the Monte Carlo chains of each satellite.
In both samples there are `excesses' of satellites near pericenter and apocenter as the radial velocity is zero there and thus satellites spend relatively more time there. 
}
\label{fig:pericenter_excess}
\end{figure}

Once systemic proper motions of most MW satellites were measured with {\it Gaia} DR2, subsequent orbital analysis revealed an excess of  satellites near their orbital pericenter  \citep{Simon2018ApJ...863...89S, Fritz2018A&A...619A.103F}. 
This is unexpected as a satellite spends more time near its orbital apocenter than its pericenter.
To assess this issue with our data set, we follow \citet{Fritz2018A&A...619A.103F} and evaluate the ratio, $f_{\rm peri}= (r_{GC} - r_{\rm peri} )/ (r_{\rm apo} - r_{\rm peri})$, which is a proxy for the orbital phase in the radial direction.
 $f_{\rm peri}=0,1$  corresponds to the satellite being at its pericenter or apocenter, respectively. 
With DR2 measurements, roughly half the dSph sample had $f_{\rm peri}<0.1$, however, the $f_{\rm peri}$ distribution becomes less extreme with  a heavier MW  \citep{Fritz2018A&A...619A.103F}.

We explore the $f_{\rm peri}$ distribution from our orbit modeling in Figure~\ref{fig:pericenter_excess}.
We see an excess  of MW satellites with  $f_{\rm peri}\sim0$ and we see a secondary peak at $f_{\rm peri}\sim1$.
We compute the fraction directly from the Monte Carlo samples.
We exclude LMC satellites (Car~II, Car~III, Hor~I, Hyi~I, Phx~II, Ret~II) and dSphs with large tangential velocity errors (Col~I, Eri~II, Hyd~II, Leo~IV, Leo~V, Peg~III, Psc~II, and Ret~III) following  \citet{CorreaMagnus2022MNRAS.511.2610C}.
We note that in a number of cases there are satellites that are either closer than their previous pericenter\footnote{The LMC satellites, Car~III, Hyi~I, Phx~II, and Ret~II suffer from this issue but are already excluded from the analysis.} (CB and UMa~II)  or more distant than their previous apocenter (Leo~II, Sgr~II, and Wil~1) and for these objects we use their current Galactocentric distance instead for the pericenter or apocenter. We note that this issue arises since we have defined the pericenter and apocenter as the local minimum and local maximum, respectively, and the dSphs have since had their orbits perturbed by the LMC.
We note that both excluded samples (LMC satellites and large tangential errors) are preferentially near either their pericenter or apocenter.

To further examine this issue, we explore the $f_{\rm peri}$ distribution of subhalos in high-resolution cosmological $N$-body zoom-in simulations of MW-like halos. These simulations are described in detail in \citet{Jethwa2018MNRAS.473.2060J}.
These simulations were run with the $N$-body part of \textsc{gadget-3} which is similar to \textsc{gadget-2} \citep{Springel2005MNRAS.364.1105S}. These simulations resolve Milky Way-like dark matter haloes with a particle mass of $2.27\times10^5 M_\odot$. From comparing with higher resolution runs, \citet{Jethwa2018MNRAS.473.2060J} found that the subhaloes in these simulations are complete down to a mass of $10^{7.5} M_\odot$, which is sufficient for comparing with the dwarfs in this work. Although these simulations are dark matter only, they include an analytical disk potential which is grown adiabatically from $z=3$ to $z=1$ (11 Gyr to 8 Gyr ago). This technique has been show to mimic the depletion of subhaloes by baryonic disks \citep[e.g.,][]{GarrisonKimmel2017MNRAS.471.1709G}.

The subhaloes in this simulation are identified with \textsc{rockstar} \citep{Behroozi2013ApJ...762..109B} and the merger trees are constructed with \textsc{consistent trees} \citep{Behroozi2013ApJ...763...18B}. 
We measure the orbit of each subhalo by taking its position and velocity relative to the Milky Way in the final snapshot and computing the angular momentum and energy of its orbit. We then compute the turning points of the effective potential, $\phi_{\rm eff}(r) = \frac{L^2}{2r^2}+\phi(r)$, where $L$ is the total angular momentum and $\phi(r)$ is the gravitational potential of the Milky Way halo in the final snapshot.
In total, there are 1576 subhalos with $r_{\rm GC}<r_{\rm vir}$ from six MW-like simulations. 

The  $f_{\rm peri}$ distribution of subhalos of MW-like dSphs is included in  Figure~\ref{fig:pericenter_excess}.
From the simulations, we see that the subhalo distribution peaks at both $f_{\rm peri}\sim0$ and $f_{\rm peri}\sim1$ which matches the dSph population. 
A pile-up at $f_{\rm peri}\sim 0$ is not unexpected. 
These  peaks occurs as $\frac{d}{dt} f_{\rm peri}\propto \frac{d}{dt} r_{\rm GC} =0$ at both pericenter and apocenter  \citep{Li2022ApJ...928...30L}.
This is not a one-to-one comparison between the simulated subhalos and the observed dSph population, as there is no selection function applied to the simulated subhalos \citep[e.g.][]{DrlicaWagner2020ApJ...893...47D},  and there are no observational errors applied to the simulations.
Regardless, we find a general agreement between the $f_{\rm peri}$ distribution of the  dSph and subhalo population. 
We leave a more detailed comparison between the orbital properties of the MW dSph population and simulated MW subhalo population to a future work.

The potential excess of satellites near their pericenter has been addressed by several other {\it Gaia} EDR3 based analyses.
\citet{Li2021ApJ...916....8L} examined the ratio of  the time to reach or leave a satellite's pericenter compared to half the total orbital period. 
Their analysis inferred an excess at lower values of this ratio while they expected a uniform distribution and they concluded that there remains a proximity-to-pericenter issue. 
 \citet{CorreaMagnus2022MNRAS.511.2610C} examined the  radial phase angle,  the canonically conjugate variable to the radial action, and found that the dSph population is distributed uniformly in the radial phase angle.  
\citet{CorreaMagnus2022MNRAS.511.2610C} concluded that there is no proximity-to-pericenter issue due to the  dSph population being well-mixed in radial phase angle.
\citet{Li2022ApJ...928...30L} examined  $f_{\rm peri}$ for the dSph population in a MW only potential and found a  $f_{\rm peri}$ distribution similar to our analysis (see their Figure~5). They analyzed the globular cluster population and a dozen stellar streams and found peaks at  $f_{\rm peri}\sim0,1$,   similar to the dSph population.
Furthermore, \citet{Li2022ApJ...928...30L} sampled  dSph-like orbits uniformly in time at eccentricities of 0.2, 0.4, and 0.6 and found the  $f_{\rm peri}$ distribution of these sampled orbits have peaks  at $f_{\rm peri}\sim0,1$.
In summary, there is not an  excess of satellites near their  pericenter and the perceived excess was due to how the orbital phase was computed.

\subsection{On the Anti-Correlation Between Pericenter and Density for Bright Satellites}

\begin{figure}
\includegraphics[width=\columnwidth]{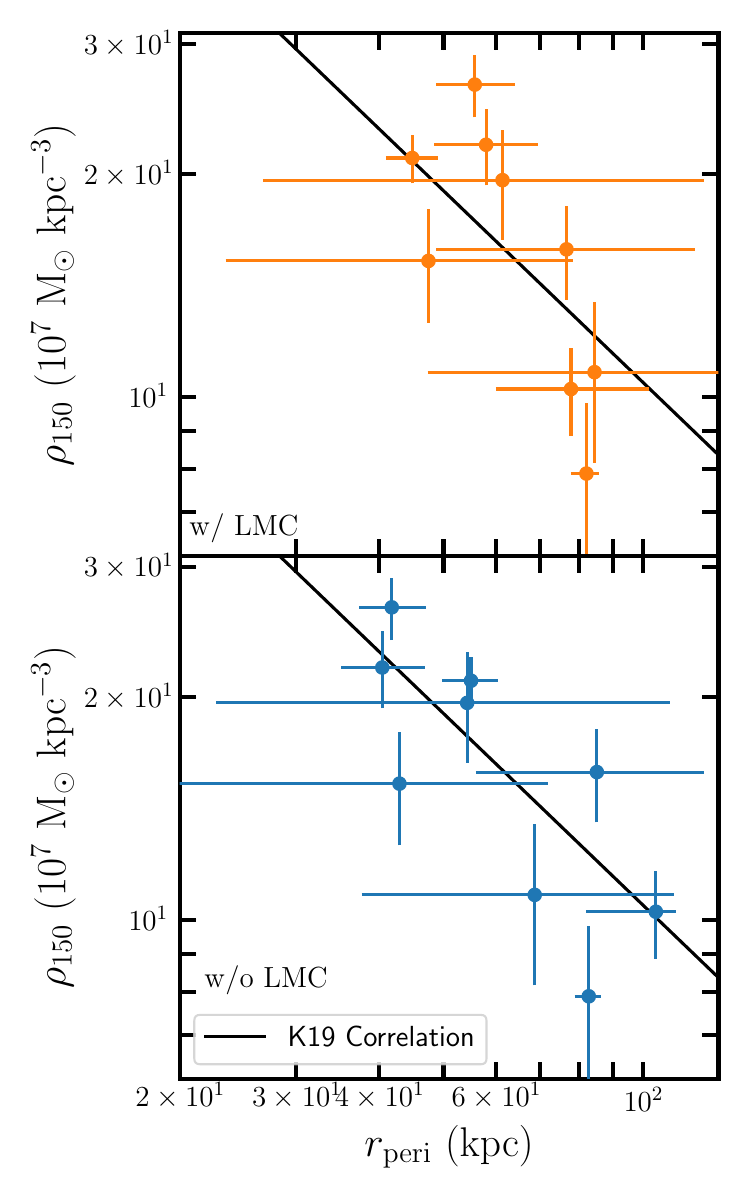}
\caption{Total density within 150 pc ($\rho_{150}$) versus the pericenter ($r_{\rm peri}$) for the classical satellites (Car, Dra, For, Leo~I, Leo~II, Scl, Sxt, and UMi.) and CVn~I. The top and bottom panels show pericenter with and without the influence of the LMC.  
Overlaid is the correlation for a cuspy dark matter  from \citet{Kaplinghat2019MNRAS.490..231K}. 
}
\label{fig:density_pericenter_correlation}
\end{figure}

There is a reported anti-correlation between the average dark matter density within 150 pc ($\rho_{150}$) and the pericenter ($r_{\rm peri}$) in the classical dSphs + CVn~I\footnote{This includes Car, CVn~I, Dra, For, Leo~I, Leo~II, Scl, Sxt, and UMi.}  based on orbits computed with  {\it Gaia} DR2 data \citep{Kaplinghat2019MNRAS.490..231K}. 
This correlation implies that only the densest dSphs can survive at small pericenters.
In Figure~\ref{fig:density_pericenter_correlation}, we update the pericenters with our results from EDR3 proper motions and include the effect of the LMC. 
For the dark matter distribution we use results from spherical Jeans dynamical models from \citet{Pace2019MNRAS.482.3480P}, which assume an NFW dark matter distribution \citep{Navarro1996ApJ...462..563N}. 
Without the LMC, we find a general agreement with the previously reported correlation between $r_{\rm peri}$ and $\rho_{150}$. 
For the models with the LMC, we find the previously reported correlation is much weaker and steeper.

The  dSphs, Ant~II and Cra~II, are in the same stellar mass range and should be included as they were likely initially hosted by similar mass haloes (given their similar stellar masses $>10^5~\Msun$) as the other classical dwarfs.
They have much lower densities than the classical dwarfs  pericenter  which is likely a signature of their tidal disruption \citep[][]{Ji2021ApJ...921...32J, Vivas2022ApJ...926...78V} and their inclusion would similarly weaken the $\rho_{150}$-$r_{\rm peri}$ anti-correlation. 
However, the question remains, where are the dSphs with high density and a large pericenter.

A similar trend is observed in $N$-body simulations between the subhalo maximum circular velocity ($V_{\rm max}$) and $r_{\rm peri}$  \citep{Robles2021MNRAS.503.5232R}.  At a fixed $V_{\rm max}$, subhalos with smaller  $r_{\rm peri}$ are more concentrated and more dense.  Less concentrated halos are less resilient to tidal disruption. 
A similar trend has been observed when considering the distance to the host in that the closer subhalos have higher concentrations on average than more distant subhalos  \citep{Moline2021arXiv211002097M}.  
This observed trend between pericenter and dark matter density has been used to  probe self-interacting dark matter  \citep[e.g.,][]{Jiang2021arXiv210803243J}. 
Pericenter and orbital analysis will be useful priors for dynamical analysis \citep{Robles2021MNRAS.503.5232R} however, we caution that the LMC needs to be included for MW orbital analyses.

\subsection{Dominant Source of Orbital Uncertainty} \label{sec:future_improvements}

\begin{figure*}
    \centering
    \includegraphics[width = .45\textwidth]{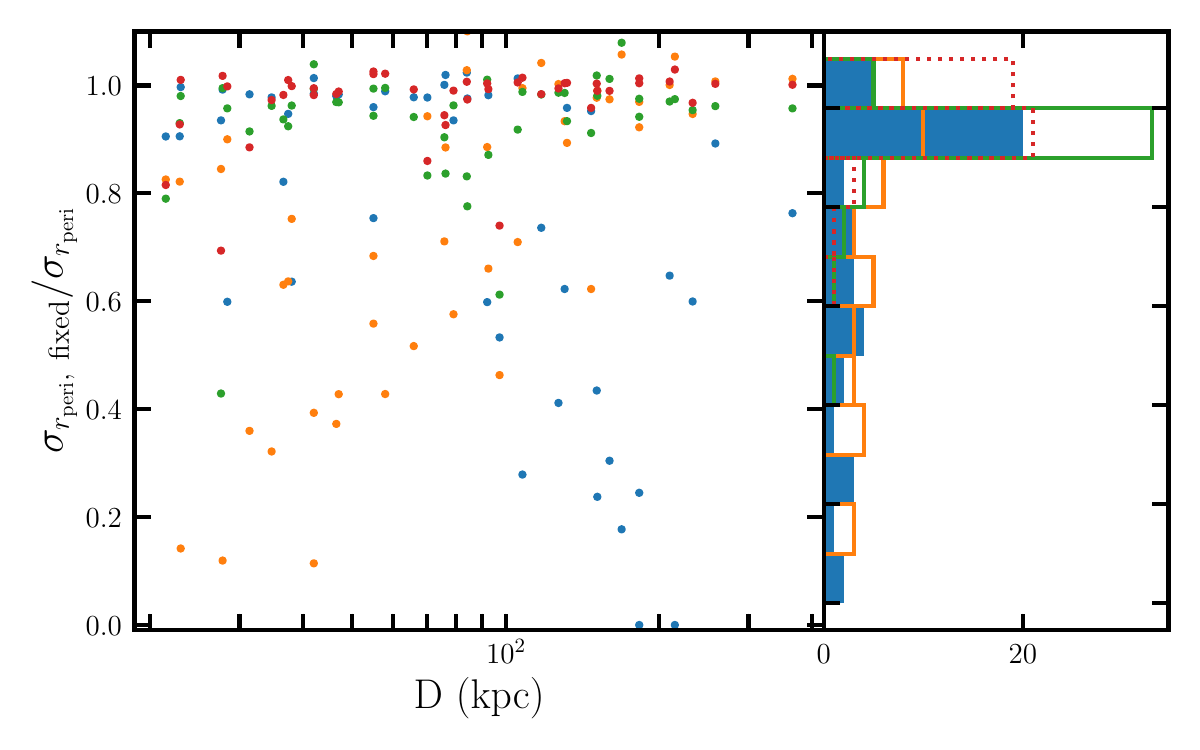}
    \includegraphics[width = .45\textwidth]{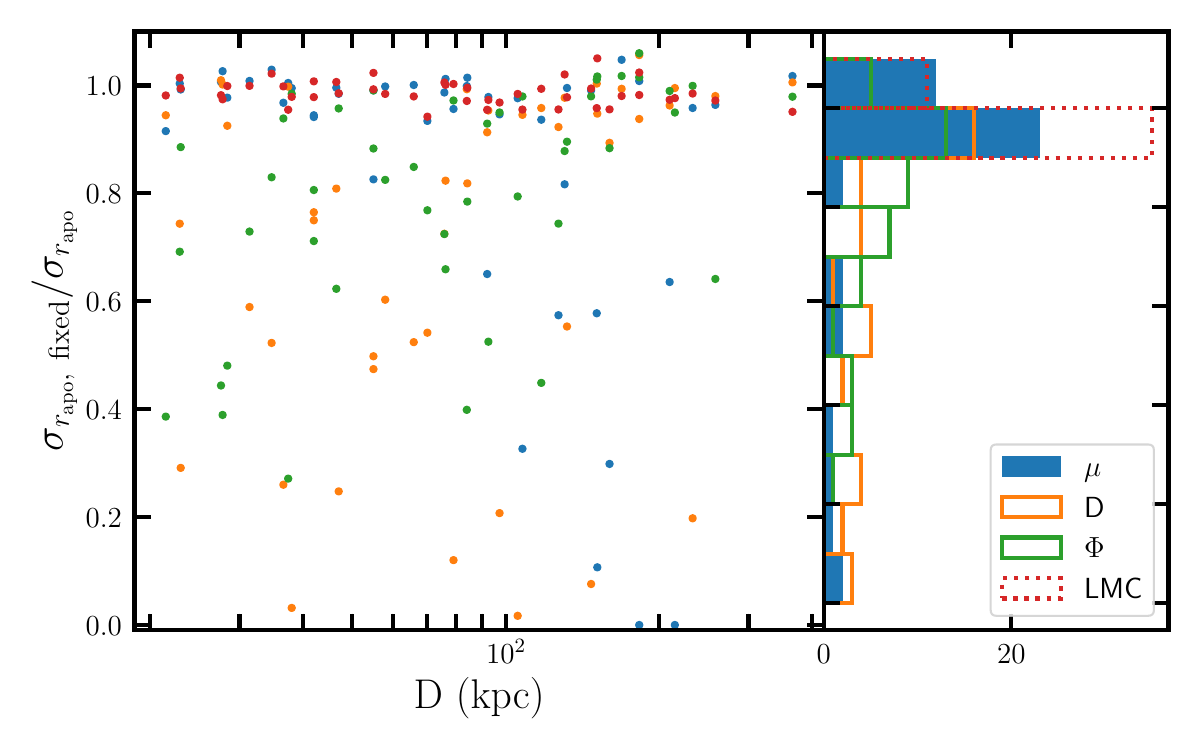}
    \caption{
    Relative fractional  error  compared to the distance for the pericenter ($\sigma_{r_{\rm peri}}$; left panels) and apocenter ($\sigma_{r_{\rm apo}}$; right panels) due to the fixing the error to zero for  the systemic proper motion (blue points), distance (orange points), MW potential (green points), and LMC mass (red points).
    For each galaxy there are four entries and in each entry three of the previously mentioned errors are set to the current value and one is set to zero to analyze its impact. 
Low factions correspond to the majority of current error being due to the fixed parameter and high fractions indicate that current error in the fixed  property does not significantly affect the total error. 
The current error in the pericenter is dominated by the distance and/or proper motion uncertainties whereas the current error in the apocenter is dominated by to  the distance and/or potential uncertainties. To aid in interpreting the sources of error, we show a histogram of the fractional uncertainties next to each panel. }
    \label{fig:current_errors}
\end{figure*}

\begin{figure*}
    \centering
    \includegraphics[width = .45\textwidth]{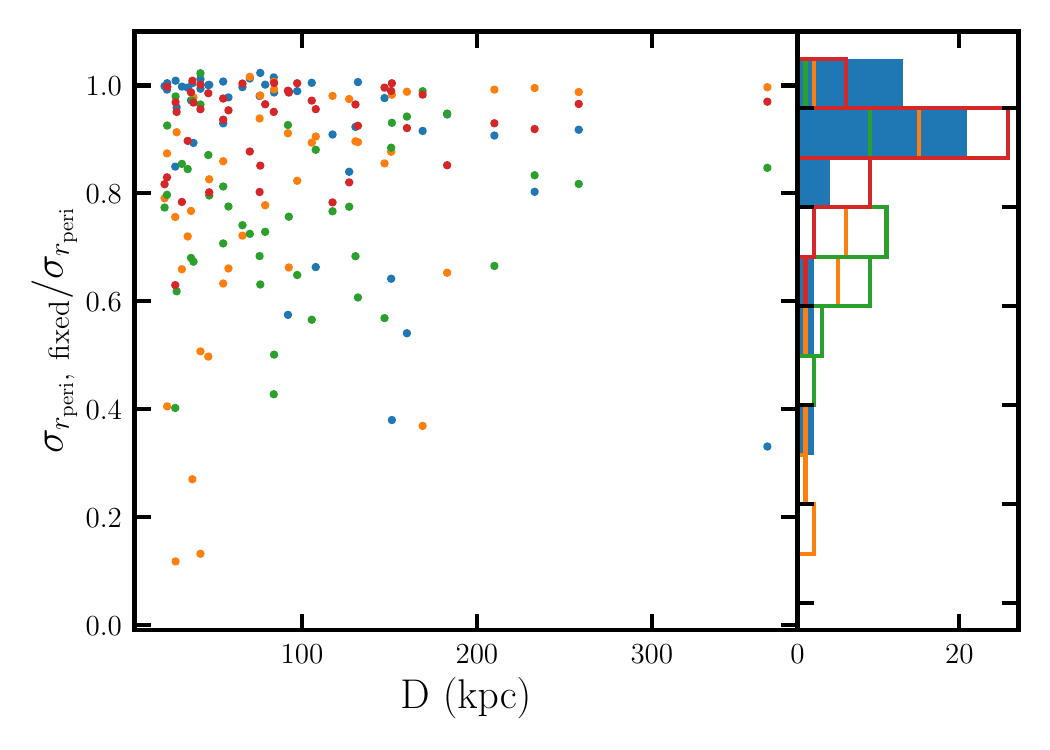}
    \includegraphics[width = .45\textwidth]{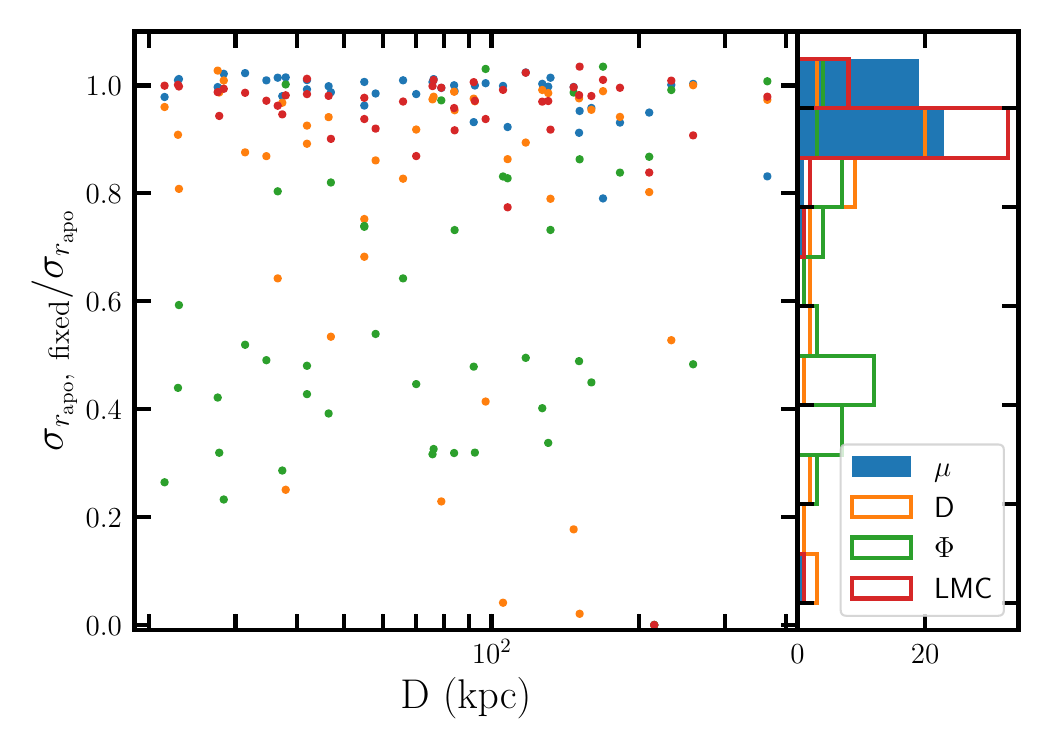}
    \caption{Similar to Figure~\ref{fig:current_errors} except here we start with the projected future errors and go to zero error for each component. The proper motion errors are based on {\it Gaia} DR5 projections, the errors on distance are assumed to be 2\%, and the Milky Way and LMC uncertainties are left at their present-day values. The largest source of error in the future will be due to the distance and MW potential uncertainty which emphasizes the need to better measure the MW potential. Interestingly, some satellites near the LMC will be sensitive to improvements in the LMC. }
    \label{fig:future_errors}
\end{figure*}

\begin{figure*}
    \centering
    \includegraphics[width = .45\textwidth]{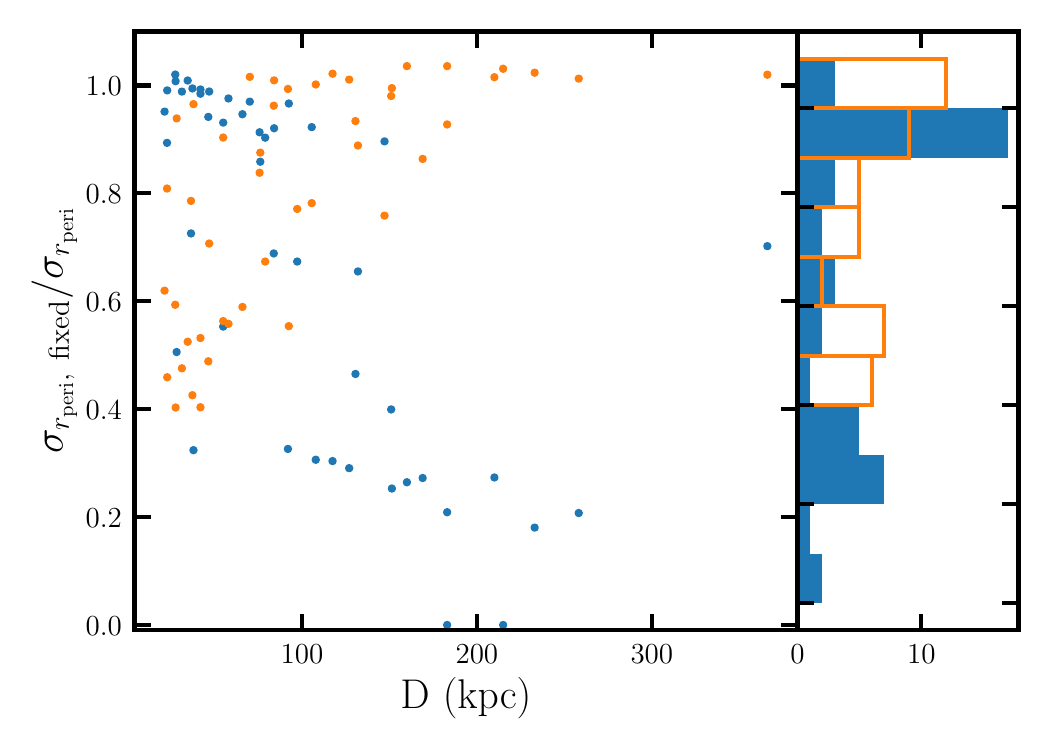}
    \includegraphics[width = .45\textwidth]{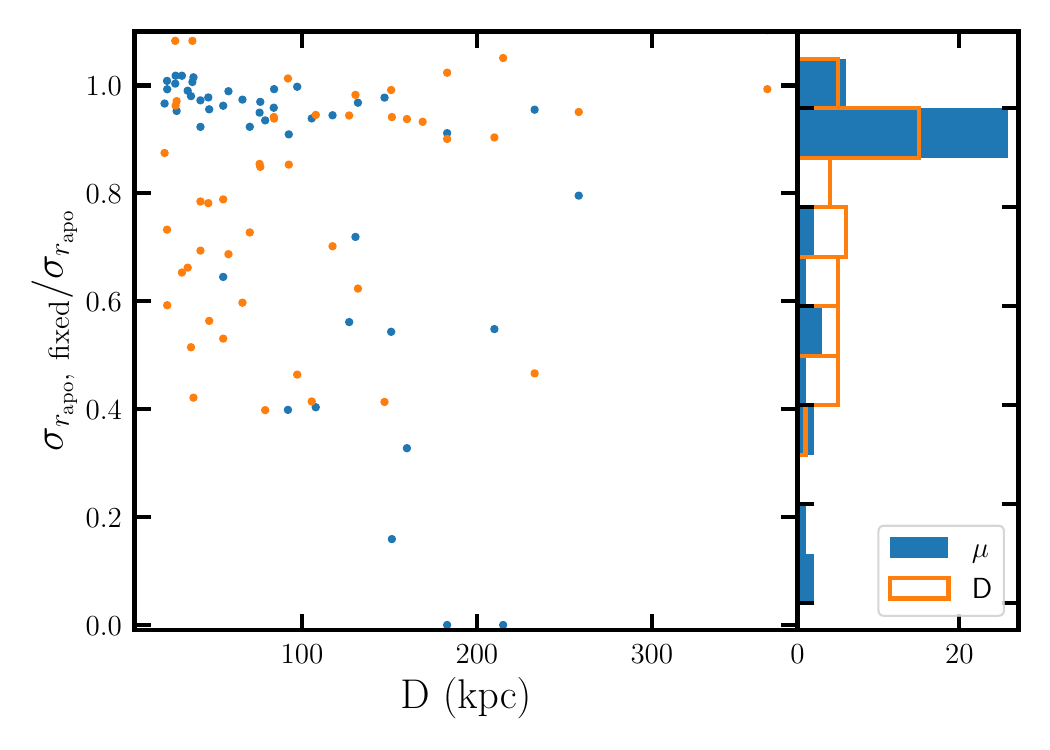}
    \caption{Similar to Figure~\ref{fig:current_errors} except the here we compare current proper motion errors to {\it Gaia} DR5 errors and compare 5\% to 2\% errors in the distance.  For some objects  5\% error in the distance is larger than their current error.}
    \label{fig:improved_errors}
\end{figure*}

In this work we have considered many sources of error when evaluating the dwarf orbits: uncertainties in the present-day phase space coordinates of the dwarfs (i.e. proper motions, distance, and radial velocity\footnote{The errors on the line-of-sight velocity and $\alpha$, $\delta$ are minuscule compared to the listed properties.}), in the LMC model (i.e. its present day proper motions, distance, radial velocity, and mass), and the Milky Way potential. In order to explore which of these sources dominates the orbital uncertainty (i.e. in $r_{\rm peri}$ and $r_{\rm apo}$), and thus which would be the most helpful to improve, we repeat our analysis and build three different suites with different assumptions about the uncertainties. We note that in this analysis, we use the Milky Way potential and associated uncertainties from \citet{McMillan2017MNRAS.465...76M}. In reality, there are larger uncertainties depending on what tracers and modelling techniques are used \citep[e.g.][]{Wang2020SCPMA..6309801W} and thus our Milky Way potential uncertainties should be seen as a conservative.  

For the first suite, we start with the present day errors (i.e. the fiducial analysis of this work), and sequentially turn off each individual source of error, leaving the other errors at their present day values. We dub this the `present day' suite. This results in 4 simulations where we (respectively) turn off the uncertainty in proper motion, distance, LMC, and MW potential. We show the present-day distance of each satellite compared to the relative reduction in orbital uncertainty when each source of uncertainty is fixed to zero (e.g., $\sigma_{r_{\rm peri,~fixed}}/\sigma_{r_{\rm peri}}$) in Figure~\ref{fig:current_errors}. The relative reduction in error on the pericenter and apocenter shows how much of the current error is due to the fixed quantity (i.e. if $\sigma_{r_{\rm peri},~{\rm fixed\,distance}}/\sigma_{r_{\rm peri}}$ is close to 0 most of the error on the pericenter is due to the distance uncertainty and if it is close to 1 the uncertainty is due to other properties). 
In general for our dSph sample, the error in the pericenter is dominated by either the distance or systemic proper motion uncertainty. Whereas for the apocenter, the error is dominated by the distance or potential uncertainties. We note that we choose to plot this relative reduction in error versus distance to give a sense of where in the Milky Way each uncertainty dominates.
We include the relative reduction in error for each dwarf in Table~\ref{tab:future_error}. We note that previous works have also explored the dominant source of uncertainty, but these have examined the uncertainty in the transverse velocity \citep{Battaglia2022A&A...657A..54B} instead of the orbital uncertainty as in this work. 

In the second suite, we consider a future version of the first suite where the proper motions and distances are improved to a level we expect in the next 
5 years. For the proper motions, we assume uncertainties based on 10 years of \textit{Gaia} data (DR5). For the distances, we assume a 2\% error which is an obtainable projection given current systematics in the period-Wesenheit-metallicity relations of RR Lyrae stars \citep[e.g.,][]{Nagarajan2021arXiv211106899N, Garofalo2022arXiv220307435G}. We leave the uncertainties in the Milky Way potential and LMC potential at their present-day uncertainties to assess whether these need to be improved to make use of  upcoming data. The  results are shown in Figure~\ref{fig:future_errors}.
Due to improvements in the systemic proper motion in future {\it Gaia} data releases, the dominant errors in the future will be due to distance and/or MW potential uncertainties. This motivates the need for more precise measurements of the Milky Way potential in order to make optimal use of {\it Gaia} DR5 data. {Measurements with stellar streams are excellent for this since they span a large range of radii and can disentangle the influence of the LMC \citep[e.g.][]{Kupper2015ApJ...803...80K,Erkal2019MNRAS.487.2685E,Vasiliev2021MNRAS.501.2279V}. Similarly, measurements based on distant tracers which include the effect of the LMC are also promising \citep[e.g.][]{Deason2021MNRAS.501.5964D,CorreaMagnus2022MNRAS.511.2610C}.} Interestingly, we also see that there are some dwarfs which have a substantial uncertainty ($\gtrsim 20\%$) in their orbital properties due to the LMC uncertainties. We note that dwarfs with a small number of members in the {\it Gaia} data will be dominated by proper motion measurements. 

For the final suite, we take a slightly different approach where we start with current proper motion errors and 5\% distance errors and (one at a time) improve these to projected 10-yr {\it Gaia} errors and 2\% distance errors. 
We only include present-day uncertainties in the Milky Way potential in this test as it is difficult to make projections for the future uncertainties.
The results of the orbital parameters are shown in Figure~\ref{fig:improved_errors}.
For both the pericenter and apocenter errors, the closer dwarfs are dominated by distance errors whereas the more distant dwarfs are dominated by proper motion errors. 
We note that there are some dwarfs that currently have distance errors that are less than 5\% and this exercise is done to compare how uniform improvement in distance and proper motion will affect different dSphs.

Overall, our analysis shows that while proper motions are currently one of the dominant sources of orbital uncertainty, once we have \textit{Gaia} DR5 data, the main uncertainties will come from the distance and Milky Way potential. This motivates the need of improving our measurement of the Milky Way potential so that we can make optimal use of the upcoming data sets, especially in the outskirts since it makes a significant contribution to the uncertainty in apocenters.

\section{Conclusion}
\label{section:conclusion}

We have presented a method to measure the systemic proper motion of MW satellites and applied it 54 MW dSphs. 
Our methodology builds on previous work utilizing mixture models to cleanly separate the MW foreground from the dSph stars and uses  {\it Gaia} astrometry combined with either DECam, {\it Gaia}, or Pan-Starrs photometry \citep{Pace2019ApJ...875...77P, McConnachie2020AJ....160..124M}.
Our primary results are:

\begin{itemize}

\item We have measured the systemic proper motion of 52 dSphs with two different background models and have identified likely members.   We publicly release  our membership catalogs to enable  spectroscopic follow-up analysis (see  Appendix~\ref{appendix:membership}).

\item Our systemic proper motion measurements are in excellent agreement with other EDR3 analysis \citep[e.g., ][]{McConnachie2020RNAAS...4..229M, Li2021ApJ...916....8L, Battaglia2022A&A...657A..54B}. We have compared our candidate dSph members to spectroscopic catalogs and found that  high probable  proper motion members are confirmed with velocity measurements. In addition, future spectroscopic measurements in the following dSphs can significantly expand spectroscopic samples by factors of 2-3: Aqu~II, Boo~II, Boo~III, Car~II, Dra~II, Hyd~II, Hyi~I, Ret~II, Tuc~II, and UMa~II. Larger spectroscopic samples can improve our knowledge of the dynamical properties of the ultra-faint dwarfs (Table~\ref{tab:future_targets}).

\item For the 46 dSphs with literature line-of-sight velocities, we have simulated their orbits in the Milky Way and Milky Way+LMC system. For 16 of the dSphs, we found that including the LMC changes the pericenter or apocenter by $>25\%$, showing that the LMC must be included for precise orbits (Figure~\ref{fig:orbit_differences}). 

\item We have compared the orbital information here to some previously used diagnostics for searching for tidal influences on dSphs.
Most directly we have compared the average dSph density to the average MW density at the dSph's pericenter (Figure~\ref{fig:density_pericenter}).  DSphs that are clearly undergoing tidal disruption (Ant~II,  Sgr, and Tuc~III) fall below the average MW density at their pericenter whereas other dSphs near  the average MW density (Boo~I, Boo~III, Cra~II, Gru~II, Seg~2, and Tuc~IV) are potentially tidally disrupting although future work is required to confirm this. 
We do not observe any trends between ellipticity, pericenter, and susceptibility to tidal disruption suggesting that not all large elongation is due to tidal disruption (Figure~\ref{fig:other_tidal}). 
At a fixed pericenter, the dSphs with the smallest values of the ratio between the average dSph density and the average MW density at the dSph's pericenter have larger half-light radii than the general dSph population.

\item We have explored alignments between the spatial orientation of a dSph and its orbital direction (via it reflex corrected proper motion) and there is an excess of dSphs aligned with their orbital motion (Figure~\ref{fig:angular_difference}). 
Moreover, the most elliptical dSphs ($\epsilon > 0.4$) are preferentially aligned with their orbital direction.  This may be evidence of large scale tidal torques on the dSph population from the MW.

\item We have identified six dSphs that were likely associated with the LMC:  Car~II, Car~III, Hor~I, Hyi~I, Phx~II, and Ret~II  (Figure~\ref{fig:lmc_rad_dist}). 
Our association results agree with previous orbital analysis with {\it Gaia} DR2 and updated EDR3 results \citep{Kallivayalil2018ApJ...867...19K,Erkal2020MNRAS.495.2554E, Patel2020ApJ...893..121P, Battaglia2022A&A...657A..54B, CorreaMagnus2022MNRAS.511.2610C}.

\item Our analysis does not suggest there is an excess of satellites near their orbital pericenter in contrast to some previous {\it Gaia} DR2 results   (Figure~\ref{fig:pericenter_excess}).  
We have examined the ratio $f_{\rm peri}$ and we have found  pileups  near $f_{\rm peri}\sim0, 1$ (i.e., pericenter and apocenter), similar to previous analyses.
We have applied the same analysis to subhalos of MW-like halos in $N$-body simulations and find  that the subhalo $f_{\rm peri}$ distribution agrees with the observed dSph distribution.
This agrees with other analyses directly examining the orbital phase \citep[e.g.,][]{CorreaMagnus2022MNRAS.511.2610C}

\item We have examined how the orbital uncertainties of these dwarfs are affected by the observational uncertainties for each dwarf as well as the uncertainties in the MW and LMC potentials. This allows us to determine which of these sources currently dominates the dSph population and what the largest source of error will be at the end of the {\it Gaia} mission (Figures~\ref{fig:current_errors}-\ref{fig:improved_errors}). 
In general, the current orbital pericenters are dominated by either distance and/or systemic proper motion errors  whereas the current orbital apocenters  are  dominated by the distance and/or potential uncertainty.  
In the future, both the orbital pericenters and apocenters will be dominated by distance and/or potential uncertainties except for dSphs with very small {\it Gaia} sample sizes.

\end{itemize}

The {\it Gaia} astrometric data sets (DR2, EDR3) have transformed our understanding of the orbital motion of the MW dSph population and enabled new analyses.
The study of proper motions and internal tangential kinematics of MW dSphs and more distant dSphs is promising with the future {\it Gaia} data releases and future space based astrometry (e.g., Nancy Grace Roman Space Telescope).

\acknowledgments
We thank Matt Walker and Sergey Kosopov for helpful  discussions.
{We thank the referee for their constructive report. We thank Josh Simon and Dmitry Makarov for their helpful comments.}
ABP is supported by NSF grant AST-1813881.
TSL acknowledges financial support from Natural Sciences and Engineering Research Council
of Canada (NSERC) through grant
RGPIN-2022-04794.

For the purpose of open access, the author has applied a Creative Commons Attribution (CC BY) licence to any Author Accepted Manuscript version arising from this submission.

This research has made use of the SIMBAD database, operated at CDS, Strasbourg, France \citep{Simbad2000A&AS..143....9W}.
This research has made use of NASA’s Astrophysics Data System Bibliographic Services.

This paper made use of the Whole Sky Database (wsdb) created by Sergey Koposov and maintained at the Institute of Astronomy, Cambridge by Sergey Koposov, Vasily Belokurov and Wyn Evans with financial support from the Science \& Technology Facilities Council (STFC) and the European Research Council (ERC).

This work has made use of data from the European Space Agency (ESA) mission
{\it Gaia} (\url{https://www.cosmos.esa.int/gaia}), processed by the {\it Gaia}
Data Processing and Analysis Consortium (DPAC,
\url{https://www.cosmos.esa.int/web/gaia/dpac/consortium}). Funding for the DPAC
has been provided by national institutions, in particular the institutions
participating in the {\it Gaia} Multilateral Agreement.

This project used public archival data from the Dark Energy Survey (DES). Funding for the DES Projects has been provided by the U.S. Department of Energy, the U.S. National Science Foundation, the Ministry of Science and Education of Spain, the Science and Technology FacilitiesCouncil of the United Kingdom, the Higher Education Funding Council for England, the National Center for Supercomputing Applications at the University of Illinois at Urbana-Champaign, the Kavli Institute of Cosmological Physics at the University of Chicago, the Center for Cosmology and Astro-Particle Physics at the Ohio State University, the Mitchell Institute for Fundamental Physics and Astronomy at Texas A\&M University, Financiadora de Estudos e Projetos, Funda{\c c}{\~a}o Carlos Chagas Filho de Amparo {\`a} Pesquisa do Estado do Rio de Janeiro, Conselho Nacional de Desenvolvimento Cient{\'i}fico e Tecnol{\'o}gico and the Minist{\'e}rio da Ci{\^e}ncia, Tecnologia e Inova{\c c}{\~a}o, the Deutsche Forschungsgemeinschaft, and the Collaborating Institutions in the Dark Energy Survey.
The Collaborating Institutions are Argonne National Laboratory, the University of California at Santa Cruz, the University of Cambridge, Centro de Investigaciones Energ{\'e}ticas, Medioambientales y Tecnol{\'o}gicas-Madrid, the University of Chicago, University College London, the DES-Brazil Consortium, the University of Edinburgh, the Eidgen{\"o}ssische Technische Hochschule (ETH) Z{\"u}rich,  Fermi National Accelerator Laboratory, the University of Illinois at Urbana-Champaign, the Institut de Ci{\`e}ncies de l'Espai (IEEC/CSIC), the Institut de F{\'i}sica d'Altes Energies, Lawrence Berkeley National Laboratory, the Ludwig-Maximilians Universit{\"a}t M{\"u}nchen and the associated Excellence Cluster Universe, the University of Michigan, the National Optical Astronomy Observatory, the University of Nottingham, The Ohio State University, the OzDES Membership Consortium, the University of Pennsylvania, the University of Portsmouth, SLAC National Accelerator Laboratory, Stanford University, the University of Sussex, and Texas A\&M University.
Based in part on observations at Cerro Tololo Inter-American Observatory, National Optical Astronomy Observatory, which is operated by the Association of Universities for Research in Astronomy (AURA) under a cooperative agreement with the National Science Foundation.

The Pan-STARRS1 Surveys (PS1) and the PS1 public science archive have been made possible through contributions by the Institute for Astronomy, the University of Hawaii, the Pan-STARRS Project Office, the Max-Planck Society and its participating institutes, the Max Planck Institute for Astronomy, Heidelberg and the Max Planck Institute for Extraterrestrial Physics, Garching, The Johns Hopkins University, Durham University, the University of Edinburgh, the Queen's University Belfast, the Harvard-Smithsonian Center for Astrophysics, the Las Cumbres Observatory Global Telescope Network Incorporated, the National Central University of Taiwan, the Space Telescope Science Institute, the National Aeronautics and Space Administration under Grant No. NNX08AR22G issued through the Planetary Science Division of the NASA Science Mission Directorate, the National Science Foundation Grant No. AST-1238877, the University of Maryland, Eotvos Lorand University (ELTE), the Los Alamos National Laboratory, and the Gordon and Betty Moore Foundation.

The Legacy Surveys consist of three individual and complementary projects: the Dark Energy Camera Legacy Survey (DECaLS; Proposal ID \#2014B-0404; PIs: David Schlegel and Arjun Dey), the Beijing-Arizona Sky Survey (BASS; NOAO Prop. ID \#2015A-0801; PIs: Zhou Xu and Xiaohui Fan), and the Mayall z-band Legacy Survey (MzLS; Prop. ID \#2016A-0453; PI: Arjun Dey). DECaLS, BASS and MzLS together include data obtained, respectively, at the Blanco telescope, Cerro Tololo Inter-American Observatory, NSF’s NOIRLab; the Bok telescope, Steward Observatory, University of Arizona; and the Mayall telescope, Kitt Peak National Observatory, NOIRLab. The Legacy Surveys project is honored to be permitted to conduct astronomical research on Iolkam Du’ag (Kitt Peak), a mountain with particular significance to the Tohono O’odham Nation.

NOIRLab is operated by the Association of Universities for Research in Astronomy (AURA) under a cooperative agreement with the National Science Foundation.

This project used data obtained with the Dark Energy Camera (DECam), which was constructed by the Dark Energy Survey (DES) collaboration. Funding for the DES Projects has been provided by the U.S. Department of Energy, the U.S. National Science Foundation, the Ministry of Science and Education of Spain, the Science and Technology Facilities Council of the United Kingdom, the Higher Education Funding Council for England, the National Center for Supercomputing Applications at the University of Illinois at Urbana-Champaign, the Kavli Institute of Cosmological Physics at the University of Chicago, Center for Cosmology and Astro-Particle Physics at the Ohio State University, the Mitchell Institute for Fundamental Physics and Astronomy at Texas A\&M University, Financiadora de Estudos e Projetos, Fundacao Carlos Chagas Filho de Amparo, Financiadora de Estudos e Projetos, Fundacao Carlos Chagas Filho de Amparo a Pesquisa do Estado do Rio de Janeiro, Conselho Nacional de Desenvolvimento Cientifico e Tecnologico and the Ministerio da Ciencia, Tecnologia e Inovacao, the Deutsche Forschungsgemeinschaft and the Collaborating Institutions in the Dark Energy Survey. The Collaborating Institutions are Argonne National Laboratory, the University of California at Santa Cruz, the University of Cambridge, Centro de Investigaciones Energeticas, Medioambientales y Tecnologicas-Madrid, the University of Chicago, University College London, the DES-Brazil Consortium, the University of Edinburgh, the Eidgenossische Technische Hochschule (ETH) Zurich, Fermi National Accelerator Laboratory, the University of Illinois at Urbana-Champaign, the Institut de Ciencies de l’Espai (IEEC/CSIC), the Institut de Fisica d’Altes Energies, Lawrence Berkeley National Laboratory, the Ludwig Maximilians Universitat Munchen and the associated Excellence Cluster Universe, the University of Michigan, NSF’s NOIRLab, the University of Nottingham, the Ohio State University, the University of Pennsylvania, the University of Portsmouth, SLAC National Accelerator Laboratory, Stanford University, the University of Sussex, and Texas A\&M University.

The Legacy Surveys imaging of the DESI footprint is supported by the Director, Office of Science, Office of High Energy Physics of the U.S. Department of Energy under Contract No. DE-AC02-05CH1123, by the National Energy Research Scientific Computing Center, a DOE Office of Science User Facility under the same contract; and by the U.S. National Science Foundation, Division of Astronomical Sciences under Contract No. AST-0950945 to NOAO.

\vspace{5mm}
\facilities{Gaia}

\begin{deluxetable*}{l | ccc | ccc }
\tablewidth{0pt}
\tablecaption{Dominant Error Source
\label{tab:future_error}
}
\tablehead{\colhead{Dwarf} & \colhead{$\sigma_{r_{\rm peri},~\mu}$} & \colhead{$\sigma_{r_{\rm peri},~d}$} & \colhead{$\sigma_{r_{\rm peri},~\Phi}$}
& \colhead{$\sigma_{r_{\rm apo},~\mu}$} & \colhead{$\sigma_{r_{\rm apo},~d}$} & \colhead{$\sigma_{r_{\rm apo},~\Phi}$}
}
\startdata
Ant II & 0.96 & {\bf 0.89} & 0.93 & 1.00 & {\bf 0.55} & 0.90 \\
Aqu II & {\bf 0.28} & 0.99 & 0.99 & {\bf 0.33} & 0.95 & 0.98 \\
Boo I & 0.98 & {\bf 0.52} & 0.94 & 1.00 & {\bf 0.52} & 0.85 \\
Boo II & 1.01 & {\bf 0.39} & 1.04 & 0.94 & {\bf 0.76} & 0.81 \\
Boo III & 0.98 & {\bf 0.37} & 0.97 & 1.00 & 0.81 & {\bf 0.62} \\
CVn I & {\bf 0.65} & 1.00 & 0.97 & {\bf 0.64} & 0.96 & 0.99 \\
CVn II & {\bf 0.30} & 0.97 & 1.01 & {\bf 0.30} & 0.89 & 0.88 \\
Car & 1.01 & {\bf 0.71} & 0.92 & 0.98 & {\bf 0.02} & 0.79 \\
Car II & 0.95 & {\bf 0.64} & 0.92 & 1.00 & 1.00 & {\bf 0.27} \\
Car III & 0.99 & {\bf 0.12} & 0.99 & 1.03 & 1.00 & {\bf 0.39} \\
Col I & {\bf 0.25} & 0.92 & 0.98 & {\bf 1.01} & 1.06 & 1.06 \\
CB & 0.99 & {\bf 0.11} & 0.99 & 0.94 & 0.75 & {\bf 0.71} \\
Cra II & {\bf 0.74} & 1.04 & 0.98 & 0.94 & 0.96 & {\bf 0.45} \\
Dra & 1.00 & {\bf 0.71} & 0.90 & 0.99 & 0.73 & {\bf 0.72} \\
Dra II & 0.91 & 0.83 & {\bf 0.79} & 0.92 & 0.94 & {\bf 0.39} \\
Eri II & {\bf 0.76} & 1.01 & 0.96 & 1.02 & 1.01 & {\bf 0.98} \\
For & 0.95 & {\bf 0.62} & 0.91 & 0.99 & {\bf 0.08} & 0.98 \\
Gru I & {\bf 0.41} & 1.00 & 0.99 & {\bf 0.57} & 0.92 & 0.74 \\
Gru II & 0.96 & {\bf 0.56} & 0.94 & 0.99 & {\bf 0.50} & 0.88 \\
Her & {\bf 0.62} & 0.93 & 0.99 & {\bf 0.82} & 0.98 & 0.88 \\
Hor I & 0.94 & {\bf 0.58} & 0.96 & 0.96 & {\bf 0.12} & 0.97 \\
Hyd II & {\bf 0.43} & 0.98 & 1.02 & {\bf 0.58} & 1.00 & 1.01 \\
Hyi I & 0.94 & 0.85 & {\bf 0.43} & 1.01 & 1.01 & {\bf 0.44} \\
Leo I & {\bf 0.89} & 1.01 & 0.96 & 0.96 & 0.98 & {\bf 0.64} \\
Leo II & {\bf 0.60} & 0.95 & 0.95 & 0.96 & {\bf 0.20} & 1.00 \\
Leo IV & {\bf 0.24} & 0.99 & 0.98 & {\bf 0.11} & 0.95 & 1.02 \\
Leo V & {\bf 0.18} & 1.06 & 1.08 & 1.05 & {\bf 0.99} & 1.02 \\
Peg III & {\bf 0.00} & 1.05 & 0.97 & {\bf 0.00} & 1.00 & 0.95 \\
Phx II & 0.98 & 1.10 & {\bf 0.78} & 1.01 & 0.82 & {\bf 0.78} \\
Psc II & {\bf 0.00} & 0.97 & 0.94 & {\bf 0.00} & 0.94 & 1.02 \\
Ret II & 0.98 & {\bf 0.36} & 0.91 & 1.01 & {\bf 0.59} & 0.73 \\
Ret III & {\bf 0.60} & 0.89 & 1.01 & {\bf 0.65} & 0.91 & 0.93 \\
Sgr II & 0.98 & 0.94 & {\bf 0.83} & 0.93 & {\bf 0.54} & 0.77 \\
Scl & 1.02 & 1.03 & {\bf 0.83} & 1.00 & 0.99 & {\bf 0.40} \\
Seg 1 & 1.00 & {\bf 0.14} & 0.98 & 0.99 & {\bf 0.29} & 0.89 \\
Seg 2 & 0.82 & {\bf 0.63} & 0.94 & 0.97 & {\bf 0.26} & 0.94 \\
Sxt & 0.98 & {\bf 0.66} & 0.87 & 0.98 & 0.95 & {\bf 0.53} \\
Tri II & {\bf 0.60} & 0.90 & 0.96 & 0.98 & 0.93 & {\bf 0.48} \\
Tuc II & 0.99 & {\bf 0.43} & 1.00 & 1.00 & {\bf 0.60} & 0.82 \\
Tuc III & 0.91 & {\bf 0.82} & 0.93 & 1.00 & 0.74 & {\bf 0.69} \\
Tuc IV & 0.98 & {\bf 0.43} & 0.97 & 0.98 & {\bf 0.25} & 0.96 \\
Tuc V & 0.75 & {\bf 0.68} & 0.99 & 0.83 & {\bf 0.47} & 0.99 \\
UMa I & 0.53 & {\bf 0.46} & 0.61 & 0.95 & {\bf 0.21} & 0.95 \\
UMa II & 0.98 & {\bf 0.32} & 0.96 & 1.03 & {\bf 0.52} & 0.83 \\
UMi & 1.02 & 0.89 & {\bf 0.84} & 1.01 & 0.82 & {\bf 0.66} \\
Wil 1 & {\bf 0.64} & 0.75 & 0.96 & 1.00 & {\bf 0.03} & 0.98 \\
\enddata
\end{deluxetable*}

\software{ \texttt{astropy} \citep{Astropy2013A&A...558A..33A, Astropy2018AJ....156..123A},
\texttt{matplotlib} \citep{matplotlib}, 
\texttt{NumPy} \citep{numpy},
\texttt{iPython} \citep{ipython},
\texttt{SciPy} \citep{2020SciPy-NMeth}
\texttt{corner.py} \citep{corner}, 
\texttt{emcee} \citep{ForemanMackey2013PASP..125..306F}  ,
\texttt{Q3C} \citep{2006ASPC..351..735K}
          }

\bibliography{main_bib_file}{}
\bibliographystyle{aasjournal}

\appendix

\section{Membership Catalogs}
\label{appendix:membership}

We provide catalogs of our membership along with select {\it Gaia} EDR3 columns {on Zenodo under a Creative Commons Attribution license: \dataset[10.5281/zenodo.6533295]{https://doi.org/10.5281/zenodo.6533295}}. We further include a diagnostic plot (similar to Figure~\ref{fig:example_diagnostic}) and a plot comparing our systemic proper motion measurement to literature values for each dSph and a machine readable compilation  of Tables~\ref{tab:overview}-\ref{tab:future_error}.

\section{Comments on Individual Dwarf Spheroidal Galaxies and Special Cases}
\label{appendix:comments}

\noindent{\bf Ant~II}---Due to the low surface density of Ant~II and the higher MW foreground due to the lower galactic latitude we only analyze the ``clean'' sample and use a magnitude limit, $G_{\rm max}=20$, that is much higher than suggested from the depth of astrometric solutions in the Ant~II region ($G_{\rm max}=20.85$).  Accurate photometry will assist with improving foreground separation in future measurements. 
Regardless our EDR3 measurement is consistent with other measurements \citep{McConnachie2020RNAAS...4..229M, Battaglia2022A&A...657A..54B} and spectroscopic based measurements \citep{Ji2021ApJ...921...32J}.
This dSph is included in our clear tidally disrupting sample based on the small pericenter and velocity gradient \citep{Torrealba2019MNRAS.488.2743T, Ji2021ApJ...921...32J}. 
While this analysis was in preparation a new analysis measuring the distance to Ant~II with RRL stars which slightly improved and updated the distance measurement \citep{Vivas2022ApJ...926...78V}.

\noindent{\bf Boo~I}---This dSph is included in our potentially tidally disrupting sample due to literature analysis  \citep{Longeard2021arXiv210710849L, Filion2021ApJ...923..218F} and a low value of $\rho_{1/2}/\rho_{MW}(r=r_{\rm peri})\sim 25$.  {We exclude the region around Boo~II ($R_{\rm Boo~II}=5\times r_{h,~Boo~II}\sim 15\arcmin$) when constructing the fixed background model.}

\noindent { {\bf Boo~II}---We exclude a region around  Boo~I ($R_{\rm Boo~I}=60\arcmin\sim6\times r_{h,~Boo~I}$) when constructing the fixed background model.}

\noindent{\bf Boo~III}---This dSph is included in our potentially  tidally disrupting based on its small pericenter and low average density  \citep{Carlin2018ApJ...865....7C}. We do not include it in our clear tidally disrupting sample because of the lack of detailed tidal stripping models or clear observational evidence (e.g., deep photometry or kinematic evidence).
There are a large number of candidate targets ($N_{\rm expected} \sim130$) that will significantly increase the spectroscopic sample size and can be used to further assess dynamical equilibrium. However, the large angular size and low luminosity makes follow-up difficult.  
{We exclude the region around the globular clusters, NGC~5466 ($R_{\rm NGC~5466}=7\times r_{h,~NGC~5466}= 16.1\arcmin$) and NGC~5272 ($R_{\rm NGC~5272}=30\arcmin \sim 13 \times r_{h,~NGC~5272}$), when constructing the fixed background model.}

\noindent{\bf Boo~IV}---A systemic proper motion measurement of this recently discovered dwarf has only been possible with EDR3. Our measurement disagrees with \citet{McConnachie2020RNAAS...4..229M} due to their choice of prior on the tangential velocity.  

\noindent {\bf Car}---
While other works have considered Car as a potential LMC satellite \citep{Pardy2020MNRAS.492.1543P}, we do not favor this scenario as it was closer to the MW at its most recent LMC pericenter.
There is larger background contamination around Car than other bright satellites, due to its low relative Galactic latitude and the LMC foreground stars.

\noindent{\bf Car~II and Car~III}---Because of the small angular separation between Car~II and Car~III  we model these two dwarfs simultaneously.  In particular, the angular separation, $\sim18\arcmin$, is roughly $2\times r_{h,~Car~II}$ and spatial overlap of members of the two dSphs is possible.
There is DECam g, r coverage in the NSC catalog, however, we find there is a large color difference ($(g-r)_0\sim0.075$) between spectroscopic members \citep{Li2018ApJ...857..145L} and a old-metal-poor isochrone. In addition,  a direct comparison to the reported magnitudes in the \citet{Li2018ApJ...857..145L} catalog to the NSC catalog finds a similar offset. This suggests that there are potential calibration issues in  this region. 
When we apply our standard g, r isochrone filter and include a color offset, we find that some  Car~II and Car~III spectroscopic members are not included.  Due to this issue, we opt to  use {\it Gaia} photometry in this region. 

We include a second dSph term in the likelihood to represent the Car~III population and we modify the prior volume on $\mu_{\alpha \star}$ and $\mu_{\delta}$ due to the large MW and LMC background.
With the addition of the Car~III component we are able to determine the systemic proper motion of both UFDs simultaneously for the fixed MW proper motion model.
When we model the MW proper motion with a Gaussian model, the Car~III component also  models the MW foreground. In particular, the number of Car~III `members' is much larger than expected, the proper motion completely disagrees with known spectroscopic members, and the $r_h$ spatial parameter takes on the largest possible value in the prior distribution to mimic a flat spatial distribution.  This attests to the complexity of the MW and LMC background model in this region.
Due to this issue, we only include the fixed background model results. 
Both Car~II and Car~III are highly likely to associated with the LMC in our analysis and  agrees with previous work \citep[e.g.,][]{Erkal2020MNRAS.495.2554E, Battaglia2022A&A...657A..54B}.

\noindent {\bf Cet~II}---There was not a signal observed in {\it Gaia} DR2 \citep{Pace2019ApJ...875...77P, McConnachie2020AJ....160..124M} but there is a clear signal observed in {\it Gaia} EDR3.  

\noindent { {\bf  Cet~III}---We exclude a region around the globular cluster Whiting~1 ($R_{\rm Whi~1}=5\times r_{h,~Whi~1}=1.1 \arcmin$) when constructing the fixed background model.}

\noindent {\bf Cra~II}---While there is not clear photometric (tidal tails) or kinematic evidence (e.g., gradients) \citep{Ji2021ApJ...921...32J} the small pericenter, large size, and small velocity dispersion suggests that Cra~II is undergoing tidal disruption \citep{Sanders2018MNRAS.478.3879S, Borukhovetskaya2022MNRAS.512.5247B} and we classify it as potentially tidally disrupting.

\noindent{\bf Dra~II}---One of the few objects in the sample that may be faint star cluster and not a dwarf galaxy. \citet{Baumgardt2022MNRAS.510.3531B} concluded that Dra~II is a star cluster  based on  evidence of mass segregation. 
There is not a resolved velocity dispersion or metallicity dispersion that would indicate there is a dark matter halo \citep{Longeard2018MNRAS.480.2609L}. 
There are potential of tidal features in smoothed stellar density distribution  \citep{Longeard2018MNRAS.480.2609L}. Due to the unresolved velocity dispersion and small size the density upper limit is quite large. Dra~II has a relatively small pericenter and if  the velocity dispersion was low ($\sim 1 ~\kms$) the satellite would likely be undergoing tidal disruption.   

\noindent{\bf Eri~II}---The photomtric selection window was increased for Eri~II due to its larger stellar mass relative to the other UFDs. Without this increase some spectroscopic members would be  excluded.  In addition, there is one spectroscopic member that is missing DECam photometry due to a nearby bright star that is manually included in the model. 
Only $\sim15\%$ of the orbital samples are bound which  have  $r_{\rm peri}\sim 100\kpc$ ($r_{\rm peri,~nL}\sim 200\kpc$) while the reminder of the chain is unbound. 

\noindent{\bf For}---
The proper motion is anti-parallel to the solar motion and the orbit is more sensitive to the distance uncertainty than other dwarfs \citep{Borukhovetskaya2022MNRAS.509.5330B}. 
The orbital pole aligns with the LMC and this has been used as an argument for For to be associated with the LMC \citep{Pardy2020MNRAS.492.1543P}.  We do not find a potential association with the LMC and note that For was outside the LMC escape velocity at its most recent closest approach.

\noindent{\bf Gru~II}---There is a bright star ($G\sim1.7$) near Gru~II. In DES DR2, a large portion of the region around this star is masked.  We opted to instead use DES DR1 \citep{DES2018ApJS..239...18A_DR1} as the mask is smaller. It is possible that sources near the bright star have biased photometry/colors due to presence of the bright star.

Gru~II  spatially overlaps with the Chenab/Orphan stream and it has been suggested to be connected to the stream \citep{Koposov2019MNRAS.485.4726K}.  
Gru~II and the Chenab/Orphan stream  are found to have the same proper motion \citep{Shipp2019ApJ...885....3S} but the radial velocities of Gru~II is $\sim90 \kms$ offset from the Chenab/Orphan stream predictions \citep{Simon2020ApJ...892..137S} and Gru~II is $\sim10~\kpc$ more distant than the stream \citep{MartinezVazquez2019MNRAS.490.2183M}. 
There will  be overlap between the two structures  in the color-magnitude diagram and both the signal and background region of Gru~II will be contaminated with the Chenab/Orphan stream members that have similar a proper motion to the  Gru~II proper motion.  Regardless we are able to successfully identify all known spectroscopic members.
The candidate members from our mixture model likely include some Chenab/Orphan stream members and the spectroscopic success rate might be lower than expected. 

This is one of the six dSphs that we have classified as likely tidally disrupted based on its low average density relative to the MW at pericenter ($\rho_{1/2}/\rho_{MW}(r=r_{\rm peri}) \lesssim 10$). This agrees with \citet{Simon2020ApJ...892..137S}, which found that the tidal radius was just larger than the Gru~II's physical size. As there is overlap in spatial and proper motion position but not radial velocity, searching for potential tidal tails and other signs of tidal disruption will be challenging. 

We do not find an association with the LMC but note that it is one of the closest MW satellites in phase space.  Other studies have considered Gru~II recently captured by the LMC \citep[e.g.,][]{Battaglia2022A&A...657A..54B}. 
We consider Gru~II as potentially tidally disrupting based on its low value of $\rho_{1/2}/\rho_{MW}(r=r_{\rm peri})$. 

\noindent{\bf Her}---
The  large elongation of Her has long been used as evidence for tidal disruption in her \citep[e.g., ][]{Martin2010ApJ...721.1333M, Kupper2017ApJ...834..112K, Fu2019ApJ...883...11F}. 
Extra-tidal photometric overdensities have been identified \citep{Sand2009ApJ...704..898S, Roderick2015ApJ...804..134R}, however, follow-up efforts have been unsuccessful \citep{Fu2019ApJ...883...11F, MutluPakdil2020ApJ...902..106M}. 
The large pericenter, $r_{\rm peri}\sim60~\kpc$ we find with the LMC+MW orbit modeling suggests that the tidal shocking at pericenter is small however, we note that the orbital motion (via the reflex corrected proper motion) is aligned with the Her major axis ($| \theta_{\mu}-\theta_{\rm xy} |\sim5\degree$)  

\noindent{\bf Hor~I}---Likely LMC satellite.  {We exclude the region around the globular cluster NGC~1261 ($R_{\rm NGC~1261}=5\times r_{h,~NGC~1261}\sim 3.4\arcmin$) when constructing the fixed background model.}

\noindent{\bf Hor~II}---Due to the low number of members and the lack of {\it Gaia} matches more observations are needed. Of the 3 spectroscopic members \citep{Fritz2019A&A...623A.129F} with astrometric solutions in DR2 only 2 have astrometric solutions in EDR3. We note that the same four member stars were identified in \citet{Pace2019ApJ...875...77P} and in EDR3 the strength of the signal has increased.

\noindent{\bf Hyd~II}---The orbital pole of Hyd~II aligns with the LMC  \citep{Kallivayalil2018ApJ...867...19K} but we do not find a large probability ($p_{\rm LMC}\sim 6\%$) for them to be associated.  At the most recenter minimum LMC distance ($r_{\rm LMC,~min}\sim125~\kpc$) Hyd~II has a large relative velocity and is outside the LMC escape velocity (($v_{\rm LMC,~min}\sim250~\kms$)) and unlikely to be associated.

\noindent{\bf Hyi~I}---For the photometry of Hyi~I we use the NSC catalog. Similar to Car~II and Car~III we find that there is a color offset between spectroscopic members and stellar isochrones. The offset is smaller than in Car~II, $(g-r)_0\sim0.05$, and we find that after applying this offset to the photometry our standard isochrone selection is able to select all known spectroscopic members.  Hyi~I is a likely LMC satellite.

\noindent{\bf Leo~V}---In the clean sample, our selection remove all MW foreground members and only known Leo~V spectroscopic members are left.  This is not the case for the complete sample. This is the only dSph for which we do not find new candidate members in.

\noindent{\bf Peg~III}---All known spectroscopic members \citep{Kim2016ApJ...833...16K} are below the  {\it Gaia} magnitude limit. 
While this analysis was in the later stages of preparation, there was a new analysis of the structural parameters and stellar distribution of Peg~III with with deep {\it HST} photometry  \citep{Richstein2022arXiv220401917R}. In particular, they found the center of Peg~II shifted by $\sim0.5\arcmin$ and the  half-light radius roughly doubled. With these updated structural parameters we are able to make a  detection of the systemic proper motion of Peg~III. In our previous analysis with the \citet{Kim2016ApJ...833...16K} structural parameters, there were 4 candidate members with $p>0.1$ but all had large errors such that $\sum p = 3.0_{-3.0}^{+0.9}$. With the updated parameters the same stars are identified as members with $\sum p = 3.9_{-0.4}^{+0.3}$.  The  candidate members are {located} $R\sim1-2~r_h$   whereas with {the} previous literature values they were $R\sim2-4~r_h$. There is no detection of Peg~III in \citet{Battaglia2022A&A...657A..54B} and the small errors in \citet{McConnachie2020RNAAS...4..229M} are likely due to their prior that the satellite be bound to the MW. 
Our measurement is just past the threshold for detection based on the small number of members. 
This highlights the next for accurate structural parameters to identify candidates dSph members without spectroscopic information.

\noindent{\bf Phx~II}---Likely LMC satellite.

\noindent{\bf Pic~II}---Similar to the other two NSC dSphs, we find a color offset for Pic~II.  This has the largest offset of the three NSC dSphs, $(g-r)_0\sim0.1$. We have modeled this object with both {\it Gaia} and NSC photometry and found that there is little difference between them.  We opt to use the NSC photometry as we are able to exclude more background stars.  
When the structural  parameters are varied, the half-light radius and ellipticity are found to be larger and more elongated than the literature values and the overall  signal is weak  at best.  The overall UFD membership from the our standard  analysis is $\sum p \sim 8 \pm 8$ and all identified stars contain large errors. 
To  determine a more confident Pic~II signal, we fixed the spatial parameters to the best fit literature values.   
When the structural parameters are fixed we find $\sum p \sim 6.1_{-1.5}^{+3.5}$ and the systemic proper motion has asymmetric error bars. 
We suspect this issues are caused by the large LMC+MW background and the overlap between the Pic~II and LMC proper motion. Other EDR3 work have not had this issue \citep{McConnachie2020RNAAS...4..229M, Battaglia2022A&A...657A..54B} and a color-magnitude term in the likelihood may address this issue. Future spectroscopic observations and membership will  assist in measuring the systemic proper motion. 

\noindent {\bf Psc~II}---Our mixture models did not return a confident signal in Psc~II, however, there are known spectroscopic members identified in the {\it Gaia} catalog \citep{Fritz2018A&A...619A.103F}. 
Two of the known spectroscopic stars were identified as members but they were assigned enormous errors ($\sim0.99, 0.5$).
With {\it Gaia} EDR3, there is one more spectroscopic member that has an astrometric solution compared to two in the {\it Gaia} DR2 catalog from the \citet{Kirby2015ApJ...810...56K} spectroscopic sample. 

As an alternative model, we rerun the mixture model but fix the three  spectroscopic members in the dSph component.  With this change, a proper motion signal is measured but no new members are identified. 
As the mixture model does not increase the number of members, we opt to calculate the systemic proper motion from the three known spectra members. 
Our result disagrees with \citet{McConnachie2020RNAAS...4..229M} likely due to their prior on tangential velocity that assumes the satellite is bound.

\noindent {\bf Ret~II}---This is one of the only objects with a well measured proper motions where our results differ from a result in the literature \citep{MartinezGarcia2021MNRAS.505.5884M}.
Ret~II is a likely LMC satellite. 

\noindent {{\bf Sxt}---We exclude the region around the globular cluster Palomar~3 ($R_{\rm Pal~3}=5\times r_{h,~Pal~3}= 3.25\arcmin$) and the local field galaxy Sextans~A ($R_{\rm Sxt~A}=3 \times r_{h,~Sxt~A}\sim 7.4\arcmin $), when constructing the fixed background model.}

\noindent {\bf Sgr}---Sgr is clearly tidally disrupting and has a prominant stellar stream \citep{Vasiliev2021MNRAS.501.2279V}.  Sgr was not included in our primary sample due to its large angular size.  We include the parameters we use for this analysis for reference: $r_{\rm peri}=16 ~\kpc$ \citep{Vasiliev2021MNRAS.501.2279V}, $r_{1/2}=2500~{\rm pc}$, $\epsilon=0.64$, and $d=26.5~\kpc$  \citep{McConnachie2012AJ....144....4M}. 

\noindent {\bf Sgr~II}---The second system in our sample that is likely to be a globular cluster \citep{Longeard2021MNRAS.503.2754L, Baumgardt2022MNRAS.510.3531B}.

\noindent {\bf Tuc~III}---We consider this dSph to be clearly tidally disrupting based on its tidal tails \citep{DrlicaWagner2015ApJ...813..109D} and a velocity gradient \citep{Li2018ApJ...866...22L}.

\noindent {\bf Tuc~IV}---Tuc~IV is above the LMC escape velocity at its closest approach to LMC and is unlikely to be LMC satellite. It is currently within the LMC escape velocity and may have been recently captured by the LMC. 
It had a close encounter with the LMC \citep[see also ][]{Simon2020ApJ...892..137S}. 

\noindent {\bf Tuc~V}---There was not a signal observed in similar {\it Gaia} DR2 \citep{Pace2019ApJ...875...77P} but  a clear signal observed in {\it Gaia}  EDR3. {We are able to identify a signal without using of spectroscopic membership as prior information but note that the systemic proper motion can be measured more precisely including this information \citep[e.g.,][]{Battaglia2022A&A...657A..54B}. }

\noindent {\bf Wil~1}---Roughly half of the stars (9/16) identified as candidate members in Table~2 of \citet{Willman2011AJ....142..128W} with matches to the {\it Gaia} EDR3 catalog are clearly MW foreground stars based on their parallax and/or proper motions. It is clear that any future kinematic analysis should include {\it Gaia} astrometry to improve the remove of foreground contamination.



\section{Extra Figures}

\begin{figure*}[b]
\includegraphics[height=4.5cm]{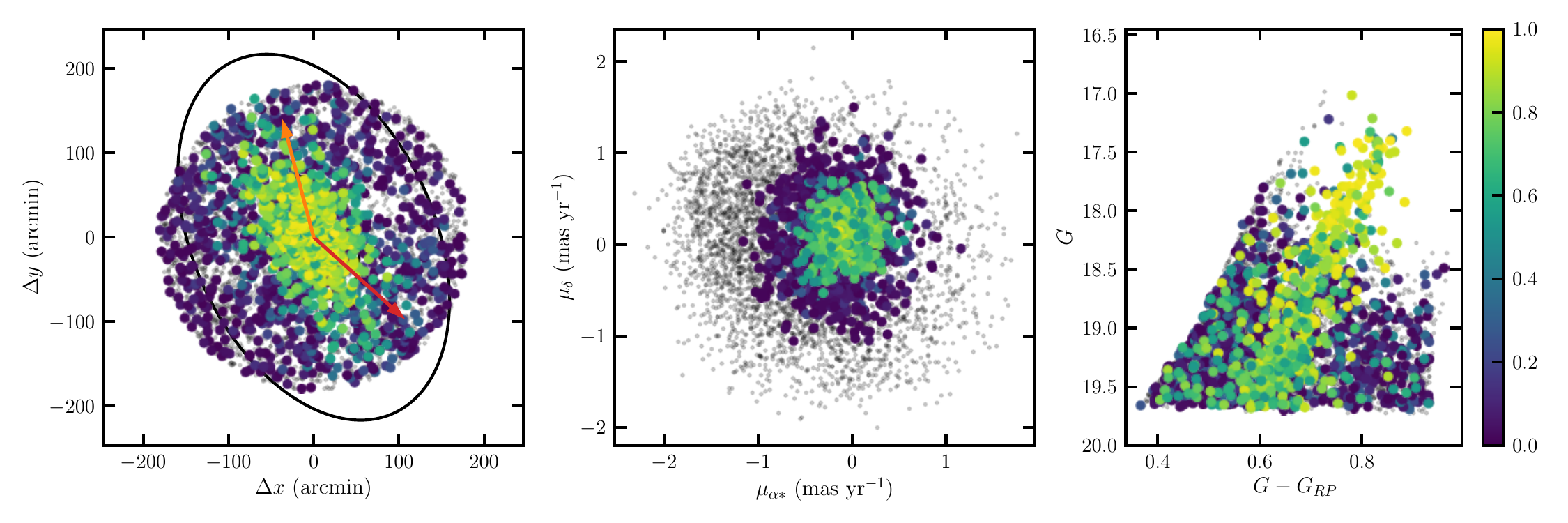}
\caption{Diagnostic plot for Antlia~II similar to Figure~\ref{fig:example_diagnostic}.  The three rows are the spatial distribution (tangent plane), the proper motion (vector point diagram),  and a {\it Gaia} color-magnitude diagram. Points with membership probability $p > 0.01$ are colored according to their probability; the rest are considered as MW stars and are shown as grey points. The red arrow points toward the Galactic center and the orange arrow is the direction of the reflex-corrected proper motion which is approximately equal to the orbital motion. 
}
\label{fig:diagnostic_ant2}
\end{figure*}

\begin{figure*}[b]
\includegraphics[height=4.5cm]{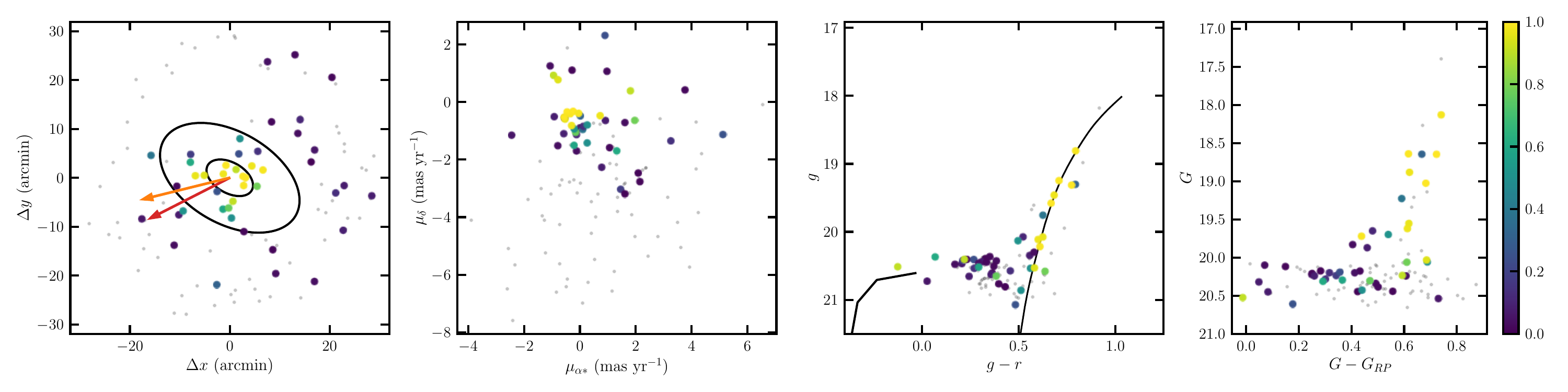}
\caption{Similar to Figure~\ref{fig:diagnostic_ant2} but for Aquarius~II.  
The center-right panel is a DECam color-magnitude diagram. 
}
\label{fig:diagnostic_aqu2}
\end{figure*}

\begin{figure*}[b]
\includegraphics[height=4.5cm]{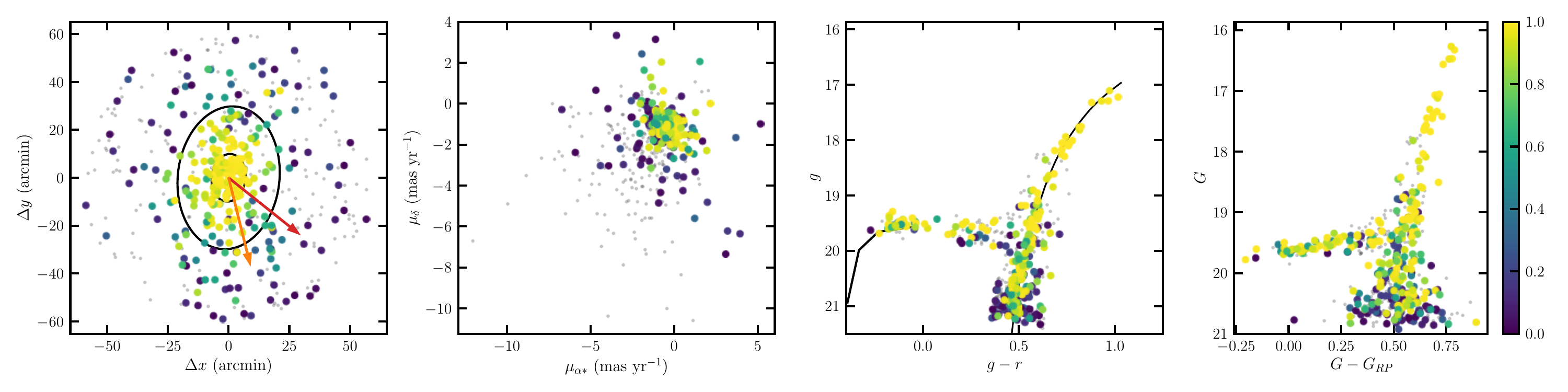}
\caption{Same as Figure~\ref{fig:diagnostic_aqu2} but for Bo\"{o}tes~I.  
}
\end{figure*}


\begin{figure*}
\includegraphics[height=4.5cm]{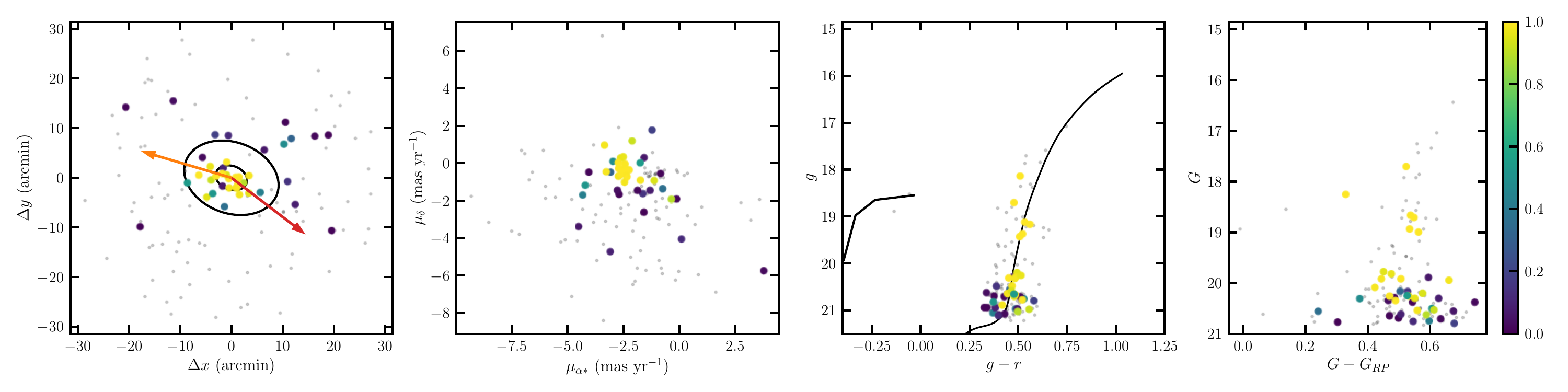}
\caption{Same as Figure~\ref{fig:diagnostic_aqu2} but for Bo\"{o}tes~II.  
}
\end{figure*}

\begin{figure*}
\includegraphics[height=4.5cm]{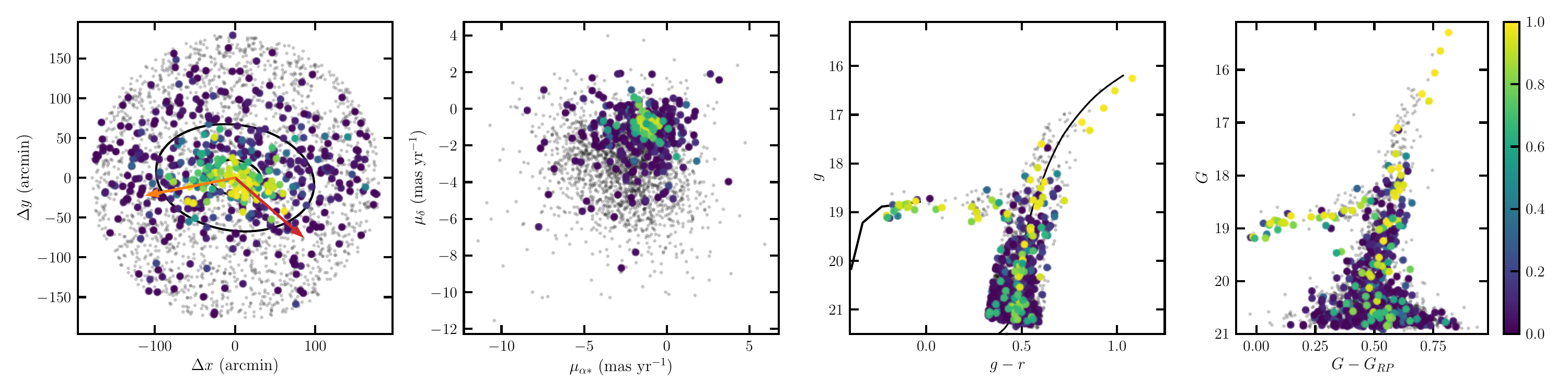}
\caption{Same as Figure~\ref{fig:diagnostic_aqu2} but for Bo\"{o}tes~III. 
}
\end{figure*}

\begin{figure*}
\includegraphics[height=4.5cm]{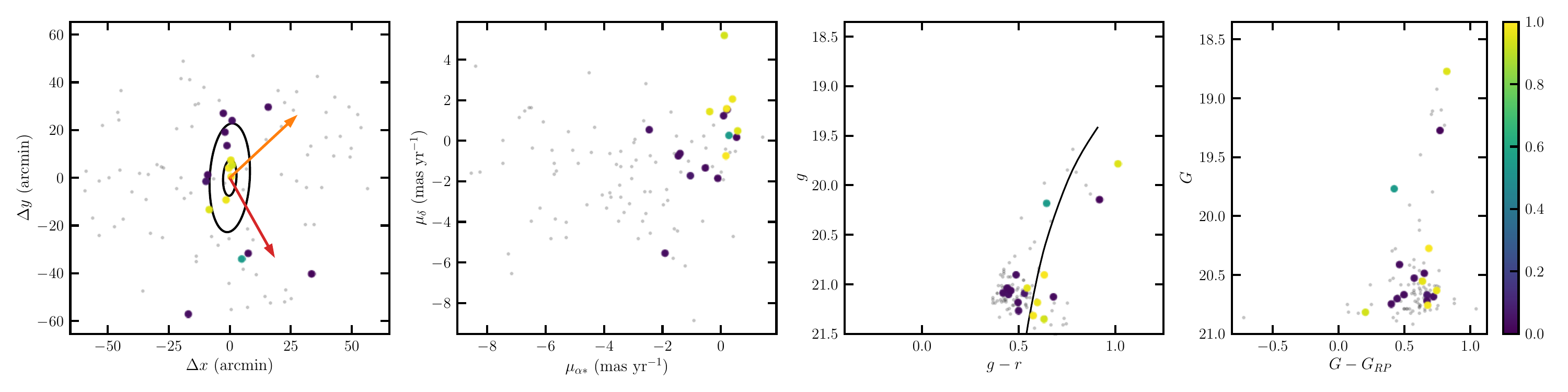}
\caption{Same as Figure~\ref{fig:diagnostic_aqu2} but for Bo\"{o}tes~IV.  
Includes PS1 photometry instead of DECam. 
}
\end{figure*}

\begin{figure*}
\includegraphics[height=4.5cm]{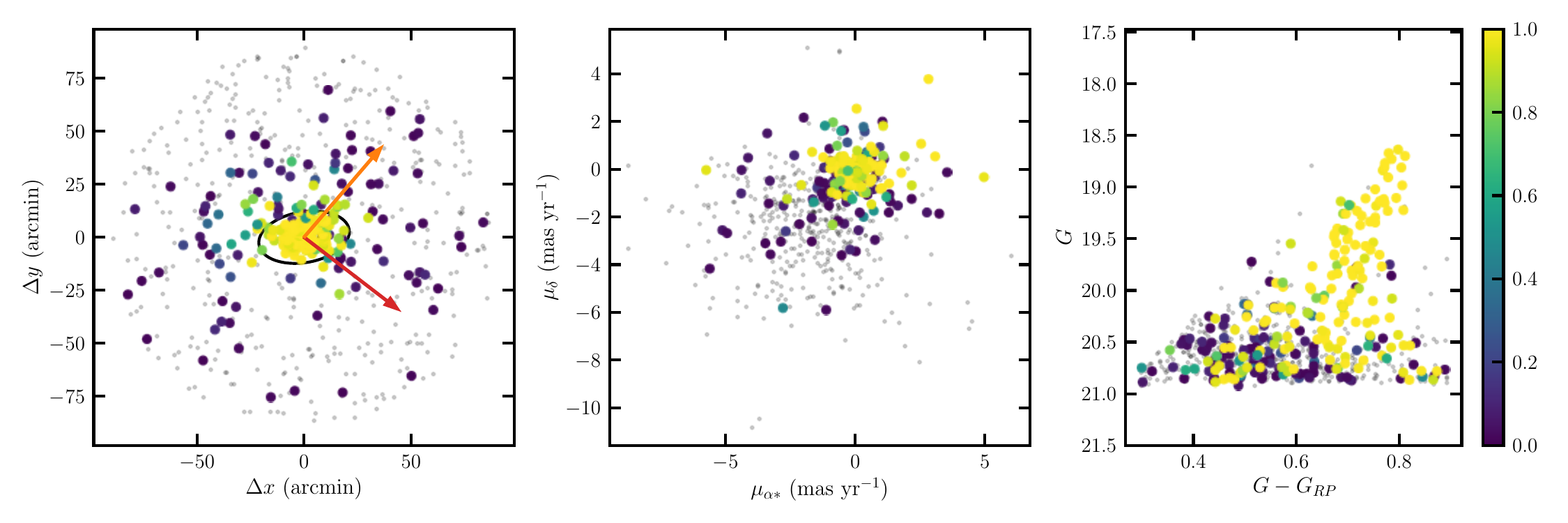}
\caption{Same as Figure~\ref{fig:diagnostic_ant2} but for Canes Venatici~I.  
}
\end{figure*}


\begin{figure*}
\includegraphics[height=4.5cm]{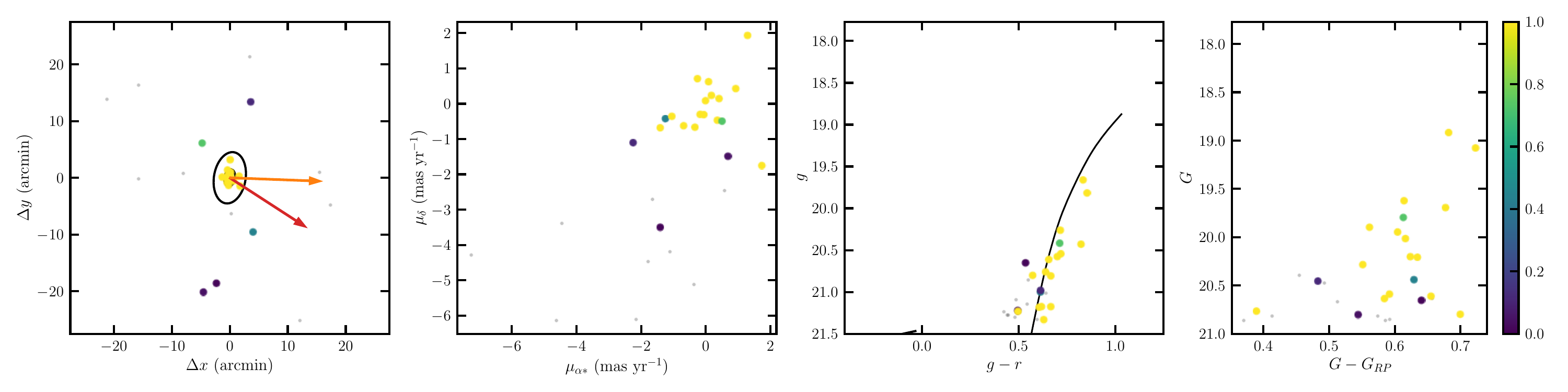}
\caption{Same as Figure~\ref{fig:diagnostic_aqu2} but for Canes Venatici~II.  
}
\end{figure*}

\begin{figure*}
\includegraphics[height=4.5cm]{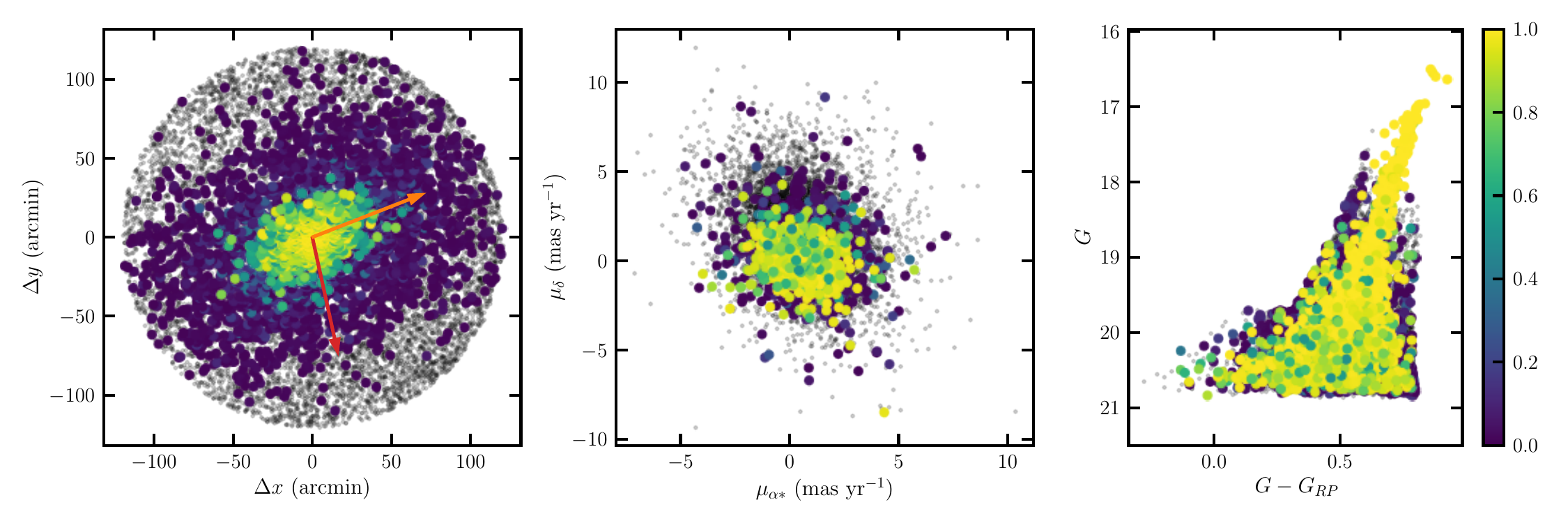}
\caption{Same as Figure~\ref{fig:diagnostic_ant2} but for  Carina.  
}
\end{figure*}

\begin{figure*}
\includegraphics[height=4.5cm]{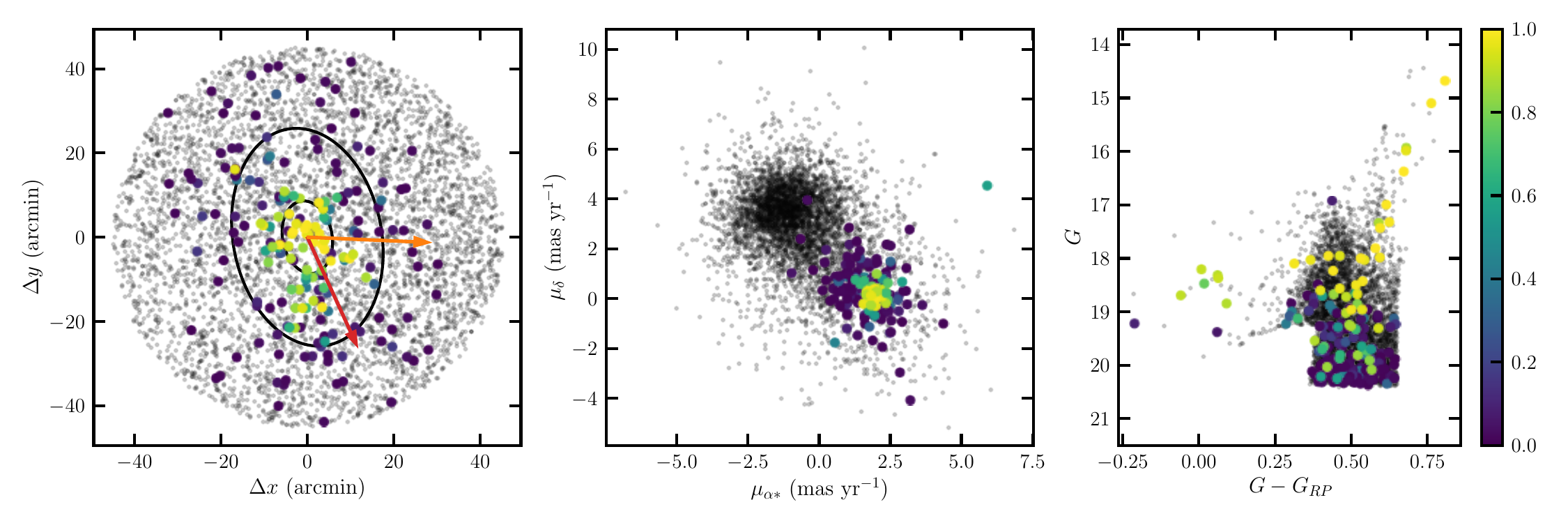}
\caption{Same as Figure~\ref{fig:diagnostic_ant2} but for Carina II.
The region overlaps with Carina~III.
}
\end{figure*}

\begin{figure*}
\includegraphics[height=4.5cm]{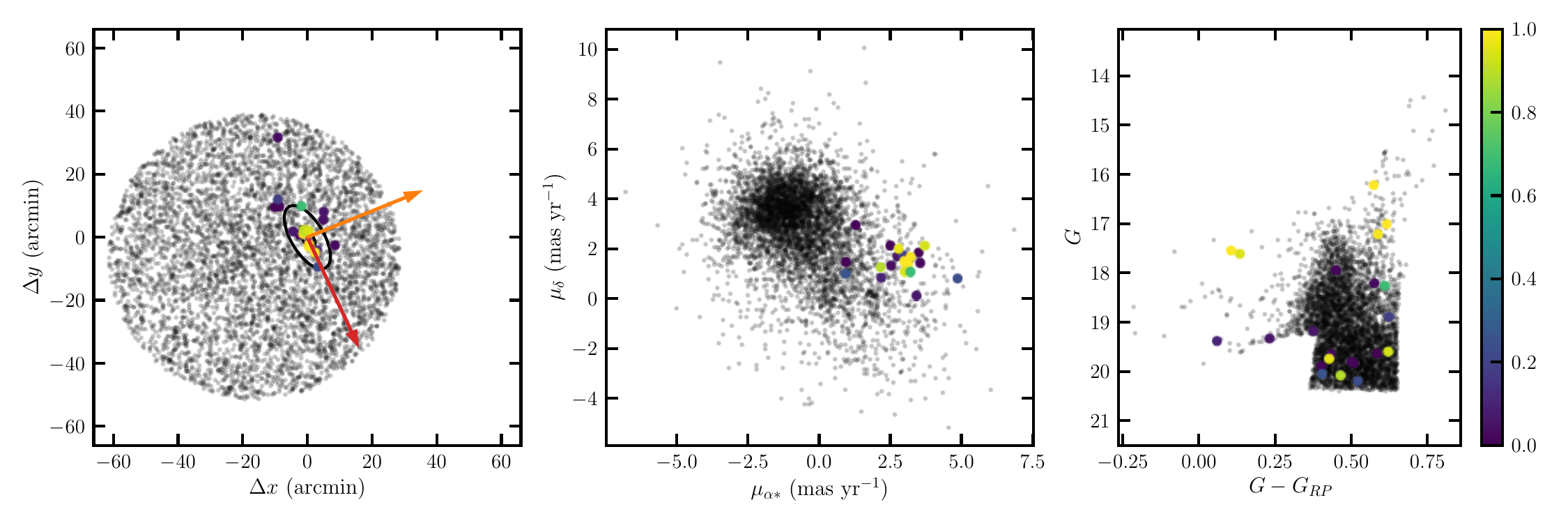}
\caption{Same as Figure~\ref{fig:diagnostic_ant2} but for Carina III.  
The area is center on Carina~II. 
}
\end{figure*}

\begin{figure*}
\includegraphics[height=4.5cm]{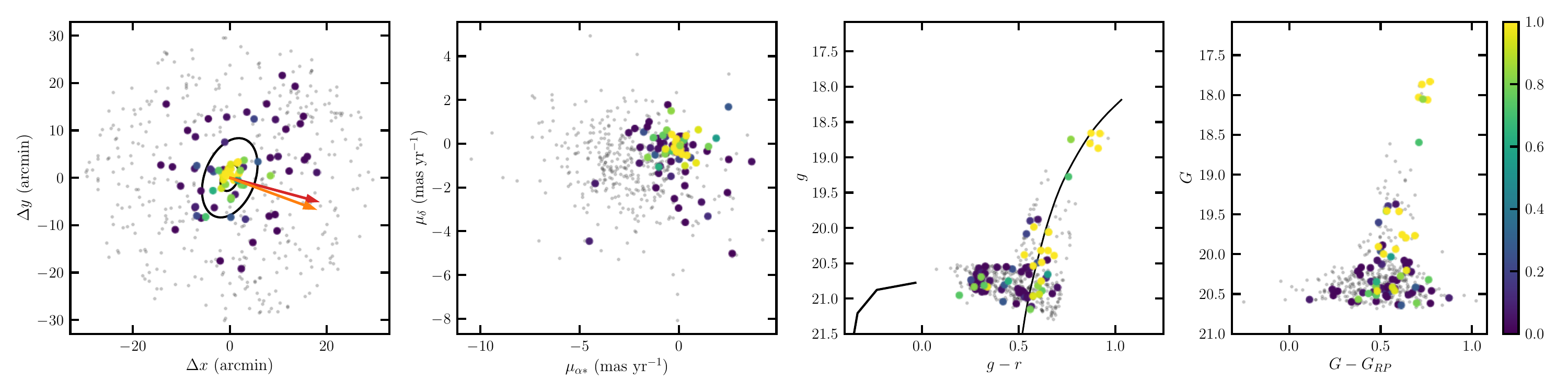}
\caption{Same as Figure~\ref{fig:diagnostic_aqu2} but for Centaurus I.  
}
\end{figure*}

\begin{figure*}
\includegraphics[height=4.5cm]{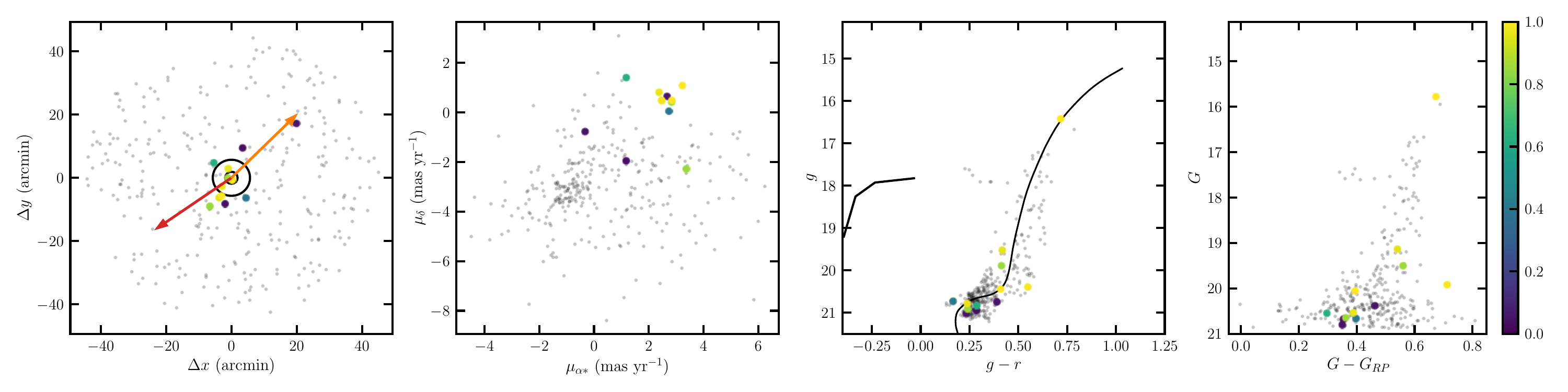}
\caption{Same as Figure~\ref{fig:diagnostic_aqu2} but for Cetus II.  
}
\end{figure*}

\begin{figure*}
\includegraphics[height=4.5cm]{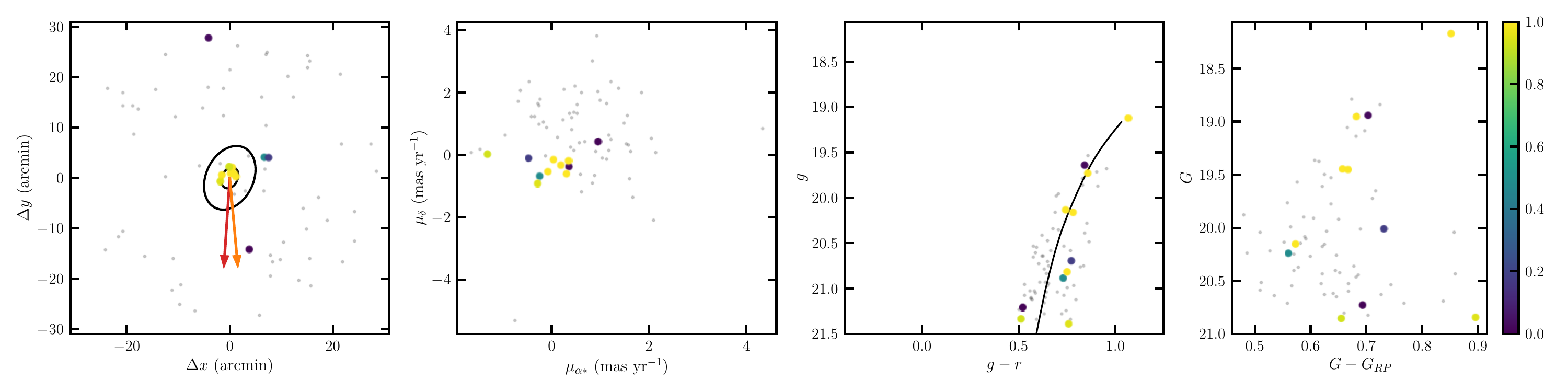}
\caption{Same as Figure~\ref{fig:diagnostic_aqu2} but for Columba I.  
}
\end{figure*}

\begin{figure*}
\includegraphics[height=4.5cm]{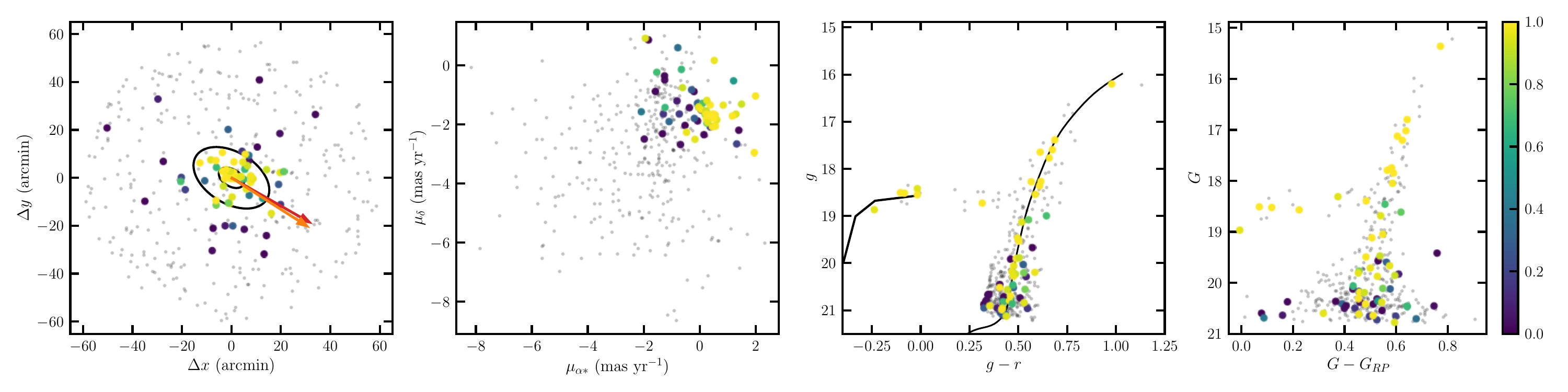}
\caption{Same as Figure~\ref{fig:diagnostic_aqu2} but for Coma Berenices.  
}
\end{figure*}

\begin{figure*}
\includegraphics[height=4.5cm]{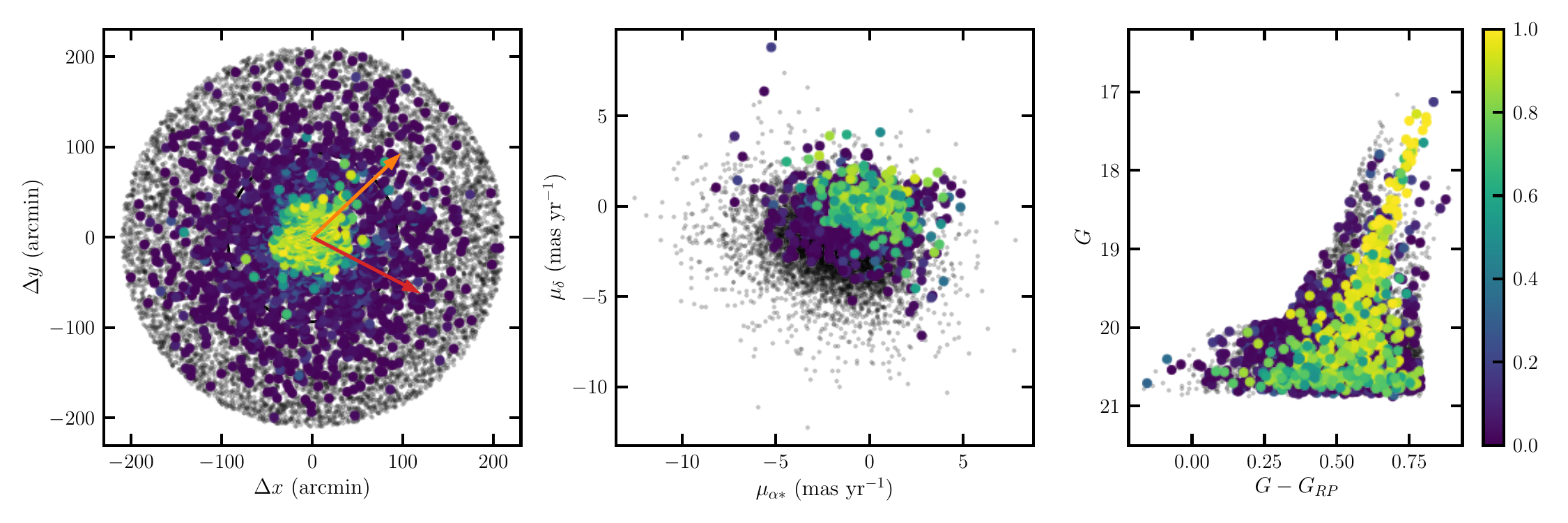}
\caption{Same as Figure~\ref{fig:diagnostic_ant2} but for Crater II.  
}
\end{figure*}

\begin{figure*}
\includegraphics[height=4.5cm]{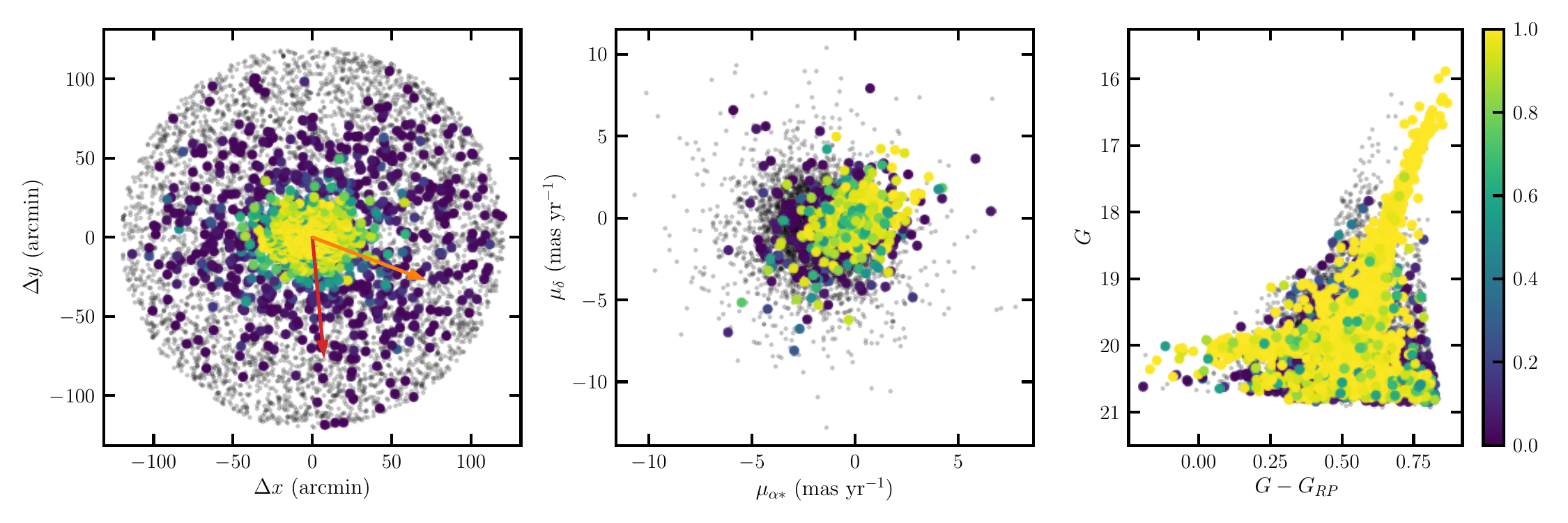}
\caption{Same as Figure~\ref{fig:diagnostic_ant2} but for  Draco.
}
\end{figure*}

\begin{figure*}
\includegraphics[height=4.5cm]{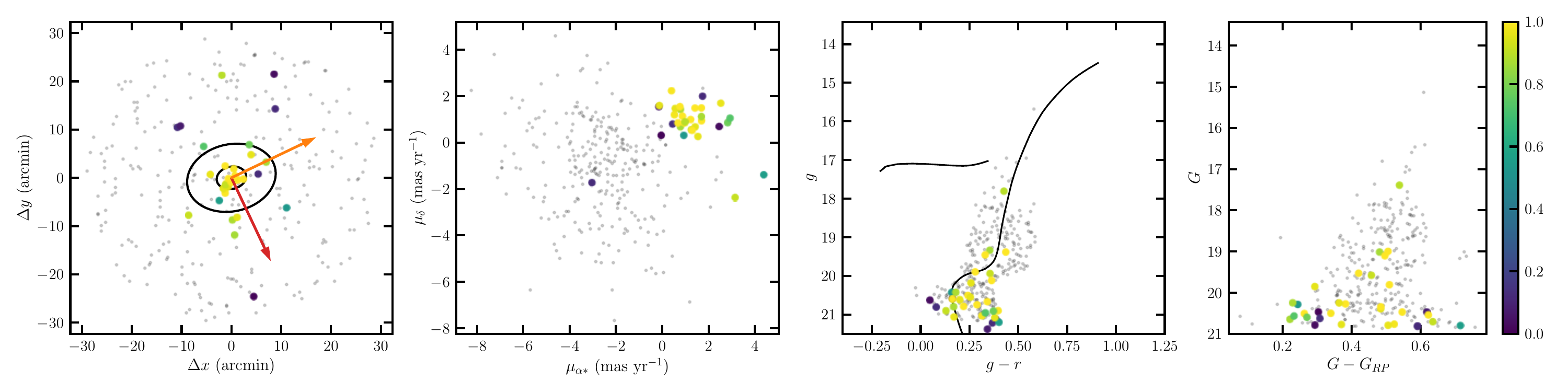}
\caption{Same as Figure~\ref{fig:diagnostic_aqu2} but for Draco II.  
}
\end{figure*}

\begin{figure*}
\includegraphics[height=4.5cm]{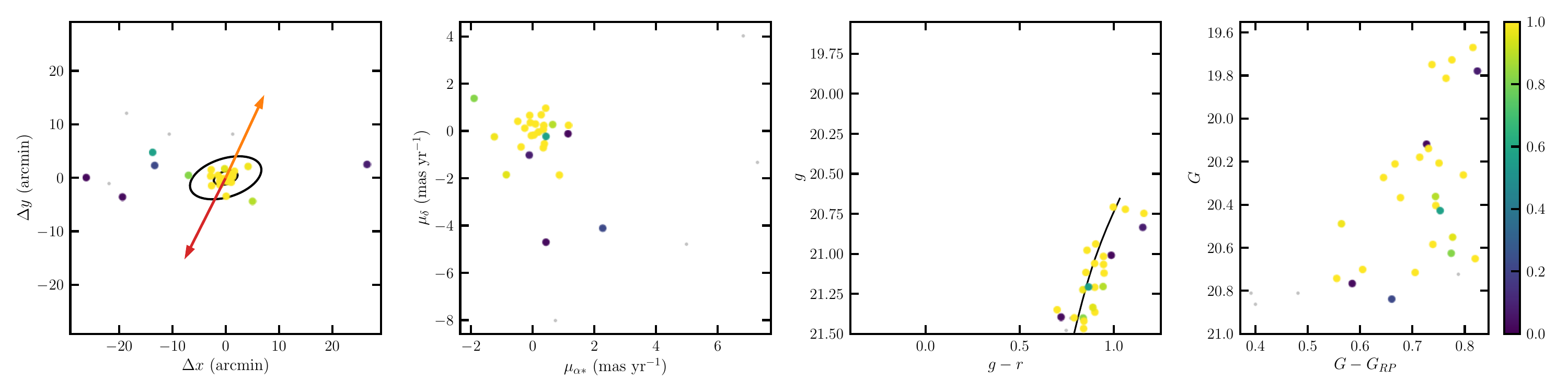}
\caption{Same as Figure~\ref{fig:diagnostic_aqu2} but for Eridanus II.  
}
\end{figure*}

\begin{figure*}
\includegraphics[height=4.5cm]{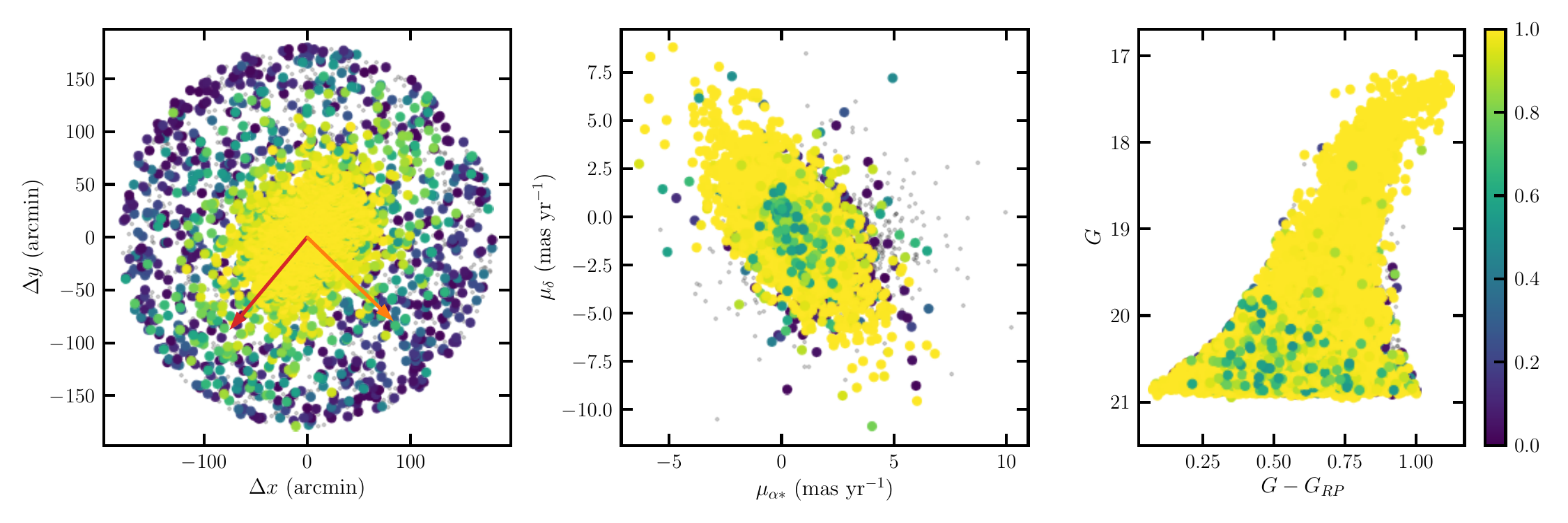}
\caption{Same as Figure~\ref{fig:diagnostic_ant2} but for  Fornax.  
}
\end{figure*}

\begin{figure*}
\includegraphics[height=4.5cm]{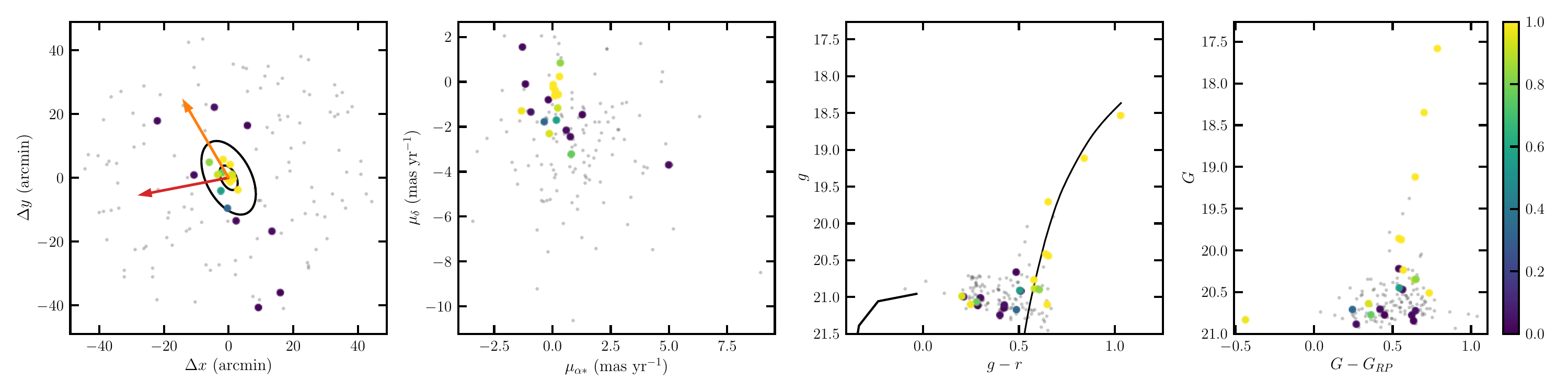}
\caption{Same as Figure~\ref{fig:diagnostic_aqu2} but for Grus I.  
}
\end{figure*}

\begin{figure*}
\includegraphics[height=4.5cm]{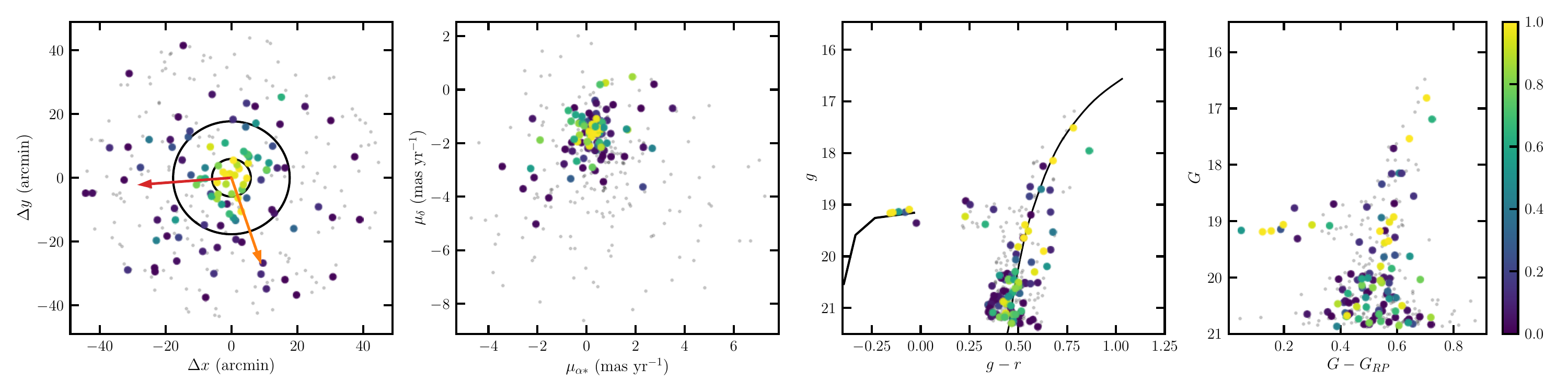}
\caption{Same as Figure~\ref{fig:diagnostic_aqu2} but for Grus II.  
}
\end{figure*}

\begin{figure*}
\includegraphics[height=4.5cm]{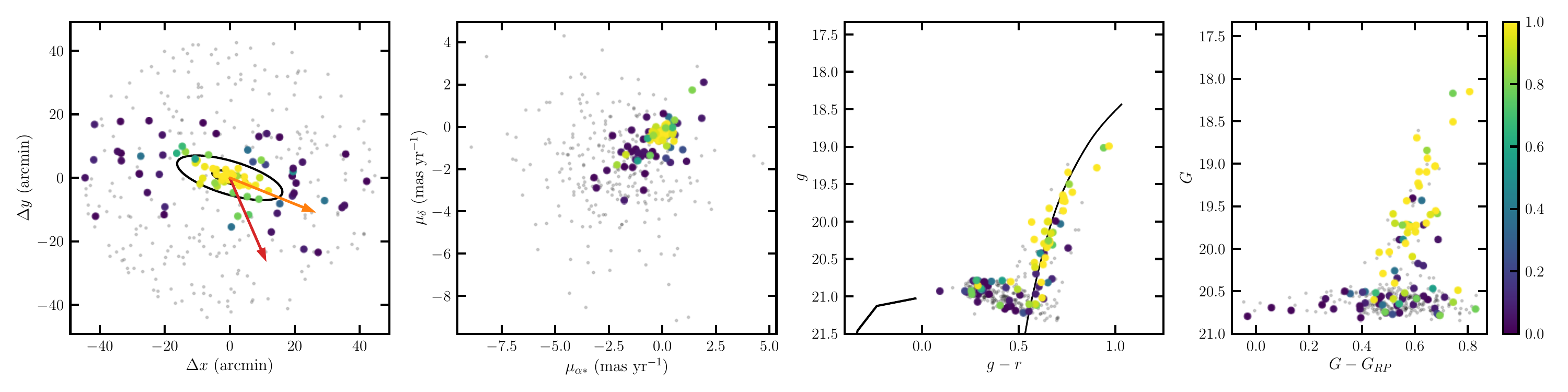}
\caption{Same as Figure~\ref{fig:diagnostic_aqu2} but for Hercules.  
}
\end{figure*}

\begin{figure*}
\includegraphics[height=4.5cm]{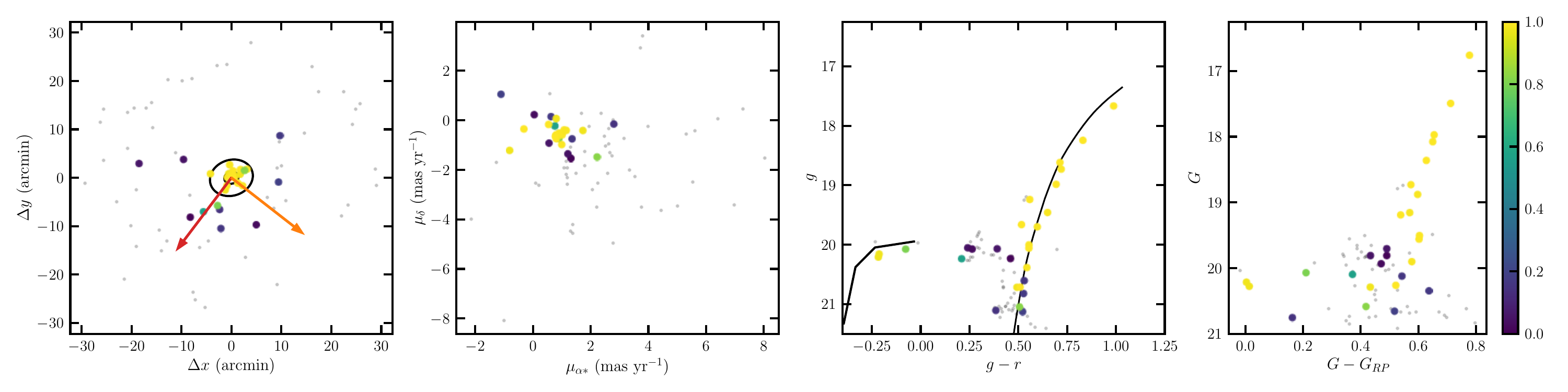}
\caption{Same as Figure~\ref{fig:diagnostic_aqu2} but for Horologium I.  
}
\end{figure*}

\begin{figure*}
\includegraphics[height=4.5cm]{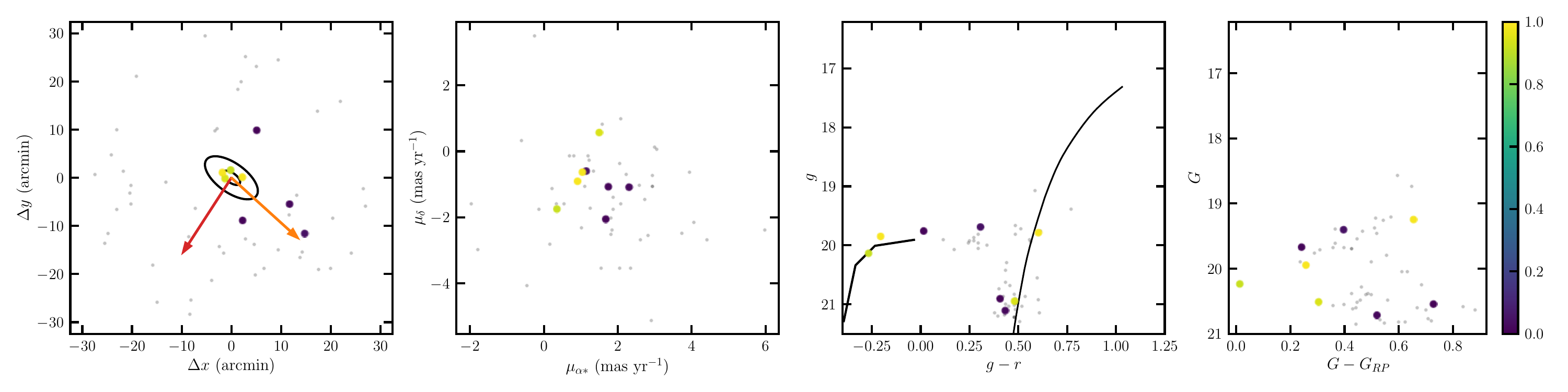}
\caption{Same as Figure~\ref{fig:diagnostic_aqu2} but for Horologium II.  
}
\end{figure*}

\begin{figure*}
\includegraphics[height=4.5cm]{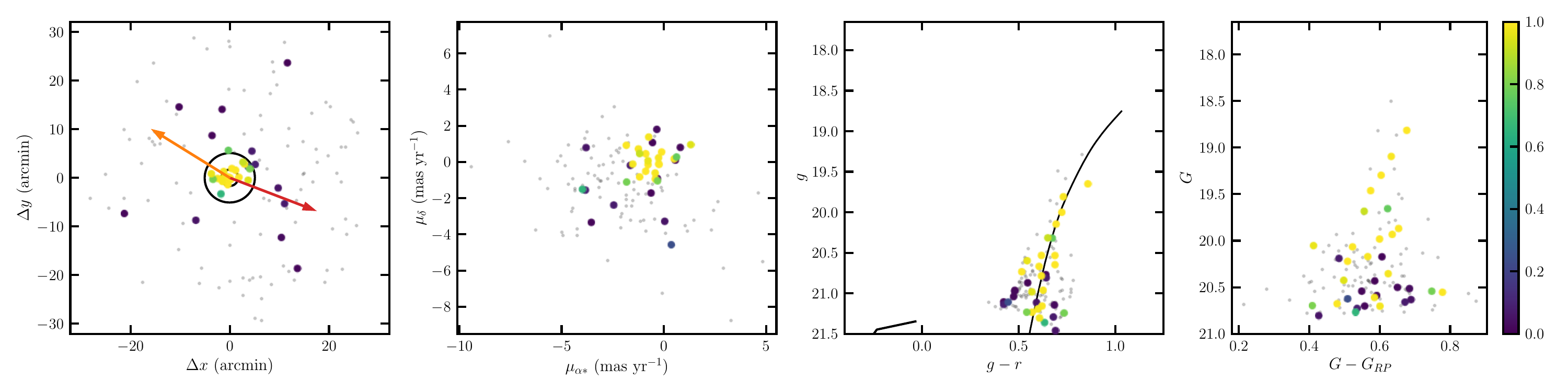}
\caption{Same as Figure~\ref{fig:diagnostic_aqu2} but for Hydra II.  
}
\end{figure*}

\begin{figure*}
\includegraphics[height=4.5cm]{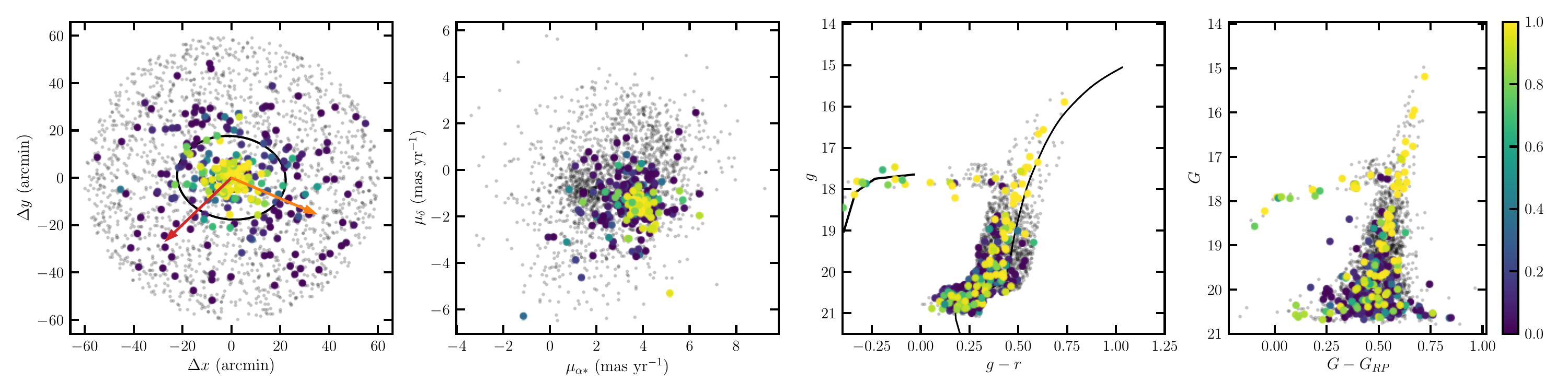}
\caption{Same as Figure~\ref{fig:diagnostic_aqu2} but for Hydrus I.  
}
\end{figure*}


\begin{figure*}
\includegraphics[height=4.5cm]{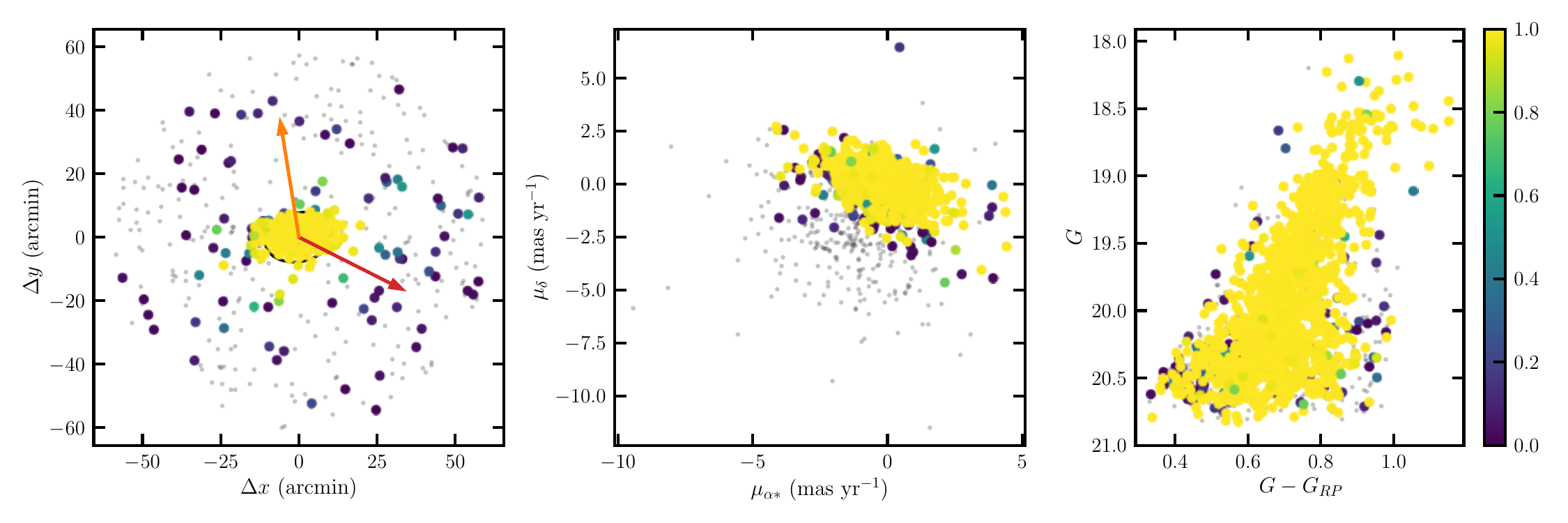}
\caption{Same as Figure~\ref{fig:diagnostic_ant2} but for Leo I.  
}
\end{figure*}

\begin{figure*}
\includegraphics[height=4.5cm]{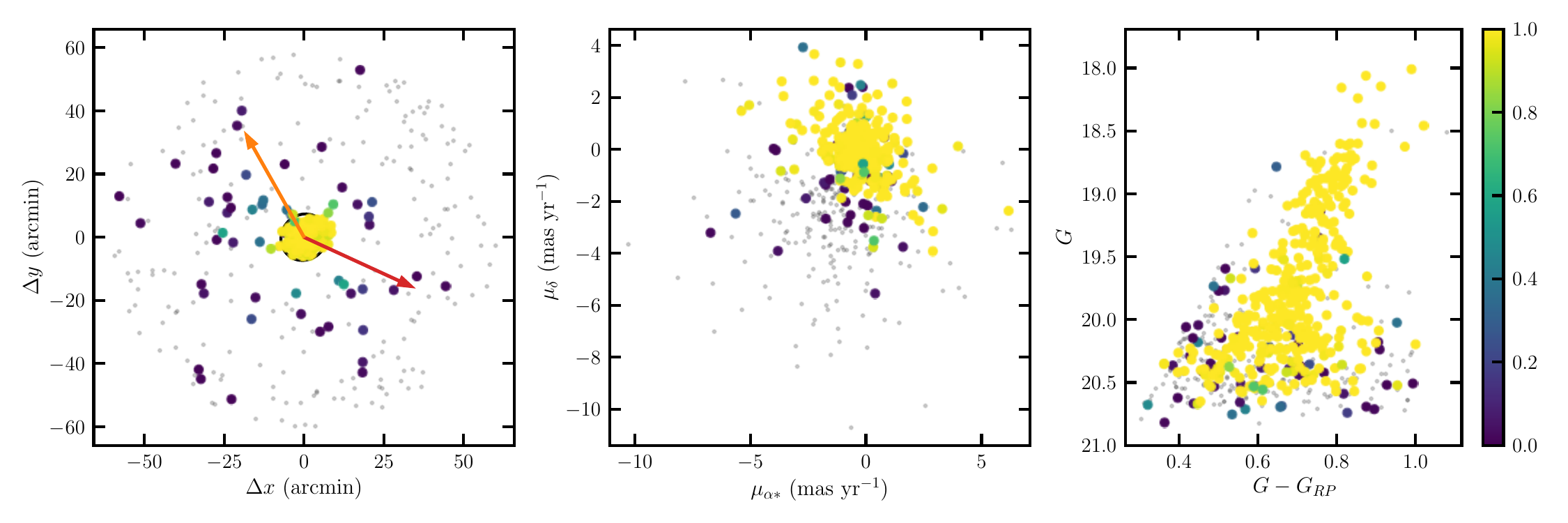}
\caption{Same as Figure~\ref{fig:diagnostic_ant2} but for  Leo II.  }
\end{figure*}

\begin{figure*}
\includegraphics[height=4.5cm]{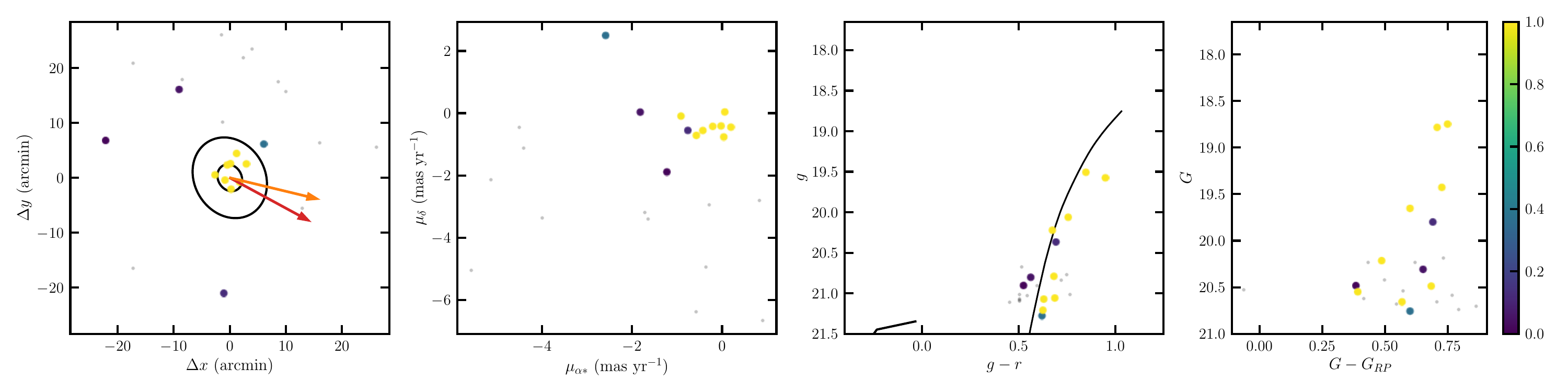}
\caption{Same as Figure~\ref{fig:diagnostic_aqu2} but for Leo IV.  
}
\end{figure*}

\begin{figure*}
\includegraphics[height=4.5cm]{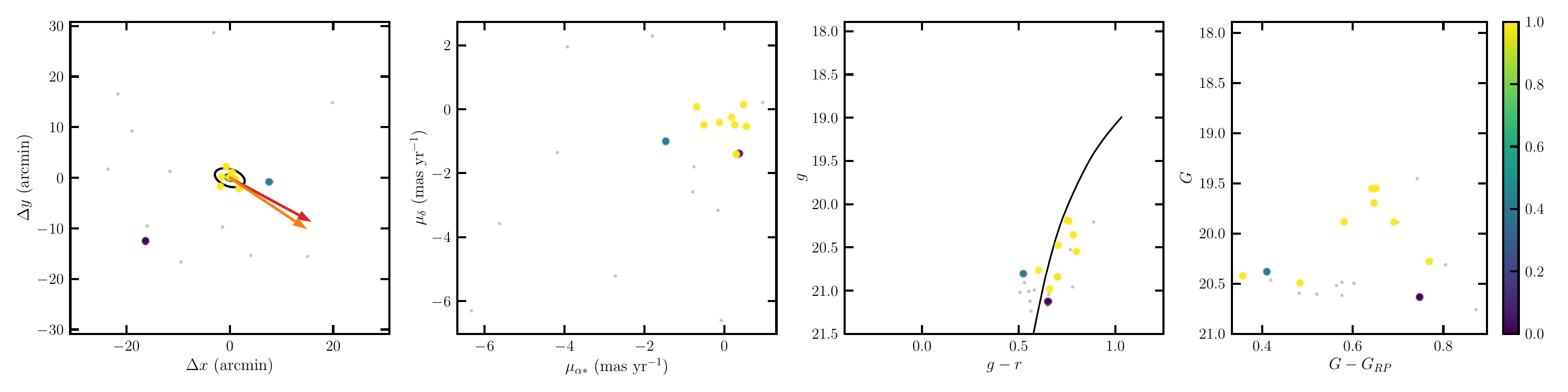}
\caption{Same as Figure~\ref{fig:diagnostic_aqu2} but for Leo V.  
}
\end{figure*}

\begin{figure*}
\includegraphics[height=4.5cm]{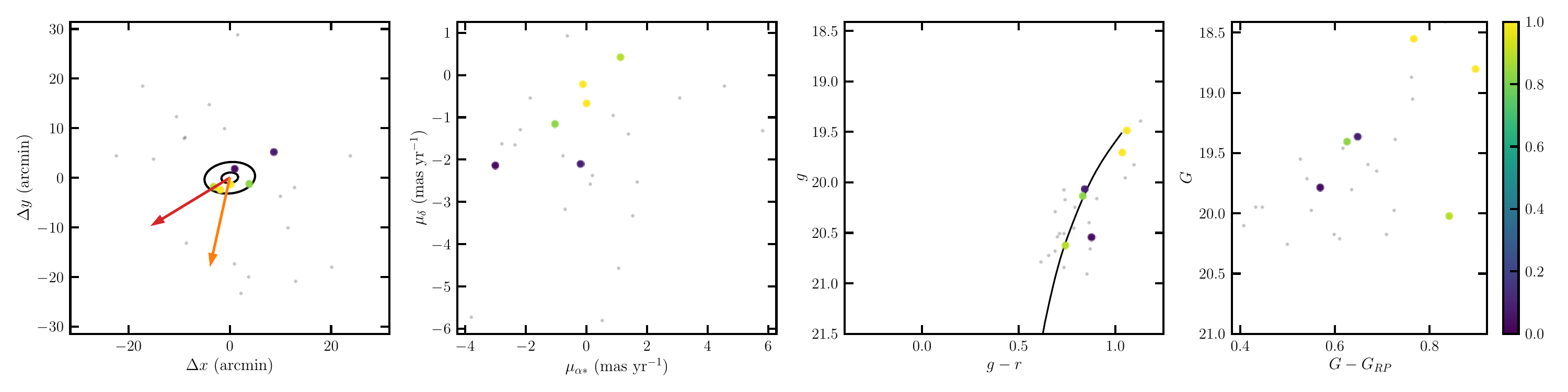}
\caption{Same as Figure~\ref{fig:diagnostic_aqu2} but for Pegasus III.  
}
\end{figure*}

\begin{figure*}
\includegraphics[height=4.5cm]{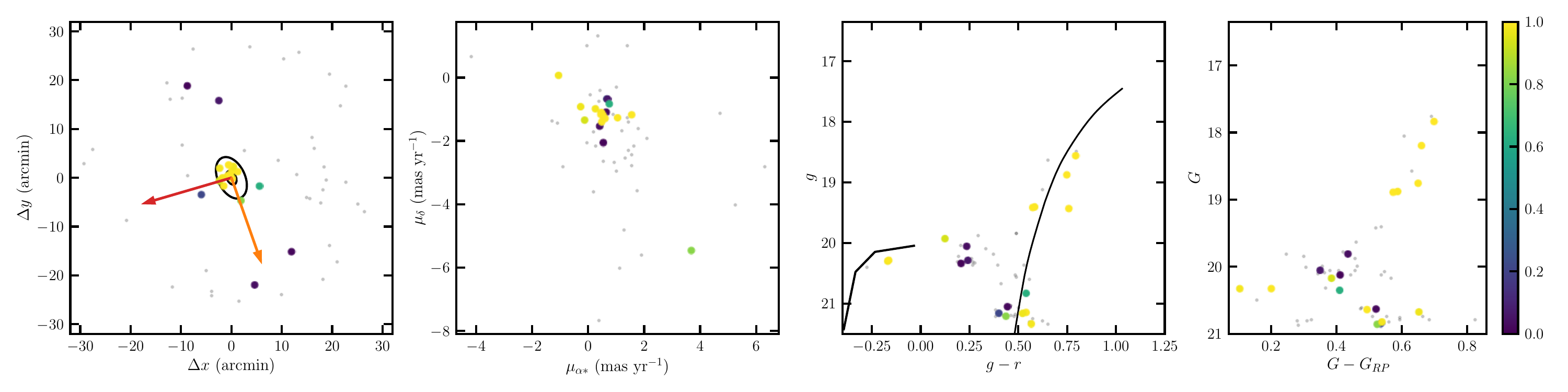}
\caption{Same as Figure~\ref{fig:diagnostic_aqu2} but for Phoenix II.  
}
\end{figure*}

\begin{figure*}
\includegraphics[height=4.5cm]{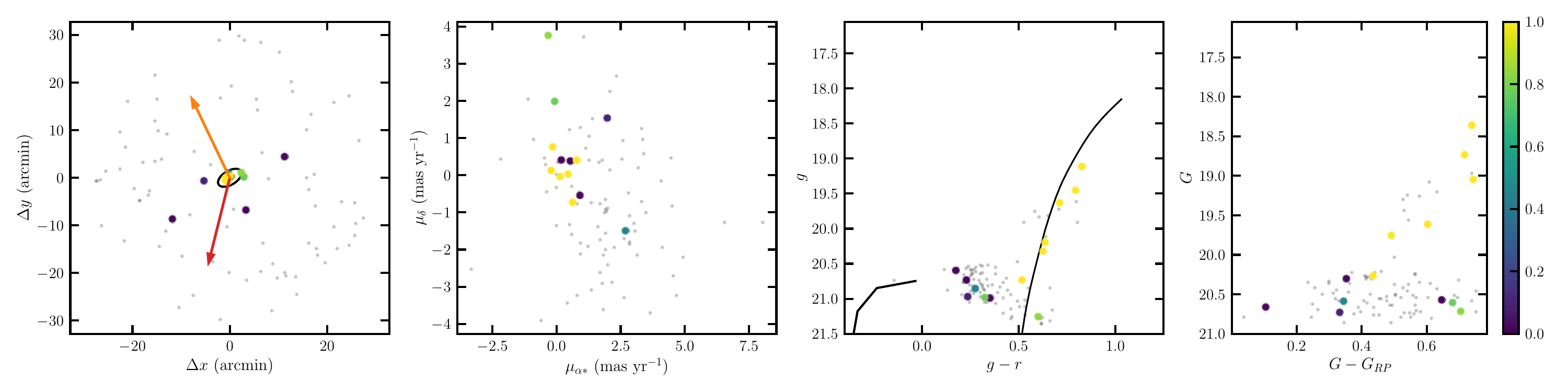}
\caption{Same as Figure~\ref{fig:diagnostic_aqu2} but for Pictor I.  
}
\end{figure*}

\begin{figure*}
\includegraphics[height=4.5cm]{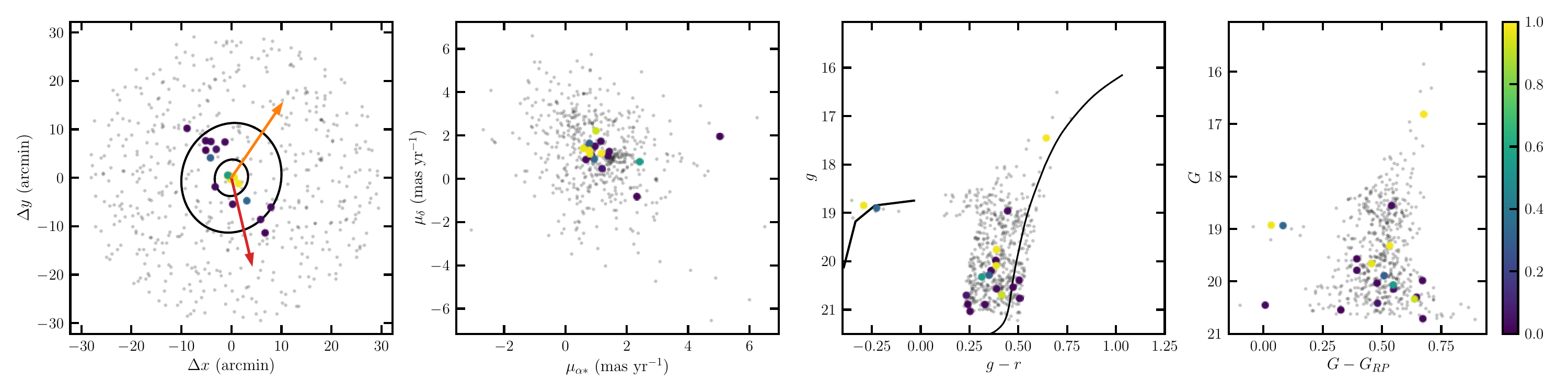}
\caption{Same as Figure~\ref{fig:diagnostic_aqu2} but for Pictor II. There is color offset in the NSC data that is unaccounted for in the isochrone. 
}
\end{figure*}

\begin{figure*}
\includegraphics[height=4.5cm]{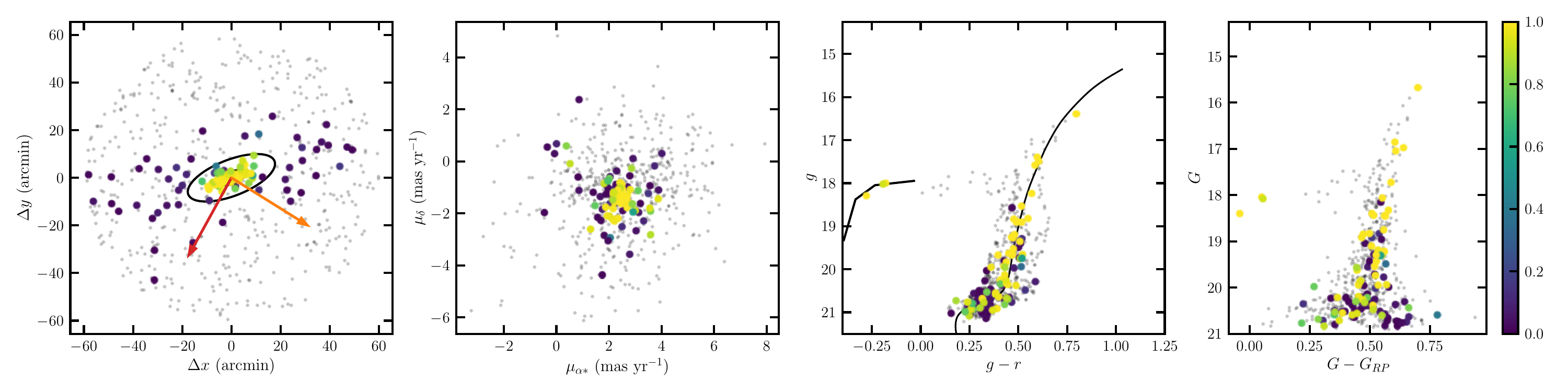}
\caption{Same as Figure~\ref{fig:diagnostic_aqu2} but for Reticulum II.  
}
\end{figure*}

\begin{figure*}
\includegraphics[height=4.5cm]{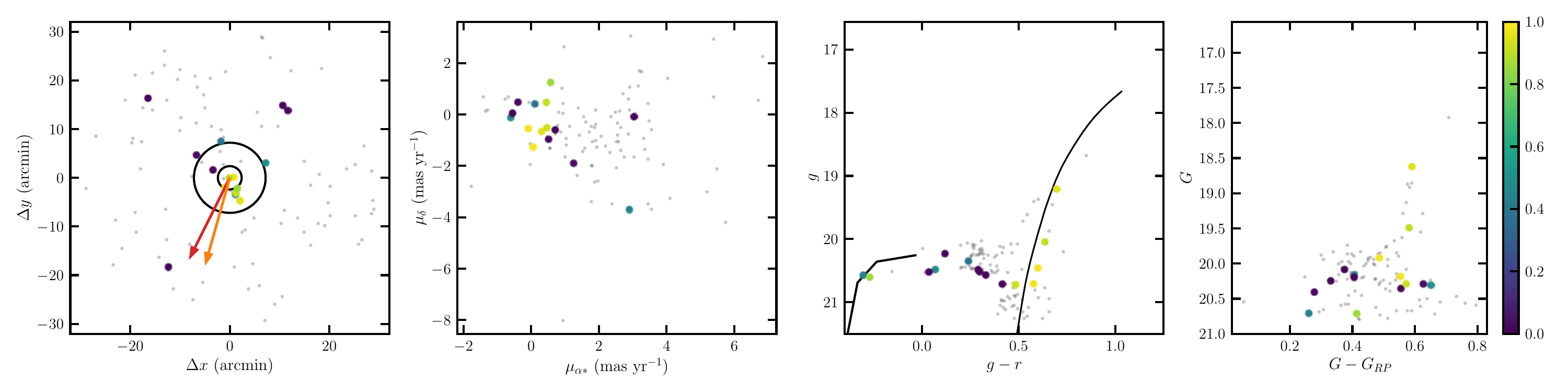}
\caption{Same as Figure~\ref{fig:diagnostic_aqu2} but for Reticulum III.  
}
\end{figure*}

\begin{figure*}
\includegraphics[height=4.5cm]{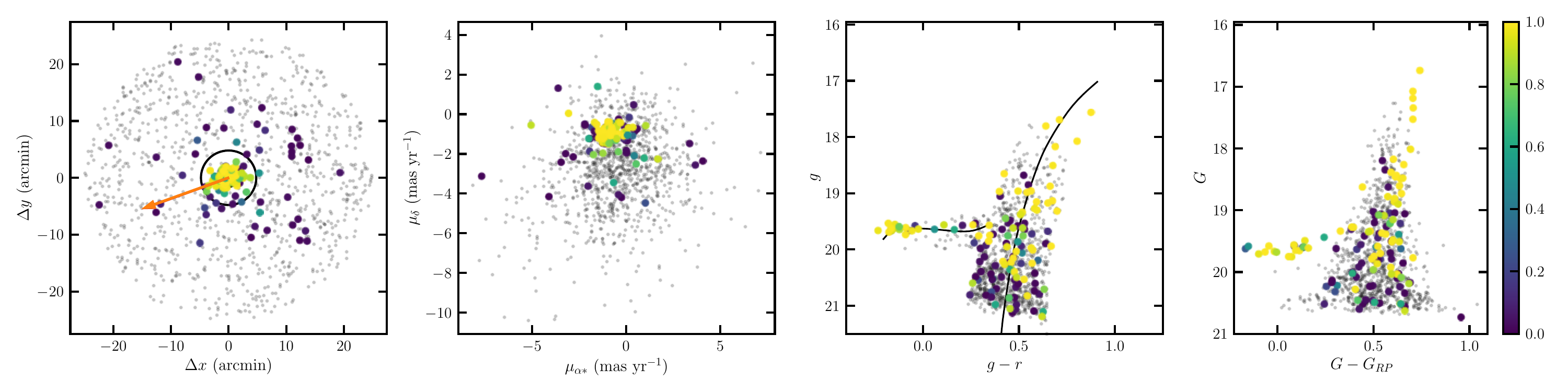}
\caption{Same as Figure~\ref{fig:diagnostic_aqu2} but for Sagittarius II.  
}
\end{figure*}

\begin{figure*}
\includegraphics[height=4.5cm]{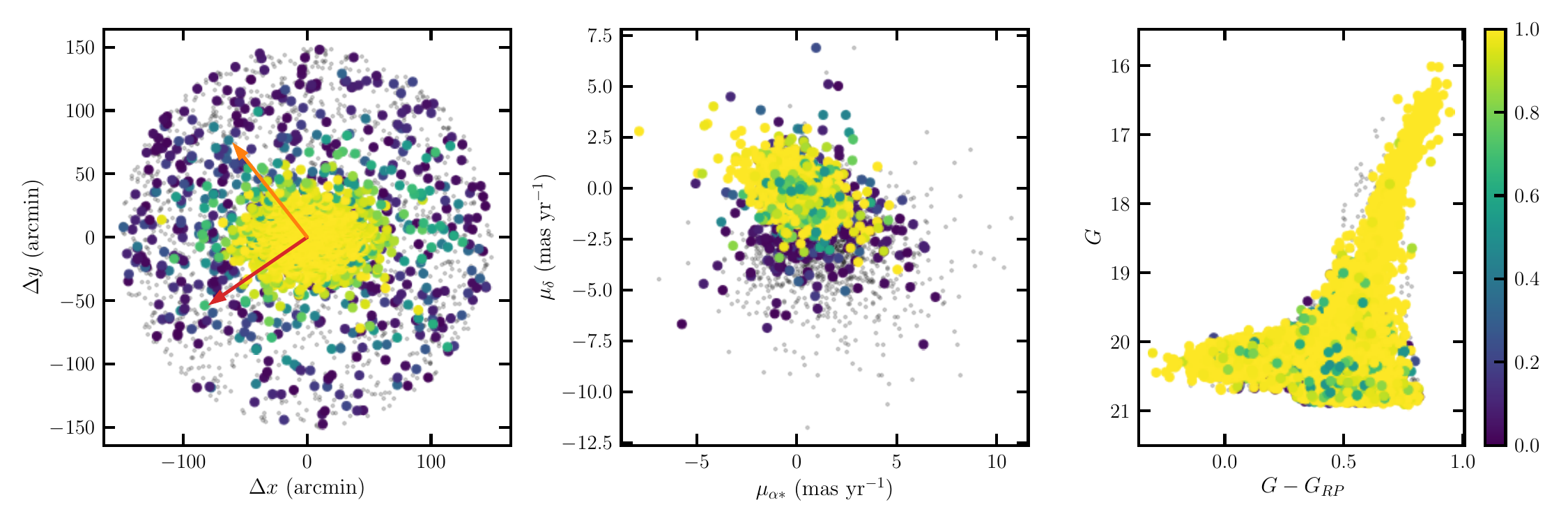}
\caption{Same as Figure~\ref{fig:diagnostic_ant2} but for Sculptor.  
}
\end{figure*}

\begin{figure*}
\includegraphics[height=4.5cm]{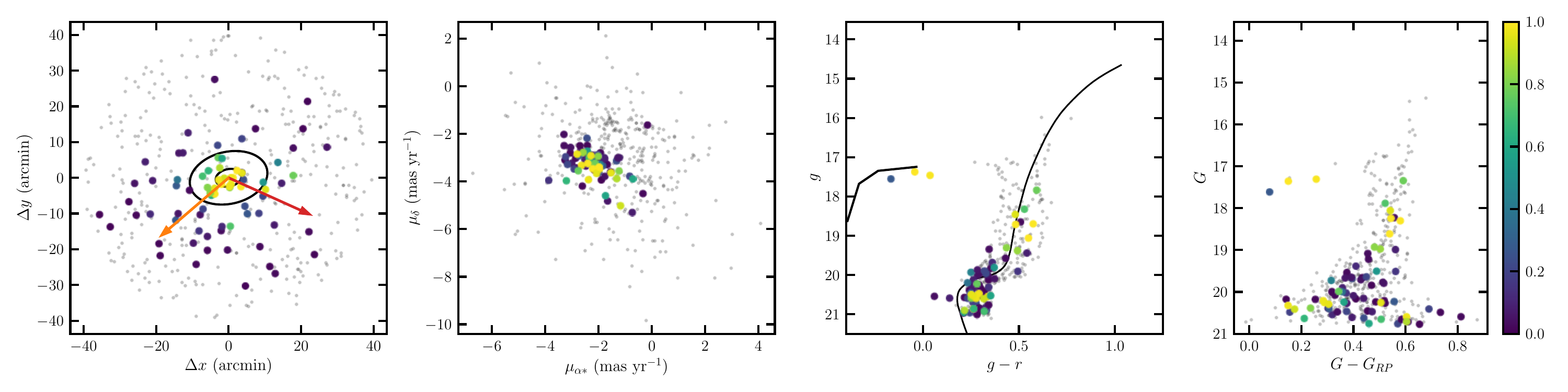}
\caption{Same as Figure~\ref{fig:diagnostic_aqu2} but for Segue 1.  
}
\end{figure*}

\begin{figure*}
\includegraphics[height=4.5cm]{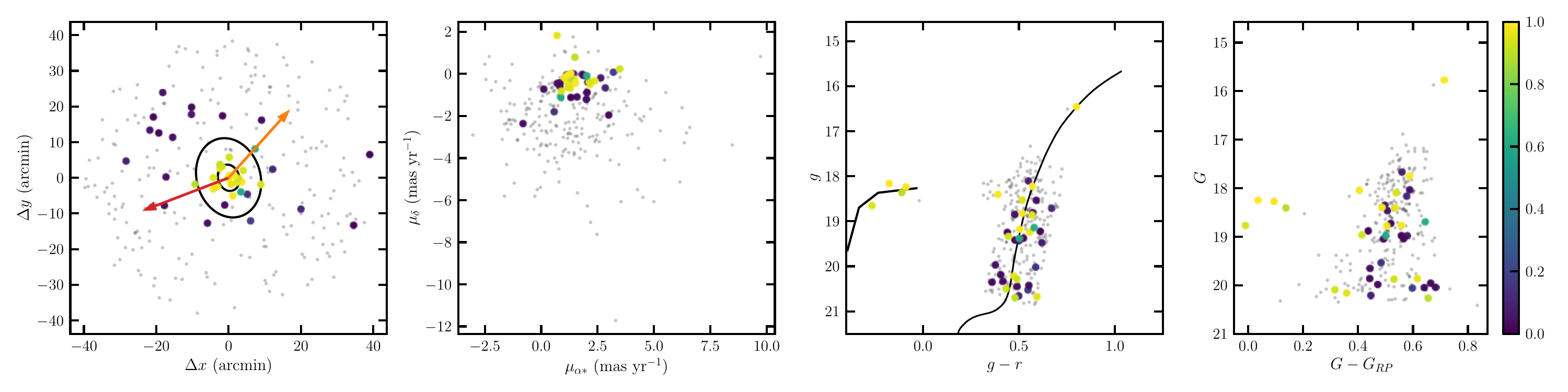}
\caption{Same as Figure~\ref{fig:diagnostic_aqu2} but for Segue 2.  
}
\end{figure*}

\begin{figure*}
\includegraphics[height=4.5cm]{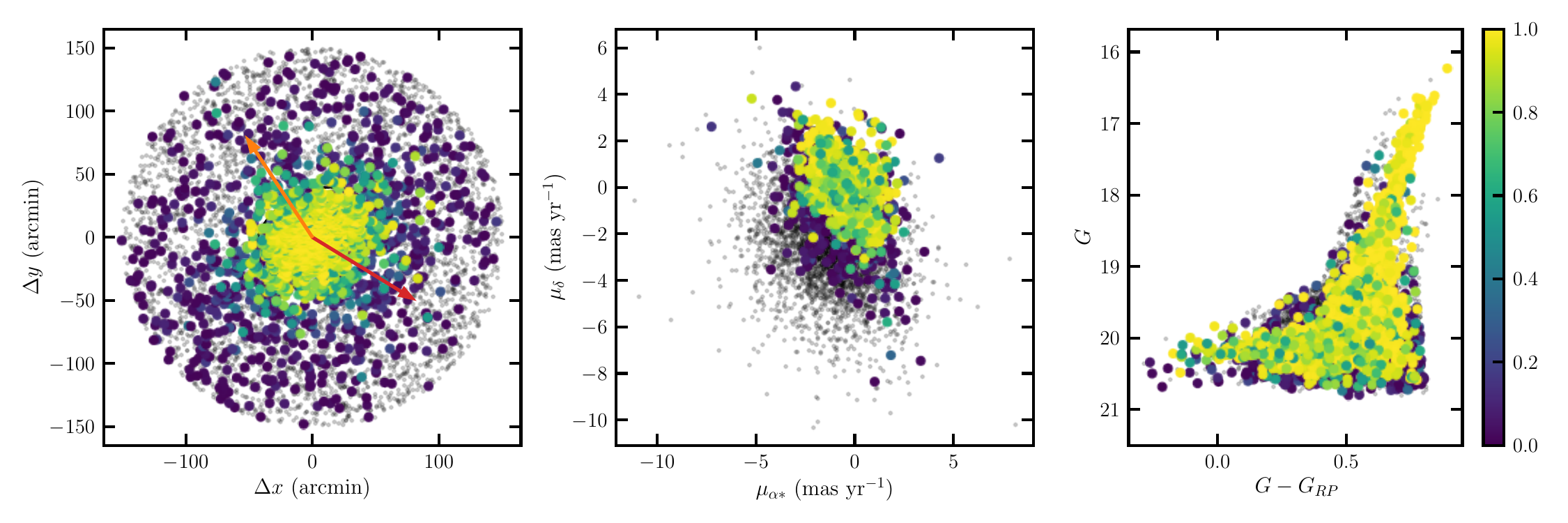}
\caption{Same as Figure~\ref{fig:diagnostic_ant2} but for Sextans.  
}
\end{figure*}

\begin{figure*}
\includegraphics[height=4.5cm]{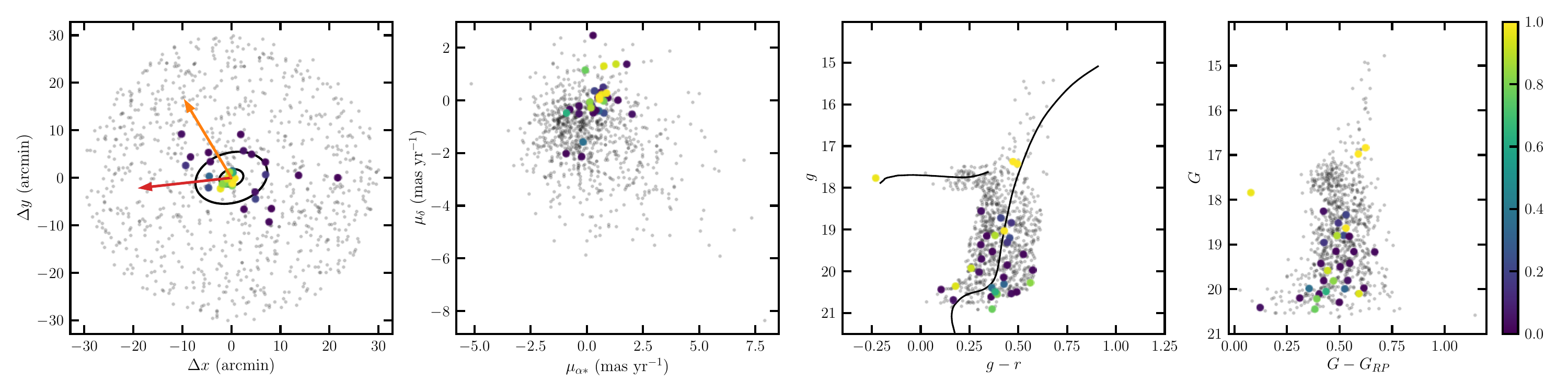}
\caption{Same as Figure~\ref{fig:diagnostic_aqu2} but for Triangulum II.  
}
\end{figure*}


\begin{figure*}
\includegraphics[height=4.5cm]{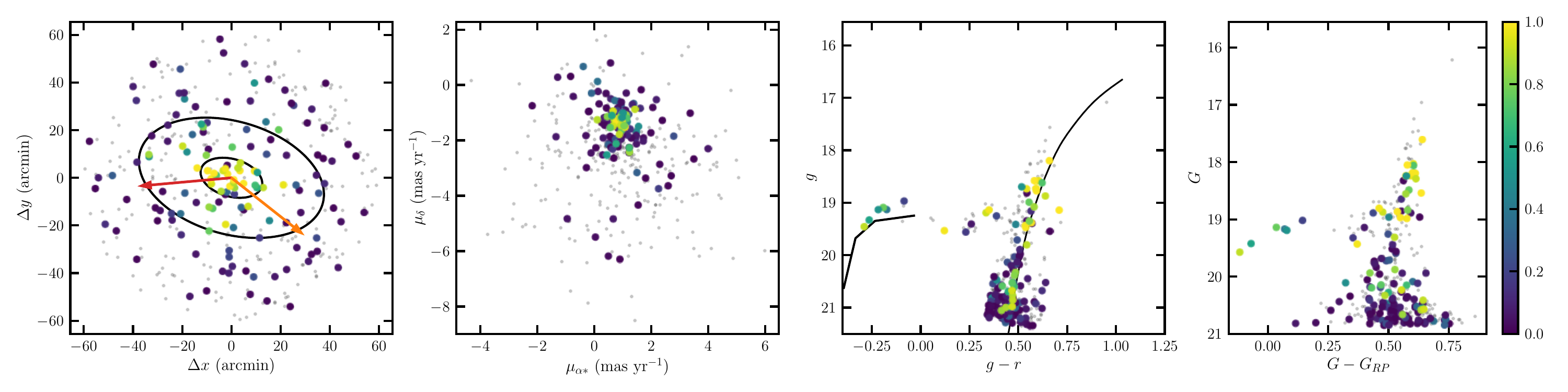}
\caption{Same as Figure~\ref{fig:diagnostic_aqu2} but for Tucana II.  
}
\end{figure*}

\begin{figure*}
\includegraphics[height=4.5cm]{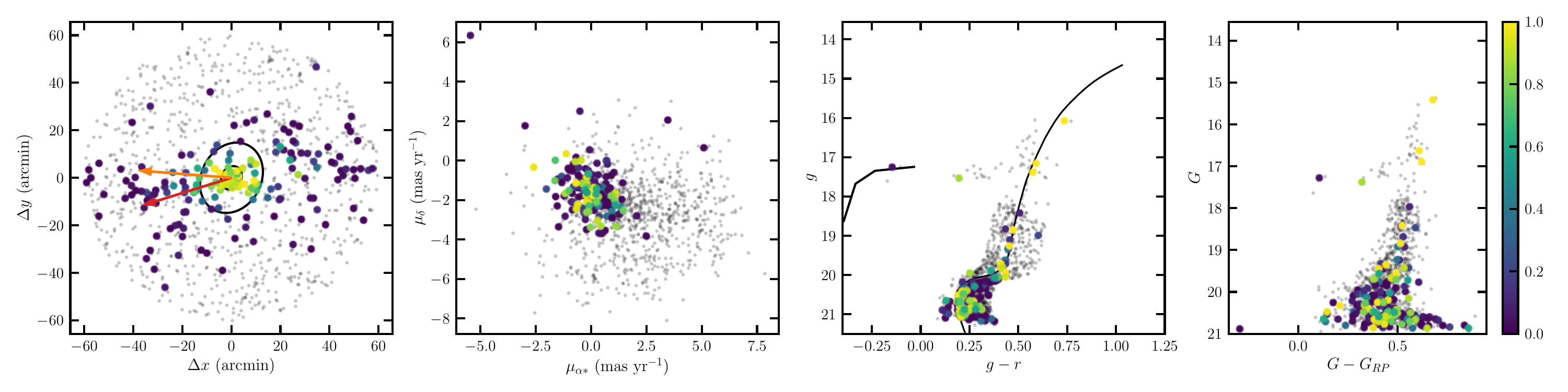}
\caption{Same as Figure~\ref{fig:diagnostic_aqu2} but for Tucana III.  
}
\end{figure*}

\begin{figure*}
\includegraphics[height=4.5cm]{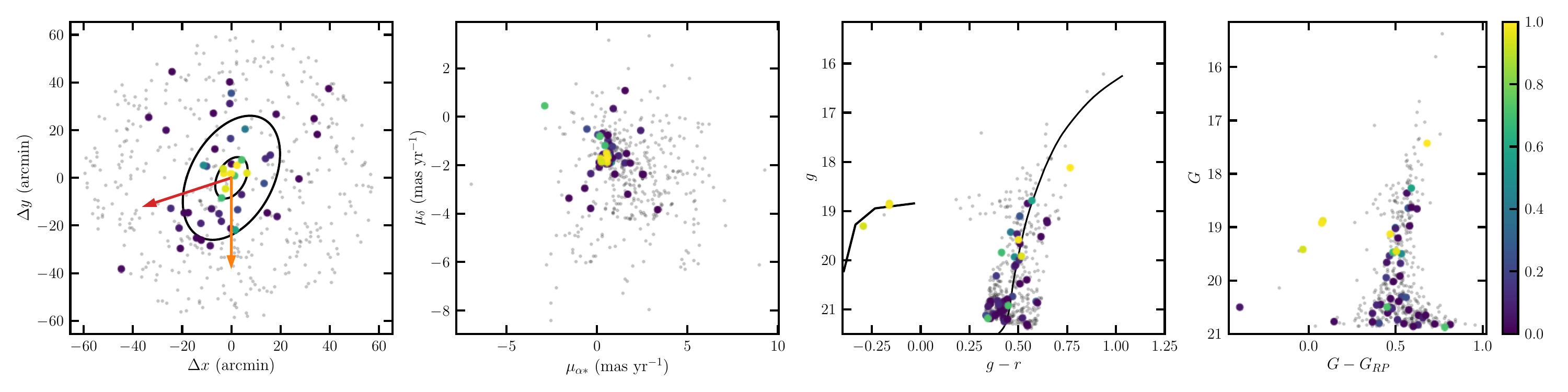}
\caption{Same as Figure~\ref{fig:diagnostic_aqu2} but for Tucana IV.  
}
\end{figure*}

\begin{figure*}
\includegraphics[height=4.5cm]{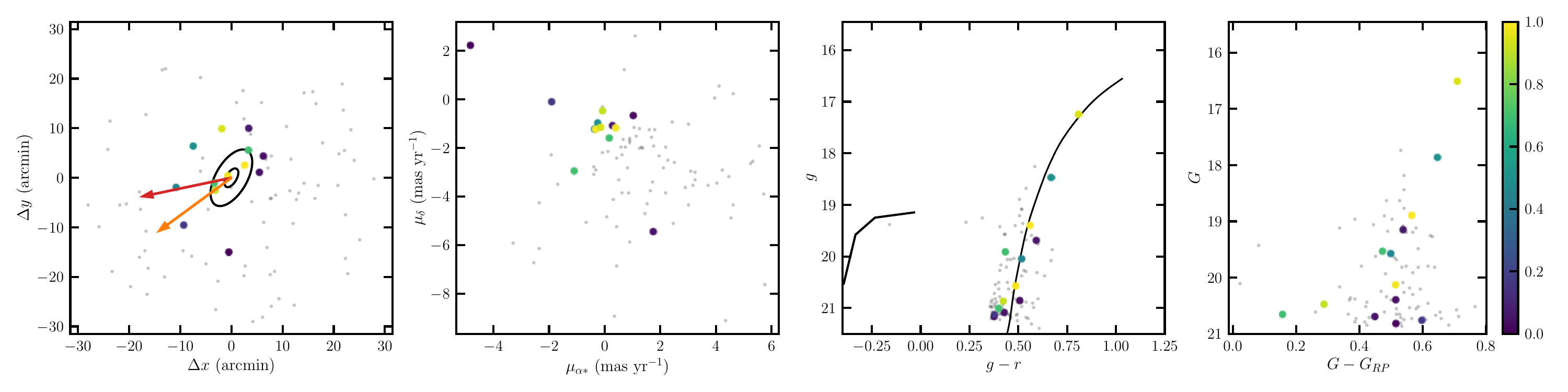}
\caption{Same as Figure~\ref{fig:diagnostic_aqu2} but for Tucana V.  
}
\end{figure*}

\begin{figure*}
\includegraphics[height=4.5cm]{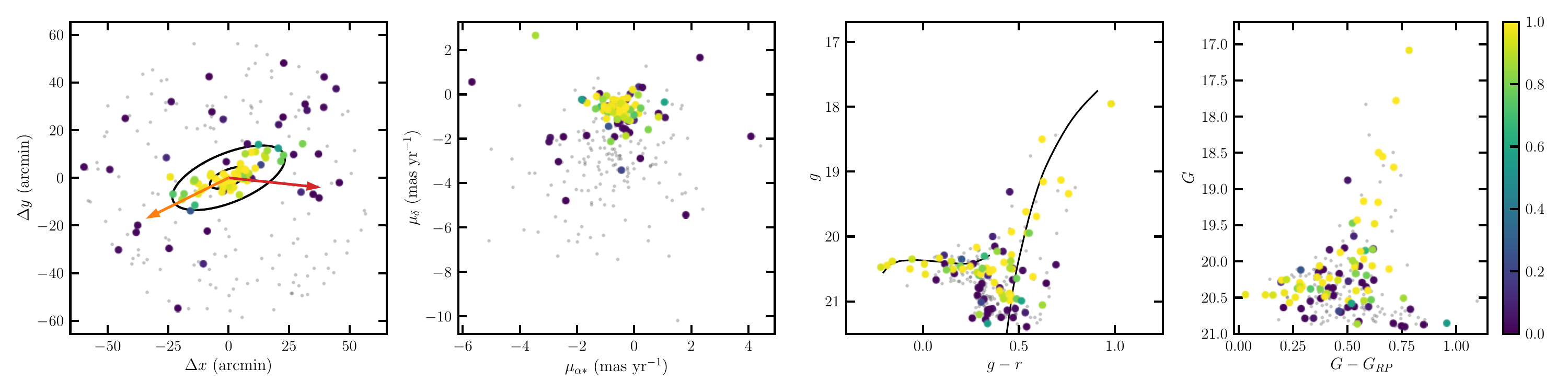}
\caption{Same as Figure~\ref{fig:diagnostic_aqu2} but for Ursa Major I.  
}
\end{figure*}

\begin{figure*}
\includegraphics[height=4.5cm]{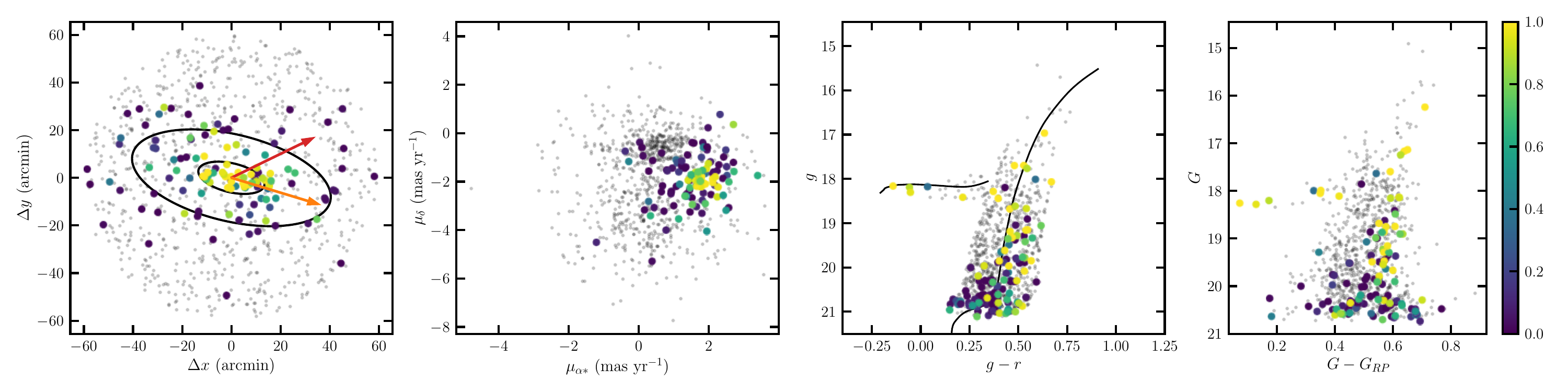}
\caption{Same as Figure~\ref{fig:diagnostic_aqu2} but for Ursa Major II.  
}
\end{figure*}

\begin{figure*}
\includegraphics[height=4.5cm]{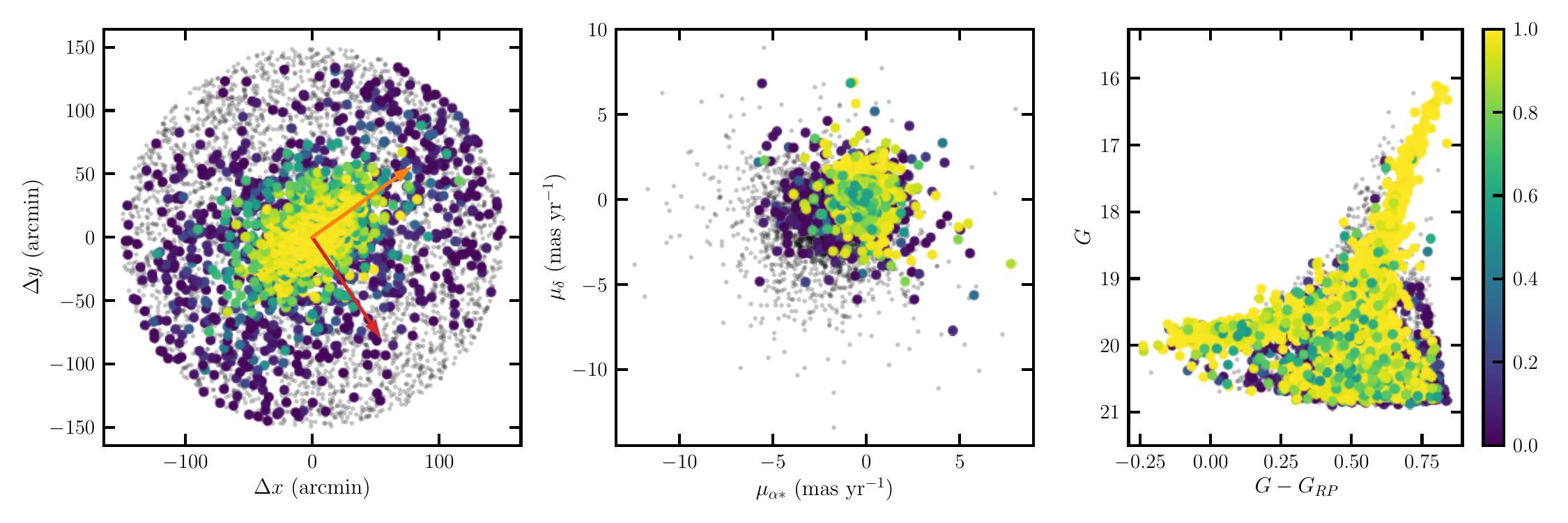}
\caption{ Same as Figure~\ref{fig:diagnostic_ant2} but for Ursa Minor.  
}
\end{figure*}

\begin{figure*}
\includegraphics[height=4.5cm]{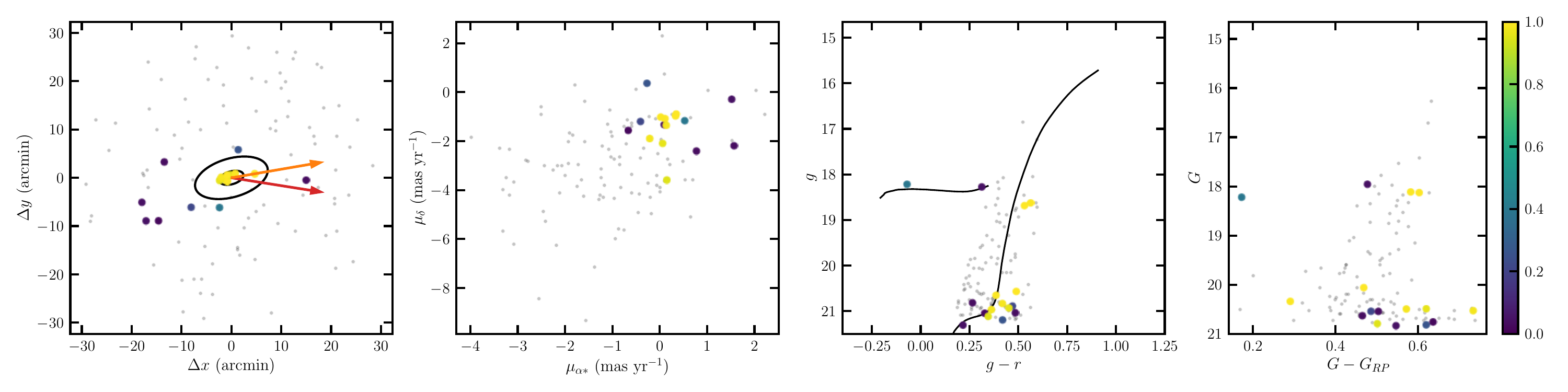}
\caption{Same as Figure~\ref{fig:diagnostic_aqu2} but for Willman 1.  
}
\end{figure*}

\clearpage

\begin{figure*}
\includegraphics[height=4.5cm]{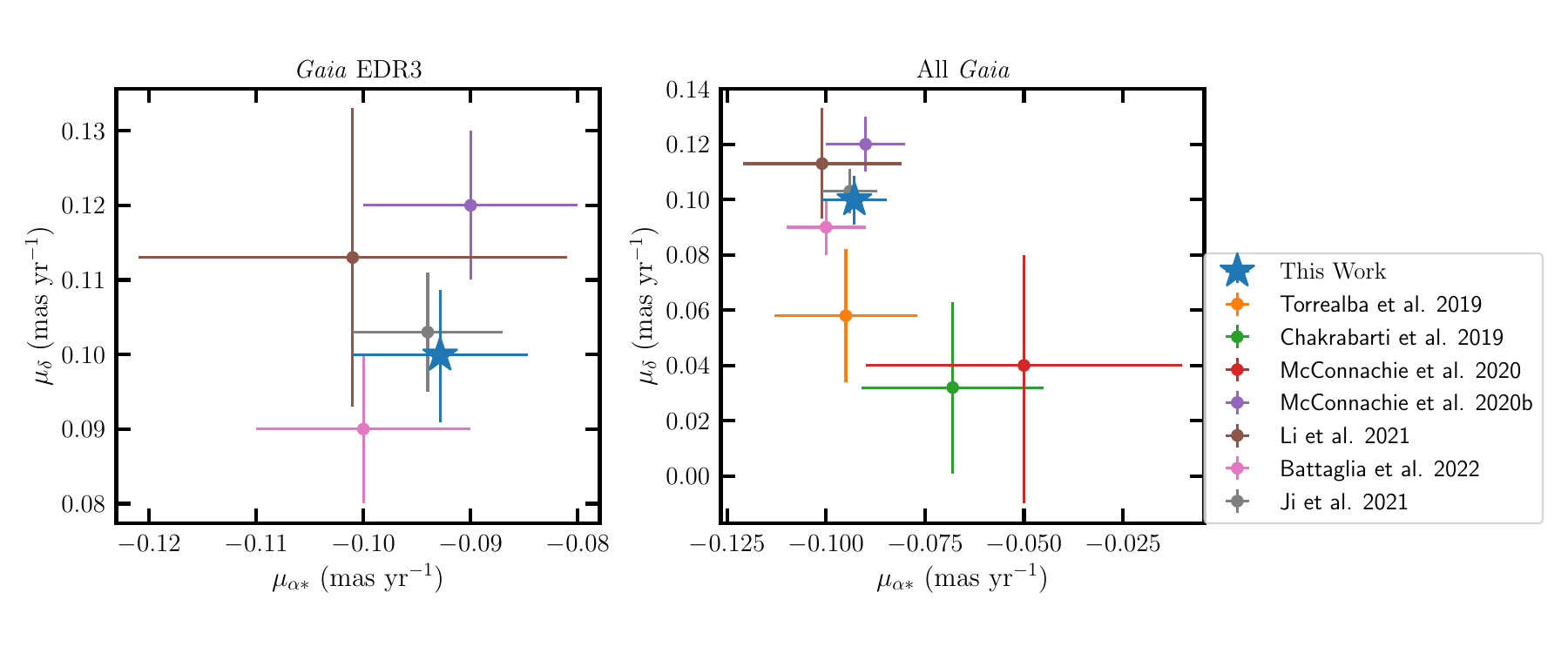}
\caption{Comparison between our systemic  proper motion measurement and literature measurements  for Antlia II.}
\label{fig:comparsion}
\end{figure*}

\begin{figure*}
\includegraphics[height=4.5cm]{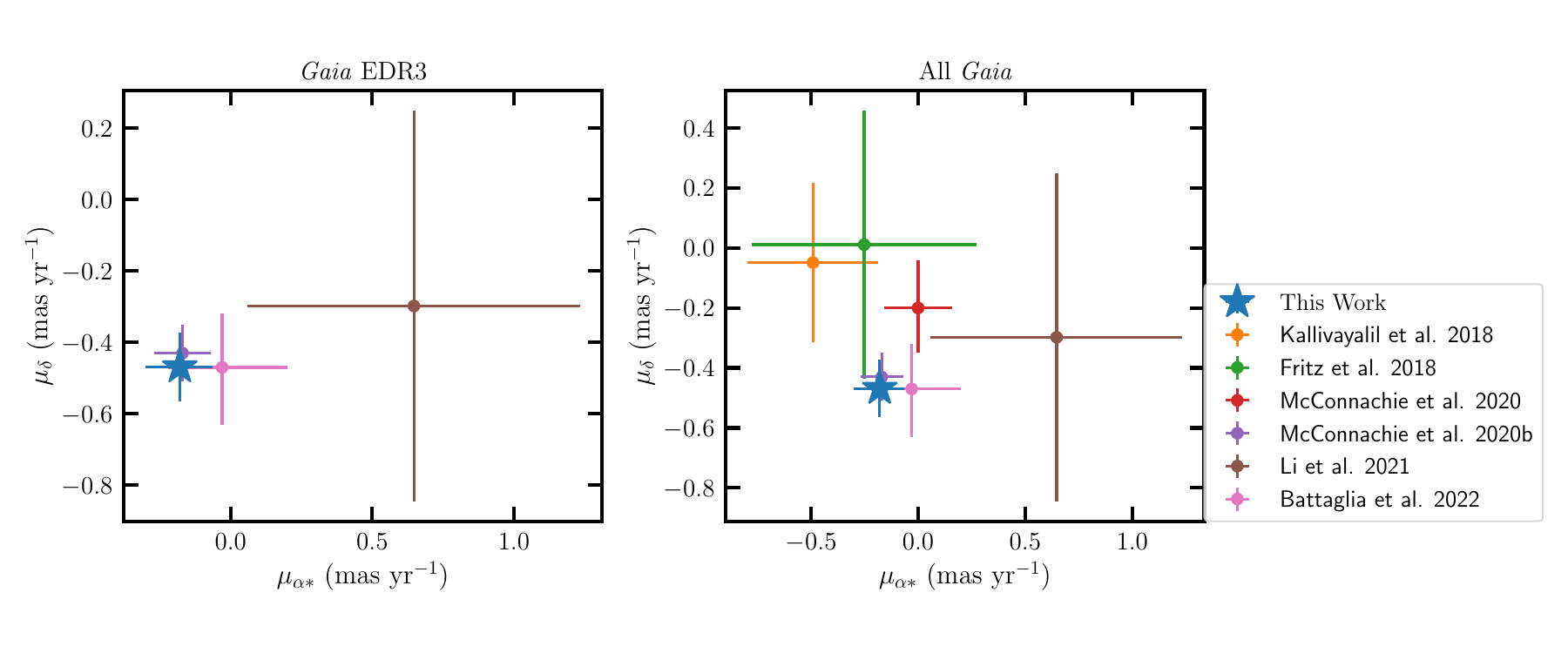}
\caption{Same as Figure~\ref{fig:comparsion} but for Aquarius II.}
\end{figure*}

\begin{figure*}
\includegraphics[height=4.5cm]{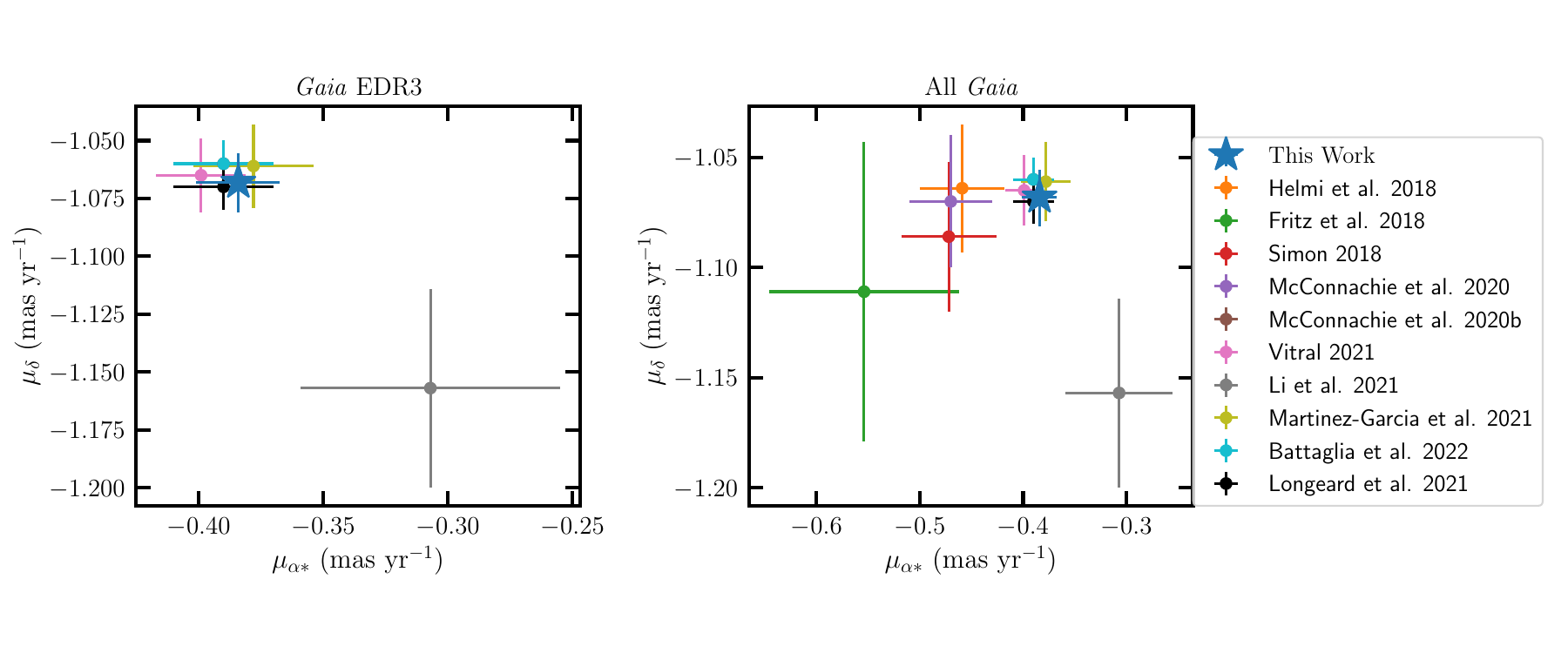}
\caption{Same as Figure~\ref{fig:comparsion} but for Bo\"{o}tes~I.}
\end{figure*}

\begin{figure*}
\includegraphics[height=4.5cm]{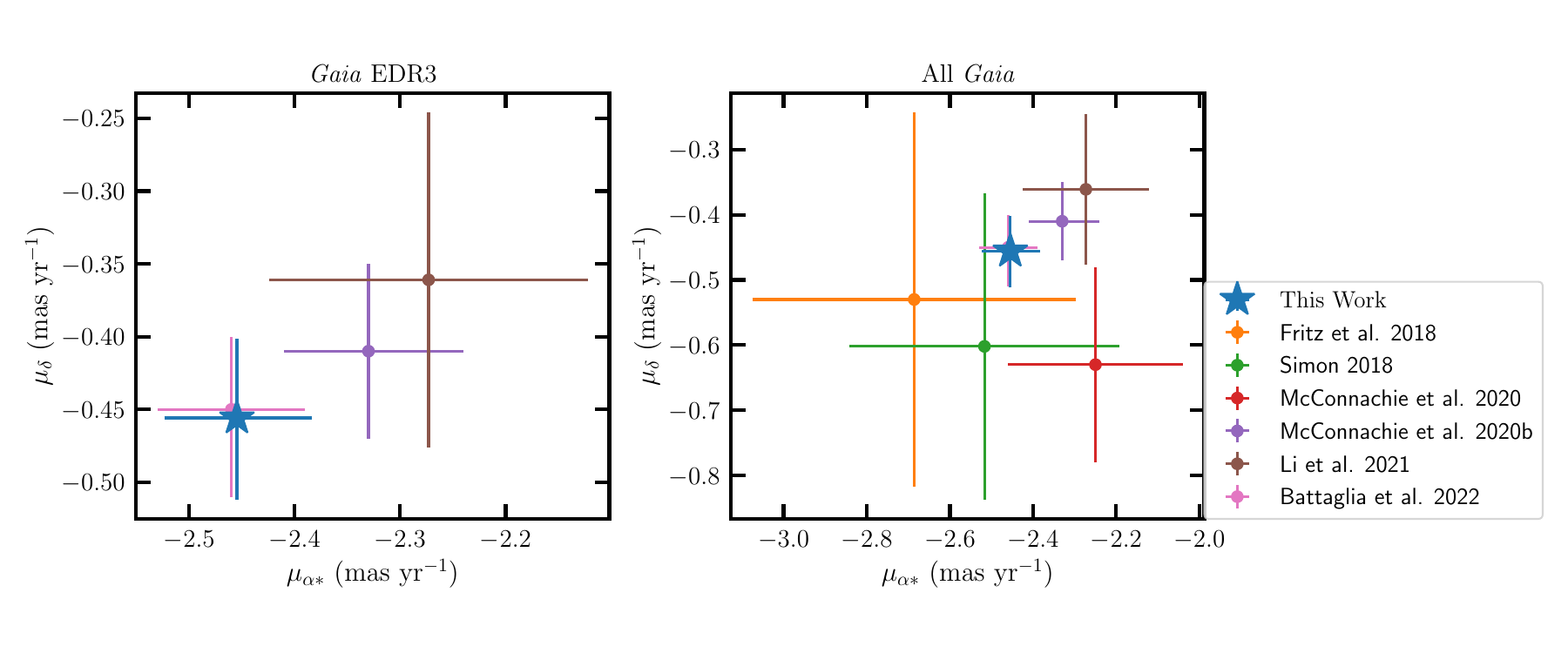}
\caption{Same as Figure~\ref{fig:comparsion} but for Bo\"{o}tes~II.}
\end{figure*}


\begin{figure*}
\includegraphics[height=4.5cm]{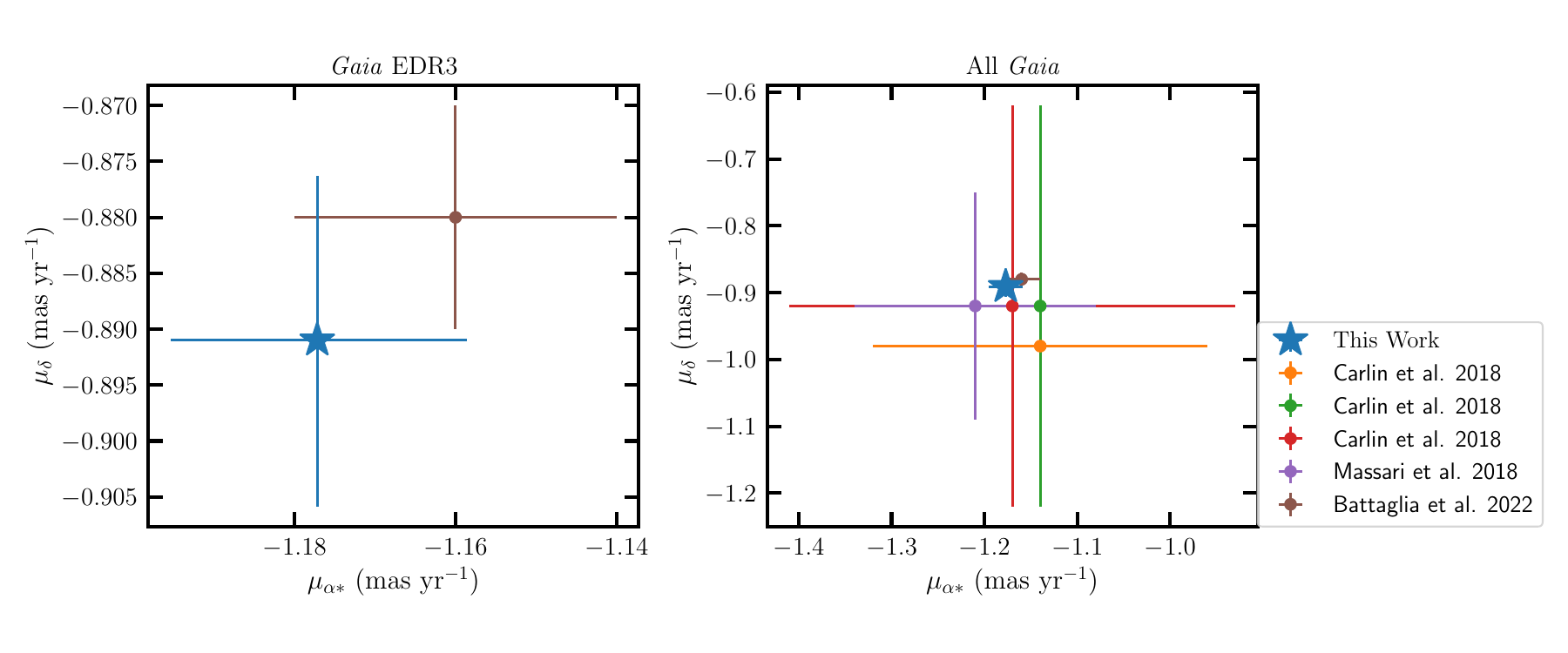}
\caption{Same as Figure~\ref{fig:comparsion} but for Bo\"{o}tes~III.}
\end{figure*}

\begin{figure*}
\includegraphics[height=4.5cm]{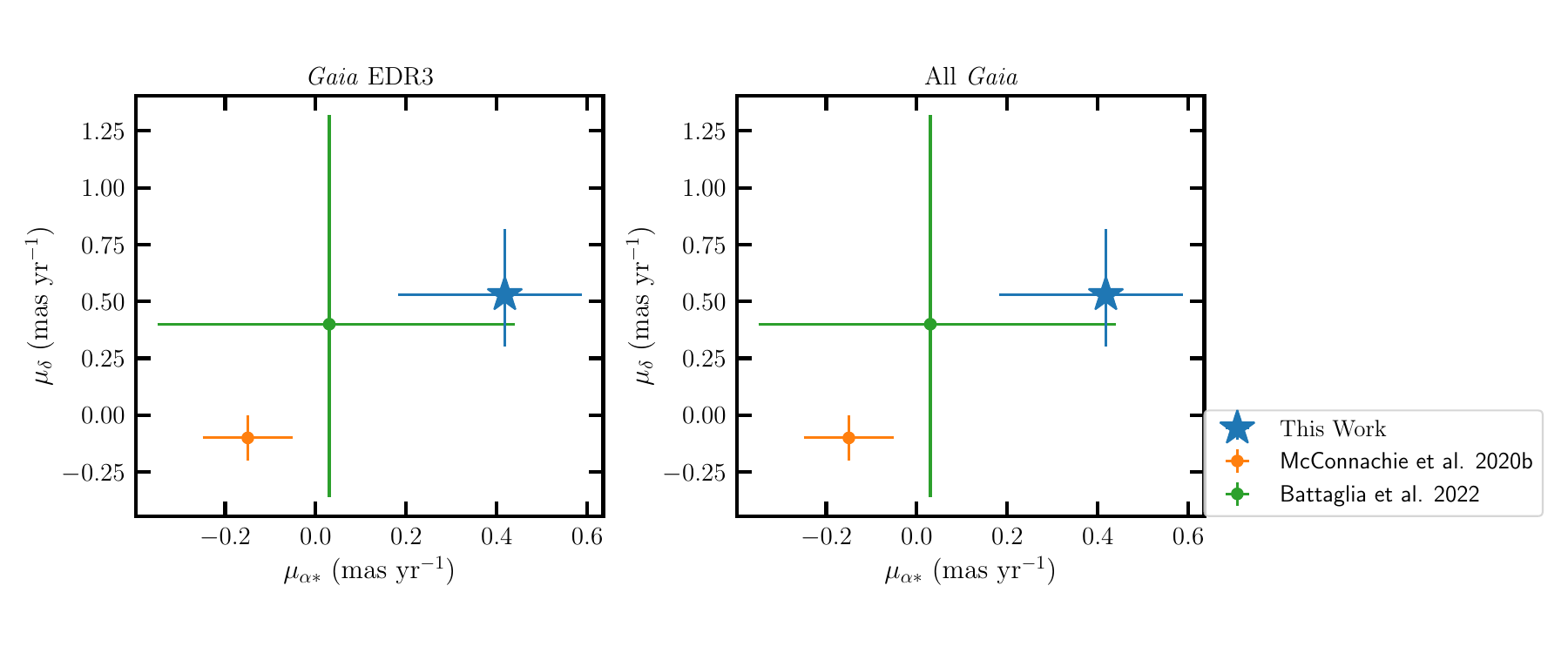}
\caption{Same as Figure~\ref{fig:comparsion} but for Bo\"{o}tes~IV.}
\end{figure*}

\begin{figure*}
\includegraphics[height=4.5cm]{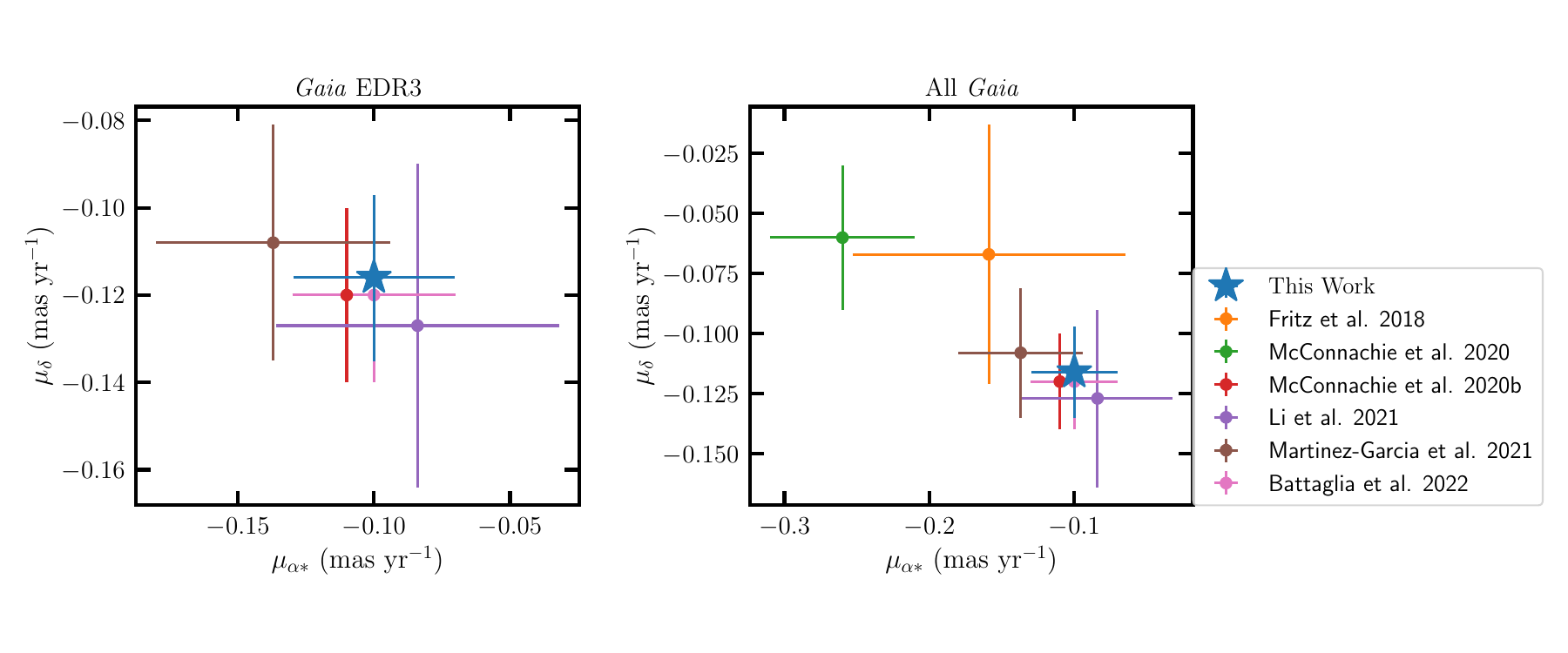}
\caption{Same as Figure~\ref{fig:comparsion} but for Canes Venatici I.}
\end{figure*}

\begin{figure*}
\includegraphics[height=4.5cm]{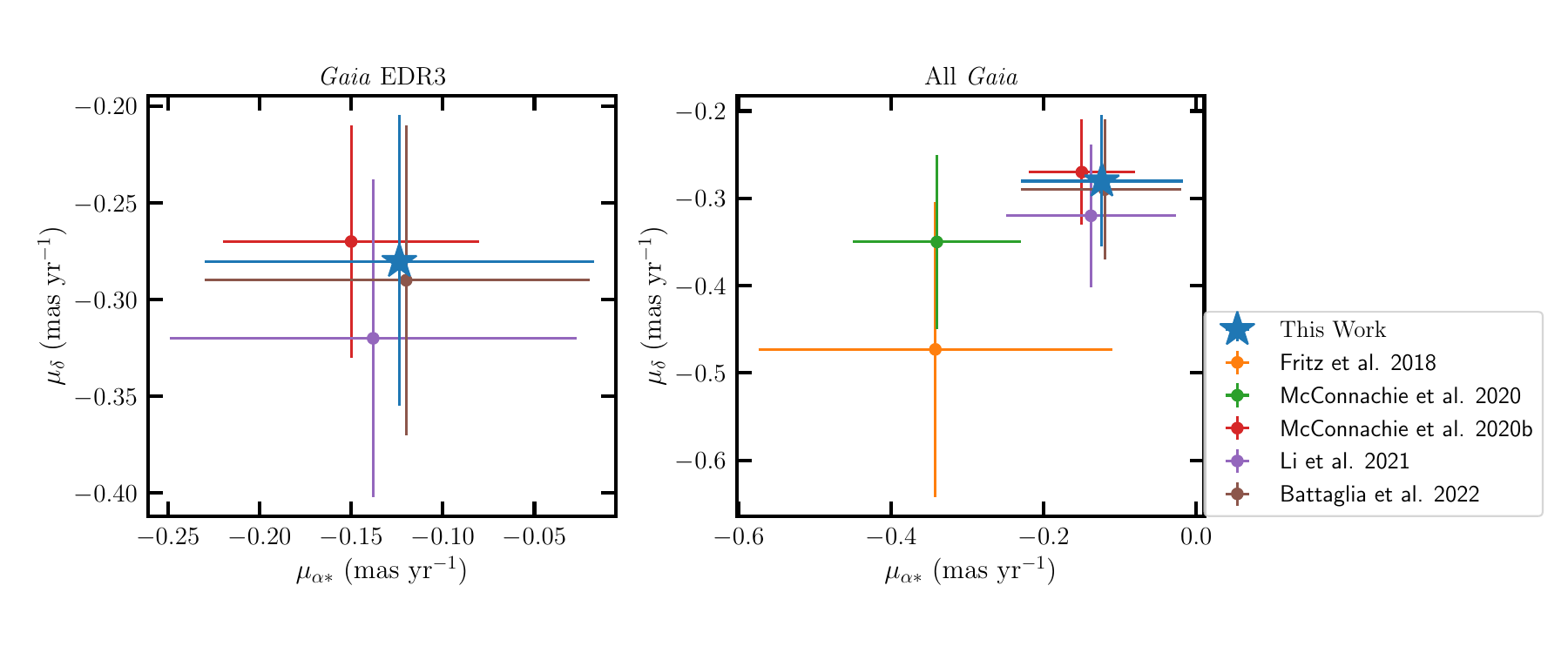}
\caption{Same as Figure~\ref{fig:comparsion} but for Canes Venatici II.}
\end{figure*}


\begin{figure*}
\includegraphics[height=4.5cm]{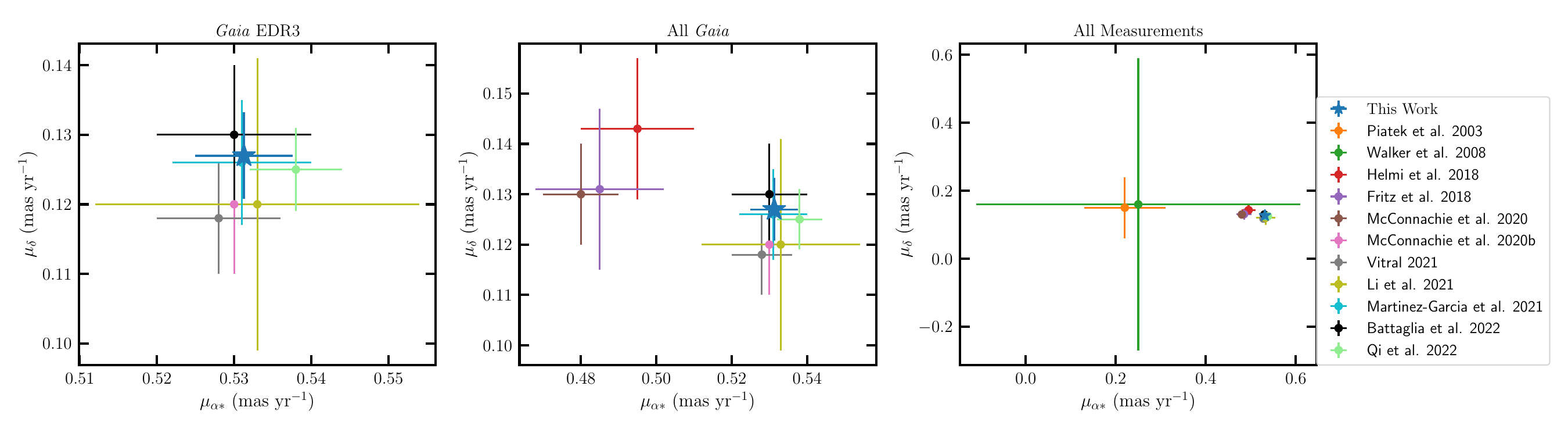}
\caption{Same as Figure~\ref{fig:comparsion} but for Carina.}
\end{figure*}

\begin{figure*}
\includegraphics[height=4.5cm]{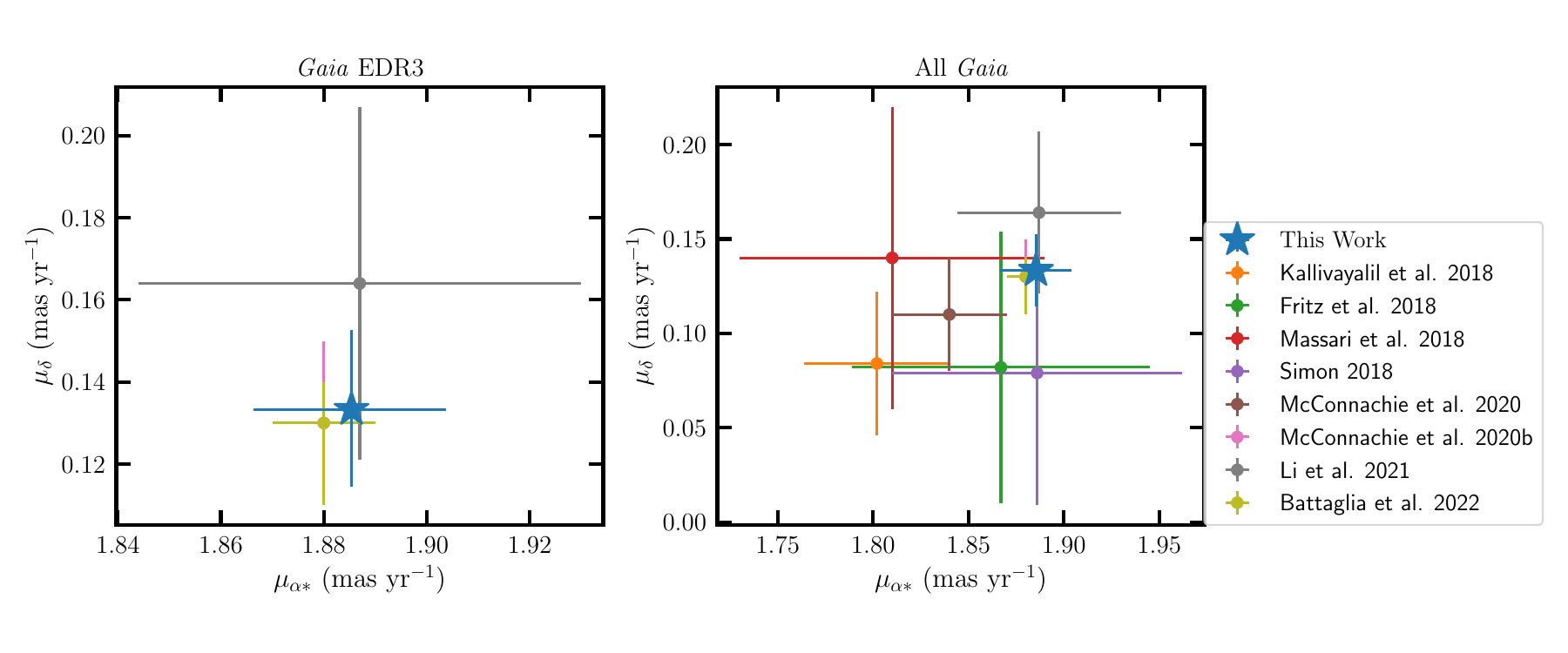}
\caption{Same as Figure~\ref{fig:comparsion} but for Carina II.}
\end{figure*}

\begin{figure*}
\includegraphics[height=4.5cm]{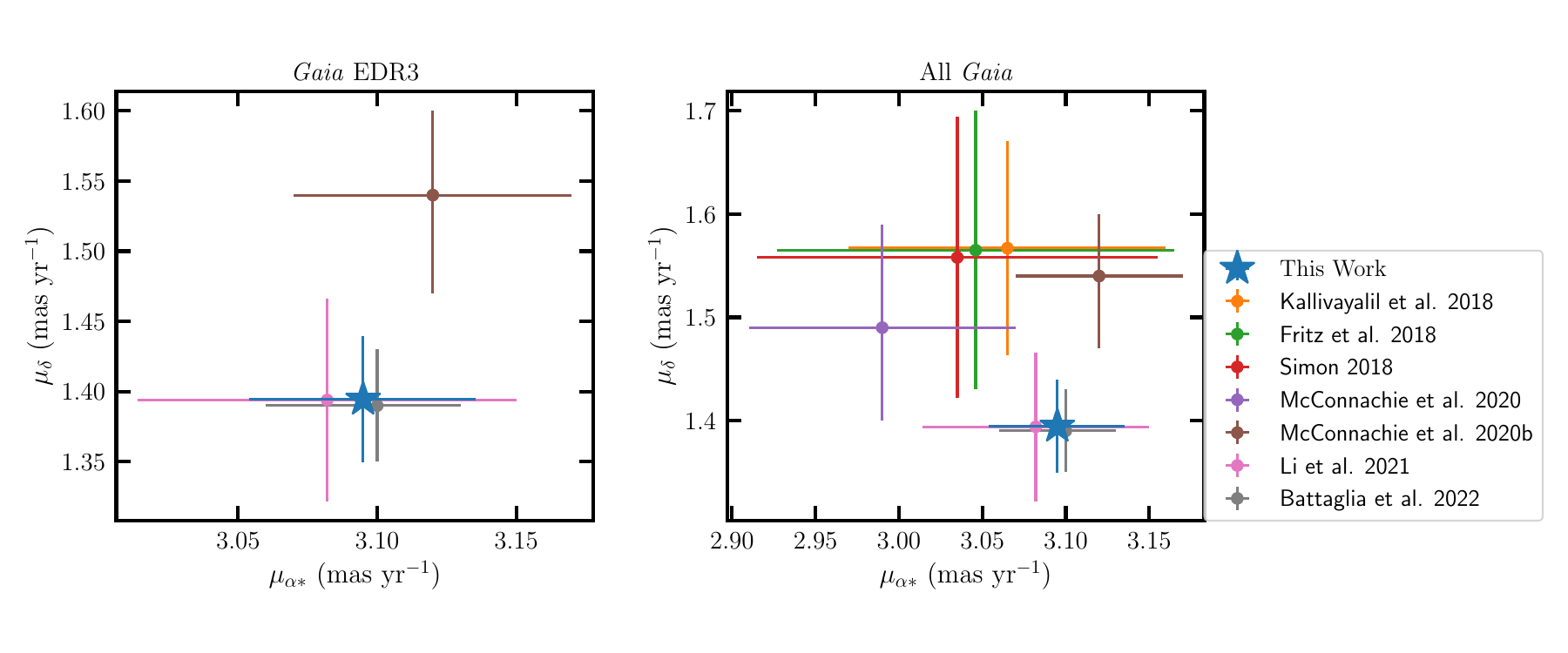}
\caption{Same as Figure~\ref{fig:comparsion} but for Carina III.}
\end{figure*}

\begin{figure*}
\includegraphics[height=4.5cm]{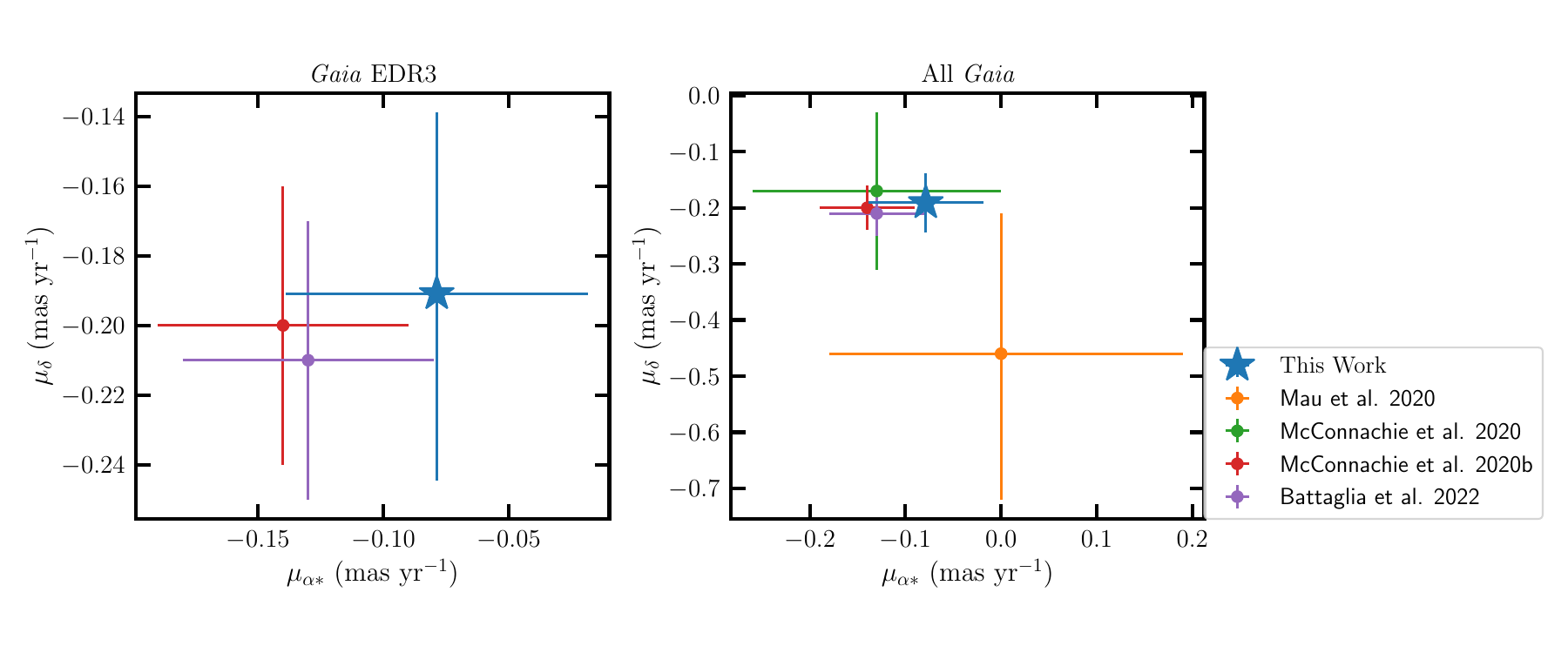}
\caption{Same as Figure~\ref{fig:comparsion} but for Centaurus I.}
\end{figure*}


\begin{figure*}
\includegraphics[height=4.5cm]{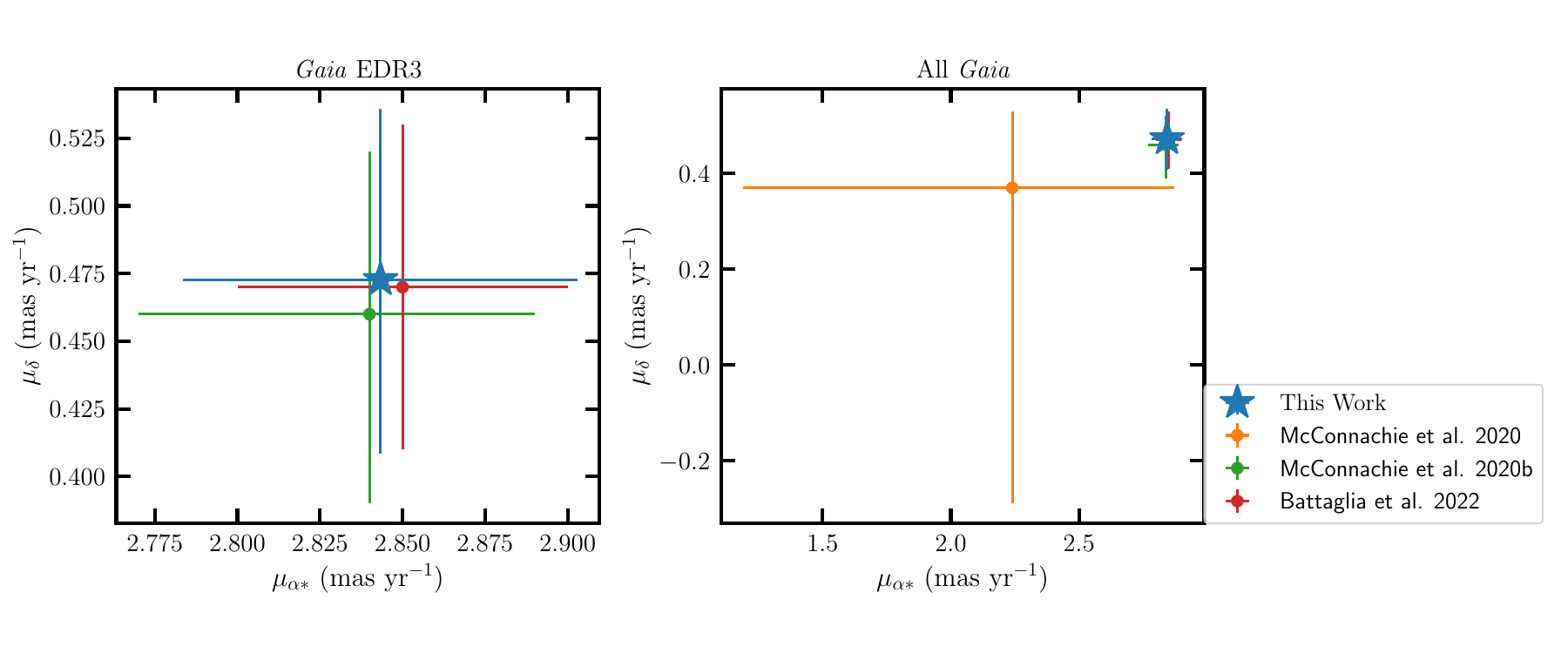}
\caption{Same as Figure~\ref{fig:comparsion} but for Cetus II.}
\end{figure*}

\begin{figure*}
\includegraphics[height=4.5cm]{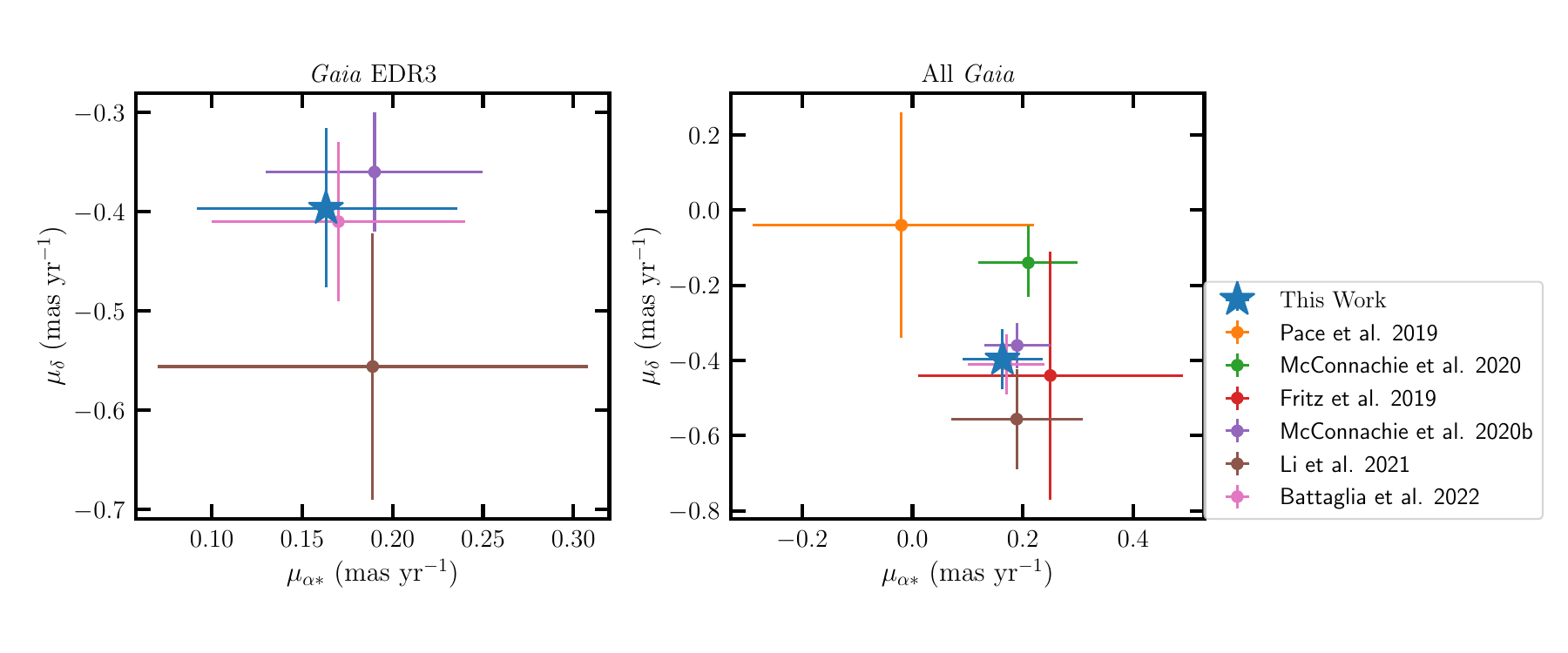}
\caption{Same as Figure~\ref{fig:comparsion} but for Columba I.}
\end{figure*}

\begin{figure*}
\includegraphics[height=4.5cm]{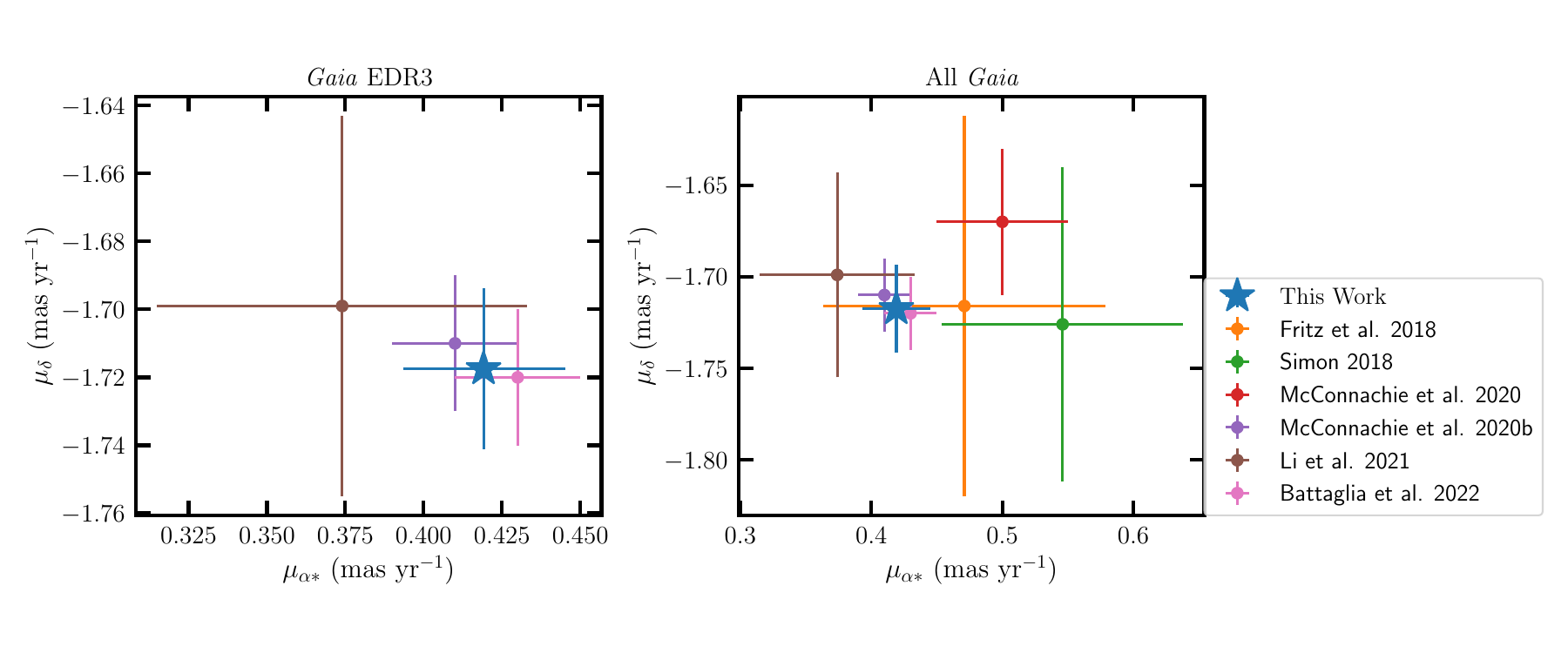}
\caption{Same as Figure~\ref{fig:comparsion} but for Coma Berenices.}
\end{figure*}


\begin{figure*}
\includegraphics[height=4.5cm]{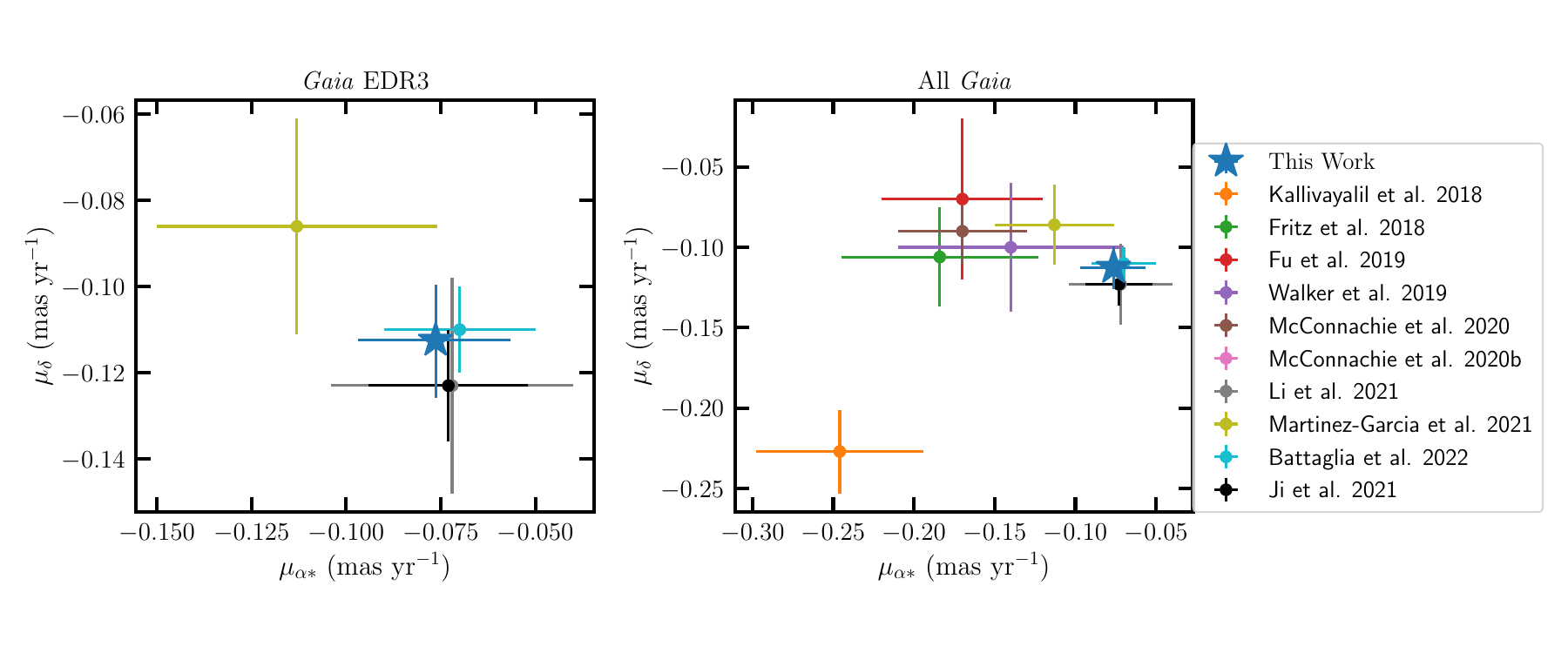}
\caption{Same as Figure~\ref{fig:comparsion} but for Crater II.}
\end{figure*}

\begin{figure*}
\includegraphics[height=4.5cm]{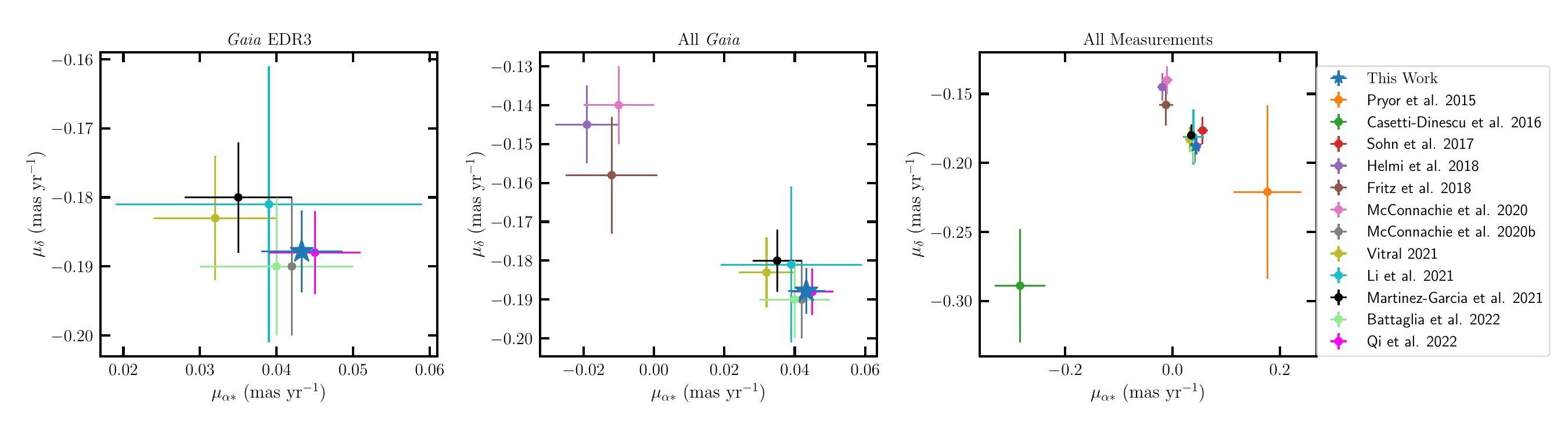}
\caption{Same as Figure~\ref{fig:comparsion} but for Draco.}
\end{figure*}

\begin{figure*}
\includegraphics[height=4.5cm]{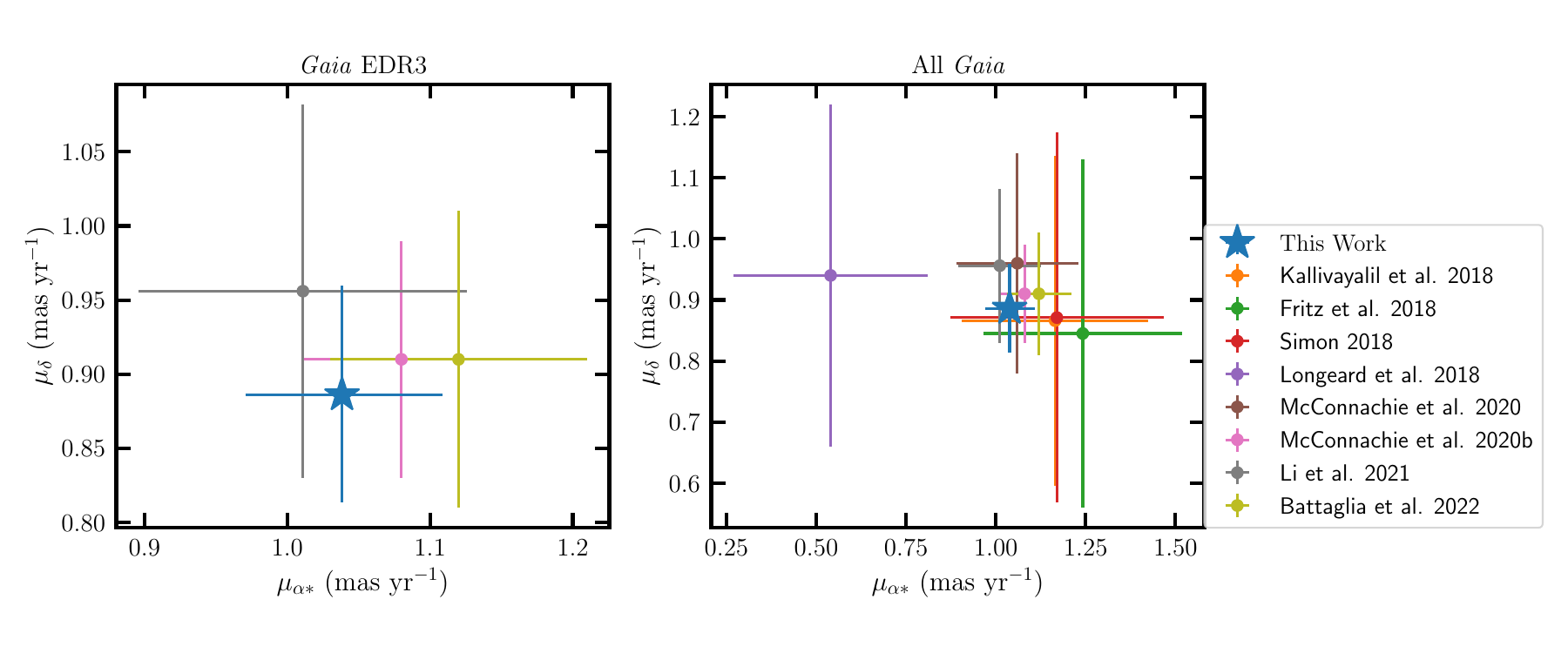}
\caption{Same as Figure~\ref{fig:comparsion} but for Draco II.}
\end{figure*}

\begin{figure*}
\includegraphics[height=4.5cm]{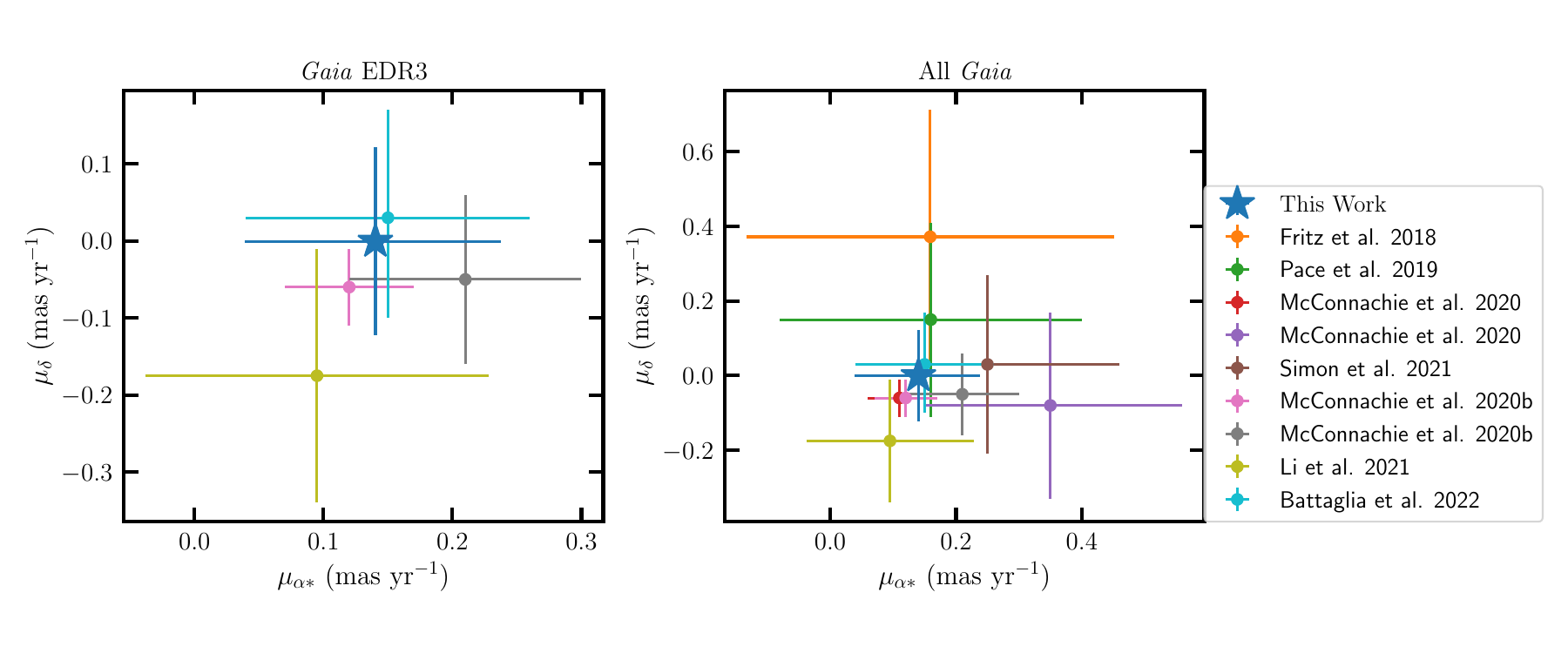}
\caption{Same as Figure~\ref{fig:comparsion} but for Eridanus II.}
\end{figure*}


\begin{figure*}
\includegraphics[height=4.5cm]{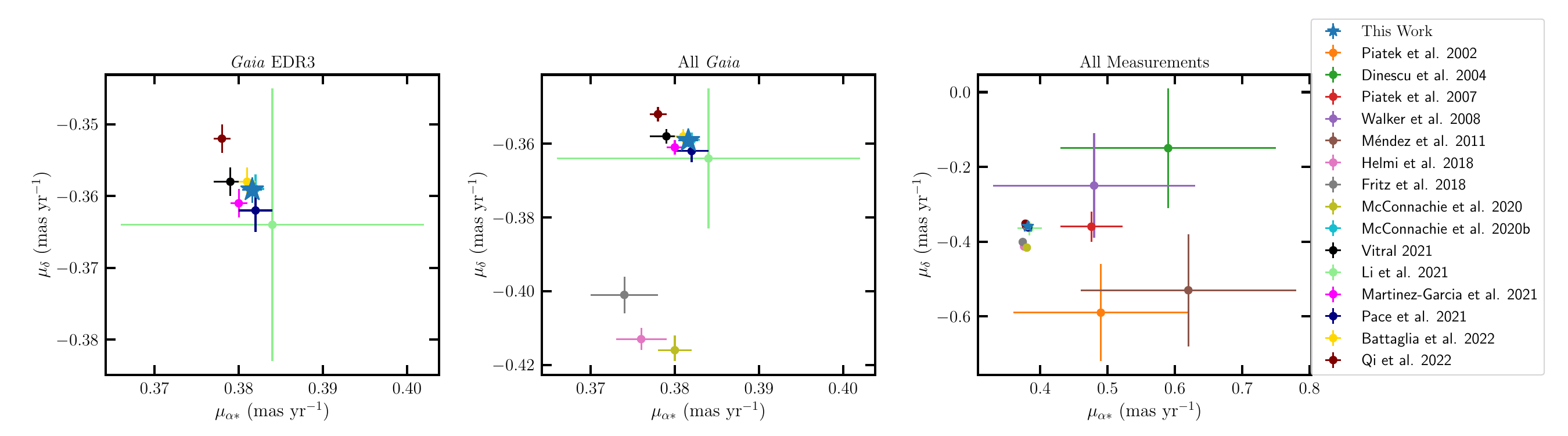}
\caption{Same as Figure~\ref{fig:comparsion} but for Fornax.}
\end{figure*}

\begin{figure*}
\includegraphics[height=4.5cm]{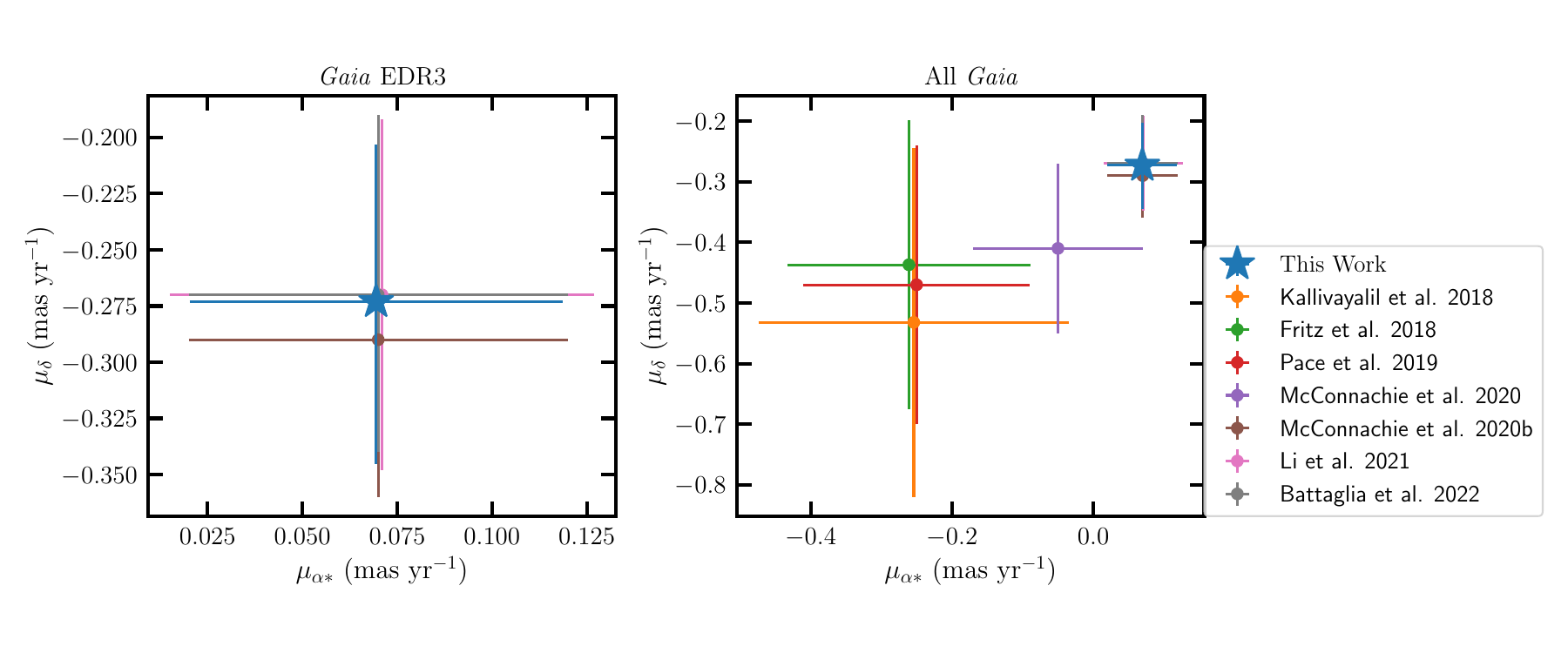}
\caption{Same as Figure~\ref{fig:comparsion} but for Grus I.}
\end{figure*}

\begin{figure*}
\includegraphics[height=4.5cm]{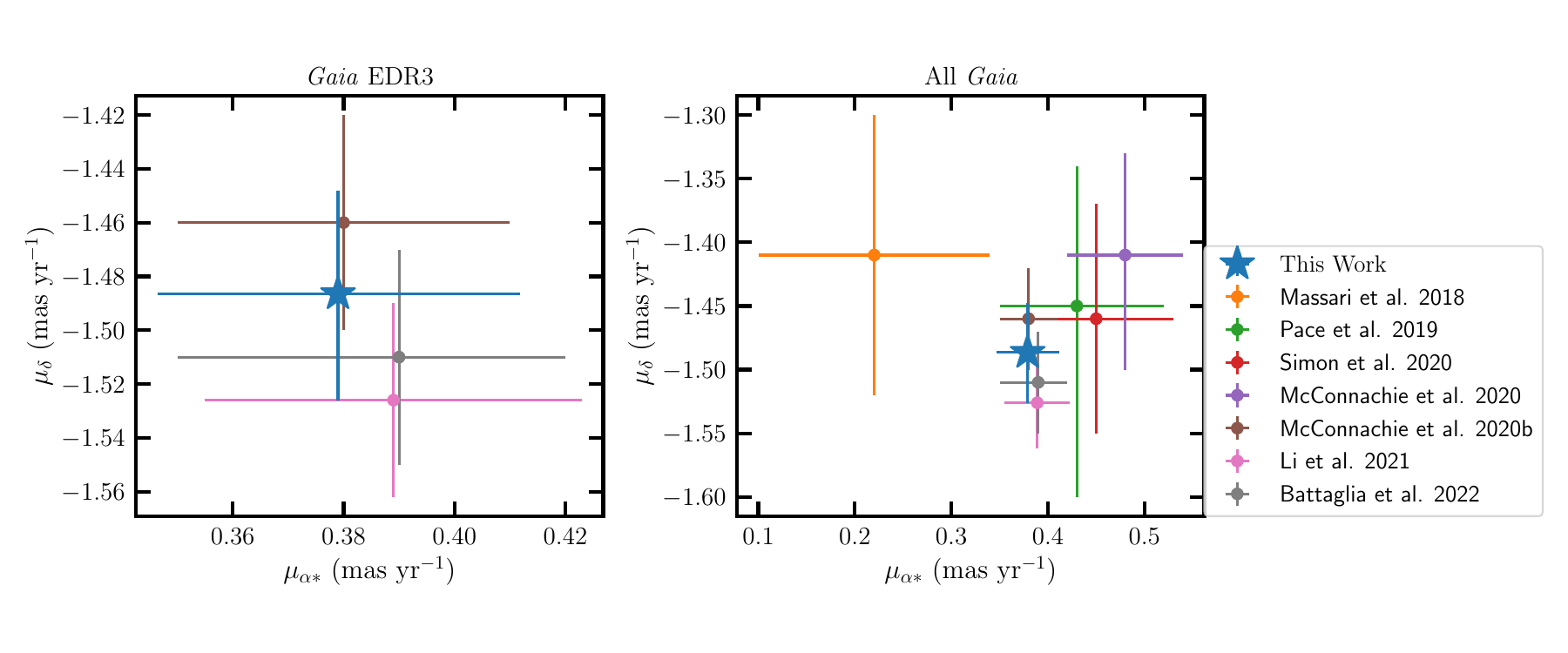}
\caption{Same as Figure~\ref{fig:comparsion} but for Grus II.}
\end{figure*}

\begin{figure*}
\includegraphics[height=4.5cm]{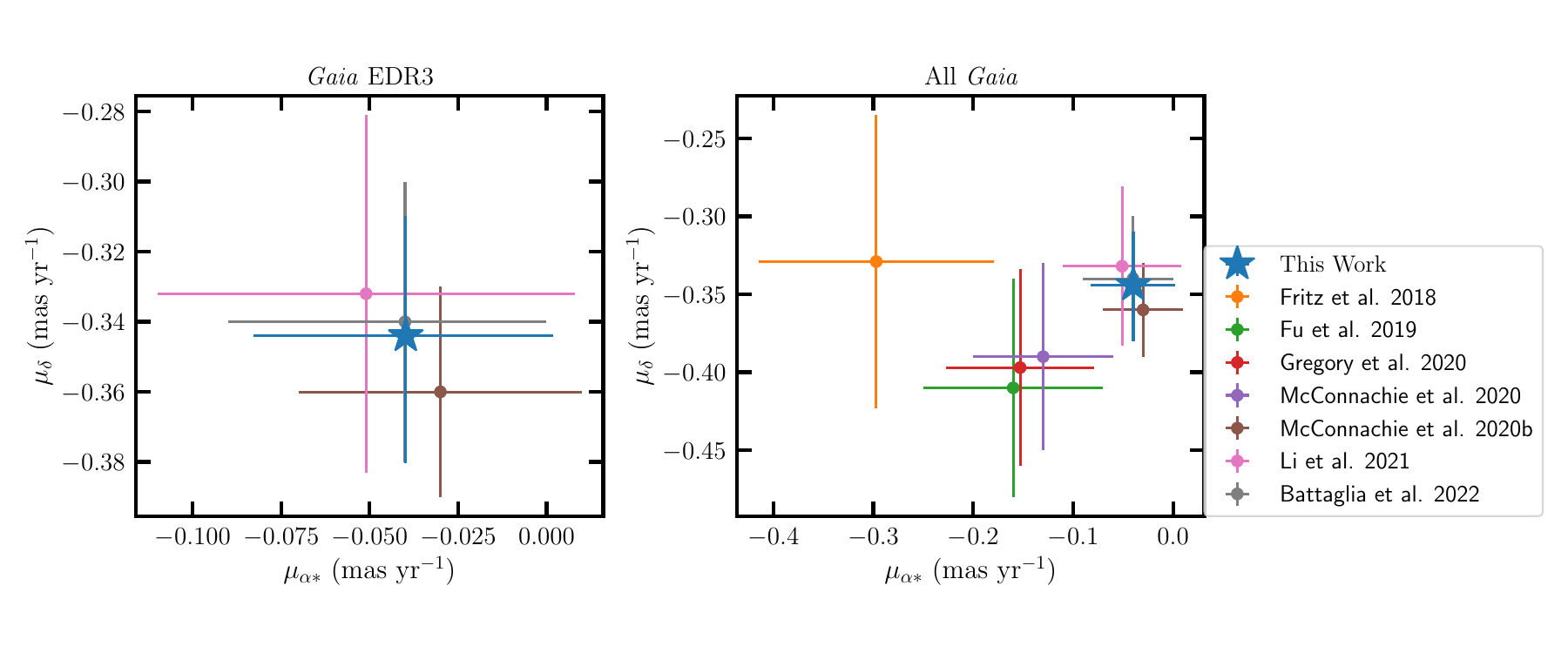}
\caption{Same as Figure~\ref{fig:comparsion} but for Hercules.}
\end{figure*}


\begin{figure*}
\includegraphics[height=4.5cm]{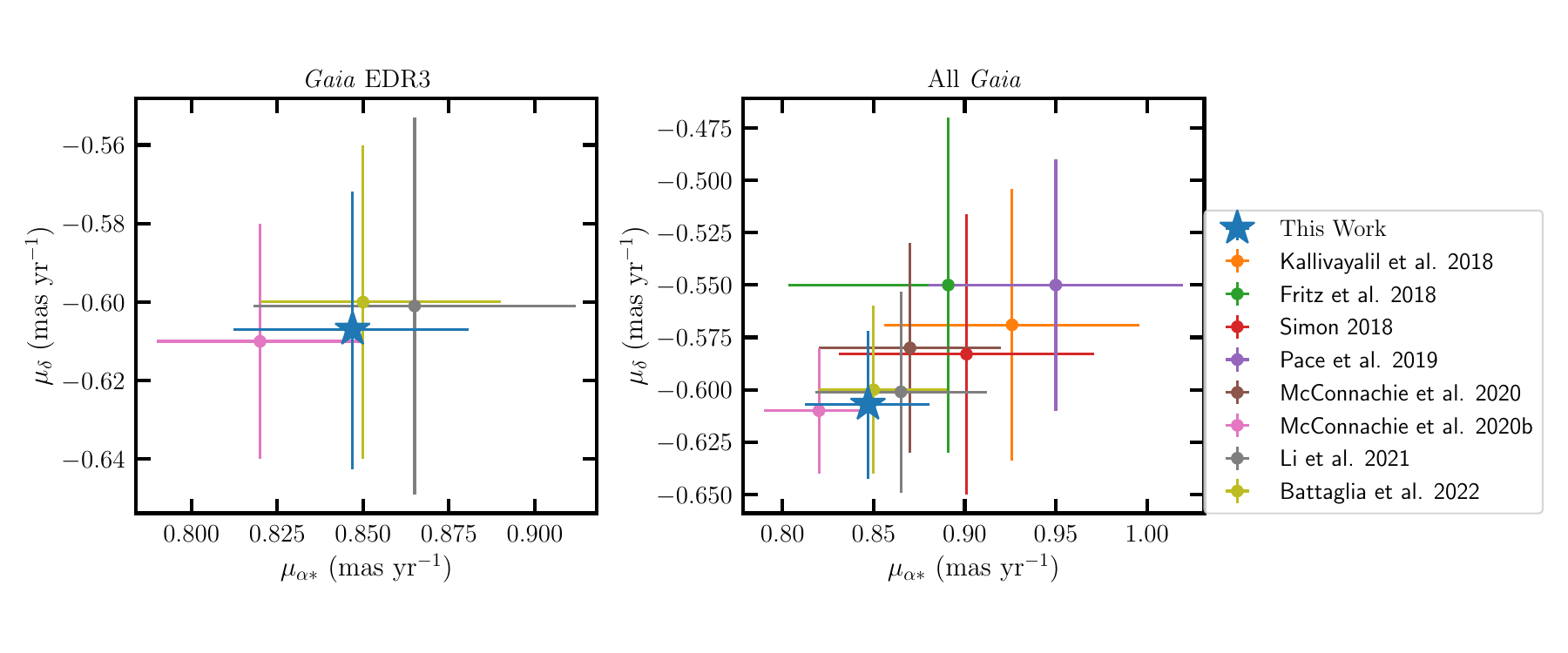}
\caption{Same as Figure~\ref{fig:comparsion} but for Horologium I.}
\end{figure*}

\begin{figure*}
\includegraphics[height=4.5cm]{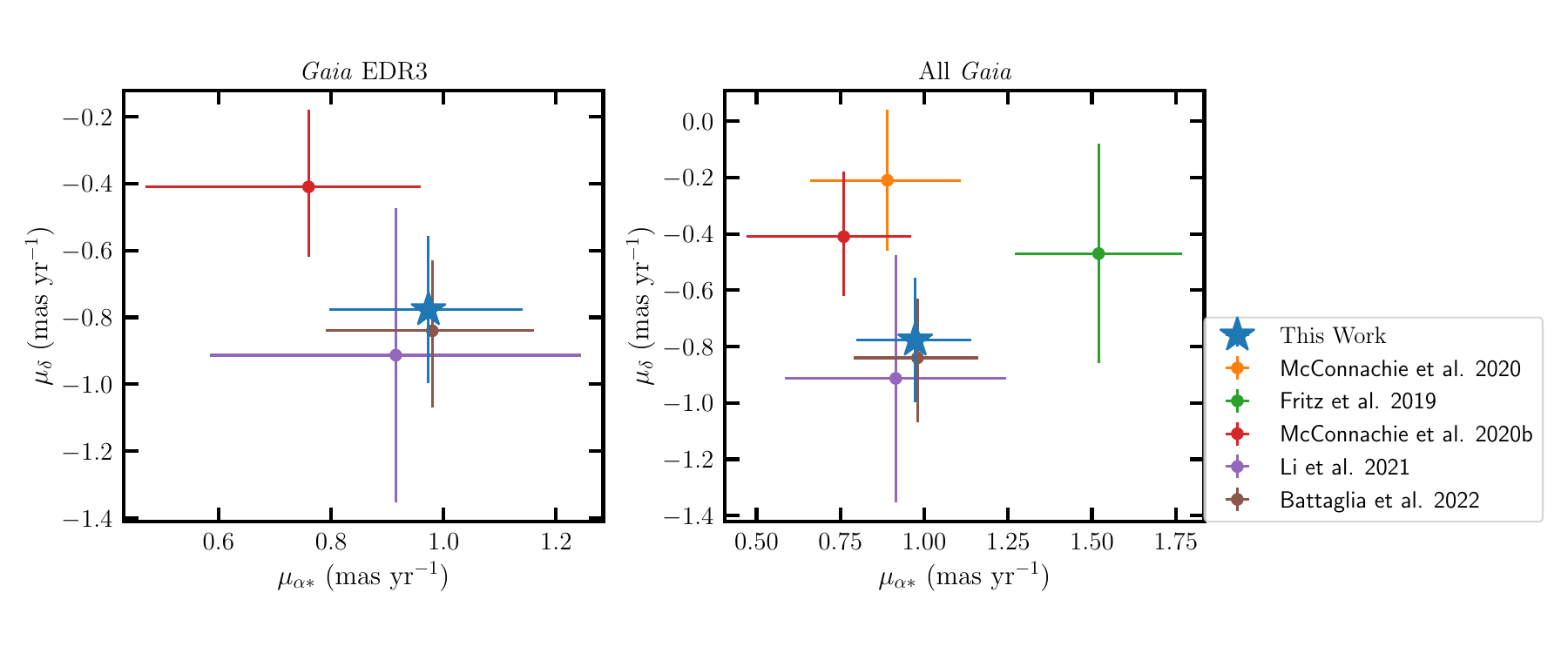}
\caption{Same as Figure~\ref{fig:comparsion} but for Horologium II.}
\end{figure*}

\begin{figure*}
\includegraphics[height=4.5cm]{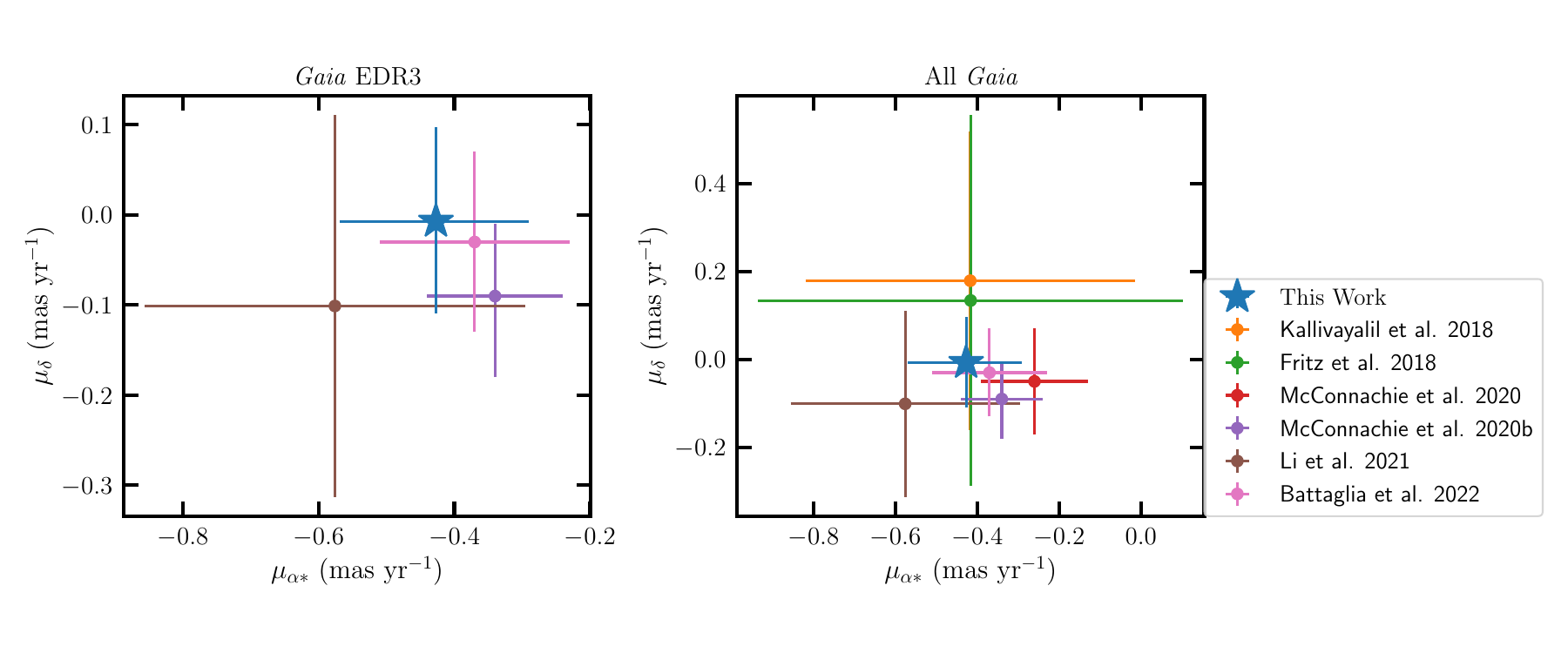}
\caption{Same as Figure~\ref{fig:comparsion} but for Hydra II.}
\end{figure*}

\begin{figure*}
\includegraphics[height=4.5cm]{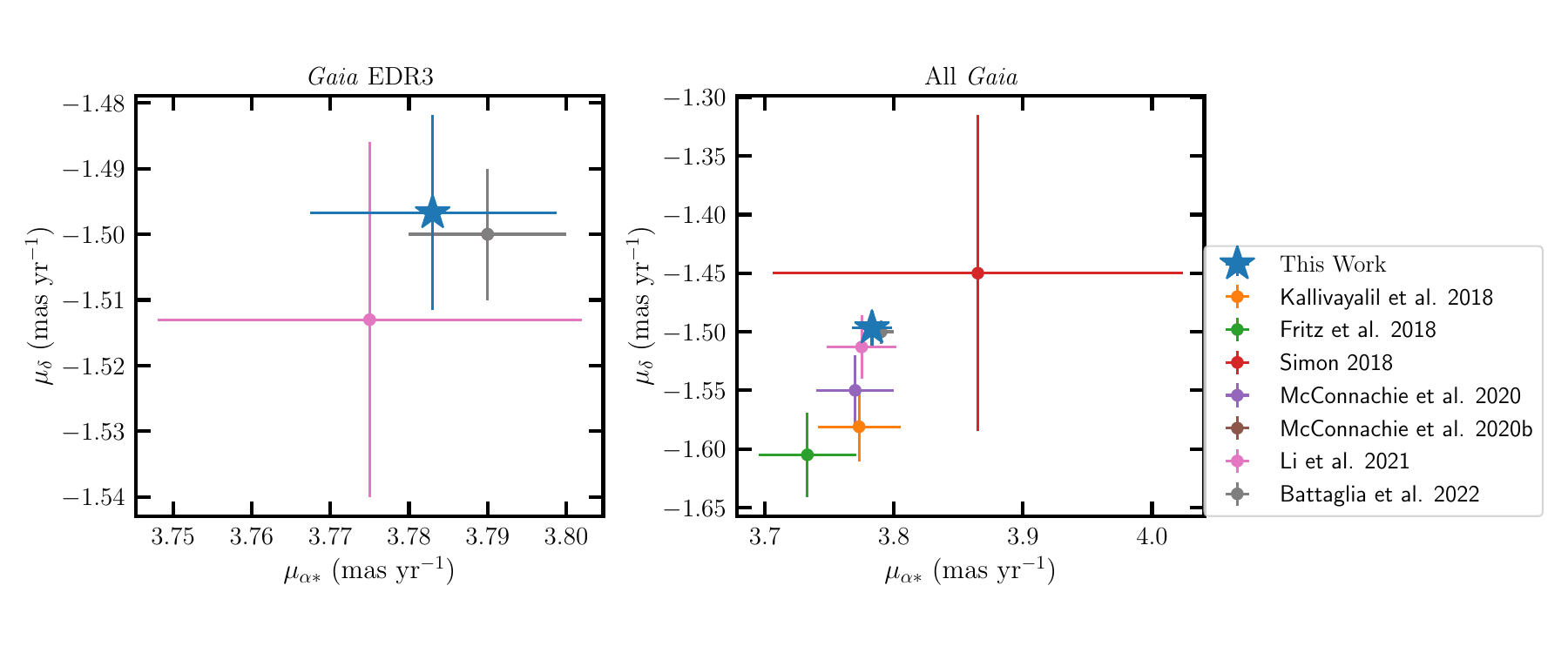}
\caption{Same as Figure~\ref{fig:comparsion} but for Hydrus I.}
\end{figure*}


\begin{figure*}
\includegraphics[height=4.5cm]{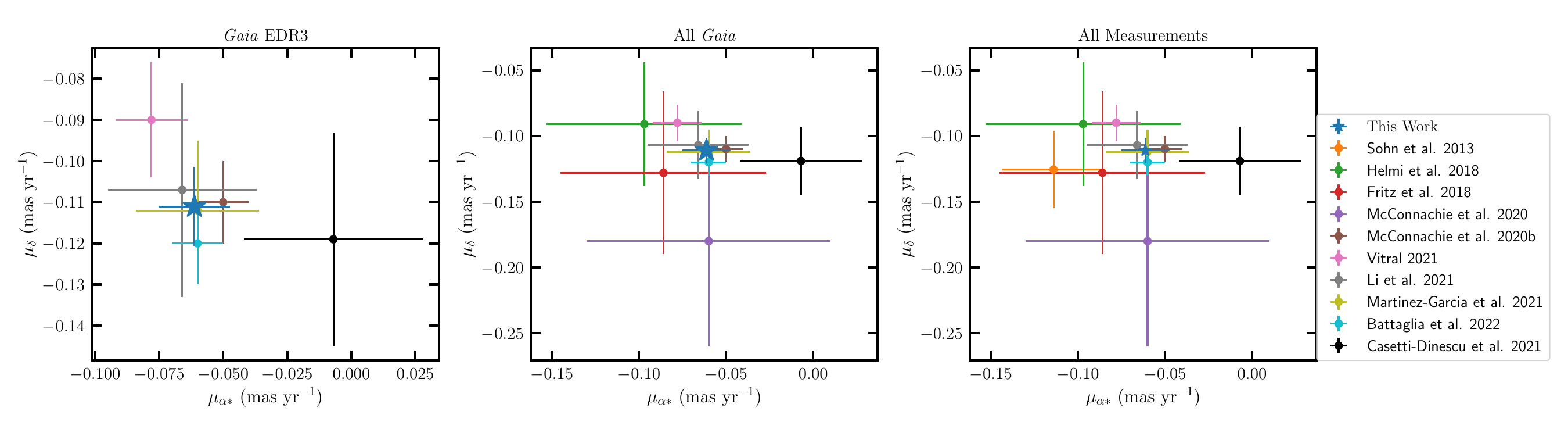}
\caption{Same as Figure~\ref{fig:comparsion} but for Leo I.}
\end{figure*}

\begin{figure*}
\includegraphics[height=4.5cm]{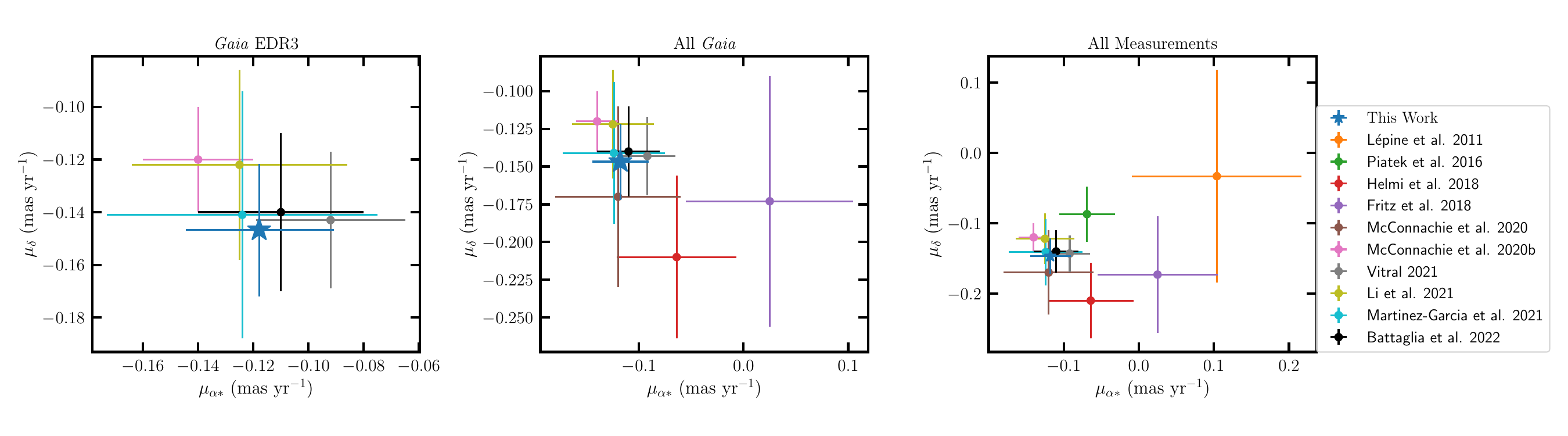}
\caption{Same as Figure~\ref{fig:comparsion} but for Leo II.}
\end{figure*}

\begin{figure*}
\includegraphics[height=4.5cm]{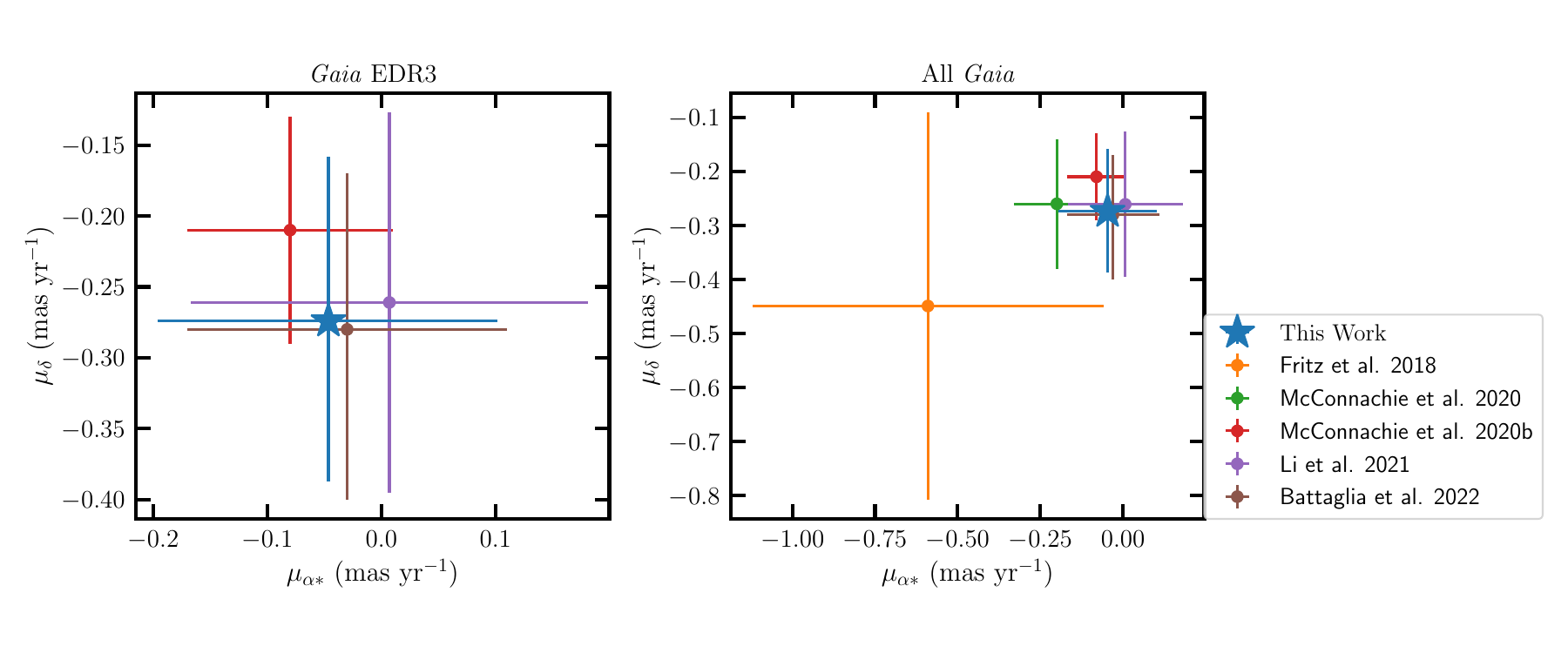}
\caption{Same as Figure~\ref{fig:comparsion} but for Leo IV.}
\end{figure*}

\begin{figure*}
\includegraphics[height=4.5cm]{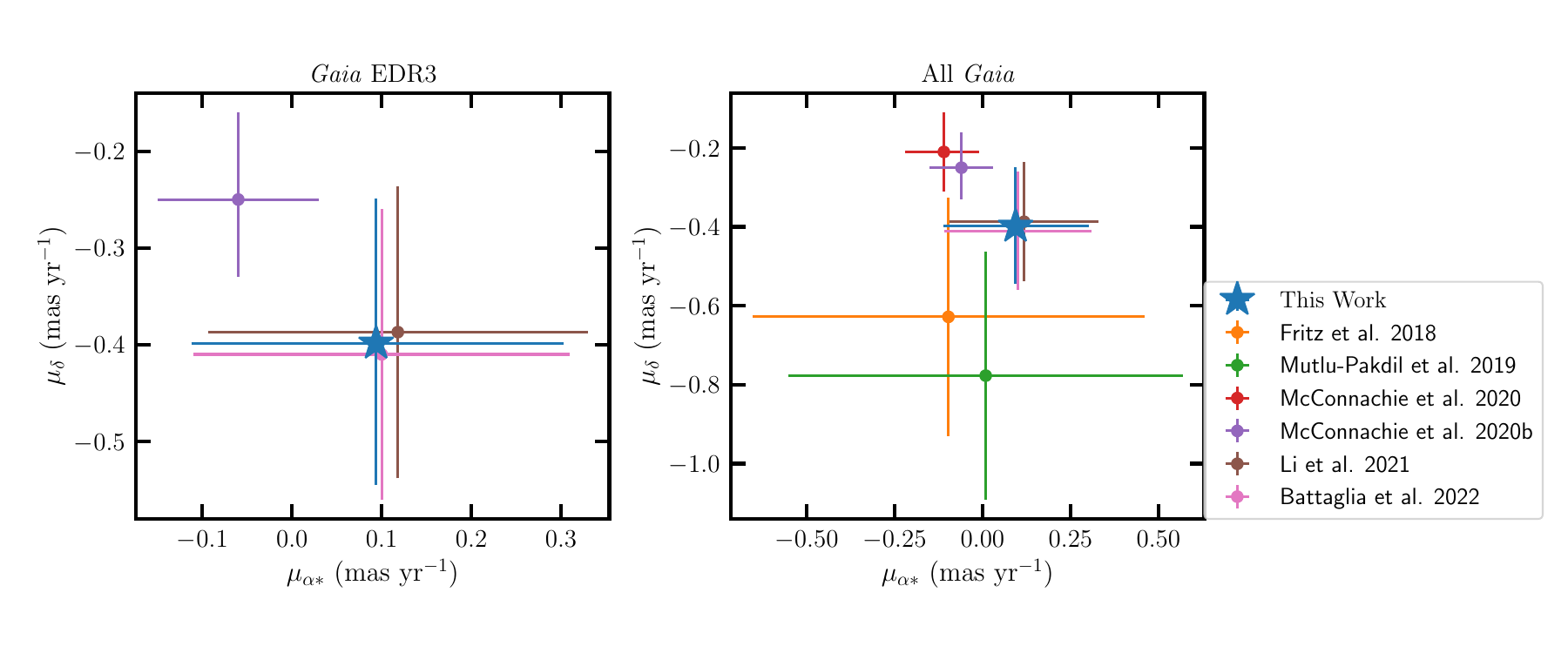}
\caption{Same as Figure~\ref{fig:comparsion} but for Leo V.}
\end{figure*}

\begin{figure*}
\includegraphics[height=4.5cm]{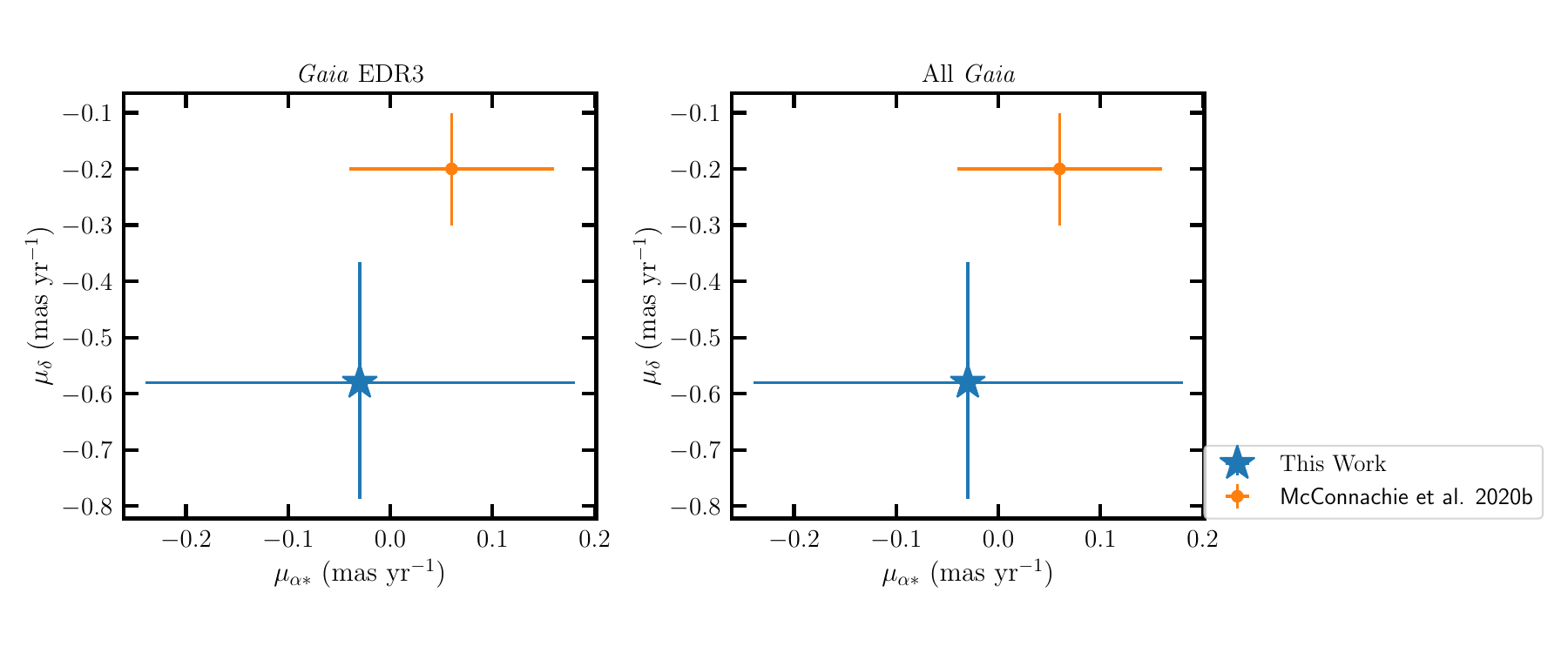}
\caption{Same as Figure~\ref{fig:comparsion} but for Pegasus III.}
\end{figure*}

\begin{figure*}
\includegraphics[height=4.5cm]{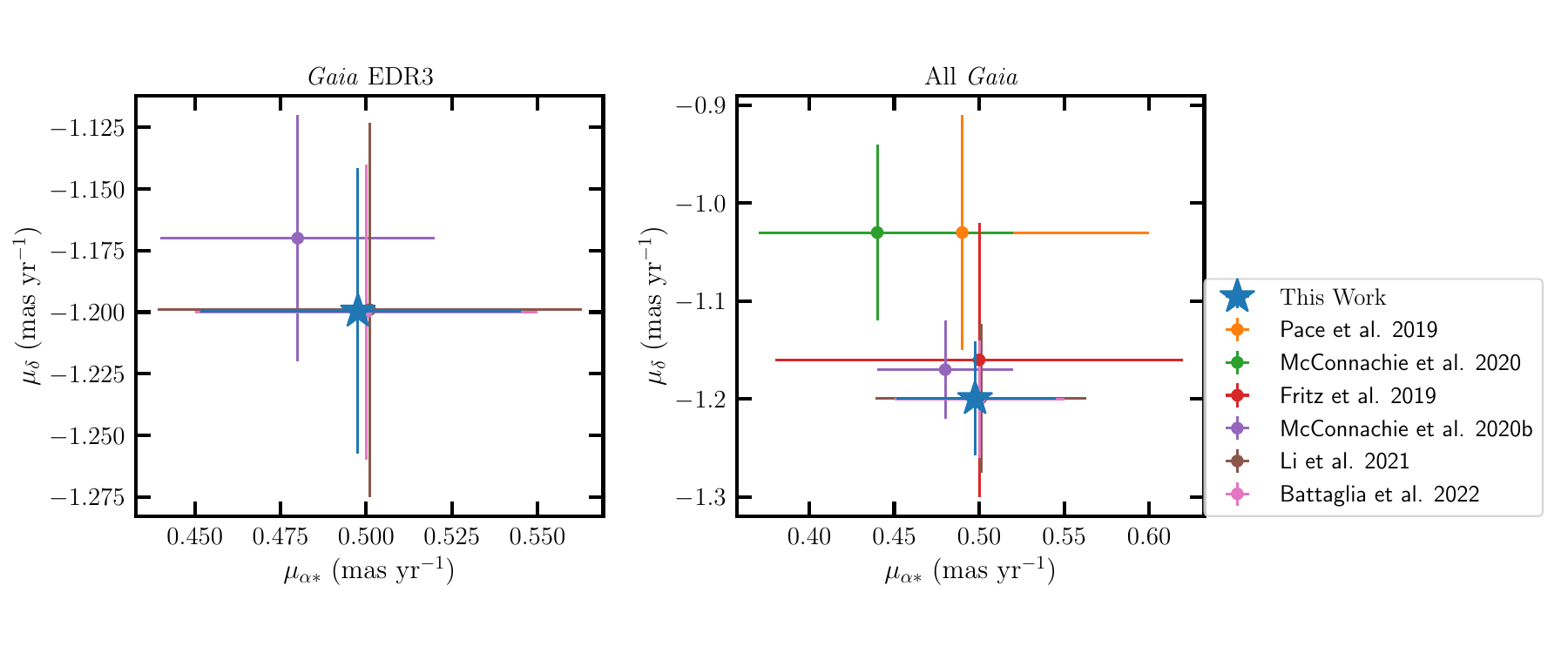}
\caption{Same as Figure~\ref{fig:comparsion} but for Phoenix II.}
\end{figure*}

\begin{figure*}
\includegraphics[height=4.5cm]{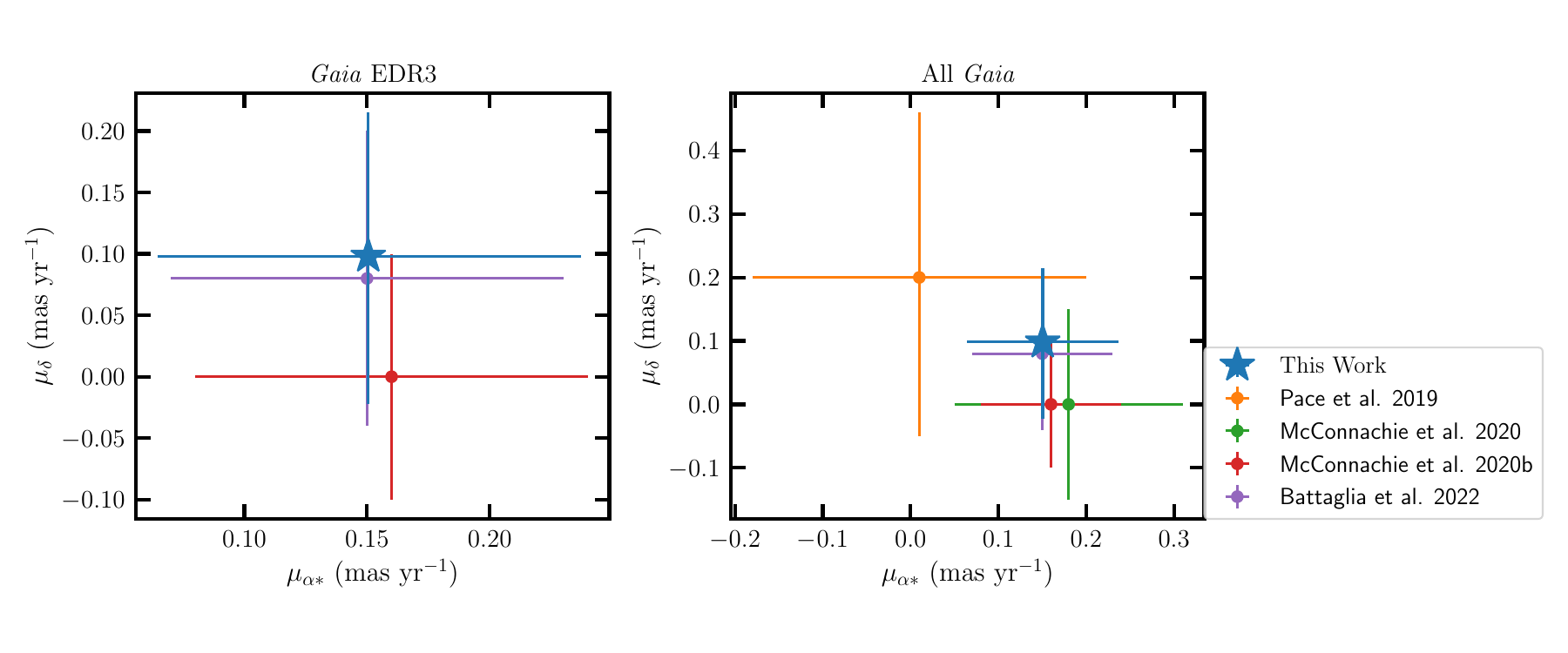}
\caption{Same as Figure~\ref{fig:comparsion} but for Pictor I.}
\end{figure*}

\begin{figure*}
\includegraphics[height=4.5cm]{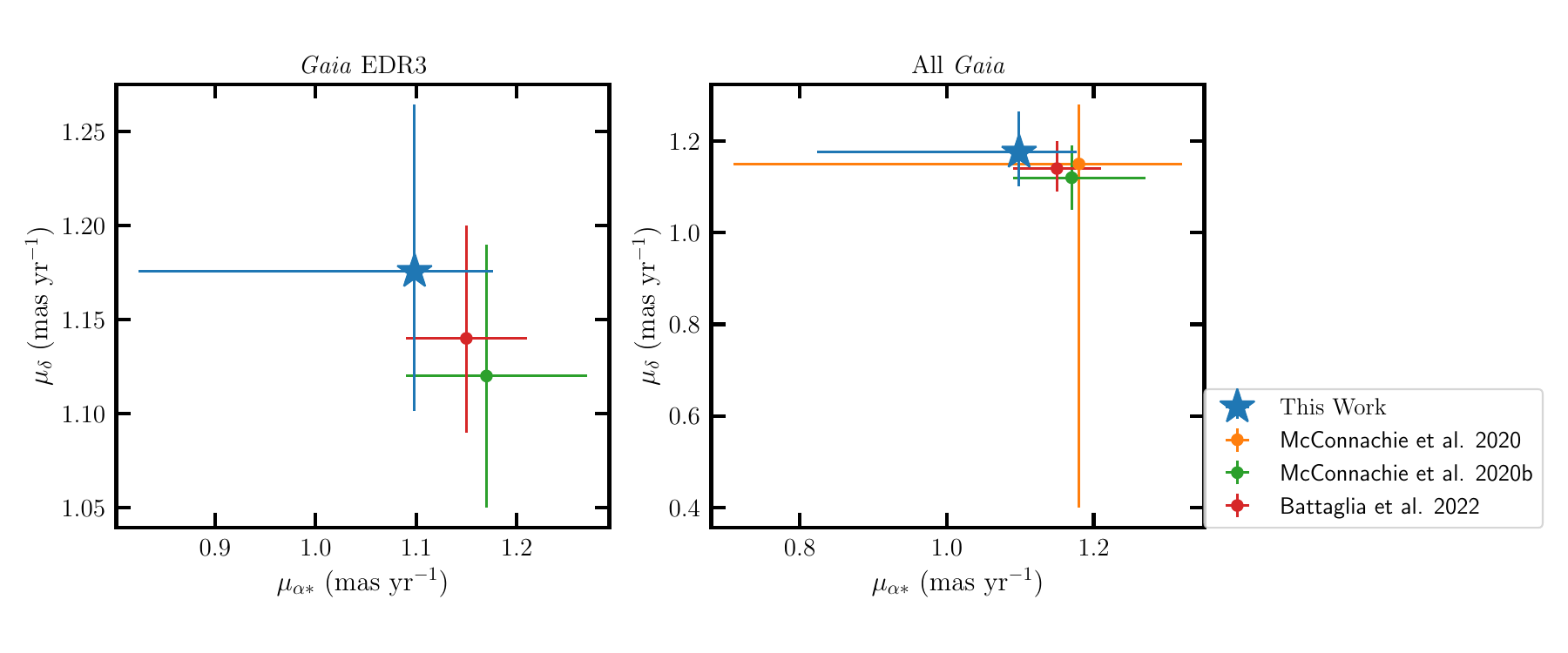}
\caption{Same as Figure~\ref{fig:comparsion} but for Pictor II.}
\end{figure*}


\begin{figure*}
\includegraphics[height=4.5cm]{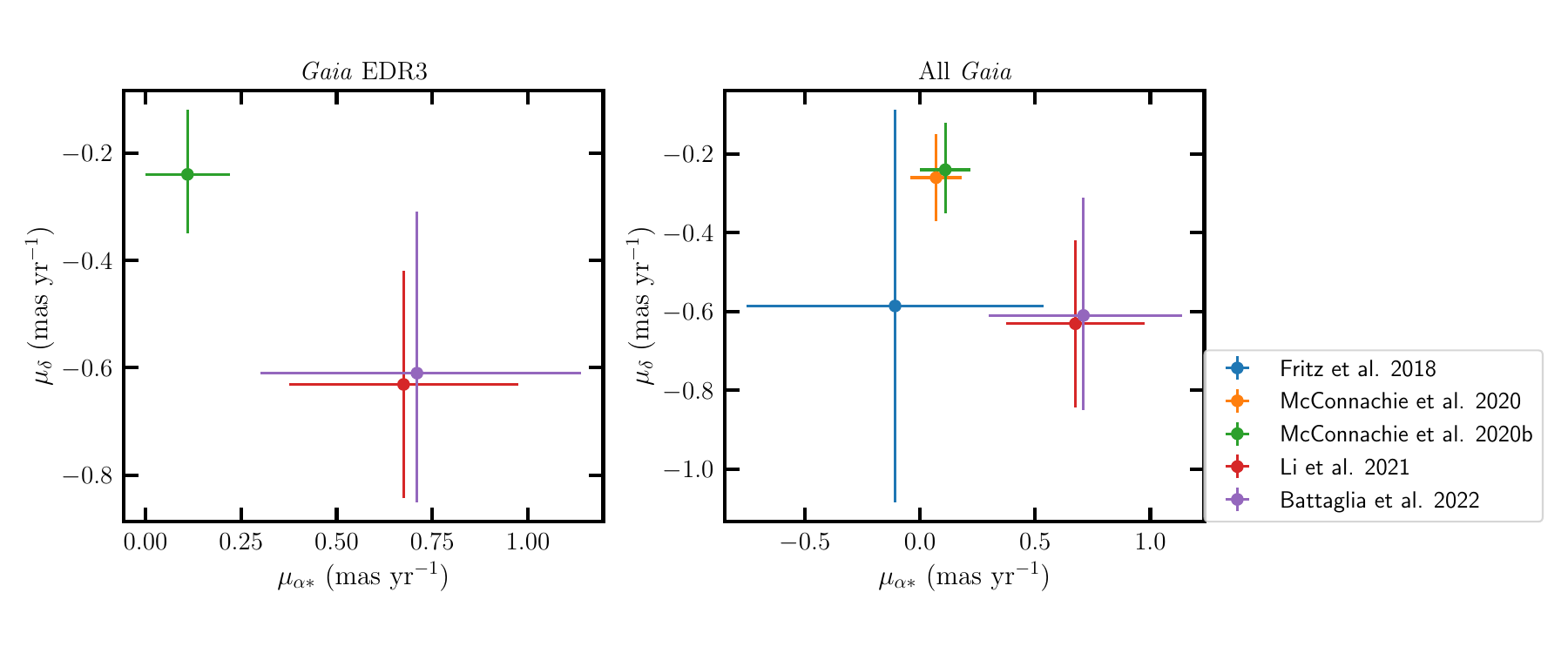}
\caption{Same as Figure~\ref{fig:comparsion} but for Pisces II.}
\end{figure*}

\begin{figure*}
\includegraphics[height=4.5cm]{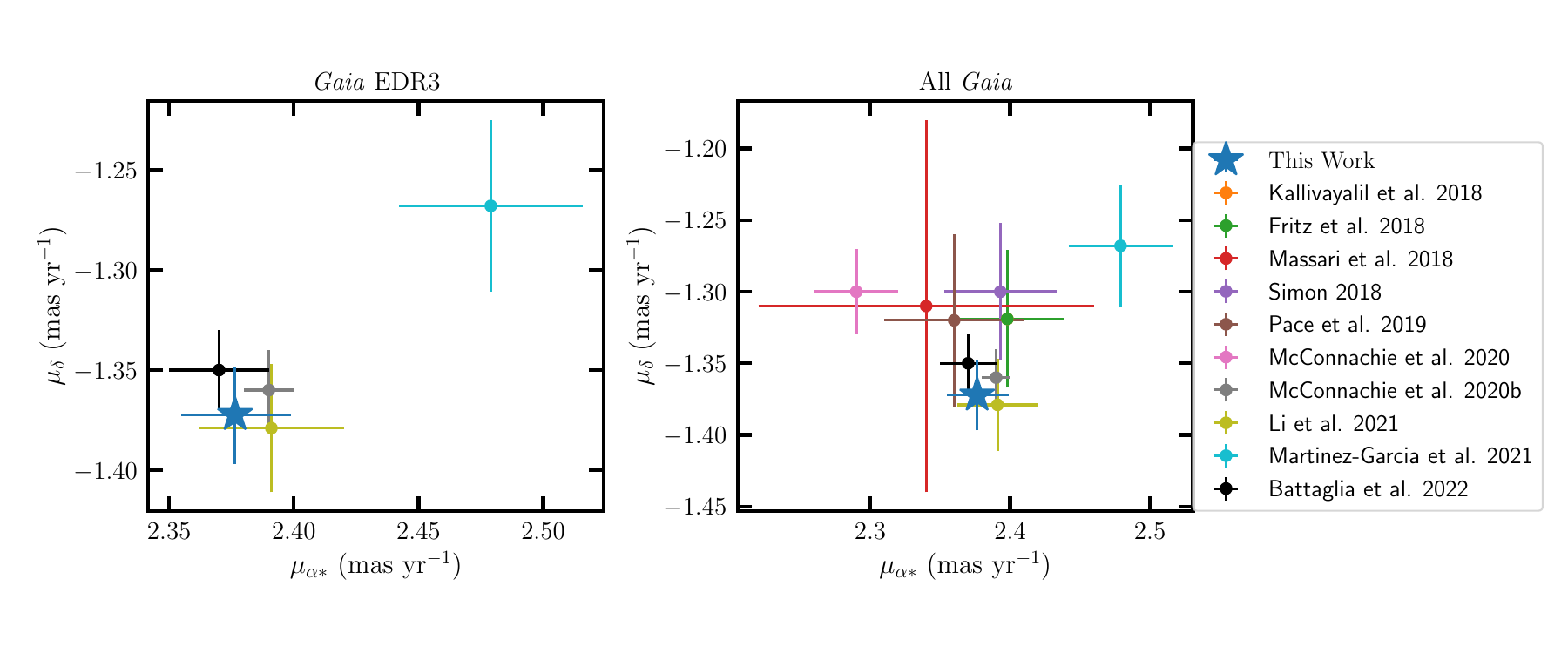}
\caption{Same as Figure~\ref{fig:comparsion} but for Reticulum II.}
\end{figure*}

\begin{figure*}
\includegraphics[height=4.5cm]{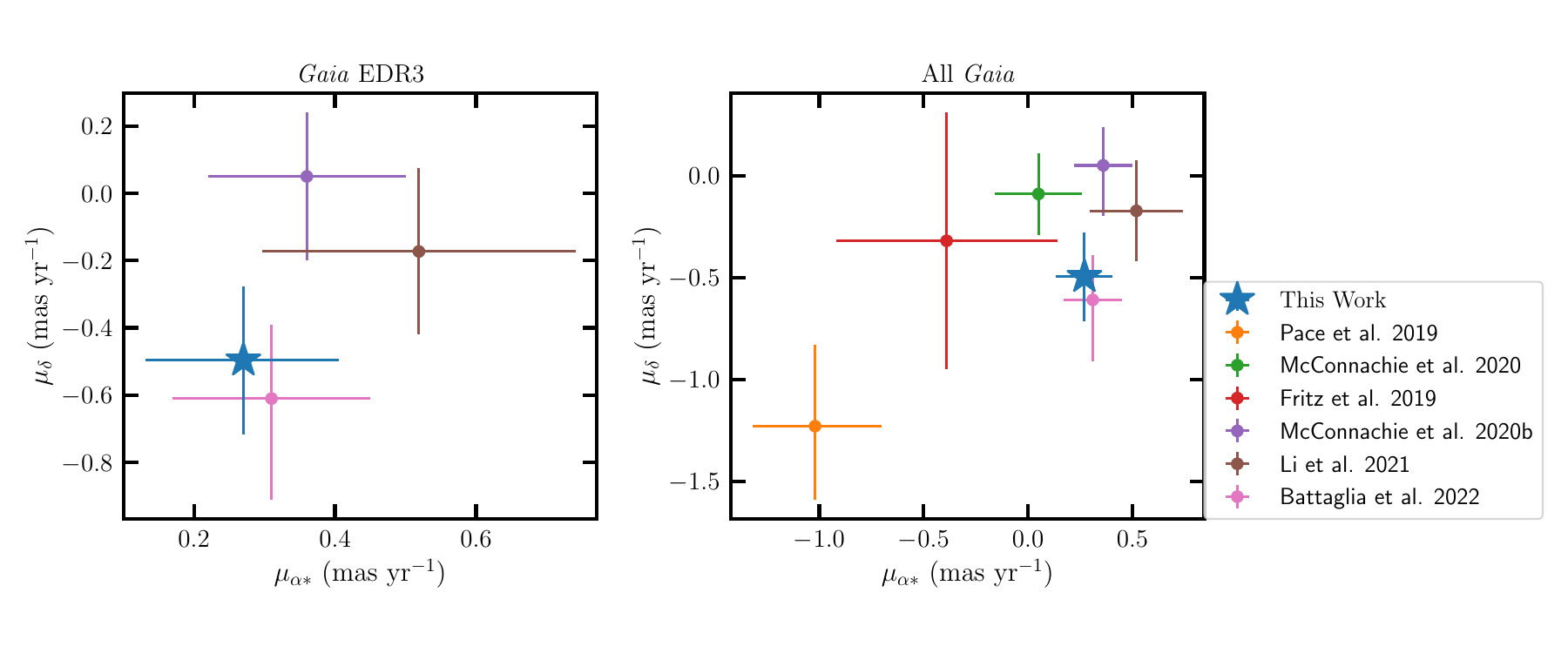}
\caption{Same as Figure~\ref{fig:comparsion} but for Reticulum III.}
\end{figure*}

\begin{figure*}
\includegraphics[height=4.5cm]{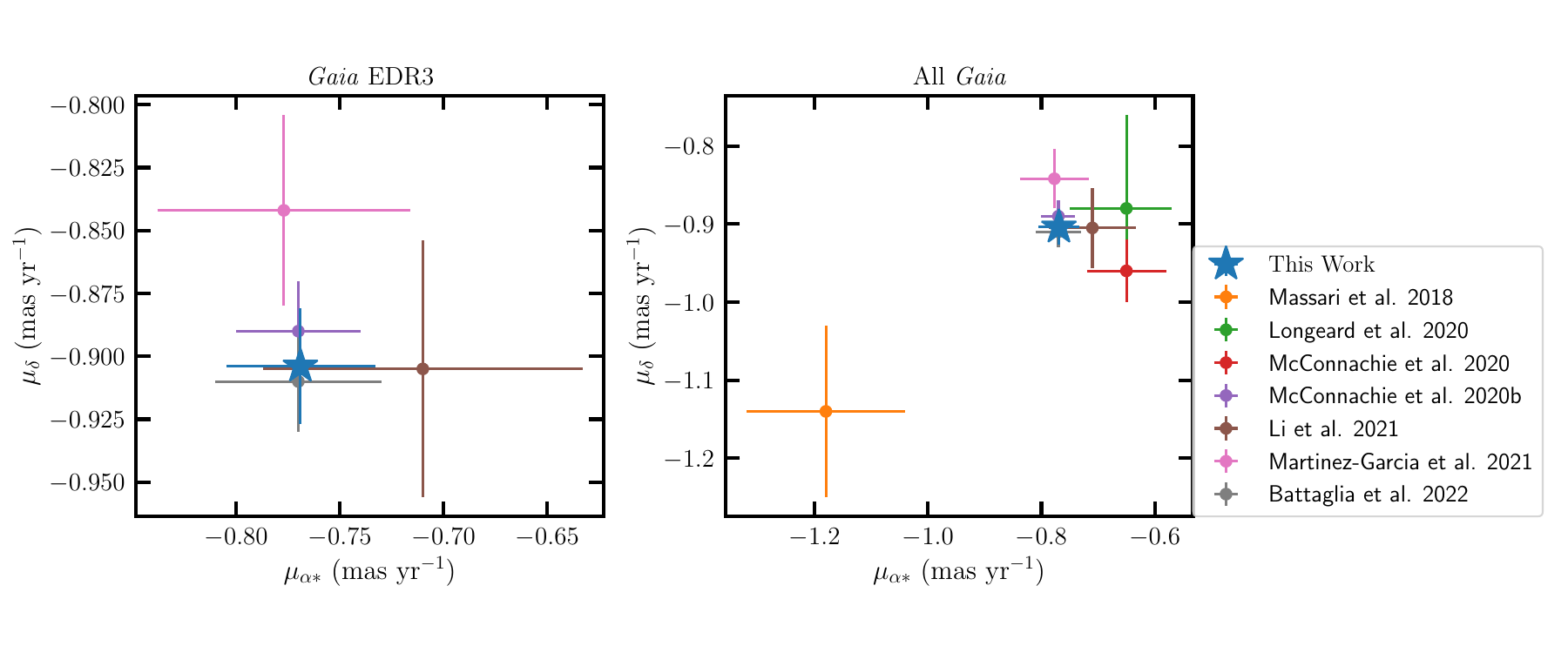}
\caption{Same as Figure~\ref{fig:comparsion} but for Sagittarius II.}
\end{figure*}


\begin{figure*}
\includegraphics[height=4.5cm]{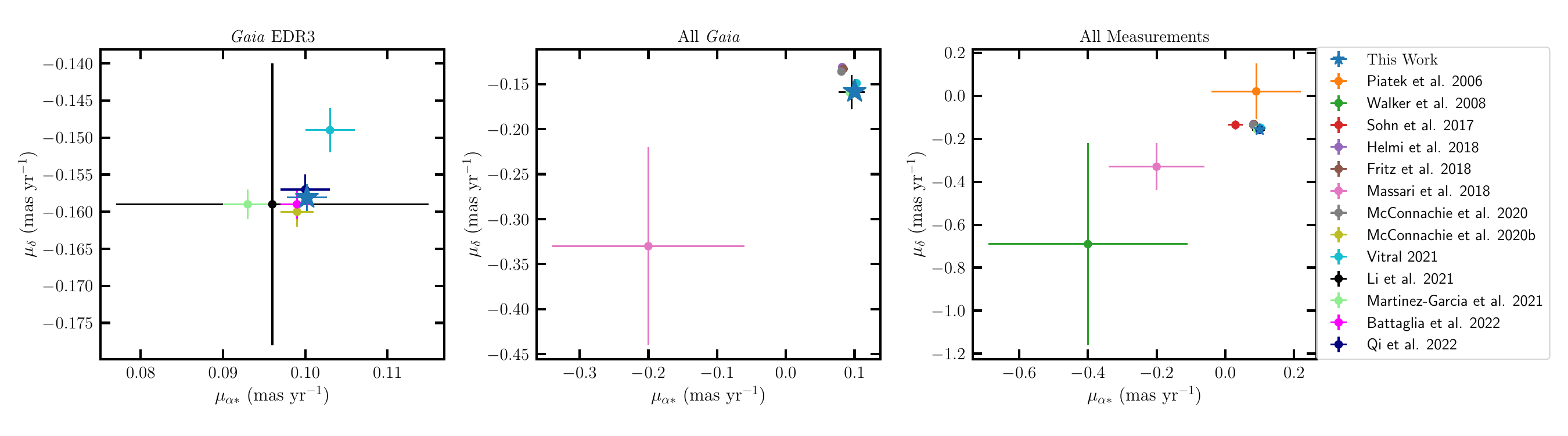}
\caption{Same as Figure~\ref{fig:comparsion} but for Sculptor.}
\end{figure*}

\begin{figure*}
\includegraphics[height=4.5cm]{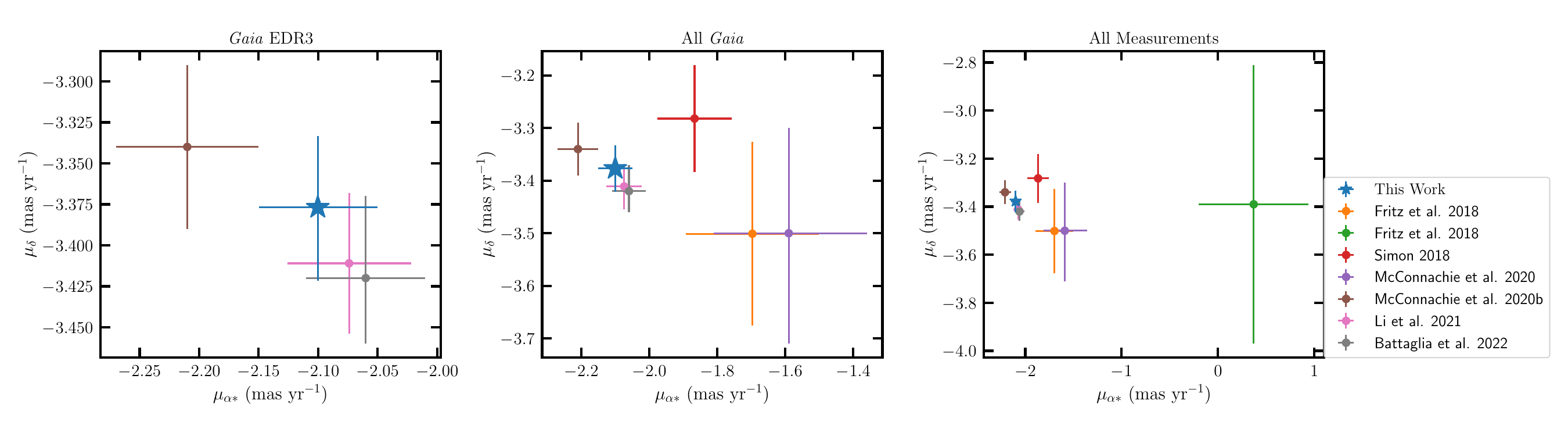}
\caption{Same as Figure~\ref{fig:comparsion} but for Segue 1.}
\end{figure*}

\begin{figure*}
\includegraphics[height=4.5cm]{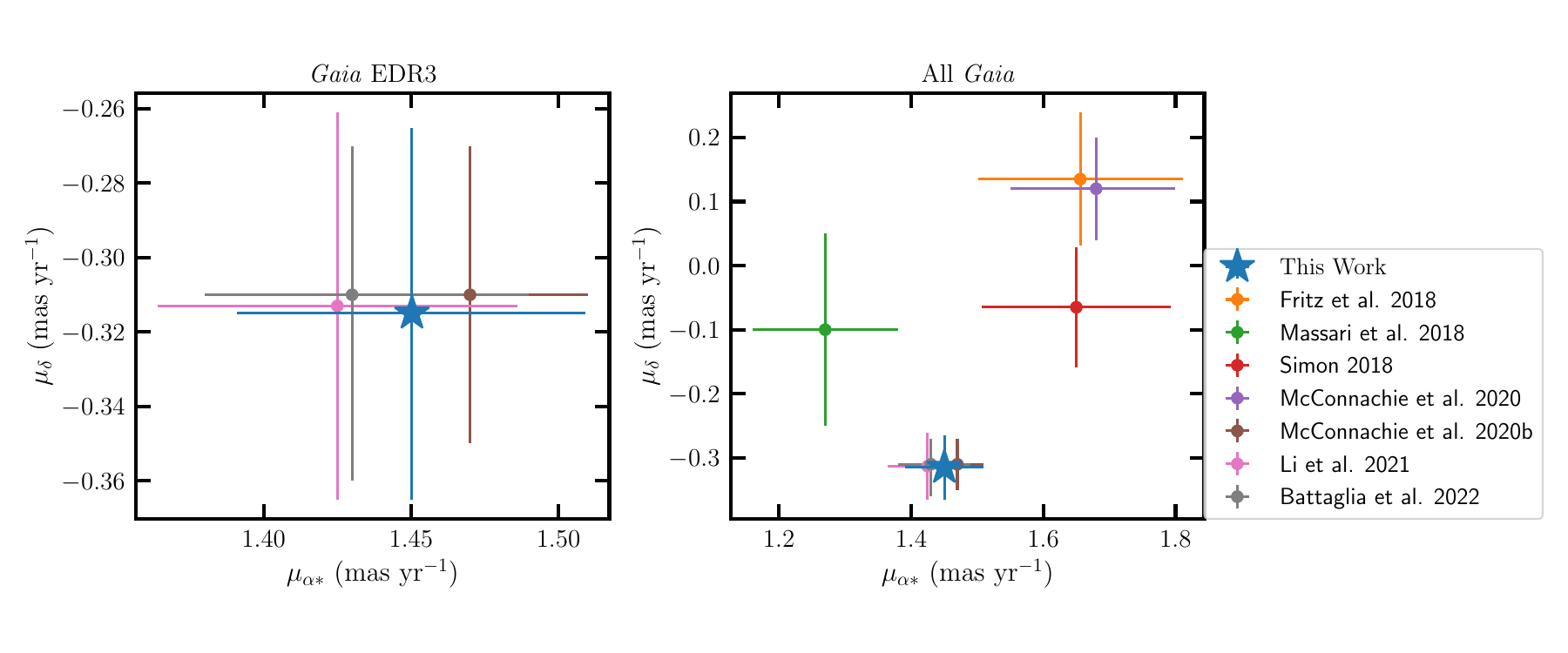}
\caption{Same as Figure~\ref{fig:comparsion} but for Segue 2.}
\end{figure*}

\begin{figure*}
\includegraphics[height=4.5cm]{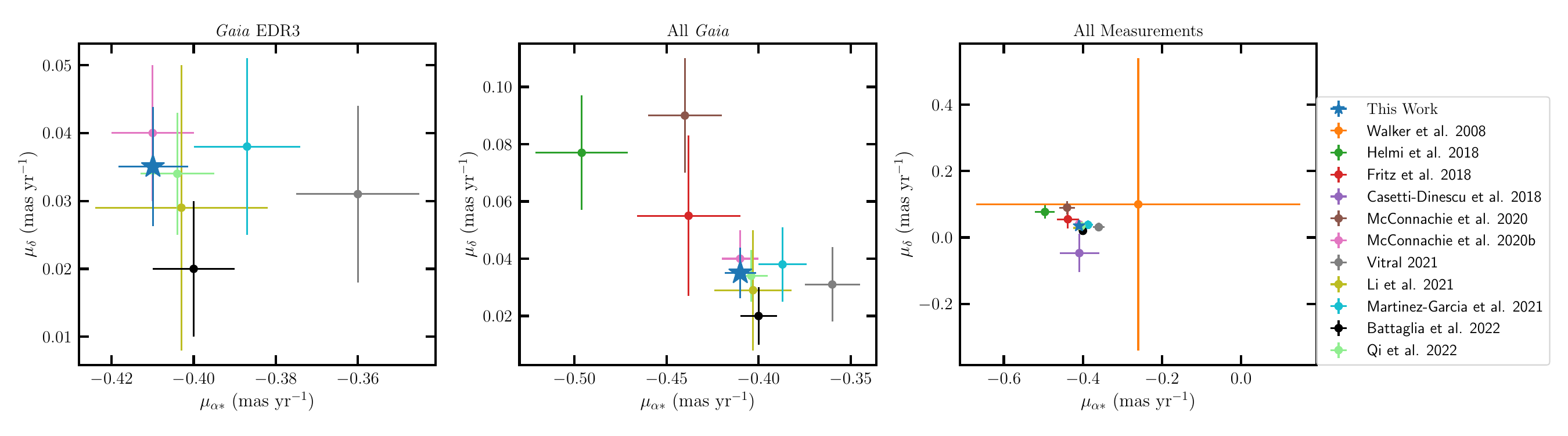}
\caption{Same as Figure~\ref{fig:comparsion} but for Sextans.}
\end{figure*}


\begin{figure*}
\includegraphics[height=4.5cm]{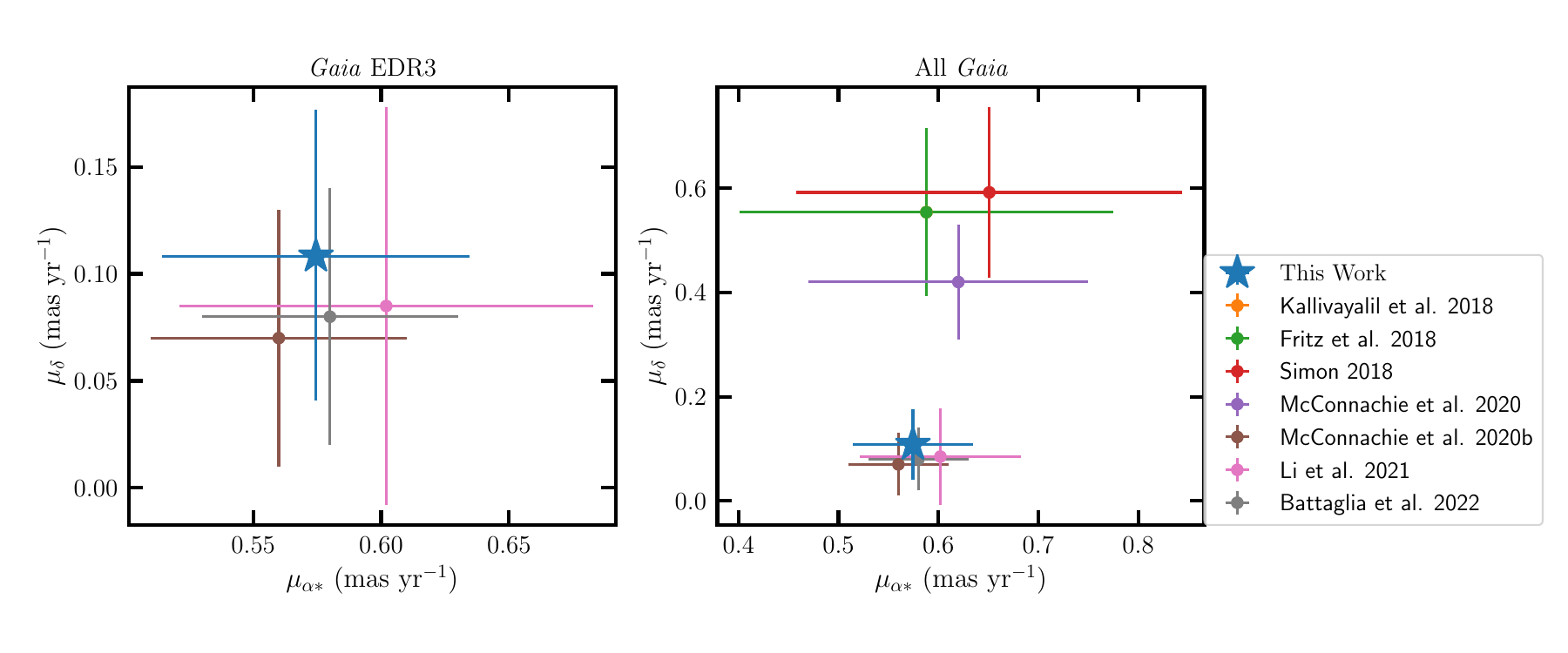}
\caption{Same as Figure~\ref{fig:comparsion} but for Triangulum II.}
\end{figure*}

\begin{figure*}
\includegraphics[height=4.5cm]{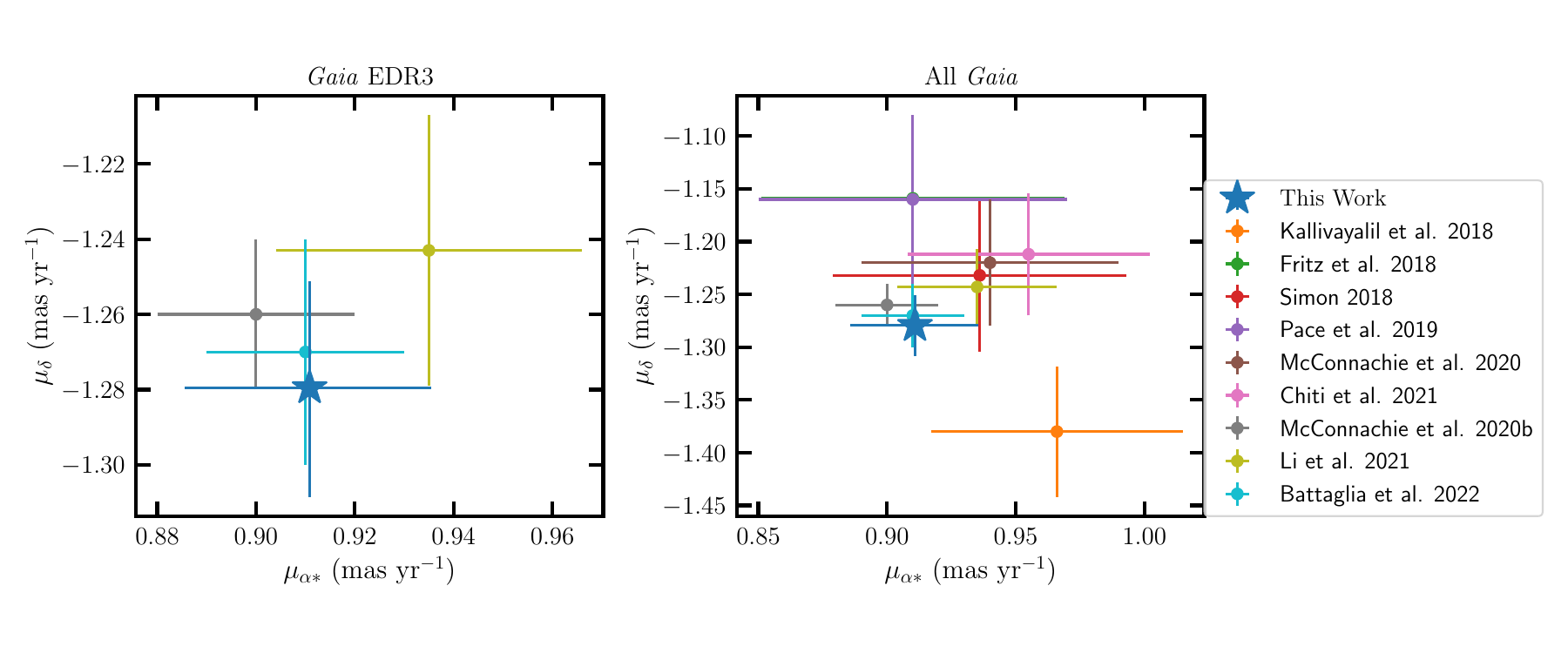}
\caption{Same as Figure~\ref{fig:comparsion} but for Tucana II.}
\end{figure*}

\begin{figure*}
\includegraphics[height=4.5cm]{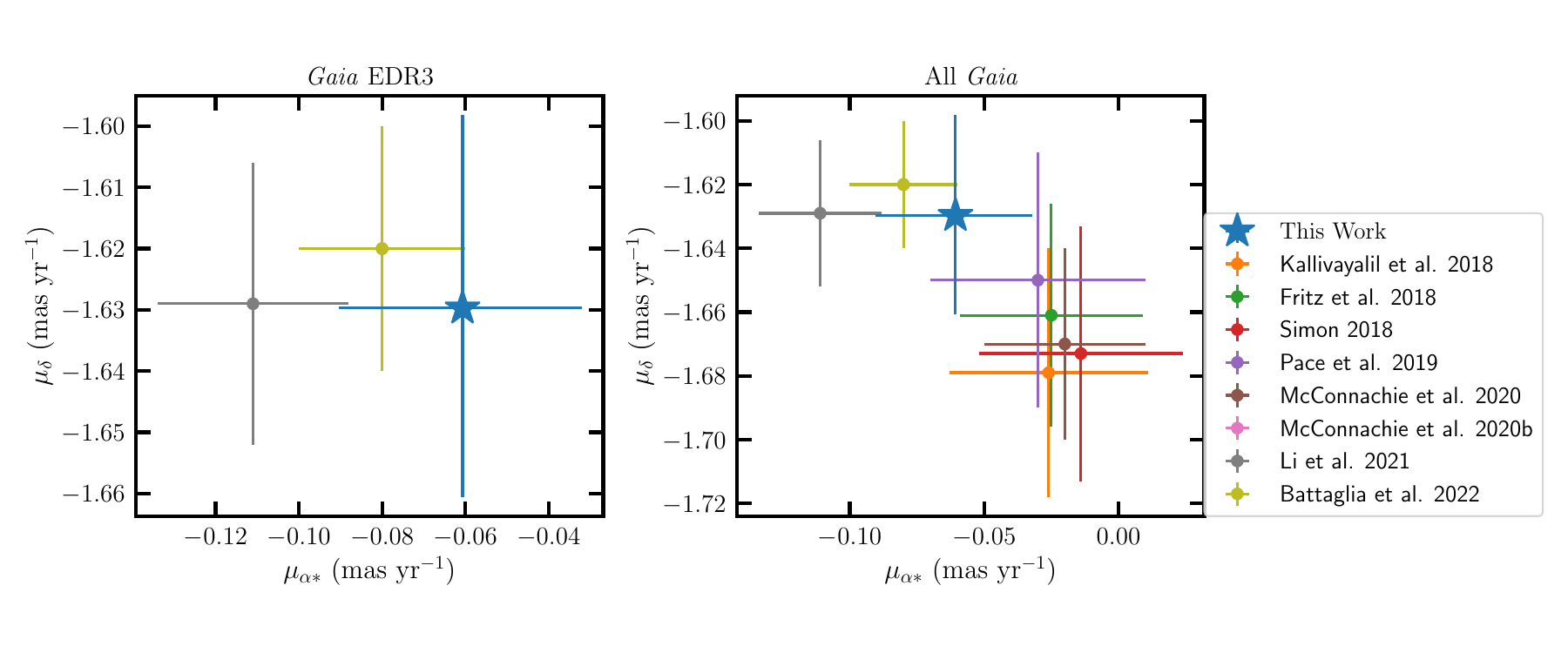}
\caption{Same as Figure~\ref{fig:comparsion} but for Tucana III.}
\end{figure*}

\begin{figure*}
\includegraphics[height=4.5cm]{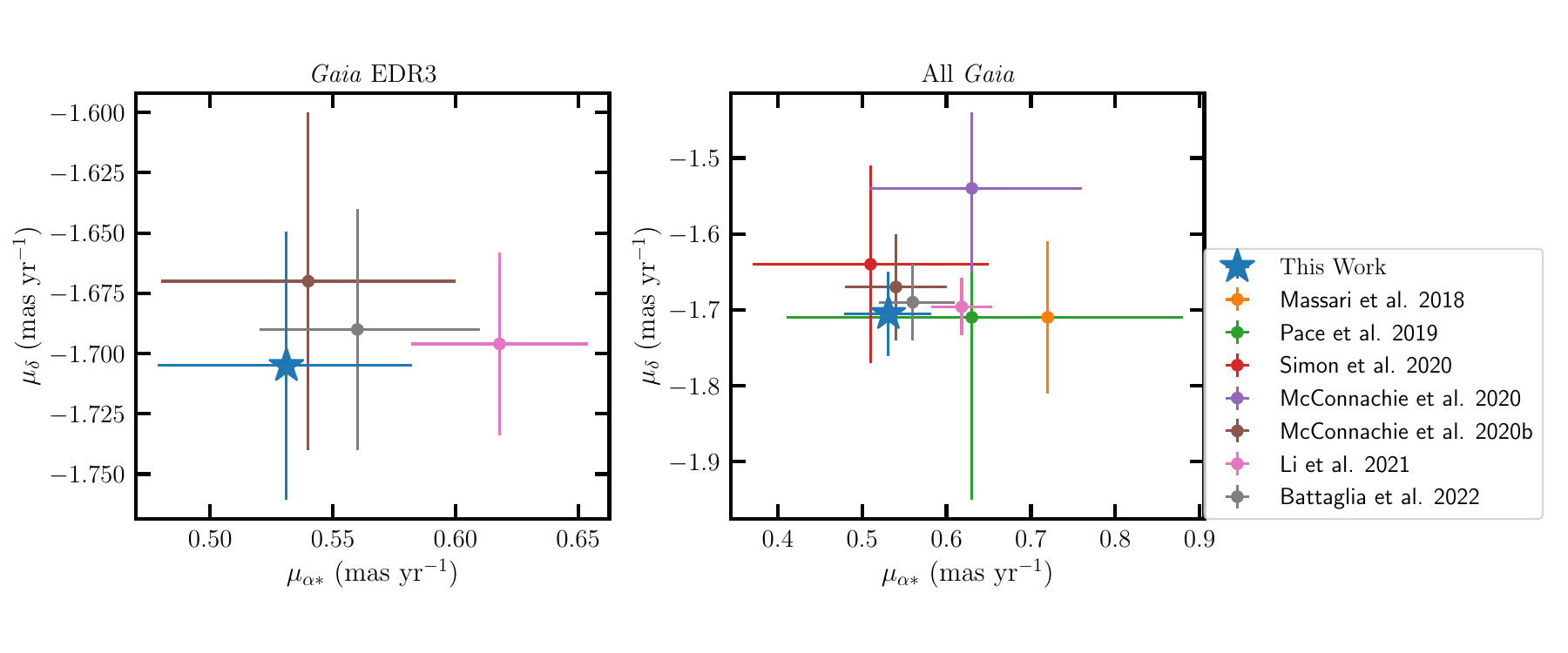}
\caption{Same as Figure~\ref{fig:comparsion} but for Tucana IV.}
\end{figure*}


\begin{figure*}
\includegraphics[height=4.5cm]{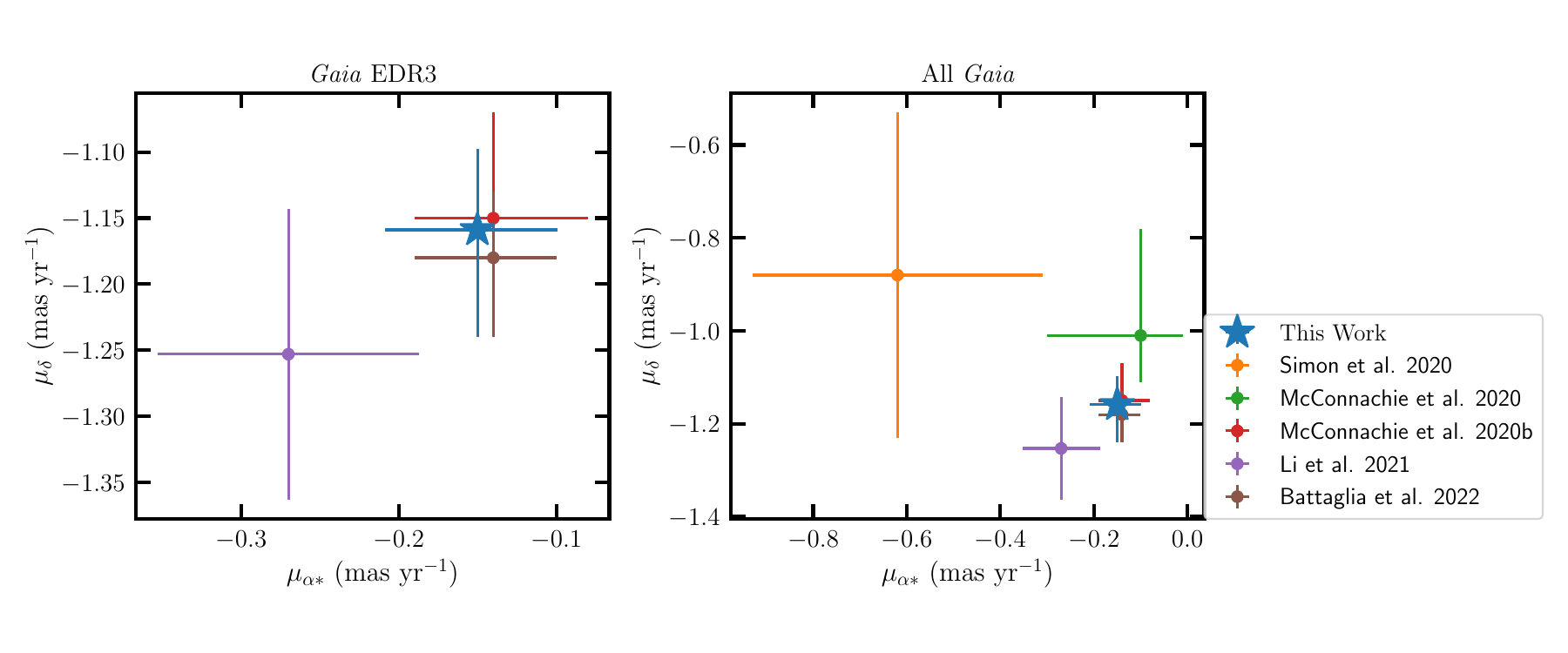}
\caption{Same as Figure~\ref{fig:comparsion} but for Tucana V.}
\end{figure*}

\begin{figure*}
\includegraphics[height=4.5cm]{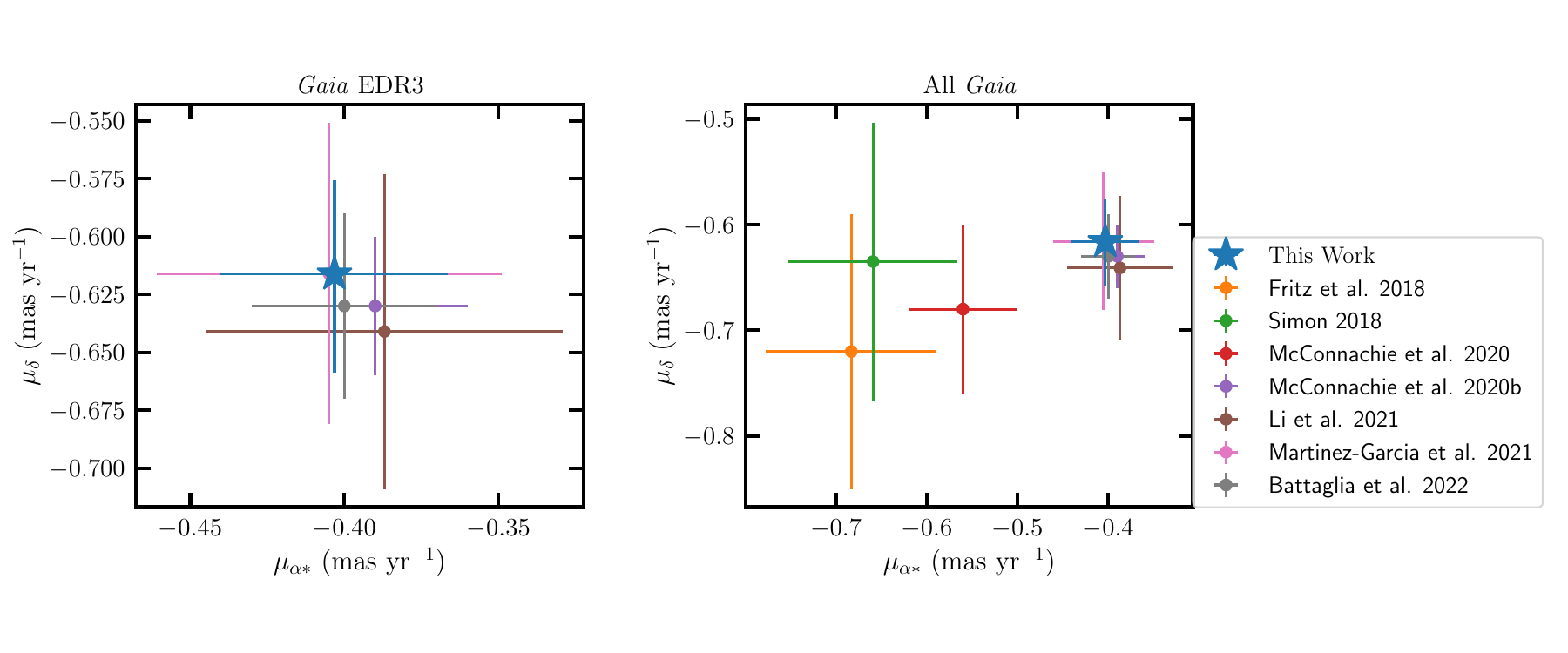}
\caption{Same as Figure~\ref{fig:comparsion} but for Ursa Major I.}
\end{figure*}

\begin{figure*}
\includegraphics[height=4.5cm]{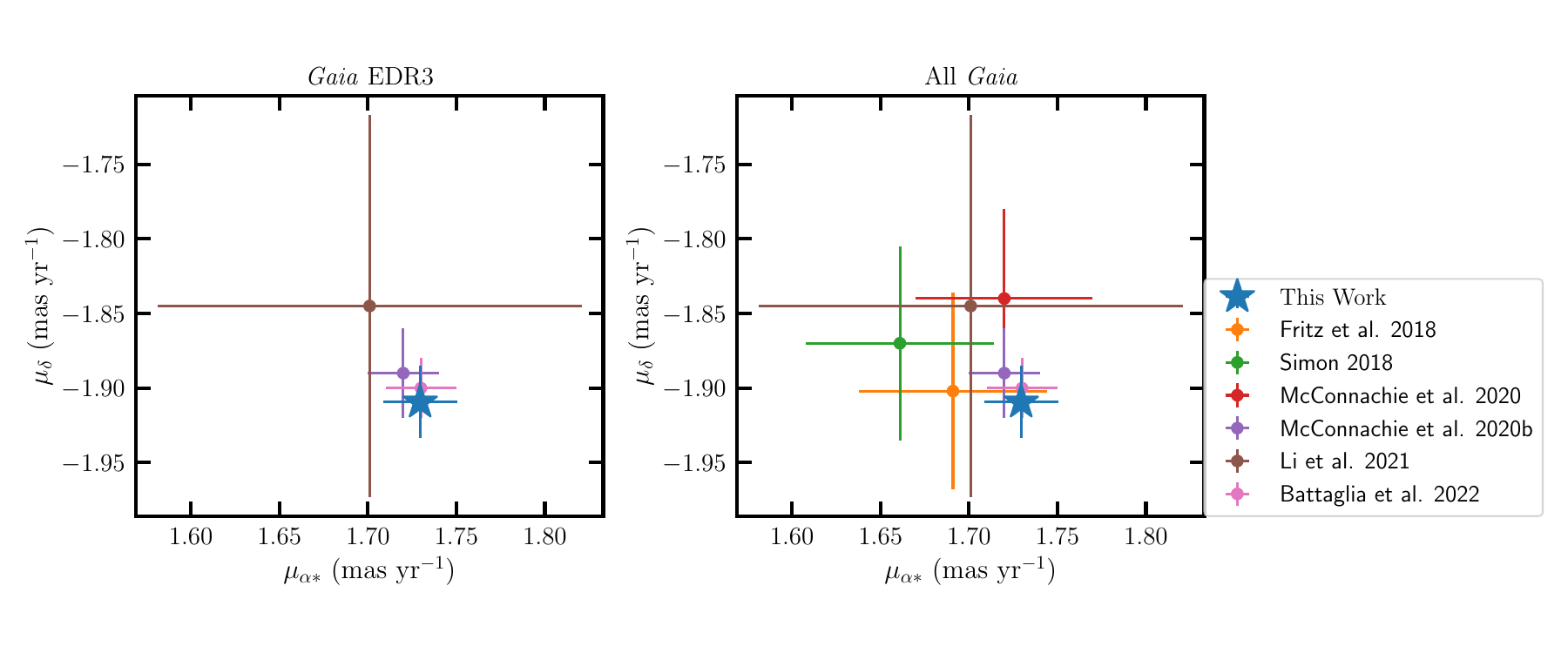}
\caption{Same as Figure~\ref{fig:comparsion} but for Ursa Major II.}
\end{figure*}

\begin{figure*}
\includegraphics[height=4.5cm]{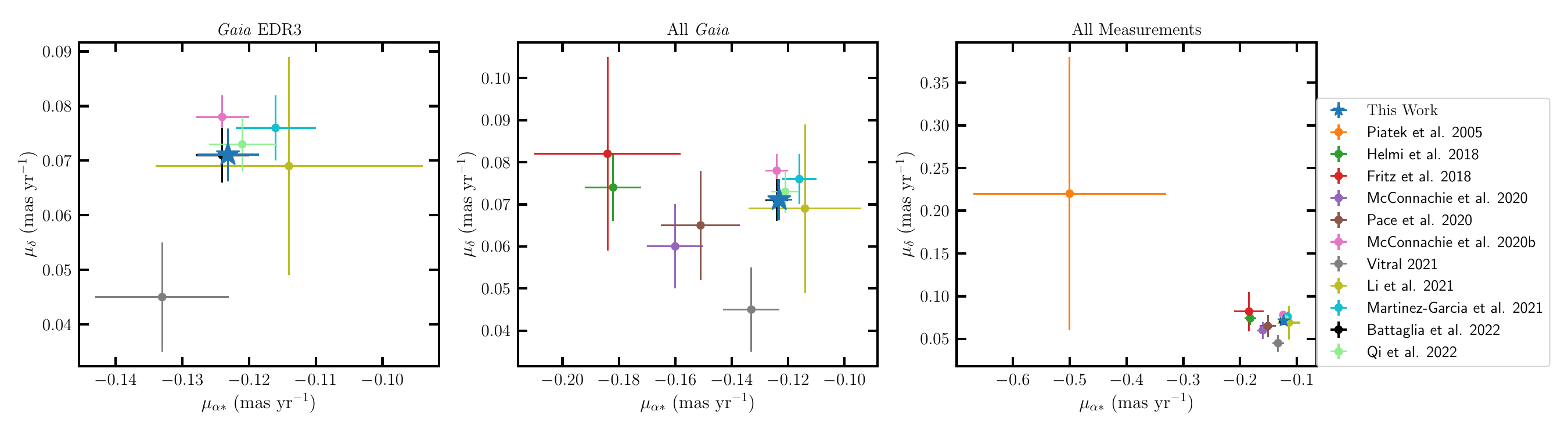}
\caption{Same as Figure~\ref{fig:comparsion} but for Ursa Minor.}
\end{figure*}

\begin{figure*}
\includegraphics[height=4.5cm]{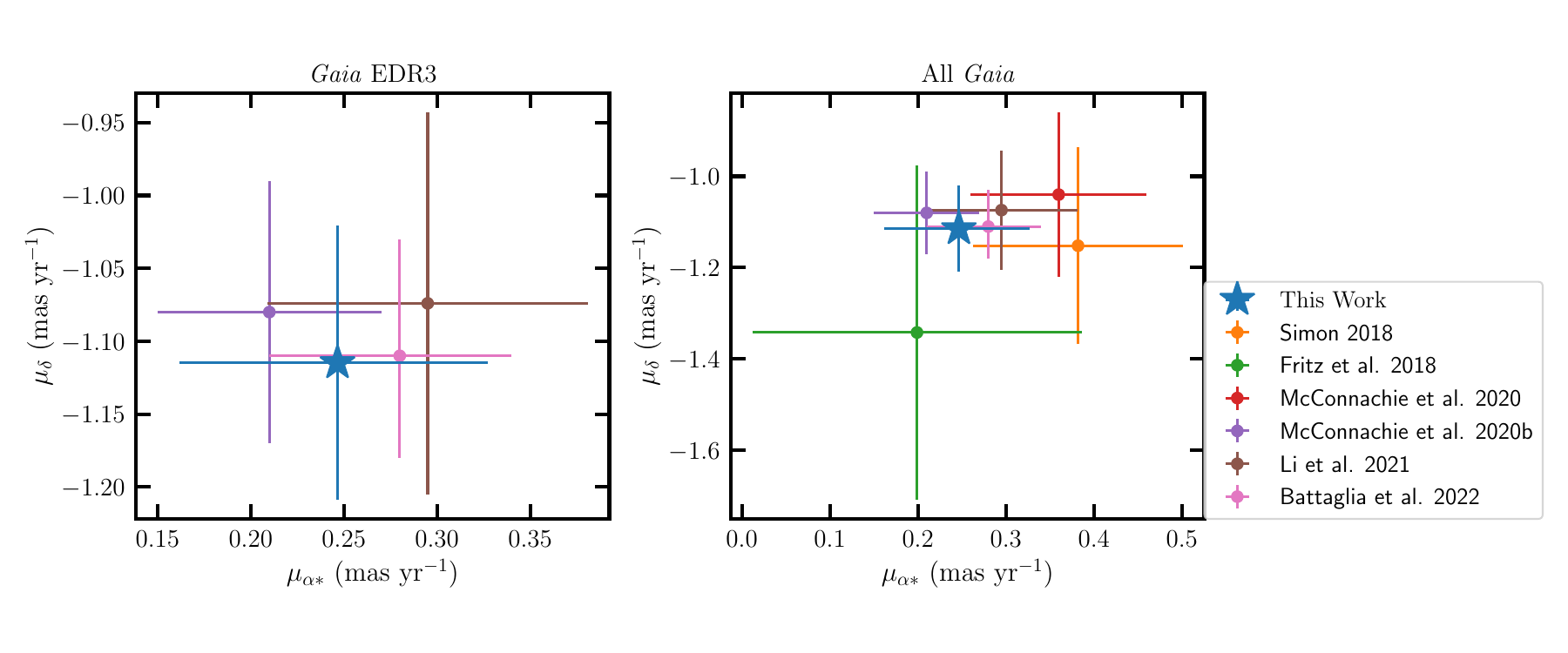}
\caption{Same as Figure~\ref{fig:comparsion} but for Willman 1.}
\end{figure*}

\end{document}